\documentclass[12pt]{scrartcl}    

\usepackage{silence}
\usepackage{a4} 
\usepackage{amsmath}    
\usepackage{paralist}
\usepackage{latexsym} 
\usepackage{amssymb}  
\usepackage{amsfonts}  
\usepackage{mathrsfs}  
\usepackage{dsfont}
\usepackage{xcolor}
\usepackage{bbm,exscale}
\definecolor{Myblue}{rgb}{0,0,0.6}  
\usepackage[colorlinks,citecolor=Myblue,linkcolor=Myblue,urlcolor=Myblue,pdfpagemode=None]{hyperref}
\usepackage{amsthm}
\usepackage{accents}
\usepackage[square,numbers,sort&compress]{natbib} 
\usepackage[all,cmtip]{xy}
\usepackage{ifthen} 
\usepackage{bbding}
\usepackage{stmaryrd}  
\usepackage{verbatim}
\usepackage{bbding} 
\usepackage{wasysym}  
\usepackage{soul}  
	\setuldepth{Berlin}
\usepackage[yyyymmdd,hhmmss]{datetime}
\usepackage{booktabs}
\usepackage{hhline}
\usepackage{enumitem}
\usepackage{color}
\usepackage{textcomp}
\usepackage{gensymb}
\usepackage{pdfpages}
\usepackage{tcolorbox}
\usepackage[bb=boondox]{mathalfa} 
\usepackage{tikz}
\usepackage{tikz-cd}
\usetikzlibrary{calc}
\usetikzlibrary{decorations.markings}
\usetikzlibrary{fadings,decorations.pathreplacing}
\usetikzlibrary{matrix,arrows}
\usetikzlibrary{patterns}
\usetikzlibrary{arrows,calc,decorations.pathreplacing,decorations.markings,shapes.geometric,shadows}

\DeclareSymbolFont{bbold}{U}{bbold}{m}{n}
\DeclareSymbolFontAlphabet{\mathbbold}{bbold}

\tikzset{
	string/.style={draw=#1, postaction={decorate}, decoration={markings,mark=at position .51 with {\arrow[draw=#1]{>}}}},
	costring/.style={draw=#1, postaction={decorate}, decoration={markings,mark=at position .51 with {\arrow[draw=#1]{<}}}},
	ostring/.style={draw=#1, postaction={decorate}, decoration={markings,mark=at position .47 with {\arrow[draw=#1]{>}}}},
	ustring/.style={draw=#1, postaction={decorate}, decoration={markings,mark=at position .56 with {\arrow[draw=#1]{>}}}},
	oostring/.style={draw=#1, postaction={decorate}, decoration={markings,mark=at position .43 with {\arrow[draw=#1]{>}}}},
	uustring/.style={draw=#1, postaction={decorate}, decoration={markings,mark=at position .59 with {\arrow[draw=#1]{>}}}},
	directed/.style={string=blue!50!black}, 
	odirected/.style={ostring=blue!50!black}, 
	udirected/.style={ustring=blue!50!black}, 
	oodirected/.style={oostring=blue!50!black}, 
	uudirected/.style={uustring=blue!50!black},     
	redirected/.style={costring= blue!50!black},
	redirectedgreen/.style={costring= green!50!black},
	directedgreen/.style={string= green!50!black},
}

\tikzset{-dot-/.style={decoration={
			markings,
			mark=at position 0.5 with {\fill circle (2pt);}},postaction={decorate}}}

\tikzset{
	Fdot/.style={circle, draw, fill, inner sep=0pt}, 
	Odot/.style={circle, draw, inner sep=0.1pt, minimum size=0.1cm}
}

\newcommand\tikzzbox[1]
{#1}

\vfuzz \hfuzz
\makeatletter
\newcommand{\raisemath}[1]{\mathpalette{\raisem@th{#1}}}
\newcommand{\raisem@th}[3]{\raisebox{#1}{$#2#3$}}
\makeatother

\newcommand{\A}{\mathcal{A}}

\renewcommand{\leq}{\leqslant}

\newcommand{\btimes}{\mathbin{\square}} 

\newcommand{\B}{\mathcal{B}}
\newcommand{\Bfd}{\mathcal{B}^{\textrm{fd}}}

\newcommand{\C}{\mathds{C}}

\newcommand{\N}{\mathds{N}}
\newcommand{\Q}{\mathds{Q}}
\newcommand{\R}{\mathds{R}}
\newcommand{\Z}{\mathds{Z}}

\def\1{\ifmmode\mathrm{1\!l}\else\mbox{\(\mathrm{1\!l}\)}\fi}
\newcommand{\one}{\mathbbm{1}}
\newcommand{\be}{\begin{equation}}
  \newcommand{\ee}{\end{equation}}
\newcommand{\bes}{\begin{equation*}}
  \newcommand{\ees}{\end{equation*}}

\newcommand{\id}{\operatorname{id}}

\newcommand{\Hom}{\operatorname{Hom}}
\newcommand{\Aut}{\operatorname{Aut}}
\newcommand{\End}{\operatorname{End}}

\def\LG{\mathcal{LG}}

\def\RW{\mathcal{RW}}
\newcommand{\hmf}{\operatorname{hmf}}

\newcommand{\ev}{\operatorname{ev}}

\newcommand{\tev}{\widetilde{\operatorname{ev}}}
\newcommand{\coev}{\operatorname{coev}}
\newcommand{\tcoev}{\widetilde{\operatorname{coev}}}
\def\lra{\longrightarrow}

\def\lmt{\longmapsto}

\newcommand{\Fun}{\textrm{Fun}^{\mathrm{sm}}}

\newcommand{\SO}{\textrm{SO}}

\newcommand{\Bord}{\textrm{Bord}}

\newcommand{\Bordfr}{\Bord_{2,1,0}^{\textrm{fr}}}
\newcommand{\Bordor}{\Bord_{2,1,0}^{\textrm{or}}}

\newcommand{\zz}{\mathcal{Z}}
\newcommand{\zzgra}{\mathcal{Z}^\textrm{gr}}

\newcommand{\Vect}{\operatorname{Vect}}

\newcommand{\dual}{\#}
\newcommand{\Ae}{A^{\textrm{e}}}

\newcommand{\dM}{{}^\dagger\hspace{-1.8pt}M}
\newcommand{\Md}{M^\dagger}

\newcommand{\bi}{\mathcal C}
\newcommand{\bigra}{{\mathcal C}^{\textrm{gr}}}
\newcommand{\idlg}{{\textrm{Id}_{\bi}}}

\newcommand{\au}{\text{\ul{$a$}}}
\newcommand{\bu}{\text{\ul{$b$}}}
\newcommand{\cu}{\text{\ul{$c$}}}
\newcommand{\du}{\text{\ul{$d$}}}
\newcommand{\eu}{\text{\ul{$e$}}}
\newcommand{\gu}{\text{\ul{$g$}}}
\newcommand{\hu}{\text{\ul{$h$}}}

\newcommand{\pu}{\text{\ul{$p$}}}
\newcommand{\qu}{\text{\ul{$q$}}}
\newcommand{\wu}{\text{\ul{$w$}}}
\newcommand{\xu}{\text{\ul{$x$}}}
\newcommand{\yu}{\text{\ul{$y$}}}
\newcommand{\zu}{\text{\ul{$z$}}}
\newcommand{\zeru}{\text{\ul{$0$}}}
\newcommand{\Bhfp}[1]{{#1}^{\circlearrowleft}}
\newcommand{\rs}{r_\text{s}}
\newcommand{\qs}{q_\text{s}}
\newcommand{\rnh}{r_\text{nh}}
\newcommand{\rsh}{r_\text{sh}}
\newcommand{\qnh}{q_\text{nh}}
\newcommand{\qsh}{q_\text{sh}}
\newcommand{\SU}{\textrm{SU}}
\newcommand{\da}{\dot{A}}
\newcommand{\db}{\dot{B}}

\def\RW{\mathcal{RW}}

\newcommand\arxiv[2]      {\href{https://arXiv.org/abs/#1}{#2}}
\newcommand\doi[2]        {\href{https://dx.doi.org/#1}{#2}}

\allowdisplaybreaks

\deffootnote[1em]{1em}{1em}{\textsuperscript{\thefootnotemark}}

\theoremstyle{definition} 
\newtheorem{definition}{Definition}
\newtheorem{proposition}[definition]{Proposition}
\newtheorem{theorem}[definition]{Theorem}

\newtheorem{lemma}[definition]{Lemma}
\newtheorem{corollary}[definition]{Corollary}
\newtheorem{remark}[definition]{Remark}

\newtheorem{example}[definition]{Example}

\numberwithin{equation}{section}
\numberwithin{definition}{section}
\numberwithin{figure}{section}

\newcommand\void[1]{}

\begin{document}

\title{%
Truncated affine Rozansky--Witten models as extended TQFTs%
}

\author{%
	Ilka Brunner$^*$ \quad Nils Carqueville$^\#$ \quad Daniel Roggenkamp$^\vee$
	\\[0.5cm]
	\normalsize{\texttt{\href{mailto:ilka.brunner@physik.uni-muenchen.de}{ilka.brunner@physik.uni-muenchen.de}}} \\ %
	\normalsize{\texttt{\href{mailto:nils.carqueville@univie.ac.at}{nils.carqueville@univie.ac.at}}} \\  %
	\normalsize{\texttt{\href{mailto:roggenkamp@uni-mannheim.de}{roggenkamp@uni-mannheim.de}}}
	\\[0.3cm]  %
	\hspace{-1.2cm} {\normalsize\slshape $^*$Arnold Sommerfeld Center,  LMU M\"unchen, Theresienstra\ss e 37, 80333 M\"unchen, Deutschland}\\[-0.1cm]
	\hspace{-1.2cm} {\normalsize\slshape $^\#$Universit\"at Wien, Fakult\"at f\"ur Physik, Boltzmanngasse 5, 1090 Wien, \"{O}sterreich}\\[-0.1cm]
	\hspace{-1.2cm} {\normalsize\slshape $^\vee$Institut f\"ur Mathematik, Universit\"at Mannheim, B6, 26, 68131 Mannheim, Deutschland}
}

\date{}

\maketitle

\vspace{-13.8cm}
\hfill {\scriptsize LMU-ASC 01/22}
\vspace{13cm}

\begin{abstract} 
We construct extended TQFTs associated to Rozansky--Witten models with target manifolds $T^*\C^n$.
The starting point of the construction is the 3-category whose objects are such Rozansky--Witten models, and whose morphisms are defects of all codimensions. 
By truncation, we obtain a (non-semisimple) 2-category $\bi$ of bulk theories, surface defects, and isomorphism classes of line defects.
Through a systematic application of the cobordism hypothesis we construct a unique extended oriented 2-dimensional TQFT valued in~$\bi$ for every affine Rozansky--Witten model. 
By evaluating this TQFT on closed surfaces we obtain the infinite-dimensional state spaces (graded by flavour and R-charges) of the initial 3-dimensional theory. 
Furthermore, we explicitly compute the commutative Frobenius algebras that classify the restrictions of the extended theories to circles and bordisms between them. 
\end{abstract}

\newpage

\tableofcontents

\section{Introduction and summary}
\label{sec:IntroductionSummary}

In an effort to make this paper accessible to readers from both the mathematics and physics communities, we provide two independent introductions.

\subsection{Introduction for physicists}

Topological quantum field theories arise in physics for example as subsectors of honest quantum field theories. 
In the context of supersymmetric field theories, topological twists single out these subsectors as the cohomology of a differential that originates from the supercharges of the initial theory. 
The resulting theories describe simpler, often solvable, parts of the initial theory that can be treated with mathematical rigour. 

Mathematically, a TQFT can be viewed as a symmetric monoidal functor $\zz$ from a geometric category of bordisms to the category of complex vector spaces.\footnote{More general target categories are possible and will be encountered in this paper.} 
In the geometric category, objects are closed $(d-1)$-dimensional manifolds, while morphisms are represented by $d$-dimensional bordisms between them, where here and throughout we assume all manifolds to be oriented. 
The functor~$\zz$ associates to the former vector spaces that correspond to the state spaces obtained by quantisation on the $(d-1)$-dimensional manifold times the time direction, and to the latter linear maps between them. 
Disjoint unions of $(d-1)$-dimensional manifolds on the geometric side are mapped to tensor products of the respective state spaces. 
The vector space associated to the empty set is $\C$, such that one assigns a number (a linear map from $\C$ to $ \C$) to a $d$-manifold without boundary. 
The latter can be regarded as the result of evaluating the path integral. 

Composition of bordisms in the geometric category consists of gluing $d$-dimensional bordisms along their $(d-1)$-dimensional boundaries. 
Functoriality therefore means that we can cut any manifold into simpler pieces, evaluate~$\zz$ on them, and glue them back together in the target category. 
The result is guaranteed to be independent of the precise way the manifold was decomposed. Physically, the cutting and gluing procedure can be interpreted in terms of an evaluation of the path integral on the pieces. 
In case of non-empty boundaries, boundary values for local fields must be specified; the path integral depends on them, and gluing involves a summation over all such boundary values. 
Given a local action functional, a piecewise evaluation of the path integral in this way is naturally independent of the initial decomposition.

To define a TQFT, it is thus enough to specify it on simple building blocks  and let functoriality (and symmetric monoidality) do the rest. 
This is particularly powerful in $d=2$ dimensions, where one has finite data and simple relations among them: 
any closed (oriented) 2-manifold can be decomposed into pairs-of-pants and caps, subject to a finite set of sewing relations. 
This leads to the well-known result that 2-dimensional (oriented) TQFTs are classified by commutative Frobenius algebras.

In higher dimensions, the situation is more involved,  as there are infinitely many building blocks as well as relations. 
To come up with a definition that leads to sufficiently ``simple" structures, mathematicians invented the notion of fully extended TQFTs, which can be viewed as maximally local extensions of ordinary TQFTs as above. 
In this setting, the higher bordism category does not only involve $d$- and $(d-1)$-manifolds, but (oriented) manifolds of any codimension.
Objects of this category are 0-dimensional manifolds, i.\,e.\ disjoint unions of points. 
Morphisms between them are 1-dimensional lines, whose boundaries correspond to their source and target objects. 
In turn, these lines can be connected by 2-dimensional surfaces with corners, which are the 2-morphisms of the category.
One proceeds like this step by step, obtaining a higher category on the geometric side, whose $d$-morphisms are represented by $d$-dimensional bordisms between $(d-1)$-dimensional manifolds; for details see \cite{l0905.0465, CalaqueScheimbauer2015}. 

Just like ordinary TQFTs, fully extended TQFTs are given by functors, which are now functors between higher categories. 
In particular, the target categories have to be higher categories as well (and carry a symmetric monoidal structure). 
A fully extended TQFT then associates objects in the target category to points, 1-morphisms to lines, and so on. 

The cobordism hypothesis of \cite{BDpaper, l0905.0465} states that a symmetric monoidal functor from the fully extended oriented bordism category to a suitable target category is determined by the object~$u$ that the functor assigns to a point, as well as one additional datum~$\lambda$ that encodes an $\SO(d)$-symmetry on~$u$.\footnote{We stress again that here we consider only \textsl{oriented} TQFTs. The cobordism hypothesis applies more generally and also classifies TQFTs on bordisms with other tangential structures such as spin or $G$-bundles -- in which case $\SO(d)$ is replaced by $\textrm{Spin}(d)$ or $BG$, respectively.} 
Thus, the entire extended TQFT can be reconstructed from~$u$ and~$\lambda$. 

Not all objects~$u$ of the target category are necessarily potential images of the point. 
They have to satisfy certain consistency conditions, the first of which is universal, namely that they have to be ``fully dualisable''. 
This is a natural finiteness condition in the categorical setting. 
To give a flavour of it, we note that in the category of vector spaces only objects that are finite-dimensional are dualisable. 
In the setting of higher categories, this simple dualisability constraint on objects is extended in a natural way, where \textsl{full} dualisability requires that the dualising 1-morphisms (evaluation, which pairs an object with its dual, and coevaluation) must have duals as well; 
these duals in turn come with dualising 2-morphisms, which also must have duals -- and so on. 
In the case $d=2$, there is only one more consistency condition besides full dualisability: namely that the Serre automorphism~$S_u$ (a canonical 1-automorphism of~$u$ arising from full dualisability) has to be trivialisable, i.\,e.\ there has to exist a 2-isomorphism $\lambda\colon S_u \lra 1_u$. 
Then as shown in \cite{HV, Hesse} pairs consisting of a fully dualisable object~$u$ together with a trivialisation~$\lambda$ of its Serre automorphism $S_u$ give rise to fully extended oriented TQFTs (which are unique for given~$u$ iff the trivialisation~$\lambda$ is unique up to isomorphism, cf.\ \eqref{eq:IsomorphismOfExtendedTQFTs}).

A class of interesting target categories for extended TQFTs arises from topological quantum field theory with defects. 
Here, higher categorical structures arise if one regards bulk theories as objects, codimension-1 defects as 1-morphism, codimension-2 defects as 2-morphisms, etc.\footnote{Even though both involve higher categories, a priori extended TQFTs and defect TQFTs are distinct concepts. 
Higher categories are inputs to the former and outputs of the latter.} 
For instance, in \cite{CMM}, the non-semisimple 2-category of defects in Landau--Ginzburg models was used as the target category of fully extended 2-dimensional TQFTs.  

\medskip 

In this paper we aim to constructively apply the cobordism hypothesis to 3-dimensional Rozansky--Witten models introduced in \cite{RW1996}. 
These are topological twists of ${\mathcal N}=4$ supersymmetric sigma models with holomorphic symplectic target manifolds. 
Defects in these models were described in \cite{KRS, KR0909.3643}.

For simplicity we restrict our considerations to Rozansky--Witten models with the affine target manifolds $T^*\C^n$. 
These can be obtained by topologically twisting the 3-dimensional ${\mathcal N}=4$ supersymmetric theory of $n$ free hypermultiplets, and their defect 3-category ${\mathcal{RW}}^\text{aff}$ can be described very explicitly.  
Objects are Rozansky--Witten theories with target space $T^*\C^n$ for $n\in\N_0$, which we will label by lists $\xu:=(x_1,\ldots,x_n)$ of variables corresponding to the~$n$ free hypermultiplet bulk fields. 

The 1-morphisms in ${\mathcal{RW}}^\text{aff}$ are given by surface defects, separating two affine Rozansky--Witten models~$\xu$ and~$\yu$, say. 
From the perspective of the untwisted models, the surface defects preserve at least a chosen 2-dimensional ${\mathcal N}= (2,2)$ subalgebra of the bulk supersymmetry, and physics on the defects is formulated in a manifestly ${\mathcal N}=(2,2)$ supersymmetric way. 
Surface defects may support additional defect fields, chiral fields denoted by $\au$ which can interact with the bulk fields $\xu$ and $\yu$ on either side of the defect via a superpotential $W(\au,\xu,\yu)$.
Colliding two parallel surface defects, the respective superpotentials add up, and the squeezed-in bulk fields become fields on the newly created defect. 
This defect fusion is the composition of 1-morphisms in the defect 3-category.

The 2-morphisms correspond to line defects separating possibly different surface defects and are represented by matrix factorisations of the differences of the respective superpotentials. 
Again, there is a product corresponding to merging parallel defect lines, given by the relative tensor product of matrix factorisations. 
This corresponds to the composition of 2-morphisms. 

Finally, 3-morphisms in ${\mathcal{RW}}^\text{aff}$ correspond to point defects separating possibly different line defects, and are given by morphisms of matrix factorisations up to homotopy. 
Their composition describes the operator product of these local fields and defines the composition of 3-morphisms in the defect 3-category.

\medskip 

Using the cobordism hypothesis, one may try to construct a fully extended 3-dimensional TQFT by finding fully dualisable objects in ${\mathcal{RW}}^\text{aff}$ that come with the correct $\SO(3)$-symmetry. 
This is however impossible: no object $\xu=(x_1,\ldots,x_n)$ in ${\mathcal{RW}}^\text{aff}$ with $n>0$ is fully dualisable. 
In fact, such objects~$\xu$ are dualisable only up to dimension two -- the dualising 2-morphisms have no duals themselves.
This is due to the non-compactness of the target spaces $T^*\C^n$, which implies that the associated state spaces are infinite-dimensional. 
Hence, even though they are of course reasonable quantum field theories from a physics perspective, affine Rozansky--Witten models do not even satisfy the axioms of an ordinary 3-dimensional TQFT. 
They do however provide intricate mathematical as well as physical structures, some of which have been recently explored in \cite{GHNPPS,CDGG}. 
In the latter paper, affine Rozansky--Witten models serve as toy models for a larger class of theories that can be regarded as a derived and non-semisimple generalisation of 3-dimensional Chern--Simons theory.

In this paper, we avoid the problem posed by infinities by truncating to two dimensions, effectively disregarding the problematic partition functions on 3-dimensional spaces. 
More precisely, instead of the 3-category ${\mathcal{RW}}^\text{aff}$ we consider a truncation, the 2-category $\bi$ whose objects and 1-morphisms are identical to the ones in ${\mathcal{RW}}^\text{aff}$, but whose 2-morphisms are isomorphism classes of 2-morphisms in ${\mathcal{RW}}^\text{aff}$. 
In this truncated setup affine Rozansky--Witten models are fully dualisable. 
The dualising maps can be explicitly constructed from identity surface and line defects in $\mathcal{RW}^\text{aff}$. 
Moreover, their Serre automorphisms are trivialisable, and hence affine Rozansky--Witten models with target manifold $T^*\C^n$ give rise to fully extended 2-dimensional TQFTs $\mathcal{Z}_n$ with target category~$\bi$. 

The construction is very explicit, basically relying on the knowledge of the trivial surface as well as line defects from the category ${\mathcal{RW}}^\text{aff}$. 
As a consequence, we can straightforwardly evaluate~$\zz_n$ on all bordisms. 
Evaluating it on closed genus-$g$ surfaces~$\Sigma_g$ yields isomorphism classes of matrix factorisations of~$0$, whose cohomologies correspond to the state spaces $\mathcal{H}_{\Sigma_g}$ of the initial 3-dimensional theory. 
Via an explicit calculation we obtain the super vector spaces
\be 
\label{eq:StateSpacePhysIntro}
{\cal H}_{\Sigma_g}
	\cong H^\bullet\big(\zz_n(\Sigma_g)\big)
	\cong \big( \C \oplus \C[1] 
		 \big)^{\otimes 2ng}
	\otimes_\C \C[\au,\xu]  \, ,
\ee 
where $\xu=(x_1,\ldots,x_n)$, $\au=(a_1,\ldots,a_n)$, and $[1]$ denotes a shift of $\Z_2$-grading associated to fermion number. 
The spaces~\eqref{eq:StateSpacePhysIntro} indeed recover -- from the ``first principles'' of the cobordism hypothesis -- the state spaces of affine Rozansky--Witten model with target manifold $T^*\C^n$, which can be obtained by canonically quantising~$n$ free hypermultiplets on $\Sigma_g\times\R$. 
From this perspective, the~$x_i$ and~$a_i$ are the scalar components of the $n$ hypermultiplets, and the tensor factor 
	 $(\C \oplus \C[1])^{\otimes 2ng}$ 
corresponds to the fermions 
associated to the $2g$ holomorphic 1-forms on $\Sigma_g$ (see Appendix~\ref{subsec:StateSpecesAppendix} for more details). 

Beyond closed surfaces, we can also evaluate the TQFT $\mathcal{Z}_n$ on surfaces with boundary, in particular on caps and pairs-of-pants. 
As for ordinary 2-dimensional TQFTs, caps and pairs-of-pants give rise to a commutative Frobenius algebra, albeit now in the category of endodefects of~$\xu$. 
This Frobenius algebra determines the TQFT on any surface with boundary but without corners.
In our construction, it turns out to have a simple physical interpretation via its action on the endomorphisms of the 1-morphism that~$\mathcal{Z}_n$ associates to the circle. 
These endomorphisms represent isomorphism classes of line operators in Rozansky--Witten models, and the Frobenius algebra encodes their fusion properties.

\medskip

It is expected that the above programme can be carried out for other Rozansky--Witten models as well, and that theories with compact target manifolds can even be understood as fully extended 3-dimensional TQFTs, valued in the defect 3-category of \cite{KRS, KR0909.3643}. 
If the latter can be endowed with the richer structure of an $(\infty,3)$-category, the cobordism hypothesis would even allow to compute the moduli space of Rozansky--Witten models, and possibly monodromies in this space (see Remarks~\ref{rem:ModuliSpaceOfFullyExtendedTQFTs} and~\ref{rem:ModuliSpaceOfTruncatedAffineRWmodels} for some comments in this direction). 
Moreover, such 3-dimensional theories would not only provide state spaces, but also e.\,g.\ mapping class group representations on them, which could be compared with those obtained in \cite{RW1996}. 

\medskip 

This paper is organised as follows. 
In Section~\ref{sec:BicategoriesTruncatedRW} we introduce our truncation~$\bi$ of the defect 3-category $\mathcal{RW}^\text{aff}$, which we then use as a target for extended TQFTs.
We furthermore introduce a related graded version $\bi^{\textrm{gr}}$ that keeps track of flavour and R-symmetries.
We show that all objects of $\bi$ and $\bi^{\textrm{gr}}$ are fully dualisable with trivialisable Serre automorphism (Theorems~\ref{thm:FullyDualisable} and~\ref{thm:FullyDualisableGraded}).

Section \ref{sec:FullyExtendedTQFT} is devoted to the explicit construction of the extended TQFTs. 
After a brief review of the cobordism hypothesis and extended TQFTs in Section~\ref{subsec:CobordismHypothesis}, in Section~\ref{subsec:ExtendedAffineRW} we apply this formalism to construct the unique extended TQFTs~$\zz_n$ with target~$\bi$ (Theorem~\ref{cor:ExtendedTQFTZn}). 
In particular, we explicitly determine~$\zz_n$ on closed surfaces~$\Sigma_g$ (Proposition~\ref{prop:StateSpaces}), and more generally its associated commutative Frobenius algebra (Proposition~\ref{prop:CommutativeFrobeniusAlgebraFromZn}).
Hence we evaluate the functor on the circle, and we frame the result in terms of the non-semisimple category of $\Z_2$-graded $\C[\au,\xu]$-modules. 
Finally, Section~\ref{subsec:ExtendedAffineRWWithCharges} provides an extension to the graded case, which takes into account flavour and R-charges.

We include two appendices. 
Appendix~\ref{sec:applemmas} contains two technical lemmas on matrix factorisations, which we use in our explicit calculations. 
In Appendix~\ref{sec:appalgebras} we collect some facts about the 3-dimensional $\mathcal{N}=4$ supersymmetry algebra as well as results from the canonical quantisation of free hypermultiplets.

\subsection{Introduction for mathematicians}

A fully extended TQFT~$\zz$ of a given dimension~$d$ is maximally local in the sense that it is a symmetric monoidal functor on a higher bordism category that features manifolds of all dimensions $0,1,\dots,d$, and~$\zz$ is compatible with cutting and gluing along submanifolds of arbitrary codimension. 
As originally argued in \cite{BDpaper}, this compatibility imposes strong constraints on~$\zz$. 
More precisely, if one considers $d$-framed bordisms, then~$\zz$ is expected to be already determined by what it assigns to a single positively framed point~$+$, because the $d$-framed bordism category is generated as a symmetric monoidal category by the fully dualisable object which is the positively framed point~$+$. 
In the case $d=1$ this is basically the statement that the framed half-circles 
$
\ev_+ = 
\tikzzbox{%
\begin{tikzpicture}[thick,scale=0.9,color=black, baseline=0cm]
\coordinate (p1) at (0,0);
\coordinate (p2) at (0,0.25);
\coordinate (g1) at (0,-0.05);
\coordinate (g2) at (0,0.3);
\draw[ultra thick, blue!20!white] (g1) .. controls +(-0.25,0) and +(-0.25,0) ..  (g2); 
\draw[very thick, red!80!black] (p1) .. controls +(-0.2,0) and +(-0.2,0) ..  (p2); 
\fill[red!80!black] (0.4,0) circle (0pt) node[left] {{\tiny$-$}};
\fill[red!80!black] (0.4,0.25) circle (0pt) node[left] {{\tiny$+$}};
\end{tikzpicture}
}
$ 
and 
$
\coev_+ 
= 
\tikzzbox{%
\begin{tikzpicture}[thick,scale=0.9,color=black, baseline=0cm, xscale=-1]
\coordinate (p1) at (0,0);
\coordinate (p2) at (0,0.25);
\coordinate (g1) at (0,-0.05);
\coordinate (g2) at (0,0.3);		%
\draw[ultra thick, blue!20!white] (g1) .. controls +(-0.25,0) and +(-0.25,0) ..  (g2); 
\draw[very thick, red!80!black] (p1) .. controls +(-0.2,0) and +(-0.2,0) ..  (p2); 
\fill[red!80!black] (-0.1,0) circle (0pt) node[left] {{\tiny$+$}};
\fill[red!80!black] (-0.1,0.25) circle (0pt) node[left] {{\tiny$-$}};
\end{tikzpicture}
}
$ are the adjunction morphisms that witness the negatively framed point~$-$ as the dual of~$+$, that all framed compact 1-manifolds are disjoint unions of framed circles and intervals, and that changes in framing can be described in terms of $\ev_+$, $\coev_+$ and their adjoints, see e.\,g.\ \cite{l0905.0465, DouglasSchommerPriesSnyder2013}. 

In the more general case of bordisms with $G$-structure for a Lie group~$G$, the cobordism hypothesis states that fully extended TQFTs are classified by fully dualisable objects in the target category that come with a $G$-homotopy fixed point structure. 
In the $(\infty,d)$-categorical setting, an extended proof sketch was given in \cite{l0905.0465}, which was completed up to a conjecture on factorisation homology in \cite{AyalaFrancis2017CH}, and a further generalisation appeared in \cite{GradyPavlov2021}. 

In dimension $d=2$, the cobordism hypothesis has been formulated and proven independently in the setting of (weak) 2-categories. 
More precisely, in the framed case, i.\,e.\ when~$G$ is trivial, 2-dimensional extended TQFTs with values in a given symmetric monoidal 2-category~$\B$ are classified by fully dualisable objects of~$\B$ as shown in \cite{Pstragowski}. 
In the oriented case, $G = \SO(2)$, the classification was described earlier in \cite{spthesis} in terms of fully dualisable objects $u\in\B$ together with a trivialisation $S_u \stackrel{\textrm{{\tiny $\cong$}}}{\longrightarrow} 1_u$ of their Serre automorphisms. 
We recall the definition of~$S_u$ in Section~\ref{subsubsec:SerreAutomorphisms}, and we review the 2-dimensional oriented cobordism hypothesis in Section~\ref{subsec:CobordismHypothesis}.\footnote{The framed and oriented cobordism hypotheses in two dimensions can be understood as the special cases $r=0$ and $r=1$ of the $r$-spin cobordism hypothesis proven in \cite{RSpinLorantNils}, where~$G$ is the $r$-fold cover of $\SO(2)$.} 

Clearly the choice of target category~$\B$ is crucial. 
For $d=1$ the standard choice are (super) vector spaces, in which case the classification is entirely explicit: fully dualisable objects in $\B = \Vect_\Bbbk$ (or $\B = \Vect_\Bbbk^{\Z_2}$) are precisely finite-dimensional (super) vector spaces. 

In dimension $d=2$ there are no standard choices for~$\B$, in the sense that there are various inequivalent symmetric monoidal 2-categories~$\B$ such that $\End_\B(\one)$ is equivalent to $\Vect_\Bbbk$ or $\Vect_\Bbbk^{\Z_2}$.\footnote{A natural source of candidate target categories are the higher categories associated to defect TQFTs \cite{CRS1}, i.\,e.\ symmetric monoidal 1-functors on stratified and decorated bordisms.} 
The examples that have been studied in detail in the literature are the 2-category $\B = \textrm{Alg}_\Bbbk$ (corresponding to state sum models) of finite-dimensional $\Bbbk$-algebras, bimodules and bimodule maps, and $\B = \LG$ (corresponding to Landau--Ginzburg models) of isolated singularities $\{ W=0 \}$ and homotopy categories of matrix factorisations, see \cite{spthesis} and \cite{CMM}, respectively. 
The result is that extended oriented TQFTs with values in $\textrm{Alg}_\Bbbk$ are classified by separable symmetric Frobenius $\Bbbk$-algebras, while in the case of $\LG$ the condition is that the polynomials~$W$ must depend on an even number of variables. 
Moreover, it follows from \cite{cw1007.2679, BanksOnRozanskyWitten} that there is a 2-category associated to B-twisted sigma models, and Calabi--Yau varieties classify extended oriented TQFTs with values in it. 

In dimension $d=3$ even fewer examples of target categories~$\B$ and $G$-structures have been considered in complete detail. 
One main result \cite{DouglasSchommerPriesSnyder2013, FreedTeleman2020} is that fully extended framed TQFTs with values in the 3-category $\B = \textrm{Tens}_\Bbbk$ of $\Bbbk$-linear tensor categories are classified by fusion categories of non-zero global dimension. 
It is also widely expected that spherical fusion categories classify extended oriented TQFTs with values in $\textrm{Tens}_\Bbbk$, and that they precisely extend Turaev--Viro--Barrett--Westbury models to the point. 

In addition to the above state sum models, it is natural to consider topologically twisted sigma models also in dimension three. 
Indeed, in \cite{KR0909.3643} an extended sketch of a 3-category $\RW$ associated to Rozansky--Witten models was proposed based on the path integral analysis of \cite{RW1996, KRS}. 
Objects of $\RW$ are holomorphic symplectic manifolds, and its Hom 2-categories feature a rich interplay of algebra and geometry. 
It remains a challenge to fully exhibit the formidable structure of a symmetric monoidal 3-category on $\RW$, and to study its fully dualisable objects and their homotopy fixed points. 
Related applications to homological link invariants have been made e.\,g.\ in \cite{OblomkovRozansky}, and generally it is expected that Rozansky--Witten models with compact target manifolds can be extended to the point as TQFTs with values in~$\RW$. 

\medskip 

In the present paper we explicitly exhibit certain Rozansky--Witten models as extended oriented TQFTs. 
For simplicity, we restrict our considerations to Rozansky--Witten models with affine target manifolds $T^*\C^n \cong \C^{2n}$. 
Moreover, we truncate the subcategory $\RW^{\textrm{aff}} \subset \RW$ of these types of models to the (non-semisimple) 2-category
\be 
\bi := \textrm{T}(\RW^{\textrm{aff}}) \, , 
\ee 
which by definition has the same objects and 1-morphisms as $\RW^{\textrm{aff}}$, while 2-morphisms are isomorphism classes of 2-morphisms in $\RW^{\textrm{aff}}$. 
The restriction on affine Rozansky--Witten models allows us to very explicitly construct the symmetric monoidal structure of~$\bi$ in terms of lists of variables $\xu = (x_1,\dots,x_n)$, polynomials, and isomorphism classes of matrix factorisations (carried out in detail in Sections~\ref{subsec:Definition2category}--\ref{subsec:SymmetricMonoidalStructure}). 
The truncation to two dimensions is necessary as the non-compactness of the target manifolds $T^*\C^n$ leads to infinite-dimensional state spaces, so affine Rozansky--Witten models cannot even give rise to closed TQFTs valued in $\Vect_\C$. 
	 
In this setup, we prove our first main result (cf.\ Theorem~\ref{thm:OrientedCobordismHypothesis}): 
	 
\bigskip 
\noindent
\textbf{Theorem. } 
Every object $\xu \in \bi = \textrm{T}(\RW^{\textrm{aff}})$ is fully dualisable, 
and up to isomorphism there is precisely one associated 2-dimensional extended oriented TQFT~$\zz_n$ with values in~$\bi$ for every positive integer~$n$. 
\bigskip 

\noindent
We also prove a variant of this result in the presence of gradings for polynomials and matrix factorisations (Sections~\ref{subsec:GradedCase} and~\ref{subsec:ExtendedAffineRWWithCharges}), the source of which are flavour and R-charges. 

The other main result of the present paper is to systematically apply the logic behind the cobordism hypothesis to explicitly compute the invariants associated to the TQFTs~$\zz_n$. 
Specifically, for a closed surface~$\Sigma_g$ of genus~$g$, we find (in Section~\ref{subsubsec:PartFuncStateSpaces}) that 
\be 
\label{eq:IntroStateSpace}
\zz_n(\Sigma_g) \cong \big( \C \oplus \C[1] 
	 \big)^{\otimes 2ng}
\otimes_\C \C[\au,\xu] 
\ee 
as a matrix factorisation of zero (here with differential zero), which in~$\bi$ is isomorphic to its cohomology, i.\,e.\ a $\Z_2$-graded vector space (with additional gradings in the case of flavour and R-charges, cf.\ Section~\ref{subsec:ExtendedAffineRWWithCharges}). 

From the perspective of the truncated 2-dimensional TQFT~$\zz_n$, the infinite-dimensional vector space~\eqref{eq:IntroStateSpace} is interpreted as the ``partition function'' associated to~$\Sigma_g$. 
However, from the perspective of 3-dimensional Rozansky--Witten theory with non-compact target $T^*\C^{n}$, it is a ``state space''. 
We note that state spaces of TQFTs with values in $\Vect_{\C}$ must be finite-dimensional, but infinite dimensions can and do occur for the extended TQFT~$\zz_n$ with values in~$\bi$. 

We also compute~$\zz_n$ on surfaces with non-trivial boundary, and hence determine the 2-dimensional closed TQFT with values in $\bi(\xu,\xu)$ obtained by restricting~$\zz_n$ to dimension~1 and~2. 
As explained in Section~\ref{subsubsec:ComFrobAlgGrothRing}, the result can be formulated in terms of the Grothendieck ring of the homotopy category of matrix factorisations of $\sum_{i=1}^n (a_i-d_i) \cdot (x_i-y_i)$, or equivalently of the non-semisimple category of graded $\C[\au,\xu]$-modules.

\subsubsection*{Acknowledgements} 

We are grateful to 
	Alexei Oblomkov, 
	Ingmar Saberi, 
	Pavel Safronov, 
		and 
	Gregor Schaumann
for helpful discussions. 
I.\,B.\ is supported by the Deutsche Forschungsgemeinschaft (DFG) under Germany's Excellence Strategy EXC-2094 390783311 and the DFG grant ID 17448, 
N.\,C.\ is supported by the DFG Heisenberg Programme, 
	and 
D.\,R.\ is supported by the Heidelberg Institute for Theoretical Studies.

\section{2-categories of truncated affine Rozansky--Witten models}
\label{sec:BicategoriesTruncatedRW} 

In this section we study the 2-category $\bi$ which is the truncation of the 3-category of Rozansky--Witten theories with affine target manifolds. 
The basic ingredients are recalled in Section~\ref{subsec:Definition2category}, and the symmetric monoidal structure for $\bi$ is described in Sections \ref{subsec:MonoidalStructure}--\ref{subsec:SymmetricMonoidalStructure}. 
Then in Section~\ref{subsec:FullDualisibility} we prove that every object in $\bi$ is fully dualisable, and we observe that all Serre automorphisms are trivialisable. 
Finally, in Section~\ref{subsec:GradedCase} we discuss a variation $\bigra$ of $\bi$ that keeps track of flavour and R-charge degrees, and we prove analogous results for $\bigra$. 

\medskip 

By a 2-category we mean a (possibly non-strict) bicategory in the sense of \cite[Sect.\,2.1]{JohnsonYauBook}. 
For general background on 2-categories we refer to \cite{LeinsterBasic2, JohnsonYauBook} and to~\cite[Sect.\,2.2]{2dDefectTQFTLectureNotes} for a short and casual discussion. 
We use \cite{spthesis} as our main reference for symmetric monoidal 2-categories, and \cite{DouglasSchommerPriesSnyder2013, Pstragowski} for dualisability.

\subsection{Definition of the ungraded 2-category}
\label{subsec:Definition2category}

Here we define the 2-category $\bi$. 
It is expected to describe, via trunctation, a small sector of the much more intricate 3-categorical structure of all Rozansky--Witten theories with arbitrary holomorphic symplectic target manifolds, introduced in \cite{KRS, KR0909.3643}.

Objects of $\bi$ are finite ordered sets of variables $(x_1,\dots,x_n)$ for $n\in\Z_{\geqslant 0}$. 
We abbreviate $\xu = (x_1,\dots,x_n)$ if the length~$n$ can be left implicit, and we note that for $n=0$ we have the empty list $\xu = \varnothing$. 
Equivalently, we may define objects in~$\bi$ to be the polynomial rings $\C[x_1,\dots,x_n]$. 
We will also use symbols like~$\xu'$ or~$\yu$ to denote objects. 

A 1-morphism $(x_1,\dots,x_n) \lra (y_1,\dots,y_m)$ in $\bi$ is a pair $(\au; W)$, where $\au = (a_1,\dots,a_k)$ is another (possibly empty) set of variables, and $W\in \C[\au,\xu,\yu]$ is a polynomial. 
To emphasise the roles of source and target variables, we sometimes write such 1-morphisms $(\au; W)$ as $(\au; W(\au,\xu,\yu))$. 
Using this notation, the horizontal composition of $(\au; W)\colon\xu \lra \yu$ and $(\bu; V)\colon \yu\lra\zu$ is given by
\be 
\label{eq:HorizontalComp} 
\big( \bu; \, V(\bu,\yu,\zu) \big) \circ \big( \au; \, W(\au,\xu,\yu) \big)
	= 
	\big( \au,\bu,\yu; \, V(\bu,\yu,\zu) + W(\au,\xu,\yu) \big) \, . 
\ee 
An example of a 1-endomorphism of $\xu = (x_1,\dots,x_n)$ is 
\be 
\label{eq:Unit1Morphism}
1_\xu = \big( \au ; \, \au \cdot (\xu'-\xu) \big) \, , 
\ee 
where by definition $\au \cdot (\xu'-\xu) = \sum_{i=1}^n a_i \cdot (x'_i-x_i)$. 
Note that in~\eqref{eq:Unit1Morphism} we use the two different symbols $\xu, \xu'$ to denote the same object; this is analogous to the isomorphism $\C[\xu] \otimes_\C \C[\xu] \cong \C[\xu,\xu']$, which we will often employ implicitly. 
Below we will see that, as the notation suggests, $1_\xu$ is a (weak) unit 1-morphism of~$\xu$. 

\medskip 

To describe 2-morphisms in $\bi$, we first recall some basics on matrix factorisations; see e.\,g.\ \cite{kr0401268, cm1208.1481} for more details and background. 
A \textsl{matrix factorisation} of a given polynomial $f\in\C[\xu]$ is a pair $(X,d_X)$, where $X=X^0 \oplus X^1$ is a free $\Z_2$-graded $\C[\xu]$-module and $d_X\colon X\lra X$ is an odd $\C[\xu]$-linear module map such that $d_X^2 = f\cdot \id_X$. 
The \textsl{shift} $(X,d_X)[1]$ of a matrix factorisation $(X,d_X)$ is defined by 
\be 
\label{eq:ShiftOfMF}
(X,d_X)[1] := (X^1 \oplus X^0, -d_X) \, . 
\ee 

A class of examples of matrix factorisations are those of Koszul type: 
for $k\in\Z_{\geqslant 1}$ and given polynomials $p_i,q_i\in\C[\xu]$, $i \in \{1,\dots,k\}$, the \textsl{Koszul matrix factorisation} 
\be 
\label{eq:Koszul1}
\big[\pu,\,\qu\big] = \big( K(\pu,\qu), \, d_{K(\pu,\qu)} \big)
\ee 
is given by 
\be 
\label{eq:Koszul2}
K(\pu,\qu) 
 	= 
 	\bigwedge\Big( \bigoplus_{i=1}^k \C[\xu] \cdot \theta_i \Big) \, , 
 	\quad 
d_{K(\pu,\qu)} 
	= 
	\sum_{i=1}^k \big( p_i\cdot \theta_i + q_i \cdot \theta_i^* \big) \, ,  
\ee 
where $\{\theta_i\}$ is a chosen $\C[\xu]$-basis of $\C[\xu]^{\oplus k}$. 
It is straightforward to check that $[\pu,\qu]$ is a matrix factorisation of $\sum_{i=1}^k p_i\cdot q_i$. 

For any $W\in \C[x_1,\dots,x_n]$, we may specialise to $k=n$ and choose 
\be 
\label{eq:IWpiqi}
q_i := x_i' - x_i \in \C[\xu,\xu'] \, , 
	\quad 
	p_i := \partial_{[i]}^{x',x}W \in \C[\xu,\xu'] \, , 
\ee 
where 
\be 
\label{eq:IWquotients}
\partial_{[i]}^{x',x}W 
	:=  
	\frac{W(x_1,\dots,x_{i-1}, x'_i, \dots x'_n) - W(x_1,\dots,x_i, x'_{i+1}, \dots x'_n)}{x'_i-x_i} \, . 
\ee 
The resulting matrix factorisation $[\pu,\qu]$ is a matrix factorisation of  $W(\xu')-W(\xu) \in \C[\xu,\xu'] \cong \C[\xu] \otimes_\C \C[\xu]$; we denote it by $I_W$. 

Finally, matrix factorisations of $f\in \C[\xu]$ are the objects of the \textsl{homotopy category of matrix factorisations} $\textrm{HMF}(\C[\xu],f)$. 
By definition a morphism $(X,d_X) \lra (X',d_{X'})$ in $\textrm{HMF}(\C[\xu],f)$ is a class in even cohomology of the differential $\delta_{X,X'}$ defined on $\Z_2$-homogeneous maps as 
\begin{align}
\delta_{X,X'} \colon \Hom_{\C[\xu]}(X,X') & \lra  \Hom_{\C[\xu]}(X,X') \nonumber 
\\ 
\zeta & \lmt d_{X'} \circ \zeta - (-1)^{|\zeta|} \zeta \circ d_X \, .
\label{eq:MFdiff}
\end{align}
We denote the idempotent completion of the full subcategory of finite-rank matrix factorisations of~$f$ by $\hmf(\C[\xu],f)^\omega$. 
Both $\textrm{HMF}(\C[\xu],f)$ and $\hmf(\C[\xu],f)^\omega$ are typically non-semisimple. 

With the above preparations, we can now define 2-morphisms in $\bi$ as follows. 
Let $(\au; W)$ and $(\bu;V)$ be 1-morphisms $\xu\lra\yu$. 
Then a 2-morphism $(\au;W)\lra (\bu;V)$ in $\bi$ is an isomorphism class of objects in $\textrm{HMF}(\C[\au,\bu,\xu,\yu],V-W)$ which has a representative that is a direct summand of a finite-rank matrix factorisation in $\textrm{HMF}(\C[\au,\bu,\xu,\yu],V-W)$. 
Put differently, such a 2-morphism is an isomorphism class of objects in $\hmf(\C[\au,\bu,\xu,\yu],V-W)^\omega$. 
For example, the Koszul matrix factorisation~$I_W$ from~\eqref{eq:IWpiqi} and~\eqref{eq:IWquotients} represents the unit 2-endomorphism $1_{(\xu;W)}$ on $(\xu;W)\colon\varnothing\lra\varnothing$. 
In general, the unit 2-morphism of $(\au;W)\colon \xu\lra \yu$ is represented by the analogous Koszul matrix factorisation that treats the variables $\xu,\yu$ as spectators, i.\,e.\ there are factors $q_i=a_i'-a_i$, and the variables $\xu$ and $\yu$ only appear in the $p_i$. 
We will usually not make a notational distinction between a 2-morphism in $\bi$ and a matrix factorisation that represents it, and we may abbreviate $(X,d_X)$ to~$X$. 

Our convention for the standard graphical calculus in 2-categories is that we read 1-morphisms from right to left, and 2-morphisms from bottom to top. 
Hence a 1-morphism $(\au; W) \colon \xu \lra \yu$ and a 2-morphism $(X,d_X) \colon (\au; W) \lra (\bu; V)$ are respectively presented as follows: 
\be 
\tikzzbox{%
	\begin{tikzpicture}[thick,scale=1.5, color=blue!50!black, baseline=0cm, xscale=-1]
	\coordinate (p1) at (0,0);
	\coordinate (p2) at (2,0);
	%
	\draw[very thick, red!80!black] (p1) -- (p2); 
	%
	\fill[red!80!black] ($(p2)+(0.1,0)$) circle (0pt) node {{\small $\yu\vphantom{\xu}$}};
	\fill[red!80!black] ($(p1)+(-0.1,0)$) circle (0pt) node {{\small $\xu$}};
	%
	\fill[black] ($(1,0)$) circle (0pt) node[above] {{\small $(\au; W)$}};
	%
	\end{tikzpicture}
} 
, \qquad 
\tikzzbox{%
	\begin{tikzpicture}[very thick,scale=1.0,color=blue!50!black, baseline=1.2cm]
	\coordinate (d1) at (-2,0);
	\coordinate (d2) at (+2,0);
	\coordinate (u1) at (-2,3);
	\coordinate (u2) at (+2,3);
	\coordinate (s) at ($(-0.5,-0.2)$);
	\coordinate (b1) at ($(d1)+(s)$);
	\coordinate (b2) at ($(d2)+(s)$);
	\coordinate (t1) at ($(u1)+(s)$);
	\coordinate (t2) at ($(u2)+(s)$);
	%
	\fill [orange!40!white, opacity=0.7] (b1) -- (b2) -- (t2) -- (t1); 
	%
	\draw[very thick] ($(d1)+(2,0)+(s)$) -- ($(d1)+(2,3)+(s)$);
	\fill ($(d1)+(2,1.5)+(s)$) circle (2.5pt) node[left] {{\small $X$}};
	\fill ($(d1)+(2.05,2.5)+(s)$) circle (0pt) node[left] {{\small $(\bu; V)$}};
	\fill ($(d1)+(2.05,0.5)+(s)$) circle (0pt) node[left] {{\small $(\au; W)$}};
	\fill[black] ($(d1)+(3.75,0.5)+(s)$) circle (0pt) node {{\small $\xu\vphantom{\yu'}$}};
	\fill[black] ($(d1)+(0.25,0.5)+(s)$) circle (0pt) node {{\small $\yu\vphantom{\yu'}$}};
	%
	\draw[very thick, red!80!black] (t1) -- (t2); 
	\draw[very thick, red!80!black] (b1) -- (b2); 
	\draw[thin, black] (b1) -- (t1); 
	\draw[thin, black] (b2) -- (t2);
	\end{tikzpicture}
}
\, . 
\ee 

\medskip 

The horizontal and vertical compositions of two appropriately composable 2-morphisms $(X,d_X)$ and $(Y,d_Y)$ are both given by the tensor product of matrix factorisations 
\begin{align}
& \big( X\otimes Y, \, d_{X\otimes Y} \big) 
\nonumber
\\
& 
= 
\Big( \big( (X^0\otimes Y^0) \oplus (X^1\otimes Y^1)\big) \oplus \big( (X^0\otimes Y^1) \oplus (X^1\otimes Y^0) \big) , \, d_X \otimes 1 + 1\otimes d_Y \Big) , 
\label{eq:XtensorY} 
\end{align}
where however the meaning of ``$\otimes$'' on the right-hand side is different for each case, denoting the appropriate relative tensor product over the respective intermediate polynomial ring: 
\begin{align}
\tikzzbox{%
	\begin{tikzpicture}[very thick,scale=1.0,color=blue!50!black, baseline=1.2cm]
	\coordinate (d1) at (-3,0);
	\coordinate (d2) at (+3,0);
	\coordinate (u1) at (-3,3);
	\coordinate (u2) at (+3,3);
	\coordinate (s) at ($(-0.5,-0.2)$);
	\coordinate (b1) at ($(d1)+(s)$);
	\coordinate (b2) at ($(d2)+(s)$);
	\coordinate (t1) at ($(u1)+(s)$);
	\coordinate (t2) at ($(u2)+(s)$);
	%
	\fill [orange!40!white, opacity=0.7] (b1) -- (b2) -- (t2) -- (t1); 
	%
	\draw[very thick] ($(d1)+(2.25,0)+(s)$) -- ($(d1)+(2.25,3)+(s)$);
	\fill ($(d1)+(2.25,1.5)+(s)$) circle (2.5pt) node[left] {{\small $X'$}};
	\fill ($(d1)+(2.3,2.5)+(s)$) circle (0pt) node[left] {{\small $(\bu'; V')$}};
	\fill ($(d1)+(2.3,0.5)+(s)$) circle (0pt) node[left] {{\small $(\au'; W')$}};
	\draw[very thick] ($(d1)+(3.75,0)+(s)$) -- ($(d1)+(3.75,3)+(s)$);
	\fill ($(d1)+(3.75,1.5)+(s)$) circle (2.5pt) node[right] {{\small $X\vphantom{X'}$}};
	\fill ($(d1)+(3.7,2.5)+(s)$) circle (0pt) node[right] {{\small $(\bu; V)$}};
	\fill ($(d1)+(3.7,0.5)+(s)$) circle (0pt) node[right] {{\small $(\au; W)$}};
	\fill[black] ($(d1)+(5.75,0.5)+(s)$) circle (0pt) node {{\small $\xu\vphantom{\yu'}$}};
	\fill[black] ($(d1)+(3,0.5)+(s)$) circle (0pt) node {{\small $\yu\vphantom{\yu'}$}};
	\fill[black] ($(d1)+(0.25,0.5)+(s)$) circle (0pt) node {{\small $\zu\vphantom{\yu'}$}};
	%
	\draw[very thick, red!80!black] (t1) -- (t2); 
	\draw[very thick, red!80!black] (b1) -- (b2); 
	\draw[thin, black] (b1) -- (t1); 
	\draw[thin, black] (b2) -- (t2);
	\end{tikzpicture}
}
& 
:= 
\tikzzbox{%
	\begin{tikzpicture}[very thick,scale=1.0,color=blue!50!black, baseline=1.2cm]
	\coordinate (d1) at (-3,0);
	\coordinate (d2) at (+3,0);
	\coordinate (u1) at (-3,3);
	\coordinate (u2) at (+3,3);
	\coordinate (s) at ($(-0.5,-0.2)$);
	\coordinate (b1) at ($(d1)+(s)$);
	\coordinate (b2) at ($(d2)+(s)$);
	\coordinate (t1) at ($(u1)+(s)$);
	\coordinate (t2) at ($(u2)+(s)$);
	%
	\fill [orange!40!white, opacity=0.7] (b1) -- (b2) -- (t2) -- (t1); 
	%
	\draw[very thick] ($(d1)+(4,0)+(s)$) -- ($(d1)+(4,3)+(s)$);
	\fill ($(d1)+(4,1.5)+(s)$) circle (2.5pt) node[left] {{\small $X'\otimes_{\C[\yu]} X$}};
	\fill ($(d1)+(4.05,2.5)+(s)$) circle (0pt) node[left] {{\small $(\bu, \bu', \yu; V+V')$}};
	\fill ($(d1)+(4.05,0.5)+(s)$) circle (0pt) node[left] {{\small $(\au, \au', \yu; W+W')$}};
	\fill[black] ($(d1)+(5.75,0.5)+(s)$) circle (0pt) node {{\small $\xu\vphantom{\yu'}$}};
	\fill[black] ($(d1)+(0.25,0.5)+(s)$) circle (0pt) node {{\small $\zu\vphantom{\yu'}$}};
	%
	\draw[very thick, red!80!black] (t1) -- (t2); 
	\draw[very thick, red!80!black] (b1) -- (b2); 
	\draw[thin, black] (b1) -- (t1); 
	\draw[thin, black] (b2) -- (t2);
	\end{tikzpicture}
}
\nonumber
\, , 
\\
\tikzzbox{%
	\begin{tikzpicture}[very thick,scale=1.0,color=blue!50!black, baseline=2cm]
	\coordinate (d1) at (-2,0);
	\coordinate (d2) at (+2,0);
	\coordinate (u1) at (-2,4.5);
	\coordinate (u2) at (+2,4.5);
	\coordinate (s) at ($(-0.5,-0.2)$);
	\coordinate (b1) at ($(d1)+(s)$);
	\coordinate (b2) at ($(d2)+(s)$);
	\coordinate (t1) at ($(u1)+(s)$);
	\coordinate (t2) at ($(u2)+(s)$);
	%
	\fill [orange!40!white, opacity=0.7] (b1) -- (b2) -- (t2) -- (t1); 
	%
	\draw[very thick] ($(d1)+(2.5,0)+(s)$) -- ($(d1)+(2.5,4.5)+(s)$);
	\fill ($(d1)+(2.5,3)+(s)$) circle (2.5pt) node[left] {{\small $Y$}};
	\fill ($(d1)+(2.5,1.5)+(s)$) circle (2.5pt) node[left] {{\small $X$}};
	\fill ($(d1)+(2.55,4)+(s)$) circle (0pt) node[left] {{\small $(\cu; U)$}};
	\fill ($(d1)+(2.55,2.25)+(s)$) circle (0pt) node[left] {{\small $(\bu; V)$}};
	\fill ($(d1)+(2.55,0.5)+(s)$) circle (0pt) node[left] {{\small $(\au ; W)$}};
	\fill[black] ($(d1)+(3.75,0.5)+(s)$) circle (0pt) node {{\small $\xu\vphantom{\yu'}$}};
	\fill[black] ($(d1)+(0.25,0.5)+(s)$) circle (0pt) node {{\small $\yu\vphantom{\yu'}$}};
	%
	\draw[very thick, red!80!black] (t1) -- (t2); 
	\draw[very thick, red!80!black] (b1) -- (b2); 
	\draw[thin, black] (b1) -- (t1); 
	\draw[thin, black] (b2) -- (t2);
	\end{tikzpicture}
}
& 
:= 
\tikzzbox{%
	\begin{tikzpicture}[very thick,scale=1.0,color=blue!50!black, baseline=2cm]
	\coordinate (d1) at (-2,0);
	\coordinate (d2) at (+2,0);
	\coordinate (u1) at (-2,4.5);
	\coordinate (u2) at (+2,4.5);
	\coordinate (s) at ($(-0.5,-0.2)$);
	\coordinate (b1) at ($(d1)+(s)$);
	\coordinate (b2) at ($(d2)+(s)$);
	\coordinate (t1) at ($(u1)+(s)$);
	\coordinate (t2) at ($(u2)+(s)$);
	%
	\fill [orange!40!white, opacity=0.7] (b1) -- (b2) -- (t2) -- (t1); 
	%
	\draw[very thick] ($(d1)+(2.5,0)+(s)$) -- ($(d1)+(2.5,4.5)+(s)$);
	\fill ($(d1)+(2.5,2.25)+(s)$) circle (2.5pt) node[left] {{\small $Y\otimes_{\C[\bu]} X$}};
	\fill ($(d1)+(2.55,4)+(s)$) circle (0pt) node[left] {{\small $(\cu; U)$}};
	\fill ($(d1)+(2.55,0.5)+(s)$) circle (0pt) node[left] {{\small $(\au ; W)$}};
	\fill[black] ($(d1)+(3.75,0.5)+(s)$) circle (0pt) node {{\small $\xu\vphantom{\yu'}$}};
	\fill[black] ($(d1)+(0.25,0.5)+(s)$) circle (0pt) node {{\small $\yu\vphantom{\yu'}$}};
	%
	\draw[very thick, red!80!black] (t1) -- (t2); 
	\draw[very thick, red!80!black] (b1) -- (b2); 
	\draw[thin, black] (b1) -- (t1); 
	\draw[thin, black] (b2) -- (t2);
	\end{tikzpicture}
}
\end{align} 

From the above and a standard matrix factorisation computation (see e.\,g.\ \cite[Sect.\,4]{kr0401268} or \cite[Sect.\,2.2]{cm1208.1481}) it follows that $1_{(\au;W)}$ is indeed the unit 2-morphism on $(\au;W)\colon \xu\lra\yu$. 
We also note that Koszul matrix factorisations as in \eqref{eq:Koszul1}--\eqref{eq:Koszul2} are tensor products: 
\be 
\big[ \pu, \, \qu \big] 
	= 
	\bigotimes_i \, [p_i, q_i] \, . 
\ee 

The horizontal composition~\eqref{eq:HorizontalComp} is strictly associative up to a re-ordering of the variables, which we will leave implicit. 
On the other hand, the unitors of the 2-category $\bi$ are non-trivial. 
For example, for $(\au; W)\colon \xu\lra\yu$ we have 
\begin{align}
1_{\yu} \circ \big( \au; \, W(\au,\xu,\yu) \big) 
	& = 
	\big( \bu; \, \bu\cdot (\yu'-\yu) \big) \circ \big( \au; \, W(\au,\xu,\yu) \big) 
	\nonumber 
	\\ 
	& = 
	\big( \au,\bu,\yu; \, \bu\cdot(\yu'-\yu) + W(\au,\xu,\yu) \big) \, . 
\end{align}
There is a canonical 2-morphism between this and $(\au;W)$ which is represented by the Koszul matrix factorisation with $q_i=y_i'-y_i$. 
That this is in fact a 2-isomorphism 
\be 
\label{eq:lambda}
\lambda_{(\au;W)} \colon 1_{\yu} \circ \big( \au; \, W(\au,\xu,\yu) \big) \lra \big( \au; \, W(\au,\xu,\yu) \big) 
\ee 
follows from Knörrer periodicity, see \cite[Sect.\,2.2.3\,\&\,2.3]{KR0909.3643}. 
Analogously we have right unitors 
\be 
\label{eq:rho}
\rho_{(\au;W)} \colon \big( \au; \, W(\au,\xu,\yu) \big) \circ 1_{\xu} \lra \big( \au; \, W(\au,\xu,\yu) \big) \, . 
\ee 

\medskip 

Summarising, the 2-category $\bi$ has sets of variables~$\xu$ as objects, polynomials $W(\au,\xu,\yu)$ with ``extra variables''~$\au$ as 1-morphisms $\xu\lra\yu$, and isomorphism classes of matrix factorisations of differences of such polynomials as 2-morphisms.

\subsection{Monoidal structure}
\label{subsec:MonoidalStructure}

To endow $\bi$ with a monoidal structure, we have to provide a 2-functor 
\be 
\btimes \colon \bi \times \bi \lra \bi \, , 
\ee 
a unit object $\one$, an associator~$a$ which is part of an adjoint equivalence $(a,a^-)$, as well as 1-unitors $l,r$, 2-unitors $\lambda',\mu',\rho'$ and a pentagonator~$\pi$; see \cite[Sect.\,2.3]{spthesis} for details. 

\medskip 

On objects, we define the \textsl{monoidal product} to be 
\be 
\label{eq:MonProdObj}
(x_1,\dots,x_n) \btimes (y_1,\dots,y_m) 
	= 
	(x_1,\dots,x_n,y_1,\dots,y_m) \, , 
\ee 
while on Hom categories we have 
\be 
\begin{tikzpicture}[
baseline=(current bounding box.base),
descr/.style={fill=white,inner sep=3.5pt},
normal line/.style={->}
]
\matrix (m) [matrix of math nodes, row sep=0.5em, column sep=3em, text height=2.5ex, text depth=0.1ex, cells={anchor=west}]
{%
	{} \phantom{aaaaaaaii}{\big(\bi \times \bi \big)\Big( (\xu,\xu'), (\yu,\yu') \Big)} &  \bi \big( \xu\btimes\xu', \yu\btimes\yu' \big) 
	\\[0.2cm]
	\Big( \big( \au; \, W(\au,\xu,\yu) \big) , \big(\bu; \, V(\bu,\xu',\yu') \big) \Big) & \Big( \au\btimes\bu; \, W(\au,\xu,\yu) + V(\bu,\xu',\yu') \Big) 
	\\[0.5cm]
	{} 
	& 
	{}
	\\[0.5cm]
	\Big( \big( \widetilde\au; \, \widetilde W(\widetilde\au,\xu,\yu) \big) , \big(\widetilde\bu; \, \widetilde V(\widetilde\bu,\xu',\yu') \big) \Big) & \Big( \widetilde\au\btimes\widetilde\bu; \, \widetilde W(\widetilde\au,\xu,\yu) + \widetilde V(\widetilde\bu,\xu',\yu') \Big) .
	\\
};
\path[font=\footnotesize] (m-1-1) edge[->] node[above] {$ \btimes $} (m-1-2);
\path[font=\footnotesize] (m-2-1) edge[|->] node[above] {} (m-2-2);
\path[font=\footnotesize] (m-4-1) edge[|->] node[above] {} (m-4-2);
\path[font=\footnotesize, transform canvas={yshift=14.5mm}] (m-4-1) edge[|->] node[above] {} (m-4-2);
\path[font=\footnotesize] (m-2-1) edge[->] node[right] {$ (X,Y) $} (m-4-1);
\path[font=\footnotesize] (m-2-2) edge[->] node[left] {$ X\otimes_\C Y $} (m-4-2);
\end{tikzpicture}
\ee 
In words, $\btimes$ acts as concatenation on lists of variables, as addition on polynomials, and as~$\otimes_\C$ on matrix factorisations. 
It is straightforward to check that this gives us a strict 2-functor $\btimes\colon \bi\times\bi \lra \bi$, whose \textsl{unit object} is clearly 
\be 
\label{eq:UnitObject}
\one = \varnothing \, . 
\ee 
This is captured by the 2-functor~$I$ from the trivial 2-category to $\bi$ which sends the unique object~$*$ to~$\one$. 

The \textsl{associator} $a \colon \btimes  \circ \, (\btimes \times \textrm{Id}_{\bi}) \lra \btimes \circ (\textrm{Id}_{\bi} \times \btimes)$ by definition has 1- and 2-morphism components 
\be 
\label{eq:Associator}
a_{\xu,\yu,\zu} = 1_{\xu\btimes\yu\btimes\zu} 
	\, , \quad 
	a_{(\au; W), (\bu;V), (\cu;U)} = \lambda^{-1}_{(\au; W)\btimes (\bu;V)\btimes (\cu;U)} \cdot \rho_{(\au; W)\btimes \,  (\bu;V)\btimes (\cu;U)}
\ee 
respectively. 
Here and below we suppress re-bracketing operations, and $\lambda, \rho$ are the unitor 2-isomorphisms~\eqref{eq:lambda}--\eqref{eq:rho}. 
The associator components are clearly isomorphisms, and we make the natural choice for the adjoint equivalence $(a,a^-)$. 

\medskip 

We will frequently employ the 3-dimensional graphical calculus for monoidal 2-categories developed in \cite{TrimbleSurfaceDiagrams, BMS}, using the conventions of \cite[Sect.\,3.1]{CMS}. 
As in string diagrams for 2-categories, objects, 1- and 2-morphisms are represented by 2-, 1- and 0-dimensional strata, respectively. 
In addition to horizontal composition (from right to left) and vertical composition (from bottom to top), in a monoidal 2-category we moreover read the monoidal product from front to back. 
Hence in $\bi$ for 2-morphisms $X\colon (\au;W) \lra (\widetilde\au;\widetilde W)$ and $Y\colon (\bu;V) \lra (\widetilde\bu;\widetilde V)$, we have 
\begin{align}
&
\tikzzbox{%
	\begin{tikzpicture}[very thick,scale=1.0,color=blue!50!black, baseline=1.2cm]
	\coordinate (d1) at (-2,0);
	\coordinate (d2) at (+2,0);
	\coordinate (u1) at (-2,3);
	\coordinate (u2) at (+2,3);
	\coordinate (s) at ($(-0.5,-0.2)$);
	\coordinate (b1) at ($(d1)+(s)$);
	\coordinate (b2) at ($(d2)+(s)$);
	\coordinate (t1) at ($(u1)+(s)$);
	\coordinate (t2) at ($(u2)+(s)$);
	%
	\fill [red!20!white, opacity=0.7] (b1) -- (b2) -- (t2) -- (t1); 
	%
	\draw[very thick] ($(d1)+(2,0)+(s)$) -- ($(d1)+(2,3)+(s)$);
	\fill ($(d1)+(2,1.5)+(s)$) circle (2.5pt) node[right] {{\small $Y$}};
	\fill ($(d1)+(1.95,2.5)+(s)$) circle (0pt) node[right] {{\small $(\widetilde\bu;\widetilde V)$}};
	\fill ($(d1)+(1.95,0.5)+(s)$) circle (0pt) node[right] {{\small $(\bu; V)$}};
	\fill[black] ($(d1)+(3.75,0.5)+(s)$) circle (0pt) node {{\small $\xu'\vphantom{\yu'}$}};
	\fill[black] ($(d1)+(0.25,0.5)+(s)$) circle (0pt) node {{\small $\yu'$}};
	%
	\draw[very thick, red!80!black] (t1) -- (t2); 
	\draw[very thick, red!80!black] (b1) -- (b2); 
	\draw[thin, black] (b1) -- (t1); 
	\draw[thin, black] (b2) -- (t2);
	\end{tikzpicture}
}
\btimes 
\tikzzbox{%
	\begin{tikzpicture}[very thick,scale=1.0,color=blue!50!black, baseline=1.2cm]
	\coordinate (d1) at (-2,0);
	\coordinate (d2) at (+2,0);
	\coordinate (u1) at (-2,3);
	\coordinate (u2) at (+2,3);
	\coordinate (s) at ($(-0.5,-0.2)$);
	\coordinate (b1) at ($(d1)+(s)$);
	\coordinate (b2) at ($(d2)+(s)$);
	\coordinate (t1) at ($(u1)+(s)$);
	\coordinate (t2) at ($(u2)+(s)$);
	%
	\fill [orange!40!white, opacity=0.7] (b1) -- (b2) -- (t2) -- (t1); 
	%
	\draw[very thick] ($(d1)+(2,0)+(s)$) -- ($(d1)+(2,3)+(s)$);
	\fill ($(d1)+(2,1.5)+(s)$) circle (2.5pt) node[left] {{\small $X$}};
	\fill ($(d1)+(2.05,2.5)+(s)$) circle (0pt) node[left] {{\small $(\widetilde\au;\widetilde W)$}};
	\fill ($(d1)+(2.05,0.5)+(s)$) circle (0pt) node[left] {{\small $(\au; W)$}};
	\fill[black] ($(d1)+(3.75,0.5)+(s)$) circle (0pt) node {{\small $\xu\vphantom{\yu'}$}};
	\fill[black] ($(d1)+(0.25,0.5)+(s)$) circle (0pt) node {{\small $\yu\vphantom{\yu'}$}};
	%
	\draw[very thick, red!80!black] (t1) -- (t2); 
	\draw[very thick, red!80!black] (b1) -- (b2); 
	\draw[thin, black] (b1) -- (t1); 
	\draw[thin, black] (b2) -- (t2);
	\end{tikzpicture}
}
=
\tikzzbox{%
	\begin{tikzpicture}[very thick,scale=1.0,color=blue!50!black, baseline=1.2cm]
	\coordinate (d1) at (-2,0);
	\coordinate (d2) at (+2,0);
	\coordinate (u1) at (-2,3);
	\coordinate (u2) at (+2,3);
	\coordinate (s) at ($(-0.5,-0.2)$);
	\coordinate (b1) at ($(d1)+(s)$);
	\coordinate (b2) at ($(d2)+(s)$);
	\coordinate (t1) at ($(u1)+(s)$);
	\coordinate (t2) at ($(u2)+(s)$);
	%
	\fill [red!20!white, opacity=0.8] (u1) -- (u2) -- (d2) -- (d1); 
	%
	\draw[very thick] ($(d1)+(2,0)$) -- ($(d1)+(2,3)$);
	\fill ($(d1)+(2,1.5)$) circle (2.5pt) node[right] {{\small $Y$}};
	\fill ($(d1)+(1.95,2.5)$) circle (0pt) node[right] {{\small $(\widetilde\bu;\widetilde V)$}};
	\fill ($(d1)+(1.95,0.5)$) circle (0pt) node[right] {{\small $(\bu; V)$}};
	\fill[black] ($(d1)+(3.75,0.5)$) circle (0pt) node {{\small $\xu'\vphantom{\yu'}$}};
	\fill[black] ($(d1)+(0.25,0.5)$) circle (0pt) node {{\small $\yu'$}};
	%
	\draw[very thick, red!80!black] (u1) -- (u2); 
	\draw[very thick, red!80!black] (d1) -- (d2); 
	\draw[thin, black] (d1) -- (u1); 
	\draw[thin, black] (d2) -- (u2);
	%
	%
	\fill [orange!40!white, opacity=0.7] (b1) -- (b2) -- (t2) -- (t1); 
	%
	\draw[very thick] ($(d1)+(2,0)+(s)$) -- ($(d1)+(2,3)+(s)$);
	\fill ($(d1)+(2,1.5)+(s)$) circle (2.5pt) node[left] {{\small $X$}};
	\fill ($(d1)+(2.05,2.5)+(s)$) circle (0pt) node[left] {{\small $(\widetilde\au;\widetilde W)$}};
	\fill ($(d1)+(2.05,0.5)+(s)$) circle (0pt) node[left] {{\small $(\au; W)$}};
	\fill[black] ($(d1)+(3.75,0.5)+(s)$) circle (0pt) node {{\small $\xu\vphantom{\yu'}$}};
	\fill[black] ($(d1)+(0.25,0.5)+(s)$) circle (0pt) node {{\small $\yu\vphantom{\yu'}$}};
	%
	\draw[very thick, red!80!black] (t1) -- (t2); 
	\draw[very thick, red!80!black] (b1) -- (b2); 
	\draw[thin, black] (b1) -- (t1); 
	\draw[thin, black] (b2) -- (t2);
	\end{tikzpicture}
} 
\nonumber 
\\
= \; & 
\tikzzbox{%
	\begin{tikzpicture}[very thick,scale=1.0,color=blue!50!black, baseline=1.2cm]
	\coordinate (d1) at (-3,0);
	\coordinate (d2) at (+3,0);
	\coordinate (u1) at (-3,3);
	\coordinate (u2) at (+3,3);
	\coordinate (s) at ($(-0.5,-0.2)$);
	\coordinate (b1) at ($(d1)+(s)$);
	\coordinate (b2) at ($(d2)+(s)$);
	\coordinate (t1) at ($(u1)+(s)$);
	\coordinate (t2) at ($(u2)+(s)$);
	%
	\fill [orange!40!white, opacity=0.7] (b1) -- (b2) -- (t2) -- (t1); 
	\fill [red!20!white, opacity=0.5] (b1) -- (b2) -- (t2) -- (t1); 
	%
	\draw[very thick] ($(d1)+(4,0)+(s)$) -- ($(d1)+(4,3)+(s)$);
	\fill ($(d1)+(4,1.5)+(s)$) circle (2.5pt) node[left] {{\small $Y\otimes_\C X$}};
	\fill ($(d1)+(4.05,2.5)+(s)$) circle (0pt) node[left] {{\small $(\widetilde\au,\widetilde\bu;\widetilde V + \widetilde W)$}};
	\fill ($(d1)+(4.05,0.5)+(s)$) circle (0pt) node[left] {{\small $(\au,\bu; V+W)$}};
	\fill[black] ($(d1)+(5.4,0.5)+(s)$) circle (0pt) node {{\small $\xu'\btimes\xu\vphantom{\yu'}$}};
	\fill[black] ($(d1)+(0.6,0.5)+(s)$) circle (0pt) node {{\small $\yu'\btimes\yu\vphantom{\yu'}$}};
	%
	\draw[very thick, red!80!black] (t1) -- (t2); 
	\draw[very thick, red!80!black] (b1) -- (b2); 
	\draw[thin, black] (b1) -- (t1); 
	\draw[thin, black] (b2) -- (t2);
	\end{tikzpicture}
}
\, . 
\end{align}

\medskip 

We continue to enumerate the data of the monoidal structure on $\bi$. 
The \textsl{pentagonator} is an invertible modification 
\begin{align} 
& 
\pi \colon \big(1_{\btimes} * (1_{\idlg} \times a)\big) \circ \big(a * 1_{\idlg \times \btimes \times \idlg}\big) \circ \big(1_{\btimes} * (a \times 1_{\idlg} )\big)\nonumber 
\\
&\qquad\qquad\qquad\qquad\qquad\qquad
\lra
\big( a * 1_{\idlg \times \idlg \times \btimes}\big) \circ \big(a*1_{{\btimes} \times \idlg \times \idlg}\big)
\end{align} 
which measures to what extent the associator~$a$ satisfies the pentagon axiom. 
Its components are
\be 
\label{eq:Pentagonatur} 
\pi_{\wu,\xu,\yu,\zu} 
	= 
	\lambda_{1_{\wu,\xu,\yu,\zu} \circ 1_{\wu,\xu,\yu,\zu}} \cdot \big( \btimes_{\wu;\xu,\yu,\zu} \cdot 1_{1_{\wu,\xu,\yu,\zu}} \cdot \btimes_{\wu,\xu,\yu;\zu} \big) \, . 
\ee 

The \textsl{left} and \textsl{right 1-unitors} are the pseudonatural transformations $l\colon \btimes \circ \, (I \times \textrm{Id}_\bi) \lra \textrm{Id}_\bi$ and $r\colon \btimes \circ \, ( \textrm{Id}_\bi \times I) \lra \textrm{Id}_\bi$ whose components are 
\be 
\label{eq:1unitors}
l_{*,\xu} = 1_\xu = r_{\xu,*} 
	\, , \quad 
	l_{1_*,(\au;W)} = \lambda^{-1}_{(\au;W)} \cdot \rho_{(\au;W)} = r_{(\au;W),1_*} \, . 
\ee 
Finally, the \textsl{2-unitors} are the invertible modifications 
$\lambda'\colon 1 \circ (l \times 1) \lra (l*1)\circ (a*1)$, 
$\rho'\colon r\circ 1 \lra (1*(1\times r)) \circ (a*1)$ 
and 
$\mu'\colon 1\circ (r\times 1) \lra (1*(1\times l)) \times (a*1)$ 
with components 
\be 
\label{eq:2unitors}
\lambda'_{*,\xu,\yu} = \lambda^{-1}_{1_{\xu,\yu}} \cdot \btimes^{-1}_{\xu;\yu} 
	\, , \quad 
	\rho'_{\xu,\yu,*} = \big( \btimes_{\xu;\yu} \circ 1_{1_{\xu,\yu}} \big) \cdot \lambda^{-1}_{1_{\xu,\yu}}
	\, , \quad 
	\mu'_{\xu,*,\yu} = \rho^{-1}_{1_\xu \btimes 1_\yu} \, . 
\ee 

It is now a straightforward exercise to verify that the above data satisfy the coherence axioms of a monoidal 2-category, cf.\ \cite[Def.\,C.1]{spthesis}. 
This can be done analogously to the proof of \cite[Prop.\,2.2]{CMM}, though our present situation is arguably simpler. 
Hence we have: 

\begin{proposition}
	\label{prop:MonoidalStructure} 
	The data in \eqref{eq:MonProdObj}--\eqref{eq:Associator} and \eqref{eq:Pentagonatur}--\eqref{eq:2unitors} endow $\bi$ with a monoidal structure. 
\end{proposition}

\subsection{Symmetric monoidal structure}
\label{subsec:SymmetricMonoidalStructure}

To endow the monoidal 2-category $\bi$ with a symmetric braided structure, we have to provide a braiding~$b$ which is part of an adjoint equivalence $(b,b^-)$, a syllepsis~$\sigma$, as well as invertible modifications~$R$ and~$S$ that mediate between the associator~$a$ and the braiding~$b$; see \cite[Sect.\,2.3]{spthesis} for details. 

\medskip 

The 1-morphism components 
\be 
\label{eq:Braiding1morphismComponents}
b_{\xu,\yu} \colon \xu\btimes\yu 
\lra 
\yu \btimes \xu 
\equiv \yu' \btimes \xu' 
\ee 
of~$b$ are simply identity 1-morphisms up to a re-ordering of variables, i.\,e.\ $b_{\xu,\yu} = (\cu,\du; \du \cdot (\yu'-\yu) + \cu \cdot (\xu'-\xu))$ which is identical to the pair which is the identity $1_{\xu\btimes\yu}$, but viewed as the 1-morphism~\eqref{eq:Braiding1morphismComponents} with a different yet isomorphic target. 
Analogously, the 2-morphism components are identity 2-morphisms: 
\be 
b_{(\au;W),(\bu;V)} = 
\tikzzbox{%
	\begin{tikzpicture}[thick,scale=1.0,color=black, baseline=2cm]
	\coordinate (p3) at (2,0.5);
	\coordinate (u3) at (2,3.5);
	\coordinate (ld) at (-1,1);
	\coordinate (lu) at (-1,4);
	\coordinate (rd) at (5,1);
	\coordinate (ru) at (5,4);
	\coordinate (rd2) at (5.5,0);
	\coordinate (ru2) at (5.5,3);
	\coordinate (ld2) at (-1.5,0);
	\coordinate (lu2) at (-1.5,3);
	%
	\fill [orange!25!white, opacity=0.8] (p3) -- (ld) -- (lu) -- (u3); 
	\fill [orange!25!white, opacity=0.8] (p3) -- (rd) -- (ru) -- (u3); 
	\draw[thin] (lu) --  (ld); 
	\draw[thin] (ru) --  (rd); 
	%
	\coordinate (xlb) at (0,0.85);
	\coordinate (xrb) at (3.5,3.75);
	\draw[ultra thick, color=blue!50!black] (xlb) .. controls +(0,0.25) and +(-0.2,-0.5) .. (2,2); 
	\draw[ultra thick, color=blue!50!black] (xrb) .. controls +(0,-0.75) and +(0.2,0.5) .. (2,2); 
	\fill[color=blue!50!black] (0.4,1.4) circle (0pt) node {{\small $V$}};
	\fill[color=blue!50!black] (3.75,3.5) circle (0pt) node {{\small $W$}};
	%
	\draw[very thick, red!80!black] (p3) -- (ld);
	\draw[very thick, red!80!black] (u3) -- (lu); 
	\draw[very thick, red!80!black] (p3) -- (rd);
	\draw[very thick, red!80!black] (ru) -- (u3); 
	%
	\fill [orange!30!white, opacity=0.8] (p3) -- (ld2) -- (lu2) -- (u3); 
	\fill [orange!30!white, opacity=0.8] (p3) -- (rd2) -- (ru2) -- (u3); 
	%
	\draw[ultra thick] (p3) --  (u3); 
	\fill (2.5,1) circle (0pt) node {{\small $b_{\xu,\yu}$}};
	\fill (1.5,3) circle (0pt) node {{\small $b_{\xu',\yu'}$}};\\
	%
	\coordinate (xlf) at (0.75,0.32);
	\coordinate (xrf) at (4,3.2);
	\draw[ultra thick, color=blue!50!black] (xlf) .. controls +(0,0.5) and +(-0.2,-0.5) .. (2,2); 
	\draw[ultra thick, color=blue!50!black] (xrf) .. controls +(0,-0.75) and +(0.2,0.3) .. (2,2); 
	\fill[color=blue!50!black] (0.55,0.5) circle (0pt) node {{\small $W$}};
	\fill[color=blue!50!black] (4.2,2.9) circle (0pt) node {{\small $V$}};
	\fill[color=blue!50!black] (2,2) circle (2.5pt) node[left] {{\small $b_{(\au;W),(\bu;V)}$}};
	%
	\draw[very thick, red!80!black] (ru2) -- (u3); 
	\draw[very thick, red!80!black] (rd2) -- (p3); 
	\draw[very thick, red!80!black] (lu2) -- (u3); 
	\draw[very thick, red!80!black] (ld2) -- (p3); 
	\draw[thin] (ru2) --  (rd2); 
	\draw[thin] (lu2) --  (ld2); 
	%
	%
	\fill (-0.5,3.5) circle (0pt) node {{\small $\yu'$}};
	\fill (4.5,3.5) circle (0pt) node {{\small $\xu\vphantom{\yu'}$}};
	\fill (-1,0.5) circle (0pt) node {{\small $\xu'\vphantom{\yu}$}};
	\fill (5,0.5) circle (0pt) node {{\small $\yu\vphantom{\yu'}$}};
	%
	\end{tikzpicture}
}  
\equiv 
1_{(\au,\bu;V+W)}
\, . 
\ee 
Hence~$b$ is clearly invertible, and we take its adjoint~$b^-$ to have identity components as well. 

Since~$a$ and~$b$ have only identity components, the invertible modifications 
$\sigma\colon 1_{\btimes} \lra b\circ b$, 
$R\colon a \circ b \circ a \lra (\idlg \btimes b) \circ a \circ (b \btimes \idlg)$ 
and 
$S\colon a^- \circ b \circ a^- \lra (b \btimes \idlg) \circ a^- \circ (\idlg \btimes b)$ 
have components constructed only from the unitors of the underlying 2-category $\bi$, and they satisfy their coherence axioms by the coherence theorem for 2-categories. 
Hence we find: 

\begin{proposition}
	\label{prop:SymmetricMonoidalStructure}
	The above data endow the 2-category $\bi$ with a symmetric monoidal structure. 
\end{proposition}

\subsection{Full dualisability}
\label{subsec:FullDualisibility}

\subsubsection{Duals for objects}
\label{subsubsec:DualsForObjects}

An object $\xu \in \bi$ is dualisable if there exists an object $\xu^\dual \in \bi$ together with adjunction 1-morphisms $\tev_\xu, \tcoev_\xu$ which satisfy the Zorro moves up to isomorphism, see e.\,g.\ \cite[Sect.\,2]{Pstragowski}. 
Here we show that every object in $\bi$ is dualisable and self-dual. 

\medskip 

For $\xu\in\bi$ we take the (right) \textsl{dual object} to be
\be 
\label{eq:DualObject}
\xu^\dual := \xu \, , 
\ee 
and we set 
\begin{align}
\tikzzbox{%
	\begin{tikzpicture}[thick,scale=1.5,color=black, baseline=0.1cm]
	\coordinate (p1) at (0,0);
	\coordinate (p2) at (0,0.25);
	\coordinate (g1) at (0,-0.05);
	\coordinate (g2) at (0,0.3);
	\draw[very thick, red!80!black] (p1) .. controls +(-0.3,0) and +(-0.3,0) ..  (p2); 
	\fill[red!80!black] (0.35,0) circle (0pt) node[left] {{\tiny$\xu'$}};
	\fill[red!80!black] (0.3,0.25) circle (0pt) node[left] {{\tiny$\xu\vphantom{\xu'}$}};
	\end{tikzpicture}
}
= \tev_\xu & := \big( \au; \, \au\cdot (\xu'-\xu) \big) \colon \xu \btimes \xu^\dual = (\xu,\xu') \lra \varnothing \, , 
\label{eq:Eval} 
\\ 
\tikzzbox{%
	\begin{tikzpicture}[thick,scale=1.5,color=black, baseline=0.1cm]
	\coordinate (p1) at (0,0);
	\coordinate (p2) at (0,0.25);
	\coordinate (g1) at (0,-0.05);
	\coordinate (g2) at (0,0.3);
	\draw[very thick, red!80!black] (p1) .. controls +(0.3,0) and +(0.3,0) ..  (p2); 
	\fill[red!80!black] (0,0) circle (0pt) node[left] {{\tiny$\xu\vphantom{\xu'}$}};
	\fill[red!80!black] (0.07,0.25) circle (0pt) node[left] {{\tiny$\xu'$}};
	\end{tikzpicture}
}
= \tcoev_\xu & := \big( \au; \, \au\cdot (\xu-\xu') \big) \colon \varnothing \lra \xu^\dual \btimes \xu = (\xu',\xu) \, . 
\label{eq:Coeval}
\end{align}
To show that these \textsl{adjunction 1-morphisms} indeed exhibit~$\xu$ as its own dual, we have to prove that there are \textsl{cusp isomorphisms}\footnote{or ``Zorro movies'', as reading the diagrams from bottom to top can be interpreted as straightening the (reversed) Zorro-Z} 
\begin{align}
\label{eq:cuspl}
\tikzzbox{%
	\begin{tikzpicture}[thick,scale=0.75,color=black, baseline=1.8cm, xscale=-1]
	\coordinate (p1) at (0,0);
	\coordinate (p2) at (1.5,-0.5);
	\coordinate (p2s) at (4,-0.5);
	\coordinate (p3) at (1.5,0.5);
	\coordinate (p4) at (3,1);
	\coordinate (p5) at (1.5,1.5);
	\coordinate (p6) at (-1,1.5);
	\coordinate (u1) at (0,2.5);
	\coordinate (u2) at (1.5,2);
	\coordinate (u2s) at (4,3.5);
	\coordinate (u3) at (1.5,3);
	\coordinate (u4) at (3,3.5);
	\coordinate (u5) at (1.5,4);
	\coordinate (u6) at (-1,3.5);
	%
	\fill [orange!90!white, opacity=1] 
	(p4) .. controls +(0,0.25) and +(1,0) ..  (p5) 
	-- (p5) -- (0.79,1.5) -- (1.5,2.5) 
	-- (1.5,2.5) .. controls +(0,0) and +(0,0.5) ..  (p4);
	%
	\fill [orange!30!white, opacity=0.8] 
	(1.5,2.5) .. controls +(0,0) and +(0,0.5) ..  (p4)
	-- 
	(p4) .. controls +(0,-0.25) and +(1,0) ..  (p3)
	-- (p3) .. controls +(-1,0) and +(0,0.25) ..  (p1)
	-- 
	(p1) .. controls +(0,0.5) and +(0,0) ..  (1.5,2.5);
	%
	\draw[very thick, red!80!black] (p4) .. controls +(0,0.25) and +(1,0) ..  (p5) -- (p6); 
	\draw[very thick, red!80!black] (p1) .. controls +(0,0.25) and +(-1,0) ..  (p3)
	-- (p3) .. controls +(1,0) and +(0,-0.25) ..  (p4); 
	%
	\draw[thin] (p2s) --  (u2s); 
	\draw[thin] (p6) --  (u6); 
	\fill [orange!20!white, opacity=0.8] 
	(p2s) -- (p2)
	-- (p2) .. controls +(-1,0) and +(0,-0.25) ..  (p1) 
	-- (p1) .. controls +(0,0.5) and +(0,0) ..  (1.5,2.5)
	-- (1.5,2.5) .. controls +(0,0) and +(0,0.5) .. (p4)
	-- (p4) .. controls +(0,0.25) and +(-1,0) ..  (p5) 
	-- (p5) -- (p6) -- (u6) -- (u2s)
	;
	%
	\draw[thin, dotted] (p4) .. controls +(0,0.5) and +(0,0) ..  (1.5,2.5); 
	\draw[thin] (p1) .. controls +(0,0.5) and +(0,0) ..  (1.5,2.5); 
	%
	%
	\draw[very thick, red!80!black] (u6) -- (u2s); 
	\draw[very thick, red!80!black] (p1) .. controls +(0,-0.25) and +(-1,0) ..  (p2) -- (p2s); 
	\draw[very thick, red!80!black] (0.79,1.5) -- (p6); 
	%
	\fill[color=blue!50!black] (3.5,0.2) circle (0pt) node {{\small $\xu$}};
	\fill (1.5,2.5) circle (2.5pt) node[right] {{\small $c_{\textrm{l}}$}};
	%
	\end{tikzpicture}
}     
& = c_{\textrm{l}} \colon \big(\tev_\xu \btimes 1_\xu\big) \circ \big(1_\xu \btimes \tcoev_\xu\big) \stackrel{\cong}{\lra} 1_\xu \, , 
\\ 
\tikzzbox{%
	\begin{tikzpicture}[thick,scale=0.75,color=black, baseline=1.8cm]
	\coordinate (p1) at (0,0);
	\coordinate (p2) at (1.5,-0.5);
	\coordinate (p2s) at (4,-0.5);
	\coordinate (p3) at (1.5,0.5);
	\coordinate (p4) at (3,1);
	\coordinate (p5) at (1.5,1.5);
	\coordinate (p6) at (-1,1.5);
	\coordinate (u1) at (0,2.5);
	\coordinate (u2) at (1.5,2);
	\coordinate (u2s) at (4,3.5);
	\coordinate (u3) at (1.5,3);
	\coordinate (u4) at (3,3.5);
	\coordinate (u5) at (1.5,4);
	\coordinate (u6) at (-1,3.5);
	%
	\fill [orange!90!white, opacity=1] 
	(p4) .. controls +(0,0.25) and +(1,0) ..  (p5) 
	-- (p5) -- (0.79,1.5) -- (1.5,2.5) 
	-- (1.5,2.5) .. controls +(0,0) and +(0,0.5) ..  (p4);
	%
	\fill [orange!30!white, opacity=0.8] 
	(1.5,2.5) .. controls +(0,0) and +(0,0.5) ..  (p4)
	-- 
	(p4) .. controls +(0,-0.25) and +(1,0) ..  (p3)
	-- (p3) .. controls +(-1,0) and +(0,0.25) ..  (p1)
	-- 
	(p1) .. controls +(0,0.5) and +(0,0) ..  (1.5,2.5);
	%
	\draw[very thick, red!80!black] (p4) .. controls +(0,0.25) and +(1,0) ..  (p5) -- (p6); 
	\draw[very thick, red!80!black] (p1) .. controls +(0,0.25) and +(-1,0) ..  (p3)
	-- (p3) .. controls +(1,0) and +(0,-0.25) ..  (p4); 
	%
	\draw[thin] (p2s) --  (u2s); 
	\draw[thin] (p6) --  (u6); 
	\fill [orange!20!white, opacity=0.8] 
	(p2s) -- (p2)
	-- (p2) .. controls +(-1,0) and +(0,-0.25) ..  (p1) 
	-- (p1) .. controls +(0,0.5) and +(0,0) ..  (1.5,2.5)
	-- (1.5,2.5) .. controls +(0,0) and +(0,0.5) .. (p4)
	-- (p4) .. controls +(0,0.25) and +(-1,0) ..  (p5) 
	-- (p5) -- (p6) -- (u6) -- (u2s)
	;
	%
	\draw[thin, dotted] (p4) .. controls +(0,0.5) and +(0,0) ..  (1.5,2.5); 
	\draw[thin] (p1) .. controls +(0,0.5) and +(0,0) ..  (1.5,2.5); 
	%
	%
	\draw[very thick, red!80!black] (u6) -- (u2s); 
	\draw[very thick, red!80!black] (p1) .. controls +(0,-0.25) and +(-1,0) ..  (p2) -- (p2s); 
	\draw[very thick, red!80!black] (0.79,1.5) -- (p6); 
	%
	\fill[color=blue!50!black] (3.5,0.2) circle (0pt) node {{\small $\xu^\dual$}};
	\fill (1.5,2.5) circle (2.5pt) node[right] {{\small $c_{\textrm{r}}$}};
	%
	\end{tikzpicture}
}   
& = c_{\textrm{r}} \colon \big( 1_{\xu^\dual} \btimes \tev_\xu \big) \circ \big( \tcoev_\xu \btimes 1_{\xu^\dual} \big) \stackrel{\cong}{\lra} 1_{\xu^\dual} \, . 
\end{align}
Writing $\au^{(i)}$ and $\xu^{(j)}$ for various copies of the lists of variables~$\au$ and~$\xu$, we can apply the unitors in~\eqref{eq:lambda} and~\eqref{eq:rho} to simplify the domain of~$c_{\textrm{l}}$ as follows:  
\begin{align}
& 
\tikzzbox{%
	\begin{tikzpicture}[thick,scale=1.5, color=blue!50!black, baseline=1.8cm, xscale=-1]
	\coordinate (p1) at (0,0);
	\coordinate (p2) at (1.5,-0.5);
	\coordinate (p2s) at (4,-0.5);
	\coordinate (p3) at (1.5,0.5);
	\coordinate (p4) at (3,1);
	\coordinate (p5) at (1.5,1.5);
	\coordinate (p6) at (-1,1.5);
	%
	\draw[very thick, red!80!black] (p4) .. controls +(0,0.25) and +(1,0) ..  (p5) -- (p6); 
	\draw[very thick, red!80!black] (p1) .. controls +(0,0.25) and +(-1,0) ..  (p3)
	-- (p3) .. controls +(1,0) and +(0,-0.25) ..  (p4); 
	%
	\draw[very thick, red!80!black] (p1) .. controls +(0,-0.25) and +(-1,0) ..  (p2) -- (p2s); 
	\draw[very thick, red!80!black] (0.79,1.5) -- (p6); 
	%
	\fill[black] (p1) circle (0pt) node[left] {{\small $\tcoev_{\xu}\vphantom{\au^{(4)}\tev_{\xu}}$}};
	\fill[red!80!black] ($(p1)+(0.05,0)$) circle (0pt) node[right] {{\tiny $\au^{(2)}\vphantom{\au^{(4)}\tev_{\xu}}$}};
	\fill[red!80!black] ($(p2)+(0,-0.05)$) circle (0pt) node[above] {{\tiny $\xu^{(4)}$}};
	\fill[red!80!black] ($(p3)+(0,-0.05)$) circle (0pt) node[above] {{\tiny $\xu^{(3)}$}};
	\fill[red!80!black] ($(p4)+(-0.07,0)$) circle (0pt) node[left] {{\tiny $\au^{(4)}\vphantom{\au^{(4)}\tev_{\xu}}$}};
	\fill[black] (p4) circle (0pt) node[right] {{\small $\tev_{\xu}\vphantom{\au^{(4)}\tev_{\xu}}$}};
	\fill[red!80!black] ($(p5)+(0,-0.05)$) circle (0pt) node[above] {{\tiny $\xu^{(2)}$}};
	\fill[red!80!black] ($(p6)+(-0.25,-0.2)$) circle (0pt) node[above] {{\small $\xu^{(1)}$}};
	\fill[red!80!black] ($(p2s)+(0.25,-0.2)$) circle (0pt) node[above] {{\small $\xu^{(5)}$}};
	\fill[black] ($(p6)+(0.9,-0.05)$) circle (0pt) node[above] {{\small $1_{\xu}$}};
	\fill[red!80!black] ($(p6)+(0.9,0.05)$) circle (0pt) node[below] {{\tiny $\au^{(1)}$}};
	\fill[red!80!black] ($(p6)+(1,0)$) circle (1.5pt) node[above] {};
	\fill[red!80!black] ($(p2s)+(-1,0)$) circle (1.5pt) node[above] {};
	\fill[red!80!black] ($(p2s)+(-1,0)$) circle (0pt) node[below] {{\tiny $\au^{(3)}$}};
	\fill[black] ($(p2s)+(-1.1,-0.05)$) circle (0pt) node[above] {{\small $1_{\xu}$}};
	%
	\end{tikzpicture}
} 
\nonumber \\ 
& =  
\au^{(1)} \cdot \big( \xu^{(2)} - \xu^{(1)} \big) 
	+ \au^{(2)} \cdot \big( \xu^{(4)} - \xu^{(3)} \big) 
	+ \au^{(3)} \cdot \big( \xu^{(5)} - \xu^{(4)} \big) 
	+ \au^{(4)} \cdot \big( \xu^{(3)} - \xu^{(2)} \big) 
\nonumber \\ 
& \cong 
\au^{(2)} \cdot \big( \xu^{(5)} - \xu^{(3)} \big) 
	+ \au^{(4)} \cdot \big( \xu^{(3)} - \xu^{(1)} \big)
\nonumber \\ 
& = 
\xu^{(3)} \cdot \big( \au^{(4)} - \au^{(2)} \big) 
	+ \au^{(4)} \cdot \xu^{(5)} - \au^{(4)} \cdot \xu^{(1)} 
\nonumber \\ 
& \cong 
	\au^{(2)} \cdot \big( \xu^{(5)} - \xu^{(1)} \big) \, . 
\label{eq:ZorroMovieComputation}
\end{align}
Hence the domain of~$c_{\textrm{l}}$ is isomorphic to the unit $1_\xu = (\au; \au\cdot(\xu'-\xu))$. 
The argument for~$c_{\textrm{r}}$ is analogous, and we have: 

\begin{proposition}
	Every object in $\bi$ is dualisable. 
\end{proposition}

We recall that in any symmetric monoidal 2-category, we do not have to distinguish between left and right duals for objects because the braiding allows to translate between left and right duality. 
Specifically in $\bi$, we obtain left adjunction 1-morphisms 
\be 
\ev_{\xu}  := \tev_{\xu} \circ b_{\xu^\dual,\xu} \colon \xu^\dual \btimes \xu \lra \varnothing \, , \quad 
\coev_\xu  := b_{\xu^\dual,\xu} \circ \tcoev_\xu \colon \varnothing \lra \xu \btimes \xu^\dual 
\ee 
from the right adjunction 1-morphisms $\tev_\xu, \tcoev_\xu$. 
Thus $\xu^\dual = \xu$ is both the left and right dual of~$\xu$.

\subsubsection{Full dualisability}
\label{subsubsec:FullDualisability} 

Recall that a 1-morphism $M\colon u\lra v$ in a 2-category has a \textsl{right adjoint} if there is $\Md \colon v \lra u$ together with \textsl{adjunction 2-morphisms} $\tev_M \colon M\circ \Md \lra 1_v$ and $\tcoev_M \colon 1_u \lra \Md \circ M$ which satisfy the Zorro moves 
\be 
\begin{tikzpicture}[very thick,scale=0.7,color=blue!50!black, baseline=0cm, xscale=-1]
\draw[line width=0] 
(-1,1.25) node[line width=0pt] (A) {}
(1,-1.25) node[line width=0pt] (A2) {}; 
\draw[directed] (0,0) .. controls +(0,-1) and +(0,-1) .. (-1,0);
\draw[directed] (1,0) .. controls +(0,1) and +(0,1) .. (0,0);
\draw (-1,0) -- (A); 
\draw (1,0) -- (A2); 
\fill (-0.5,-1.2) circle (0pt) node {{\tiny $\tcoev_M$}};
\fill (0.5,1.2) circle (0pt) node {{\tiny $\tev_M$}};
\end{tikzpicture}
=
1_M 
\, , \quad 
\begin{tikzpicture}[very thick,scale=0.7,color=blue!50!black, baseline=0cm, xscale=-1]
\draw[line width=0] 
(1,1.25) node[line width=0pt] (A) {}
(-1,-1.25) node[line width=0pt] (A2) {}; 
\draw[directed] (0,0) .. controls +(0,1) and +(0,1) .. (-1,0);
\draw[directed] (1,0) .. controls +(0,-1) and +(0,-1) .. (0,0);
\draw (-1,0) -- (A2); 
\draw (1,0) -- (A); 
\fill (-0.5,1.2) circle (0pt) node {{\tiny $\tev_M$}};
\fill (0.5,-1.2) circle (0pt) node {{\tiny $\tcoev_M$}};
\end{tikzpicture}
=
1_{\Md} \, . 
\ee 
Similarly, a \textsl{left adjoint} consists of $\dM\colon v\lra u$ together with $\ev_M\colon \dM \circ M \lra 1_u$ and $\coev_M\colon 1_v \lra M \circ \dM$ that satisfy analogous Zorro moves. 
If adjoints for~$M$ exist, then~$\dM$ and/or~$\Md$ are unique up to unique 2-isomorphism (compatible with the adjunction maps). 

An object in a monoidal 2-category is called \textsl{fully dualisable} if it is dualisable, and if its adjunction 1-morphisms themselves have both left and right adjoints, see e.\,g.\ \cite[Sect.\,3]{Pstragowski} for more details. 
Hence to show that every $\xu\in\bi$ is fully dualisable, we have to provide (existence of) left and right adjoints for the adjunction 1-morphisms $\tev_\xu, \tcoev_\xu$ in \eqref{eq:Eval}, \eqref{eq:Coeval} as well as associated adjunction 2-morphisms that satisfy the Zorro moves. 
We claim that those data are given by 
\begin{align}
\tikzzbox{%
	\begin{tikzpicture}[thick,scale=1.5,color=black, baseline=0.1cm]
	\coordinate (p1) at (0,0);
	\coordinate (p2) at (0,0.25);
	\coordinate (g1) at (0,-0.05);
	\coordinate (g2) at (0,0.3);
	\draw[very thick, red!80!black] (p1) .. controls +(0.3,0) and +(0.3,0) ..  (p2); 
	\fill[red!80!black] (0.08,0) circle (0pt) node[left] {{\tiny$\xu^\dual$}};
	\fill[red!80!black] (0.03,0.25) circle (0pt) node[left] {{\tiny$\xu\vphantom{\xu^\dual}$}};
	\end{tikzpicture}
} 
\equiv 
\tikzzbox{%
	\begin{tikzpicture}[thick,scale=1.5,color=black, baseline=0.1cm]
	\coordinate (p1) at (0,0);
	\coordinate (p2) at (0,0.25);
	\coordinate (g1) at (0,-0.05);
	\coordinate (g2) at (0,0.3);
	\draw[very thick, red!80!black] (p1) .. controls +(0.3,0) and +(0.3,0) ..  (p2); 
	\fill[red!80!black] (0.08,0) circle (0pt) node[left] {{\tiny$\xu'$}};
	\fill[red!80!black] (0.03,0.25) circle (0pt) node[left] {{\tiny$\xu\vphantom{\xu'}$}};
	\fill[red!80!black] (0.45,0.1) circle (0pt) node[left] {{\tiny$\au$}};
	\end{tikzpicture}
} 
= 
\coev_{\xu} = 
{}^\dagger \tev_\xu = \tev_\xu^\dagger & := \big( \au; \, \au \cdot (\xu-\xu') \big) \, , 
\label{eq:LeftRightEval}
\\ 
\tikzzbox{%
	\begin{tikzpicture}[thick,scale=1.5,color=black, baseline=0.1cm]
	\coordinate (p1) at (0,0);
	\coordinate (p2) at (0,0.25);
	\coordinate (g1) at (0,-0.05);
	\coordinate (g2) at (0,0.3);
	\draw[very thick, red!80!black] (p1) .. controls +(-0.3,0) and +(-0.3,0) ..  (p2); 
	\fill[red!80!black] (0.257,0) circle (0pt) node[left] {{\tiny$\xu\vphantom{\xu^\dual}$}};
	\fill[red!80!black] (0.4,0.25) circle (0pt) node[left] {{\tiny$\xu^\dual$}};
	\end{tikzpicture}
}  
\equiv 
\tikzzbox{%
	\begin{tikzpicture}[thick,scale=1.5,color=black, baseline=0.1cm]
	\coordinate (p1) at (0,0);
	\coordinate (p2) at (0,0.25);
	\coordinate (g1) at (0,-0.05);
	\coordinate (g2) at (0,0.3);
	\draw[very thick, red!80!black] (p1) .. controls +(-0.3,0) and +(-0.3,0) ..  (p2); 
	\fill[red!80!black] (0.257,0) circle (0pt) node[left] {{\tiny$\xu\vphantom{\xu'}$}};
	\fill[red!80!black] (0.32,0.25) circle (0pt) node[left] {{\tiny$\xu'$}};
	\fill[red!80!black] (-0.15,0.1) circle (0pt) node[left] {{\tiny$\au$}};
	\end{tikzpicture}
} 
= 
\ev_{\xu} = 
{}^\dagger \tcoev_\xu = \tcoev_\xu^\dagger & := \big( \au; \, \au \cdot (\xu'-\xu) \big) 
\label{eq:LeftRightCoeval}
\end{align}
and 
\begin{align}
\tikzzbox{%
	\begin{tikzpicture}[thick,scale=1.0,color=black, baseline=0.9cm]
	\coordinate (p1) at (0,0);
	\coordinate (p2) at (0.5,0.2);
	\coordinate (p3) at (-0.5,0.4);
	\coordinate (p4) at (2.5,0);
	\coordinate (p5) at (1.5,0.2);
	\coordinate (p6) at (2.0,0.4);
	\coordinate (s) at ($(0,1.5)$);
	\coordinate (d) at ($(3.8,0)$);
	\coordinate (h) at ($(0,2.5)$);
	\coordinate (t) at ($(0.1,0)$);
	\coordinate (u) at ($(0,-1.1)$);
	%
	\fill [orange!20!white, opacity=0.8] 
	(p3) .. controls +(0.5,0) and +(0,0.1) ..  (p2)
	-- (p2) .. controls +(0,0.9) and +(0,0.9) ..  (p5)
	-- (p5) .. controls +(0,0.1) and +(-0.3,0) .. (p6)
	-- (p6) -- ($(p6)+(s)$) -- ($(p3)+(s)$);
	%
	\fill[red!80!black] ($(p1)+(t)$) circle (0pt) node[left] {{\small $\yu'$}};
	\fill[red!80!black] ($(p2)-0.5*(t)$) circle (0pt) node[right] {{\small $\au'$}};
	\fill[red!80!black] ($(p3)+(t)$) circle (0pt) node[left] {{\small $\xu'$}};
	\fill[red!80!black] ($(p4)-0.5*(t)$) circle (0pt) node[right] {{\small $\yu$}};
	\fill[red!80!black] ($(p5)+0.5*(t)$) circle (0pt) node[left] {{\small $\au\vphantom{\au'}$}};
	\fill[red!80!black] ($(p6)-0.5*(t)$) circle (0pt) node[right] {{\small $\xu$}};
	\fill[red!80!black] ($(p4)+(s)-0.5*(t)$) circle (0pt) node[right] {{\small $\yu$}};
	\fill[red!80!black] ($(p3)+(s)+(1.25,0.2)$) circle (0pt) node {{\small $\bu$}};
	\fill[red!80!black] ($(p1)+(s)+(1.25,0.15)$) circle (0pt) node {{\small $\cu$}};
	\fill[red!80!black] ($(p1)+(s)+(t)$) circle (0pt) node[left] {{\small $\yu'$}};
	\fill[red!80!black] ($(p3)+(s)+(t)$) circle (0pt) node[left] {{\small $\xu'$}};
	\fill[red!80!black] ($(p6)+(s)-0.5*(t)$) circle (0pt) node[right] {{\small $\xu$}};
	%
	%
	\draw[very thick, red!80!black] (p2) .. controls +(0,0.1) and +(0.5,0) ..  (p3); 
	\draw[very thick, red!80!black] (p6) .. controls +(-0.3,0) and +(0,0.1) ..  (p5); 
	\draw[thin] (p6) -- ($(p6)+(s)$);
	%
	\fill [orange!30!white, opacity=0.8] 
	(p1) .. controls +(0.3,0) and +(0,-0.1) ..  (p2)
	-- (p2) .. controls +(0,0.9) and +(0,0.9) ..  (p5)
	-- (p5) .. controls +(0,-0.1) and +(-0.5,0) ..  (p4) 
	-- (p4) -- ($(p4)+(s)$) -- ($(p1)+(s)$);
	%
	\draw[very thick, red!80!black] (p1) .. controls +(0.3,0) and +(0,-0.1) ..  (p2); 
	\draw[very thick, red!80!black] (p5) .. controls +(0,-0.1) and +(-0.5,0) ..  (p4); 
	\draw[very thick, red!80!black] ($(p4)+(s)$) -- ($(p1)+(s)$);
	\draw[very thick, red!80!black] ($(p6)+(s)$) -- ($(p3)+(s)$);
	%
	\draw (p2) .. controls +(0,0.9) and +(0,0.9) ..  (p5);
	\draw[thin] (p1) -- ($(p1)+(s)$);
	\draw[thin] (p3) -- ($(p3)+(s)$);
	\draw[thin] (p4) -- ($(p4)+(s)$);
	\end{tikzpicture}
}%
& 
= 
\ev_{\tev_\xu} 
	= \tev_{\tcoev_\xu} 
 \label{eq:DualityData1}\\ 
& 
:= \big[ \cu-\au, \, \yu-\yu' \big] \otimes \big[ \bu-\au', \, \xu'-\xu \big] \otimes \big[ \au'-\au, \, \yu'-\xu \big] \, ,
\nonumber
 \\ 
\tikzzbox{%
	\begin{tikzpicture}[thick,scale=1.0,color=black, baseline=-0.9cm]
	\coordinate (p1) at (-2.75,0);
	\coordinate (p2) at (-1,0);
	\fill [orange!20!white, opacity=0.8] 
	(p1) .. controls +(0,-0.5) and +(0,-0.5) ..  (p2)
	-- (p2) .. controls +(0,0.5) and +(0,0.5) ..  (p1)
	;
	\fill [orange!30!white, opacity=0.8] 
	(p1) .. controls +(0,-0.5) and +(0,-0.5) ..  (p2)
	-- (p2) .. controls +(0,-2) and +(0,-2) ..  (p1)
	;
	\draw (p1) .. controls +(0,-2) and +(0,-2) ..  (p2); 
	\draw[very thick, red!80!black] (p1) .. controls +(0,0.5) and +(0,0.5) ..  (p2); 
	\draw[very thick, red!80!black] (p1) .. controls +(0,-0.5) and +(0,-0.5) ..  (p2); 
	%
	\fill[red!80!black] ($(p1)+(0.1,0)$) circle (0pt) node[left] {{\small $\au'$}};
	\fill[red!80!black] ($(p2)+(-0.05,0)$) circle (0pt) node[right] {{\small $\au\vphantom{\au'}$}};
	\fill[red!80!black] (-1.875,+0.38) circle (2.5pt) node[above] {{\small $\xu$}};
	\fill[red!80!black] (-1.875,-0.38) circle (2.5pt) node[below] {{\small $\yu$}};
	\end{tikzpicture}
}%
& 
= \coev_{\tev_\xu} 
	= \tcoev_{\tcoev_\xu} 
	:= \big[ \au'-\au, \, \xu-\yu \big] \, ,
	\label{eq:DualityData2}
 \\ 
\tikzzbox{%
	\begin{tikzpicture}[thick,scale=1.0,color=black, rotate=180, baseline=0.5cm]
	\coordinate (p1) at (-2.75,0);
	\coordinate (p2) at (-1,0);
	%
	\draw[very thick, red!80!black] (p1) .. controls +(0,-0.5) and +(0,-0.5) ..  (p2); 
	\fill[red!80!black] (-1.875,-0.38) circle (2.5pt) node[above] {{\small $\yu$}};
	\fill [orange!20!white, opacity=0.8] 
	(p1) .. controls +(0,-0.5) and +(0,-0.5) ..  (p2)
	-- (p2) .. controls +(0,0.5) and +(0,0.5) ..  (p1)
	;
	\fill [orange!30!white, opacity=0.8] 
	(p1) .. controls +(0,-0.5) and +(0,-0.5) ..  (p2)
	-- (p2) .. controls +(0,-2) and +(0,-2) ..  (p1)
	;
	\coordinate (q1) at (-1.875,0.38);
	\coordinate (q2) at (-2.67,-0.7);
	\draw (p1) .. controls +(0,-2) and +(0,-2) ..  (p2); 
	%
	\fill[red!80!black] ($(p1)+(0.1,0)$) circle (0pt) node[right] {{\small $\au'$}};
	\fill[red!80!black] ($(p2)+(-0.05,0)$) circle (0pt) node[left] {{\small $\au\vphantom{\au'}$}};
	\fill[red!80!black] (-1.875,+0.38) circle (2.5pt) node[below] {{\small $\xu$}};
	%
	\draw[very thick, red!80!black] (p1) .. controls +(0,0.5) and +(0,0.5) ..  (p2); 
	\end{tikzpicture}
}%
& 
= \tev_{\tev_\xu} 
	= \ev_{\tcoev_\xu} 
	:= \big[ \yu-\xu, \, \au'-\au \big] \, ,
	\label{eq:DualityData3}
 \\ 
\tikzzbox{%
	\begin{tikzpicture}[thick,scale=1.0,color=black, baseline=-0.9cm, yscale=-1]
	\coordinate (p1) at (0,0);
	\coordinate (p2) at (0.5,0.2);
	\coordinate (p3) at (-0.5,0.4);
	\coordinate (p4) at (2.5,0);
	\coordinate (p5) at (1.5,0.2);
	\coordinate (p6) at (2.0,0.4);
	\coordinate (s) at ($(0,1.5)$);
	\coordinate (d) at ($(3.8,0)$);
	\coordinate (h) at ($(0,2.5)$);
	\coordinate (t) at ($(0.1,0)$);
	\coordinate (u) at ($(0,-1.1)$);
	%
	\fill[red!80!black] ($(p1)+(t)$) circle (0pt) node[left] {{\small $\xu'$}};
	\fill[red!80!black] ($(p2)-0.5*(t)$) circle (0pt) node[right] {{\small $\au'$}};
	\fill[red!80!black] ($(p3)+(t)$) circle (0pt) node[left] {{\small $\yu'$}};
	\fill[red!80!black] ($(p4)-0.5*(t)$) circle (0pt) node[right] {{\small $\xu$}};
	\fill[red!80!black] ($(p5)+0.5*(t)$) circle (0pt) node[left] {{\small $\au\vphantom{\au'}$}};
	\fill[red!80!black] ($(p4)+(s)-0.5*(t)$) circle (0pt) node[right] {{\small $\xu$}};
	\fill[red!80!black] ($(p3)+(s)+(1.25,0.2)$) circle (0pt) node {{\small $\cu$}};
	\fill[red!80!black] ($(p1)+(s)+(1.25,0.15)$) circle (0pt) node {{\small $\bu$}};
	\fill[red!80!black] ($(p1)+(s)+(t)$) circle (0pt) node[left] {{\small $\xu'$}};
	\fill[red!80!black] ($(p3)+(s)+(t)$) circle (0pt) node[left] {{\small $\yu'$}};
	\fill[red!80!black] ($(p6)+(s)-0.5*(t)$) circle (0pt) node[right] {{\small $\yu$}};
	%
	\fill [orange!20!white, opacity=0.8] 
	(p1) .. controls +(0.3,0) and +(0,-0.1) ..  (p2)
	-- (p2) .. controls +(0,0.9) and +(0,0.9) ..  (p5)
	-- (p5) .. controls +(0,-0.1) and +(-0.5,0) ..  (p4) 
	-- (p4) -- ($(p4)+(s)$) -- ($(p1)+(s)$);
	%
	\draw[very thick, red!80!black] ($(p4)+(s)$) -- ($(p1)+(s)$);
	\draw[thin] (p1) -- ($(p1)+(s)$);
	%
	\fill [orange!30!white, opacity=0.8] 
	(p3) .. controls +(0.5,0) and +(0,0.1) ..  (p2)
	-- (p2) .. controls +(0,0.9) and +(0,0.9) ..  (p5)
	-- (p5) .. controls +(0,0.1) and +(-0.3,0) .. (p6)
	-- (p6) -- ($(p6)+(s)$) -- ($(p3)+(s)$);
	%
	\draw[very thick, red!80!black] (p1) .. controls +(0.3,0) and +(0,-0.1) ..  (p2); 
	\draw[very thick, red!80!black] (p2) .. controls +(0,0.1) and +(0.5,0) ..  (p3); 
	\draw[very thick, red!80!black] (p6) .. controls +(-0.3,0) and +(0,0.1) ..  (p5); 
	\draw[very thick, red!80!black] (p1) .. controls +(0.3,0) and +(0,-0.1) ..  (p2); 
	\draw[very thick, red!80!black] (p5) .. controls +(0,-0.1) and +(-0.5,0) ..  (p4); 
	\draw[very thick, red!80!black] ($(p6)+(s)$) -- ($(p3)+(s)$);
	%
	\draw[thin] (p6) -- ($(p6)+(s)$);
	\draw (p2) .. controls +(0,0.9) and +(0,0.9) ..  (p5);
	\draw[thin] (p3) -- ($(p3)+(s)$);
	\draw[thin] (p4) -- ($(p4)+(s)$);
	\fill[red!80!black] ($(p6)-0.5*(t)$) circle (0pt) node[right] {{\small $\yu$}};
	\end{tikzpicture}
}%
& 
= \tcoev_{\tev_\xu} 
	= \coev_{\tcoev_\xu}
\label{eq:DualityData} \\ 
& := \big[ \cu-\au, \, \yu'-\yu \big] \otimes \big[ \bu-\au, \, \xu-\xu' \big] \otimes \big[ \yu'-\xu' ,\,\au-\au' \big] \, ,
\nonumber
\end{align}
where the labels $\au, \xu, \yu, \dots$ in the diagrams either indicate objects in $\bi$, or the ``extra variables''~$\au$ in 1-morphisms $(\au;W)$. 

\begin{theorem}
	\label{thm:FullyDualisable}
	Every object in $\bi$ is fully dualisable, as witnessed by the duality data in \eqref{eq:DualObject}--\eqref{eq:Coeval} and \eqref{eq:LeftRightEval}--\eqref{eq:DualityData}. 
\end{theorem}

\begin{proof}
	We have to verify that the Zorro moves for the adjunction 2-morphisms \eqref{eq:DualityData1}--\eqref{eq:DualityData} hold. 
	We will do this in detail for the adjunction ${}^\dagger \tev_\xu \dashv \tev_\xu$. 
	
	One of the two Zorro moves for ${}^\dagger \tev_\xu \dashv \tev_\xu$ states that 
	\be 
	\label{eq:3dZorro}
	\tikzzbox{%
		\begin{tikzpicture}[thick,scale=1.0,color=black, baseline=0.9cm]
		\coordinate (p1) at (0,0);
		\coordinate (p2) at (0.5,0.2);
		\coordinate (p3) at (-0.5,0.4);
		\coordinate (p4) at (2.5,0);
		\coordinate (p5) at (1.5,0.2);
		\coordinate (p6) at (2.0,0.4);
		\coordinate (s) at ($(0,1.5)$);
		\coordinate (d) at ($(3.8,0)$);
		\coordinate (h) at ($(0,2.5)$);
		\coordinate (t) at ($(0.1,0)$);
		\coordinate (u) at ($(0,-1.1)$);
		\coordinate (u2) at ($(-3.8,-2.5)$);
		%
		\fill [orange!20!white, opacity=0.8] 
		(p3) .. controls +(0.5,0) and +(0,0.1) ..  (p2)
		-- (p2) .. controls +(0,0.9) and +(0,0.9) ..  (p5)
		-- (p5) .. controls +(0,0.1) and +(-0.3,0) .. (p6)
		-- (p6) -- ($(p6)+(s)$) -- ($(p3)+(s)$);
		%
		\fill[red!80!black] ($(p1)+(t)$) circle (0pt) node[left] {{\small $\yu'$}};
		\fill[red!80!black] ($(p2)-0.5*(t)$) circle (0pt) node[right] {{\small $\widetilde\au$}};
		\fill[red!80!black] ($(p3)+(t)$) circle (0pt) node[left] {{\small $\xu'$}};
		\fill[red!80!black] ($(p4)-0.5*(t)$) circle (0pt) node[right] {{\small $\yu$}};
		\fill[red!80!black] ($(p5)+0.5*(t)$) circle (0pt) node[left] {{\small $\au\vphantom{\widetilde\au}$}};
		\fill[red!80!black] ($(p6)-0.5*(t)$) circle (0pt) node[right] {{\small $\xu$}};
		\fill[red!80!black] ($(p4)+(s)-0.5*(t)$) circle (0pt) node[right] {{\small $\yu$}};
		\fill[red!80!black] ($(p3)+(s)+(1.25,0.2)$) circle (0pt) node {{\small $\bu$}};
		\fill[red!80!black] ($(p1)+(s)+(1.25,0.15)$) circle (0pt) node {{\small $\cu$}};
		\fill[red!80!black] ($(p1)+(s)+(t)$) circle (0pt) node[left] {{\small $\yu'$}};
		\fill[red!80!black] ($(p3)+(s)+(t)$) circle (0pt) node[left] {{\small $\xu'$}};
		\fill[red!80!black] ($(p6)+(s)-0.5*(t)$) circle (0pt) node[right] {{\small $\xu$}};
		%
		%
		\draw[very thick, red!80!black] (p2) .. controls +(0,0.1) and +(0.5,0) ..  (p3); 
		\draw[very thick, red!80!black] (p6) .. controls +(-0.3,0) and +(0,0.1) ..  (p5); 
		\draw[thin] (p6) -- ($(p6)+(s)$);
		%
		\fill [orange!30!white, opacity=0.8] 
		(p1) .. controls +(0.3,0) and +(0,-0.1) ..  (p2)
		-- (p2) .. controls +(0,0.9) and +(0,0.9) ..  (p5)
		-- (p5) .. controls +(0,-0.1) and +(-0.5,0) ..  (p4) 
		-- (p4) -- ($(p4)+(s)$) -- ($(p1)+(s)$);
		%
		\draw[very thick, red!80!black] (p1) .. controls +(0.3,0) and +(0,-0.1) ..  (p2); 
		\draw[very thick, red!80!black] (p5) .. controls +(0,-0.1) and +(-0.5,0) ..  (p4); 
		\draw[very thick, red!80!black] ($(p4)+(s)$) -- ($(p1)+(s)$);
		\draw[very thick, red!80!black] ($(p6)+(s)$) -- ($(p3)+(s)$);
		%
		\draw (p2) .. controls +(0,0.9) and +(0,0.9) ..  (p5);
		\draw[thin] (p1) -- ($(p1)+(s)$);
		\draw[thin] (p3) -- ($(p3)+(s)$);
		\draw[thin] (p4) -- ($(p4)+(s)$);
		%
		%
		%
		\fill [orange!20!white, opacity=0.8] 
		($(p2)+(d)$) .. controls +(0,0.1) and +(0.5,0) ..  ($(p3)+(d)$)
		-- ($(p3)+(d)$) -- ($(p3)+(d)+(s)$) 
		-- ($(p3)+(d)+(s)$) .. controls +(0.5,0) and +(0,0.1) ..($(p2)+(d)+(s)$);
		%
		\draw[very thick, red!80!black] ($(p2)+(d)$) .. controls +(0,0.1) and +(0.5,0) ..  ($(p3)+(d)$); 
		\draw[very thick, red!80!black] ($(p2)+(d)+(s)$) .. controls +(0,0.1) and +(0.5,0) ..  ($(p3)+(d)+(s)$); 
		%
		\fill [orange!30!white, opacity=0.8] 
		($(p1)+(d)$) .. controls +(0.3,0) and +(0,-0.1) .. ($(p2)+(d)$)
		-- ($(p2)+(d)$) -- ($(p2)+(d)+(s)$)
		-- ($(p2)+(d)+(s)$) .. controls +(0,-0.1) and +(0.3,0) .. ($(p1)+(d)+(s)$); 
		%
		\draw[thin] ($(p1)+(d)$) -- ($(p1)+(d)+(s)$);
		\draw[thin] ($(p2)+(d)$) -- ($(p2)+(d)+(s)$);
		\draw[thin] ($(p3)+(d)$) -- ($(p3)+(d)+(s)$);
		%
		\draw[very thick, red!80!black] ($(p1)+(d)$) .. controls +(0.3,0) and +(0,-0.1) ..  ($(p2)+(d)$); 
		\draw[very thick, red!80!black] ($(p1)+(d)+(s)$) .. controls +(0.3,0) and +(0,-0.1) ..  ($(p2)+(d)+(s)$); 
		%
		\fill[red!80!black] ($(p2)+(d)-0.5*(t)$) circle (0pt) node[right] {{\small $\au''$}};
		\fill[red!80!black] ($(p2)+(d)+(s)-0.5*(t)$) circle (0pt) node[right] {{\small $\au'''$}};
		\fill[red!80!black] ($(p1)+(s)+(d)+(t)$) circle (0pt) node[left] {{\small $\yu$}};
		\fill[red!80!black] ($(p1)+(d)+(t)$) circle (0pt) node[left] {{\small $\yu$}};
		\fill[red!80!black] ($(p3)+(s)+(d)+(t)$) circle (0pt) node[left] {{\small $\xu$}};
		\fill[red!80!black] ($(p3)+(d)+(t)$) circle (0pt) node[left] {{\small $\xu$}};
		%
		%
		%
		%
		\fill [orange!20!white, opacity=0.8] 
		($(p2)+(d)+(u2)$) .. controls +(0,0.1) and +(0.5,0) ..  ($(p3)+(d)+(u2)$)
		-- ($(p3)+(d)+(u2)$) -- ($(p3)+(d)+(s)+(u2)$) 
		-- ($(p3)+(d)+(s)+(u2)$) .. controls +(0.5,0) and +(0,0.1) ..($(p2)+(d)+(s)+(u2)$);
		%
		\draw[very thick, red!80!black] ($(p2)+(d)+(u2)$) .. controls +(0,0.1) and +(0.5,0) ..  ($(p3)+(d)+(u2)$); 
		\draw[very thick, red!80!black] ($(p2)+(d)+(s)+(u2)$) .. controls +(0,0.1) and +(0.5,0) ..  ($(p3)+(d)+(s)+(u2)$); 
		%
		\fill [orange!30!white, opacity=0.8] 
		($(p1)+(d)+(u2)$) .. controls +(0.3,0) and +(0,-0.1) .. ($(p2)+(d)+(u2)$)
		-- ($(p2)+(d)+(u2)$) -- ($(p2)+(d)+(s)+(u2)$)
		-- ($(p2)+(d)+(s)+(u2)$) .. controls +(0,-0.1) and +(0.3,0) .. ($(p1)+(d)+(s)+(u2)$); 
		%
		\draw[thin] ($(p1)+(d)+(u2)$) -- ($(p1)+(d)+(s)+(u2)$);
		\draw[thin] ($(p2)+(d)+(u2)$) -- ($(p2)+(d)+(s)+(u2)$);
		\draw[thin] ($(p3)+(d)+(u2)$) -- ($(p3)+(d)+(s)+(u2)$);
		%
		\draw[very thick, red!80!black] ($(p1)+(d)+(u2)$) .. controls +(0.3,0) and +(0,-0.1) ..  ($(p2)+(d)+(u2)$); 
		\draw[very thick, red!80!black] ($(p1)+(d)+(s)+(u2)$) .. controls +(0.3,0) and +(0,-0.1) ..  ($(p2)+(d)+(s)+(u2)$); 
		%
		\fill[red!80!black] ($(p2)+(d)-0.5*(t)+(u2)$) circle (0pt) node[right] {{\small $\au'$}};
		\fill[red!80!black] ($(p2)+(d)+(s)-0.5*(t)+(u2)$) circle (0pt) node[right] {{\small $\widetilde\au$}};
		\fill[red!80!black] ($(p1)+(s)+(d)+(t)+(u2)$) circle (0pt) node[left] {{\small $\yu'$}};
		\fill[red!80!black] ($(p1)+(d)+(t)+(u2)$) circle (0pt) node[left] {{\small $\yu'$}};
		\fill[red!80!black] ($(p3)+(s)+(d)+(t)+(u2)$) circle (0pt) node[left] {{\small $\xu'$}};
		\fill[red!80!black] ($(p3)+(d)+(t)+(u2)$) circle (0pt) node[left] {{\small $\xu'$}};
		%
		%
		\fill [orange!20!white, opacity=0.8] 
		($(p2)+(d)+(h)$) .. controls +(0,0.1) and +(0.5,0) ..  ($(p3)+(d)+(h)$) 
		-- ($(p3)+(h)$) -- ($(p3)+(s)+(h)$) -- ($(p3)+(d)+(s)+(h)$)
		-- ($(p3)+(d)+(s)+(h)$) .. controls +(0.5,0) and +(0,0.1) ..($(p2)+(d)+(s)+(h)$);
		%
		\fill[red!80!black] ($(p3)+(h)+(d)$) circle (2.5pt) node[below] {{\small $\xu$}};
		\fill[red!80!black] ($(p3)+(s)+(1.25,1.2)$) circle (0pt) node {{\small $\bu$}};
		\fill[red!80!black] ($(p1)+(s)+(1.25,1.15)$) circle (0pt) node {{\small $\cu$}};
		%
		\draw[very thick, red!80!black] ($(p2)+(d)+(h)$) .. controls +(0,0.1) and +(0.5,0) ..  ($(p3)+(d)+(h)$) 
		-- ($(p3)+(h)$);
		\draw[very thick, red!80!black] ($(p2)+(d)+(h)+(s)$) .. controls +(0,0.1) and +(0.5,0) ..  ($(p3)+(d)+(h)+(s)$) 
		-- ($(p3)+(h)+(s)$);
		%
		\fill [orange!30!white, opacity=0.8] 
		($(p1)+(h)$) -- ($(p1)+(d)+(h)$) .. controls +(0.3,0) and +(0,-0.1) .. ($(p2)+(d)+(h)$)
		-- ($(p2)+(d)+(h)$) -- ($(p2)+(d)+(s)+(h)$)
		-- ($(p2)+(d)+(s)+(h)$) .. controls +(0,-0.1) and +(0.3,0) .. ($(p1)+(d)+(s)+(h)$)
		-- ($(p1)+(s)+(h)$); 
		%
		\draw[thin] ($(p1)+(h)$) -- ($(p1)+(h)+(s)$);
		\draw[thin] ($(p2)+(d)+(h)$) -- ($(p2)+(d)+(s)+(h)$);
		\draw[thin] ($(p3)+(h)$) -- ($(p3)+(h)+(s)$);
		%
		\draw[very thick, red!80!black] ($(p1)+(h)$) -- ($(p1)+(d)+(h)$) .. controls +(0.3,0) and +(0,-0.1) .. ($(p2)+(d)+(h)$); 
		\draw[very thick, red!80!black] ($(p1)+(h)+(s)$) -- ($(p1)+(d)+(h)+(s)$) .. controls +(0.3,0) and +(0,-0.1) .. ($(p2)+(d)+(h)+(s)$); 
		%
		\fill[red!80!black] ($(p1)+(h)+(t)$) circle (0pt) node[left] {{\small $\yu'$}};
		\fill[red!80!black] ($(p1)+(h)+(s)+(t)$) circle (0pt) node[left] {{\small $\yu'$}};
		\fill[red!80!black] ($(p3)+(h)+(t)$) circle (0pt) node[left] {{\small $\xu'$}};
		\fill[red!80!black] ($(p3)+(h)+(s)+(t)$) circle (0pt) node[left] {{\small $\xu'$}};
		\fill[red!80!black] ($(p2)+(d)+(h)-0.5*(t)$) circle (0pt) node[right] {{\small $\au'''$}};
		\fill[red!80!black] ($(p2)+(d)+(s)+(h)-0.5*(t)$) circle (0pt) node[right] {{\small $\du$}};
		\fill[red!80!black] ($(p1)+(h)+(d)$) circle (2.5pt) node[below] {{\small $\yu$}};
		%
		%
		%
		%
		\fill [orange!20!white, opacity=0.8] 
		($(p5)+(u)$) .. controls +(0,-2) and +(0,-2) .. ($(p2)+(d)+(u)$) 
		-- ($(p2)+(d)+(u)$) .. controls +(0,0.4) and +(0,0.4) .. ($(p5)+(u)$);
		%
		\fill [orange!30!white, opacity=0.8] 
		($(p5)+(u)$) .. controls +(0,-2) and +(0,-2) .. ($(p2)+(d)+(u)$) 
		-- ($(p2)+(d)+(u)$) .. controls +(0,-0.4) and +(0,-0.4) .. ($(p5)+(u)$);
		%
		\draw ($(p5)+(u)$) .. controls +(0,-2) and +(0,-2) ..  ($(p2)+(d)+(u)$);
		\draw[very thick, red!80!black] ($(p5)+(u)$) .. controls +(0,0.4) and +(0,0.4) ..  ($(p2)+(d)+(u)$);
		\draw[very thick, red!80!black] ($(p5)+(u)$) .. controls +(0,-0.4) and +(0,-0.4) ..  ($(p2)+(d)+(u)$);
		\fill[red!80!black] ($(p5)+(u)+(1.4,0.3)$) circle (2.5pt) node[above] {{\small $\xu$}};
		\fill[red!80!black] ($(p5)+(u)+(1.4,-0.3)$) circle (2.5pt) node[below] {{\small $\yu$}};
		%
		\fill[red!80!black] ($(p5)+(u)+0.5*(t)$) circle (0pt) node[left] {{\small $\au\vphantom{\au'}$}};
		\fill[red!80!black] ($(p2)+(d)+(u)-0.5*(t)$) circle (0pt) node[right] {{\small $\au''$}};
		\end{tikzpicture}
	}%
	= 
	\tikzzbox{%
		\begin{tikzpicture}[thick,scale=1.0,color=black, baseline=0.9cm]
		\coordinate (p1) at (0,0);
		\coordinate (p2) at (0.5,0.2);
		\coordinate (p3) at (-0.5,0.4);
		\coordinate (p4) at (2.5,0);
		\coordinate (p5) at (1.5,0.2);
		\coordinate (p6) at (2.0,0.4);
		\coordinate (s) at ($(0,1.5)$);
		\coordinate (d) at ($(1.5,0)$);
		\coordinate (h) at ($(0,2.5)$);
		\coordinate (t) at ($(0.1,0)$);
		\coordinate (u) at ($(0,-1.1)$);
		\coordinate (m) at ($(0,-5)$);
		%
		\fill [orange!20!white, opacity=0.8] 
		($(p2)+(d)+(h)+(m)$) .. controls +(0,0.1) and +(0.5,0) ..  ($(p3)+(d)+(h)+(m)$) 
		-- ($(p3)+(h)+(m)$) -- ($(p3)+(s)+(h)$) -- ($(p3)+(d)+(s)+(h)$)
		-- ($(p3)+(d)+(s)+(h)$) .. controls +(0.5,0) and +(0,0.1) ..($(p2)+(d)+(s)+(h)$);
		%
		\draw[very thick, red!80!black] ($(p2)+(d)+(h)+(m)$) .. controls +(0,0.1) and +(0.5,0) ..  ($(p3)+(d)+(h)+(m)$) 
		-- ($(p3)+(h)+(m)$);
		\draw[very thick, red!80!black] ($(p2)+(d)+(h)+(s)$) .. controls +(0,0.1) and +(0.5,0) ..  ($(p3)+(d)+(h)+(s)$) 
		-- ($(p3)+(h)+(s)$);
		%
		\fill [orange!30!white, opacity=0.8] 
		($(p1)+(h)+(m)$) -- ($(p1)+(d)+(h)+(m)$) .. controls +(0.3,0) and +(0,-0.1) .. ($(p2)+(d)+(h)+(m)$)
		-- ($(p2)+(d)+(h)+(m)$) -- ($(p2)+(d)+(s)+(h)$)
		-- ($(p2)+(d)+(s)+(h)$) .. controls +(0,-0.1) and +(0.3,0) .. ($(p1)+(d)+(s)+(h)$)
		-- ($(p1)+(s)+(h)$); 
		%
		\draw[thin] ($(p1)+(h)+(m)$) -- ($(p1)+(h)+(s)$);
		\draw[thin] ($(p2)+(d)+(h)+(m)$) -- ($(p2)+(d)+(s)+(h)$);
		\draw[thin] ($(p3)+(h)+(m)$) -- ($(p3)+(h)+(s)$);
		%
		\draw[very thick, red!80!black] ($(p1)+(h)+(m)$) -- ($(p1)+(d)+(h)+(m)$) .. controls +(0.3,0) and +(0,-0.1) .. ($(p2)+(d)+(h)+(m)$); 
		\draw[very thick, red!80!black] ($(p1)+(h)+(s)$) -- ($(p1)+(d)+(h)+(s)$) .. controls +(0.3,0) and +(0,-0.1) .. ($(p2)+(d)+(h)+(s)$); 
		%
		\fill[red!80!black] ($(p1)+(h)+(t)+(m)$) circle (0pt) node[left] {{\small $\yu'$}};
		\fill[red!80!black] ($(p1)+(h)+(s)+(t)$) circle (0pt) node[left] {{\small $\yu'$}};
		\fill[red!80!black] ($(p3)+(h)+(t)+(m)$) circle (0pt) node[left] {{\small $\xu'$}};
		\fill[red!80!black] ($(p3)+(h)+(s)+(t)$) circle (0pt) node[left] {{\small $\xu'$}};
		\fill[red!80!black] ($(p2)+(d)+(h)-0.5*(t)+(m)$) circle (0pt) node[right] {{\small $\au'$}};
		\fill[red!80!black] ($(p2)+(d)+(s)+(h)-0.5*(t)$) circle (0pt) node[right] {{\small $\du$}};
		\end{tikzpicture}
	}%
	\, . 
	\ee 
	The right-hand side is $1_{{}^\dagger \tev_\xu}$, represented by the matrix factorisation $[\du-\au',\xu'-\yu']$. 
	Our task is to show that this is also true of the left-hand side. 
	By definition, three of the four lower subdiagrams are 
	\begin{align}
	\tikzzbox{%
		\begin{tikzpicture}[thick,scale=1.0,color=black, baseline=-1.5cm]
		\coordinate (p1) at (0,0);
		\coordinate (p2) at (0.5,0.2);
		\coordinate (p3) at (-0.5,0.4);
		\coordinate (p4) at (2.5,0);
		\coordinate (p5) at (1.5,0.2);
		\coordinate (p6) at (2.0,0.4);
		\coordinate (s) at ($(0,1.5)$);
		\coordinate (d) at ($(3.8,0)$);
		\coordinate (h) at ($(0,2.5)$);
		\coordinate (t) at ($(0.1,0)$);
		\coordinate (u) at ($(0,-1.1)$);
		%
		\fill [orange!20!white, opacity=0.8] 
		($(p5)+(u)$) .. controls +(0,-2) and +(0,-2) .. ($(p2)+(d)+(u)$) 
		-- ($(p2)+(d)+(u)$) .. controls +(0,0.4) and +(0,0.4) .. ($(p5)+(u)$);
		%
		\fill [orange!30!white, opacity=0.8] 
		($(p5)+(u)$) .. controls +(0,-2) and +(0,-2) .. ($(p2)+(d)+(u)$) 
		-- ($(p2)+(d)+(u)$) .. controls +(0,-0.4) and +(0,-0.4) .. ($(p5)+(u)$);
		%
		\draw ($(p5)+(u)$) .. controls +(0,-2) and +(0,-2) ..  ($(p2)+(d)+(u)$);
		\draw[very thick, red!80!black] ($(p5)+(u)$) .. controls +(0,0.4) and +(0,0.4) ..  ($(p2)+(d)+(u)$);
		\draw[very thick, red!80!black] ($(p5)+(u)$) .. controls +(0,-0.4) and +(0,-0.4) ..  ($(p2)+(d)+(u)$);
		\fill[red!80!black] ($(p5)+(u)+(1.4,0.3)$) circle (2.5pt) node[above] {{\small $\xu$}};
		\fill[red!80!black] ($(p5)+(u)+(1.4,-0.3)$) circle (2.5pt) node[below] {{\small $\yu$}};
		%
		\fill[red!80!black] ($(p5)+(u)+0.5*(t)$) circle (0pt) node[left] {{\small $\au\vphantom{\au'}$}};
		\fill[red!80!black] ($(p2)+(d)+(u)-0.5*(t)$) circle (0pt) node[right] {{\small $\au''$}};
		\end{tikzpicture}
	}%
	& = \big[ \au-\au'', \, \xu-\yu \big] 
	\label{eq:sub1} 
	\\ 
	\tikzzbox{%
		\begin{tikzpicture}[thick,scale=1.0,color=black, baseline=0.9cm]
		\coordinate (p1) at (0,0);
		\coordinate (p2) at (0.5,0.2);
		\coordinate (p3) at (-0.5,0.4);
		\coordinate (p4) at (2.5,0);
		\coordinate (p5) at (1.5,0.2);
		\coordinate (p6) at (2.0,0.4);
		\coordinate (s) at ($(0,1.5)$);
		\coordinate (d) at ($(3,0)$);
		%
		\fill [orange!20!white, opacity=0.8] 
		($(p2)+(d)$) .. controls +(0,0.1) and +(0.5,0) ..  ($(p3)+(d)$)
		-- ($(p3)+(d)$) -- ($(p3)+(d)+(s)$) 
		-- ($(p3)+(d)+(s)$) .. controls +(0.5,0) and +(0,0.1) ..($(p2)+(d)+(s)$);
		%
		\draw[very thick, red!80!black] ($(p2)+(d)$) .. controls +(0,0.1) and +(0.5,0) ..  ($(p3)+(d)$); 
		\draw[very thick, red!80!black] ($(p2)+(d)+(s)$) .. controls +(0,0.1) and +(0.5,0) ..  ($(p3)+(d)+(s)$); 
		%
		\fill [orange!30!white, opacity=0.8] 
		($(p1)+(d)$) .. controls +(0.3,0) and +(0,-0.1) .. ($(p2)+(d)$)
		-- ($(p2)+(d)$) -- ($(p2)+(d)+(s)$)
		-- ($(p2)+(d)+(s)$) .. controls +(0,-0.1) and +(0.3,0) .. ($(p1)+(d)+(s)$); 
		%
		\draw[thin] ($(p1)+(d)$) -- ($(p1)+(d)+(s)$);
		\draw[thin] ($(p2)+(d)$) -- ($(p2)+(d)+(s)$);
		\draw[thin] ($(p3)+(d)$) -- ($(p3)+(d)+(s)$);
		%
		\draw[very thick, red!80!black] ($(p1)+(d)$) .. controls +(0.3,0) and +(0,-0.1) ..  ($(p2)+(d)$); 
		\draw[very thick, red!80!black] ($(p1)+(d)+(s)$) .. controls +(0.3,0) and +(0,-0.1) ..  ($(p2)+(d)+(s)$); 
		%
		\fill[red!80!black] ($(p4)+(s)$) circle (0pt) node[right] {{\small $\yu$}};
		\fill[red!80!black] ($(p6)+(s)$) circle (0pt) node[right] {{\small $\xu$}};
		\fill[red!80!black] (p4) circle (0pt) node[right] {{\small $\yu$}};
		\fill[red!80!black] (p6) circle (0pt) node[right] {{\small $\xu$}};
		\fill[red!80!black] ($(p2)+(d)$) circle (0pt) node[right] {{\small $\au''$}};
		\fill[red!80!black] ($(p2)+(d)+(s)$) circle (0pt) node[right] {{\small $\au'''$}};
		\end{tikzpicture}
	}%
	& = \big[ \au'''-\au'', \, \yu-\xu \big] 
	\\ 
	\tikzzbox{%
		\begin{tikzpicture}[thick,scale=1.0,color=black, baseline=0.9cm]
		\coordinate (p1) at (0,0);
		\coordinate (p2) at (0.5,0.2);
		\coordinate (p3) at (-0.5,0.4);
		\coordinate (p4) at (2.5,0);
		\coordinate (p5) at (1.5,0.2);
		\coordinate (p6) at (2.0,0.4);
		\coordinate (s) at ($(0,1.5)$);
		%
		\fill [orange!20!white, opacity=0.8] 
		(p3) .. controls +(0.5,0) and +(0,0.1) ..  (p2)
		-- (p2) .. controls +(0,0.9) and +(0,0.9) ..  (p5)
		-- (p5) .. controls +(0,0.1) and +(-0.3,0) .. (p6)
		-- (p6) -- ($(p6)+(s)$) -- ($(p3)+(s)$);
		%
		\fill[red!80!black] (p1) circle (0pt) node[left] {{\small $\yu'$}};
		\fill[red!80!black] (p2) circle (0pt) node[right] {{\small $\widetilde{\au}$}};
		\fill[red!80!black] (p3) circle (0pt) node[left] {{\small $\xu'$}};
		\fill[red!80!black] (p4) circle (0pt) node[right] {{\small $\yu$}};
		\fill[red!80!black] (p5) circle (0pt) node[left] {{\small $\au\vphantom{\widetilde{\au}}$}};
		\fill[red!80!black] (p6) circle (0pt) node[right] {{\small $\xu$}};
		\fill[red!80!black] ($(p3)+(s)+(1.25,0.2)$) circle (0pt) node {{\small $\bu$}};
		\fill[red!80!black] ($(p1)+(s)+(1.25,0.15)$) circle (0pt) node {{\small $\cu$}};
		\fill[red!80!black] ($(p1)+(s)$) circle (0pt) node[left] {{\small $\yu'$}};
		\fill[red!80!black] ($(p3)+(s)$) circle (0pt) node[left] {{\small $\xu'$}};
		\fill[red!80!black] ($(p4)+(s)$) circle (0pt) node[right] {{\small $\yu$}};
		\fill[red!80!black] ($(p6)+(s)$) circle (0pt) node[right] {{\small $\xu$}};
		%
		\draw[very thick, red!80!black] (p2) .. controls +(0,0.1) and +(0.5,0) ..  (p3); 
		\draw[very thick, red!80!black] (p6) .. controls +(-0.3,0) and +(0,0.1) ..  (p5); 
		\draw[thin] (p6) -- ($(p6)+(s)$);
		%
		\fill [orange!30!white, opacity=0.8] 
		(p1) .. controls +(0.3,0) and +(0,-0.1) ..  (p2)
		-- (p2) .. controls +(0,0.9) and +(0,0.9) ..  (p5)
		-- (p5) .. controls +(0,-0.1) and +(-0.5,0) ..  (p4) 
		-- (p4) -- ($(p4)+(s)$) -- ($(p1)+(s)$);
		%
		\draw[very thick, red!80!black] (p1) .. controls +(0.3,0) and +(0,-0.1) ..  (p2); 
		\draw[very thick, red!80!black] (p5) .. controls +(0,-0.1) and +(-0.5,0) ..  (p4); 
		\draw[very thick, red!80!black] ($(p4)+(s)$) -- ($(p1)+(s)$);
		\draw[very thick, red!80!black] ($(p6)+(s)$) -- ($(p3)+(s)$);
		%
		\draw (p2) .. controls +(0,0.9) and +(0,0.9) ..  (p5);
		\draw[thin] (p1) -- ($(p1)+(s)$);
		\draw[thin] (p3) -- ($(p3)+(s)$);
		\draw[thin] (p4) -- ($(p4)+(s)$);
		\end{tikzpicture}
	}%
	& = \big[ \cu-\au, \, \yu-\yu' \big] \otimes \big[ \bu-\widetilde{\au}, \, \xu'-\xu \big]  \otimes \big[ \widetilde{\au}-\au, \, \yu'-\xu \big] \, , 
	\label{eq:sub3}
	\end{align}
	while for the top subdiagram we have 
	\be
	\label{eq:sub4}
	\tikzzbox{%
		\begin{tikzpicture}[thick,scale=1.0,color=black, baseline=3.3cm]
		\coordinate (p1) at (0,0);
		\coordinate (p2) at (0.5,0.2);
		\coordinate (p3) at (-0.5,0.4);
		\coordinate (p4) at (2.5,0);
		\coordinate (p5) at (1.5,0.2);
		\coordinate (p6) at (2.0,0.4);
		\coordinate (s) at ($(0,1.5)$);
		\coordinate (d) at ($(2,0)$);
		\coordinate (h) at ($(0,2.5)$);
		\coordinate (t) at ($(0.1,0)$);
		\coordinate (u) at ($(0,-1.1)$);
		%
		%
		\fill [orange!20!white, opacity=0.8] 
		($(p2)+(d)+(h)$) .. controls +(0,0.1) and +(0.5,0) ..  ($(p3)+(d)+(h)$) 
		-- ($(p3)+(h)$) -- ($(p3)+(s)+(h)$) -- ($(p3)+(d)+(s)+(h)$)
		-- ($(p3)+(d)+(s)+(h)$) .. controls +(0.5,0) and +(0,0.1) ..($(p2)+(d)+(s)+(h)$);
		%
		\fill[red!80!black] ($(p3)+(h)+(d)$) circle (2.5pt) node[below] {{\small $\xu$}};
		\fill[red!80!black] ($(p3)+(s)+(0.8,1.2)$) circle (0pt) node {{\small $\bu$}};
		\fill[red!80!black] ($(p1)+(s)+(0.8,1.15)$) circle (0pt) node {{\small $\cu$}};
		%
		\draw[very thick, red!80!black] ($(p2)+(d)+(h)$) .. controls +(0,0.1) and +(0.5,0) ..  ($(p3)+(d)+(h)$) 
		-- ($(p3)+(h)$);
		\draw[very thick, red!80!black] ($(p2)+(d)+(h)+(s)$) .. controls +(0,0.1) and +(0.5,0) ..  ($(p3)+(d)+(h)+(s)$) 
		-- ($(p3)+(h)+(s)$);
		%
		\fill [orange!30!white, opacity=0.8] 
		($(p1)+(h)$) -- ($(p1)+(d)+(h)$) .. controls +(0.3,0) and +(0,-0.1) .. ($(p2)+(d)+(h)$)
		-- ($(p2)+(d)+(h)$) -- ($(p2)+(d)+(s)+(h)$)
		-- ($(p2)+(d)+(s)+(h)$) .. controls +(0,-0.1) and +(0.3,0) .. ($(p1)+(d)+(s)+(h)$)
		-- ($(p1)+(s)+(h)$); 
		%
		\draw[thin] ($(p1)+(h)$) -- ($(p1)+(h)+(s)$);
		\draw[thin] ($(p2)+(d)+(h)$) -- ($(p2)+(d)+(s)+(h)$);
		\draw[thin] ($(p3)+(h)$) -- ($(p3)+(h)+(s)$);
		%
		\draw[very thick, red!80!black] ($(p1)+(h)$) -- ($(p1)+(d)+(h)$) .. controls +(0.3,0) and +(0,-0.1) .. ($(p2)+(d)+(h)$); 
		\draw[very thick, red!80!black] ($(p1)+(h)+(s)$) -- ($(p1)+(d)+(h)+(s)$) .. controls +(0.3,0) and +(0,-0.1) .. ($(p2)+(d)+(h)+(s)$); 
		%
		\fill[red!80!black] ($(p1)+(h)+(t)$) circle (0pt) node[left] {{\small $\yu'$}};
		\fill[red!80!black] ($(p1)+(h)+(s)+(t)$) circle (0pt) node[left] {{\small $\yu'$}};
		\fill[red!80!black] ($(p3)+(h)+(t)$) circle (0pt) node[left] {{\small $\xu'$}};
		\fill[red!80!black] ($(p3)+(h)+(s)+(t)$) circle (0pt) node[left] {{\small $\xu'$}};
		\fill[red!80!black] ($(p2)+(d)+(h)-0.5*(t)$) circle (0pt) node[right] {{\small $\au'''$}};
		\fill[red!80!black] ($(p2)+(d)+(s)+(h)-0.5*(t)$) circle (0pt) node[right] {{\small $\du$}};
		\fill[red!80!black] ($(p1)+(h)+(d)$) circle (2.5pt) node[below] {{\small $\yu$}};
		\end{tikzpicture}
	} 
	= \big[ \du-\cu, \, \yu-\yu' \big] \otimes \big[ \du-\au''', \, \xu-\yu \big]  \otimes \big[ \du-\bu, \, \xu'-\xu \big] \, .
	\ee 	
	This latter matrix factorisation represents the 2-isomorphism $1_{\xu,\yu}\circ {}^\dagger \tev_\xu \lra {}^\dagger \tev_\xu$, as can be checked straightforwardly. 
	
	Horizontally and vertically composing \eqref{eq:sub1}--\eqref{eq:sub4} amounts to taking the tensor product over $\C[\au,\au'',\au''',\widetilde\au,\bu,\cu,\xu,\yu]$ of those matrix factorisations. 
	Thus, the 2-morphism on the left-hand side of the Zorro move \eqref{eq:3dZorro} is represented by the matrix factorisation
\begin{align}
&\big[ \au-\au'', \, \xu-\yu \big]\otimes\big[ \au'''-\au'', \, \yu-\xu \big]
\otimes \big[ \cu-\au, \, \yu-\yu' \big] \otimes \big[ \bu-\widetilde\au, \, \xu'-\xu \big] \nonumber
\\
&\quad \otimes \big[ \widetilde\au-\au, \, \yu'-\xu \big]\otimes  \big[ \du-\cu, \, \yu-\yu' \big] \otimes \big[ \du-\au''', \, \xu-\yu \big]  \otimes \big[ \du-\bu, \, \xu'-\xu \big] \nonumber
\\
&\quad \otimes \big[ \widetilde\au-\au', \, \yu'-\xu' \big] \, . \label{eq:zorrostep1}
\end{align}
This is to be considered as a matrix factorisation over $\C[\xu',\yu',\au',\du]$. 
The remaining variables $\bu,\cu,\au,\au'',\au''',\widetilde\au,\xu,\yu$ are internal variables and can be eliminated by the following trick, which is formulated and proved as Lemma~\ref{lem:elimination} in Appendix~\ref{sec:applemmas}.
Let $(X,d_X) \equiv (X,d_X)(\xu,\au,\bu)$ in variables $\xu=(x_1,\ldots,x_n)$, $\au=(a_1,\ldots,a_k)$ and $\bu=(b_1,\ldots,b_k)$ be a matrix factorisation. 
Then under the assumption that for $\pu=(p_1,\ldots,p_k)\in\C[\xu,\au,\bu]^{\times k}$ the variables~$\bu$ are internal to $(X,d_X)\otimes [\bu-\au,\pu]$, this latter matrix factorisation is isomorphic to $(X,d_X)(\xu,\au,\au)$, i.\,e.\ the matrix factorisation obtained from $(X,d_X)$ by setting~$\bu$ equal to~$\au$. 
Thus, the internal variable $\bu$ is eliminated from the tensor product by setting the Koszul factors $\bu-\au$ to zero. 
In this way, one can use the first five tensor factors in~\eqref{eq:zorrostep1} to set $\bu=\cu=\au=\au''=\au'''=\widetilde\au=\au'$, showing that~\eqref{eq:zorrostep1} is isomorphic to 
\begin{equation}\label{eq:zorrostep2}
\big[ \du-\au', \, \yu-\yu' \big] \otimes \big[ \du-\au', \, \xu-\yu \big]  \otimes \big[ \du-\au, \, \xu'-\xu \big]\,.
\end{equation}

Next we can interchange the polynomials of Koszul matrix factorisations $[\pu,\,\qu]$ at the expense of a shift (recall~\eqref{eq:ShiftOfMF}), 
\begin{equation}
\big[\qu,\,\pu\big]\cong\big[\pu,\,\qu\big][k]\,,
\end{equation}
where $k$ is the length of the lists $\pu=(p_1,\ldots,p_k)$ and $\qu=(q_1,\ldots,q_k)$.
Since the shift of matrix factorisations can be pulled out of the tensor product, 
\begin{equation}\label{eq:tpshift}
(X,d_X)[1]\otimes (Y,d_Y)\cong \big((X,d_X)\otimes (Y,d_Y)\big)[1]
\cong(X,d_X)\otimes (Y,d_Y)[1] \, , 
\end{equation}
and since a shift by $2$ acts trivially, \eqref{eq:zorrostep2} is isomorphic to 
\begin{equation}\label{eq:zorrostep3}
\big[ \yu-\yu' ,\, \du-\au' \big] \otimes \big[ \xu-\yu,\, \du-\au'\big]  \otimes \big[ \du-\au', \, \xu'-\xu \big]\,.
\end{equation}
Applying Lemma~\ref{lem:elimination} again, one can eliminate the variables $\xu$ and $\yu$ by means of the first two tensor factors in \eqref{eq:zorrostep3}, setting them to $\yu'$. 
This yields $[ \du-\au', \xu'-\yu' ]$ which is indeed a representative of the right-hand side of the Zorro move \eqref{eq:3dZorro}.
All other Zorro moves are checked in complete analogy. 
\end{proof}

\subsubsection{Serre automorphisms}
\label{subsubsec:SerreAutomorphisms} 

Let~$u$ be a fully dualisable object in some symmetric monoidal 2-category~$\B$. 
Then by definition its adjunction 1-morphisms $\tev_u, \tcoev_u$ themselves have both left and right adjoints; but in fact $\tev_u, \tcoev_u$ have arbitrary multiple adjoints $\tev_u^{\dagger\dagger}, {}^{\dagger\dagger\dagger}\tcoev_u, \dots$, see \cite[Thm.\,3.9]{Pstragowski}. 
These are constructed only from $\tev_u, \tcoev_u$, the braiding~$b$ of~$\B$, and appropriate powers of the \textsl{Serre automorphism}
\begin{align}
S_u
:= & \; \big( 1_u \btimes \tev_u \big) \circ \big( b_{u,u} \btimes 1_{u^\dual} \big) \circ \big( 1_u \btimes \tev_u^\dagger \big) 
\nonumber
\\
{=} & \;
\tikzzbox{%
	\begin{tikzpicture}[thick,scale=1.0,color=black, baseline=3.7cm]
	\coordinate (p1) at (0,0);
	\coordinate (p2) at (2,-0.5);
	\coordinate (p3) at (2,0.5);
	\coordinate (p4) at (4,0);
	\coordinate (u1) at (0,3);
	\coordinate (u2) at (2,2.5);
	\coordinate (u3) at (2,3.5);
	\coordinate (u4) at (4,3);
	\coordinate (ld) at (-2,1);
	\coordinate (lu) at (-2,3.8);
	\coordinate (rd) at (6,1);
	\coordinate (ru) at (6,3.8);
	%
	\draw[very thick, red!80!black] (u3) .. controls +(-1,+0.2) and +(+1,0) .. (lu); 
	\draw[very thick, red!80!black] (u3) .. controls +(+1,+0.2) and +(-1,0) .. (ru); 
	\draw[very thick, red!80!black] (u1) .. controls +(0,-0.25) and +(-1,0) .. (u2) 
	--(u2) .. controls +(1,0) and +(0,-0.25) .. (u4); 
	\draw[very thick, red!80!black] (u4) .. controls +(0,0.25) and +(1,-0.2) ..  (u3); 
	\draw[very thick, red!80!black] (u1) .. controls +(0,0.25) and +(-1,-0.2) .. (u3); 
	\fill (2,3.5) circle (3pt) node[above] {{\small $b_{u,u}$}};
	\node[red!80!black] at (-2.2,3.8) {{\small $u$}};
	\node[red!80!black] at (6.2,3.8) {{\small $u$}};
	\node[red!80!black] at (2,2.8) {{\small $u^\dual$}};
	\end{tikzpicture}
} 
\, . 
\label{eq:SerreFormula}
\end{align}
For example, we have 
\begin{align}
\tev_u^\dagger 
& \cong \big( S_u \btimes 1_{u^\dual} \big) \circ b_{u^\dual,u} \circ \tcoev_u \, , 
\label{eq:RightAdjointFromSerre}
\\ 
\tev_u^{\dagger\dagger} 
& \cong \tev_u \circ b_{u^\dual,u} \circ \big( 1_{u^\dual} \btimes S_u^{-2} \big) \, , 
\\ 
\tev_u^{\dagger\dagger\dagger} 
& \cong \big( 1_{u^\dual} \btimes S_u^3 \big) \circ b_{u^\dual,u} \circ \tcoev_u \, , 
\end{align}
and similarly for left adjoints of $\tev_u$ and adjoints of $\tcoev_u$. 
In particular, it follows from \cite[Thm.\,3.9]{Pstragowski} that if~$S_u$ is its own weak inverse, then left and right adjoints coincide. 

The Serre automorphisms~$S_u$ for all $u\in\B$ are precisely the 1-morphism components of a pseudonatural transformation\footnote{As shown in \cite[Prop.\,3.2]{RSpinLorantNils}, if all 1-morphisms in~$\B$ have adjoints, then~$S$ lifts to a pseudonatural transformation $S\colon \textrm{Id}_{\Bfd} \lra \textrm{Id}_{\Bfd}$.} 
\be 
\label{eq:SerrePseudonaturalTransformationOnBfdCore}
S\colon \textrm{Id}_{(\Bfd)^\times} \lra \textrm{Id}_{(\Bfd)^\times} \, , 
\ee 
where by definition $(\Bfd)^\times$ is the maximal sub-2-groupoid of fully dualisable objects in~$\B$, see \cite[Prop.\,2.8]{HV}. 
This means that the objects of $(\Bfd)^\times$ are the fully dualisable objects in~$\B$, while the 1- and 2-morphisms in $(\Bfd)^\times$ are precisely the (weakly) invertible 1- and 2-morphisms (between fully dualisable objects) in~$\B$. 

\medskip 

According to Theorem~\ref{thm:FullyDualisable}, every object in $\B = \bi$ is fully dualisable. 
Moreover, a straightforward computation analogous to that of the cusp isomorphism in~\eqref{eq:ZorroMovieComputation} shows that every Serre automorphism in $\bi$ is trivialisable: 

\begin{proposition}
	\label{lem:SerreTrivial}
	For all $\xu \in \bi$, there are precisely two isomorphisms 
	\be 
	\label{eq:SerreTrivial}
	S_\xu \stackrel{\cong}{\longrightarrow} 1_{\xu} \, ,  
	\ee 
	represented by the matrix factorisations~$I_{1_\xu}$ and $I_{1_\xu}[1]$. 
\end{proposition} 
\begin{proof}
	Using the fact that $\xu^\dual=\xu$ for any object $\xu$ in $\bi$, and plugging the explicit formulas
	 \eqref{eq:Unit1Morphism}, \eqref{eq:Braiding1morphismComponents}, \eqref{eq:Eval}  and \eqref{eq:LeftRightEval} for $1_\xu$, $b_{\xu,\xu}$, $\tev_{\xu}$ and $\tev_{\xu}^\dagger$, respectively, into the definition \eqref{eq:SerreFormula} of the Serre automorphism, one obtains
	\begin{align}
	S_{\xu}
	& = \big( 1_{\xu} \btimes \tev_{\xu} \big) \circ \big( b_{\xu,\xu} \btimes 1_{\xu^\dual} \big) \circ \big( 1_{\xu} \btimes \tev_{\xu}^\dagger \big) 
	\\
	& = \;\;
	\tikzzbox{%
		\begin{tikzpicture}[thick,scale=0.40,color=black, baseline=0cm]
		\coordinate (p1) at (-8,5.6);
		\coordinate (p2) at (-4,4);
		\coordinate (p3) at (4,0);
		\coordinate (p4) at (4,-6);
		\coordinate (p5) at (-4,-6);
		\coordinate (p6) at (-4,0);
		\coordinate (p7) at (4,4);
		\coordinate (p8) at (8,5.6);
		\coordinate (x1) at (-4,6.5);
		\coordinate (x2) at (-4,-7.8);
		\coordinate (x3) at (4,6.5);
		\coordinate (x4) at (4,-7.8);
		\draw[very thick, red!80!black] (p1) .. controls +(1,0) and +(-1,0.5) .. (p2) .. controls +(-1,0.5) and +(-2,0) .. (p3)
		arc(90:-90:3) -- (p5) arc(270:90:3) .. controls +(2,0) and +(-1,-0.5) .. (p7) .. controls +(1,0.5) and +(-1,0) ..(p8);
		\draw[dotted, blue] (x1)--(x2);
		\draw[dotted, blue] (x3)--(x4);
		\fill[red!80!black] ($(p1)+(0.2,0.2)$) circle (0pt) node[left] {${\small \xu^{(1)}}$};
		\fill[red!80!black] (p2) circle (5pt) node[below] {${\small \xu^{(2)}}$};
		\fill[red!80!black] (p3) circle (5pt) node[below] {${\small \xu^{(3)}}$};
		\fill[red!80!black] (p4) circle (5pt) node[below] {${\small \xu^{(4)}}$};
		\fill[red!80!black] (p5) circle (5pt) node[below] {${\small \xu^{(5)}}$};
		\fill[red!80!black] (p6) circle (5pt) node[below] {${\small \xu^{(6)}}$};
		\fill[red!80!black] (p7) circle (5pt) node[below] {${\small \xu^{(7)}}$};
		\fill[red!80!black] ($(p8)+(-0.1,0.2)$) circle (0pt) node[right] {${\small \xu^{(8)}}$};
		\node[above] at (-5.3,4.8) {\color{red!80!black}$\au^{(1)}$};
		\node[above] at (-1.6,2.7) {\color{red!80!black}$\au^{(2)}$};
		\node[right] at (7,-3) {\color{red!80!black}$\au^{(3)}$};
		\node[below] at (0,-6) {\color{red!80!black}$\au^{(4)}$};
		\node[left] at (-6.8,-3) {\color{red!80!black}$\au^{(5)}$};
		\node[below] at (-1.6,1) {\color{red!80!black}$\au^{(6)}$};
		\node[above] at (5.5,4.6) {\color{red!80!black}$\au^{(7)}$};
		%
		\end{tikzpicture}
	}\nonumber\\
	&=
	\Big(\au^{(1)},\ldots,\au^{(7)},\xu^{(2)},\ldots,\xu^{(7)}\,;\;\sum_{i=1}^7\;\au^{(i)}\cdot\big(\xu^{(i+1)}-\xu^{(i)}\big)\Big)\nonumber\\
	&=\Big(\au^{(1)};\;\au^{(1)} \cdot \big(\xu^{(2)}-\xu^{(1)}\big)\Big)\,\circ\,
	\Big(\au^{(2)};\;\au^{(2)}\cdot\big(\xu^{(3)}-\xu^{(2)}\big)\Big) \nonumber
	\\
	& \qquad \,\circ\,
	\dots\, \circ\Big(\au^{(7)};\;\au^{(7)}\cdot\big(\xu^{(8)}-\xu^{(7)}\big)\Big)
	\nonumber
	\\
	&=\left(1_\xu\right)^{7}
	\, . 
	\end{align}
	 Since $1_{\xu}\circ \varphi\cong\varphi$ for any 1-morphism $\varphi\in\bi(\yu,\xu)$, we have $\left(1_\xu\right)^{7}\cong 1_\xu$, and hence $S_\xu\cong 1_\xu$.
	 This proves trivialisability of~$S_\xu$. 
	 
	 Finding all trivialisations $S_\xu \cong 1_\xu$ is equivalent to finding all automorphisms of~$1_\xu$ in~$\bi$. 
	 By definition the latter are isomorphism classes of invertible objects in the monoidal category $\hmf(\C[\au,\bu,\xu,\yu], (\au-\bu)\cdot (\xu-\yu))^\omega \cong \hmf(\C[\au,\bu,\xu,\yu], \bu\cdot \yu)^\omega$. 
	 By Knörrer periodicity this is equivalent to $\hmf(\C[\au,\xu], 0)^\omega \cong \textrm{mod}^{\Z_2}(\C[\au,\xu])$, the category of finitely generated free $\Z_2$-graded $\C[\au,\xu]$-modules. 
	 But up to isomorphism the only invertible objects in $\textrm{mod}^{\Z_2}(\C[\au,\xu])$ are $\C[\au,\xu]$ concentrated in $\Z_2$-degree~0 or~1, and we find that $\textrm{Aut}_\bi(1_\xu) = \{ [I_{1_\xu}], [I_{1_\xu}][1]\}$. 
\end{proof}

Together with the triviality of the braiding~$b$, this explains why the respective left and right adjunction 2-morphisms in \eqref{eq:DualityData1}--\eqref{eq:DualityData} are equal.

\subsection{Gradings by flavour and R-symmetries}
\label{subsec:GradedCase}

Rozansky--Witten theories carry two $\textrm{U}(1)$-symmetries: R-symmetry and flavour symmetry. 
In the following we briefly discuss a variant $\bigra$ of the 2-category~$\bi$ in which we keep track of flavour and R-charges. 
This is important in particular when comparing infinite-dimensional state spaces such as~\eqref{eq:IntroStateSpace} to results obtained by other means.
We find that all the structure exhibited above for~$\bi$ lifts to~$\bigra$. 

\medskip 

The objects of $\bigra$ are the same as the ones of $\bi$, namely finite ordered sets of variables $(x_1,\ldots,x_n)$, where however we also assign a bidegree $(1,-1)$ to all the variables $x_i$. 
The first degree corresponds to the \textsl{R-charge}, the second one to the \textsl{flavour charge}. 
Put differently, instead of thinking of the objects as polynomial rings $\C[x_1,\ldots,x_n]$, we think of them as bigraded rings, where each variable $x_i$ is homogeneous of degree $(1,-1)$. 

The bigrading is then extended to morphisms in the following way: 
A 1-morphism $(x_1,\ldots,x_n)\lra(y_1,\ldots,y_m)$ in $\bigra$ is a triple $(\au;\gu;W)$, where $\au=(a_1,\ldots,a_k)$ is a possibly empty list of additional variables which are assigned bidegrees 
\be 
\gu=\big(g_1=(r_1,q_1),\ldots, g_k=(r_k,q_k)\big) \in (\Q \times \Q)^{\times k} \, , 
\ee 
such that $W$ is a homogeneous element of the $(\Q\times\Q)$-graded ring $\C[\au,\xu,\yu]$ of bidegree $(2,0)$. 
For ease of notation we will often omit the bidegrees, in particular in cases in which the bidegrees are determined by the ones of $\xu,\yu$. 

A 2-morphism between two 
1-morphisms $(\au;\gu;W),\,(\bu;\hu;V)\colon \xu\lra \yu$ in $\bigra$ is given by an equivalence class of matrix factorisations $(X,d_X)$ of $V-W$ over $\C[\au,\bu,\xu,\yu]$, which in addition to the ordinary $\Z_2$-grading of matrix factorisations possess a bigrading: the modules~$X^i$ in $X=X^0\oplus X^1$ are bigraded modules over the bigraded ring $\C[\au,\bu,\xu,\yu]$, and the odd map~$d_X$ must be homogeneous of bidegree $(1,0)$. 
The extra bigrading induces a bigrading on the space of morphisms of matrix factorisations, and isomorphisms are required to be homogeneous of bidegree $(0,0)$. 

The definitions of the compositions of 1- and 2-morphisms in $\bigra$ agree with the corresponding definitions  in $\bi$, cf.~\eqref{eq:HorizontalComp} and~\eqref{eq:XtensorY}. 
In fact, in our entire discussion of the 2-category $\bi$, the bigrading just goes along for the ride. All constructions and arguments are compatible with the bigrading, and everything stated above for the 2-category $\bi$ carries over immediately to the graded version~$\bigra$. 
In particular, we find:

\begin{theorem}
	\label{thm:FullyDualisableGraded}
	The structure morphisms of the symmetric monoidal 2-category~$\bi$ lift to a symmetric monoidal structure on~$\bigra$, 
	and every object of $\bigra$ is fully dualisable. 
\end{theorem}

Since the arguments are basically identical, we will refrain from repeating them here for the graded case. Instead, we will briefly sketch a few aspects which differ from the discussion of $\bi$. 

Indeed, all 1-morphisms $(\au;W)$ used in the discussion of $\bi$ above are gradable, in the sense that it is possible to choose bidegrees for the additional variables $a_i$ such that $W$ is homogeneous of bidegree $(2,0)$. For instance, 
the unit 1-morphism $1_\xu\in\bi(\xu,\xu)$ defined in~\eqref{eq:Unit1Morphism}
becomes the unit 1-morphism in $\bigra$ by assigning bidegree $(1,1)$ to the variables $a_i$. 
The same holds for the evaluations $\tev_\xu$, $\ev_\xu$  in~\eqref{eq:Eval}, \eqref{eq:LeftRightCoeval}, and the coevaluations $\tcoev_\xu$, $\coev_\xu$ in~\eqref{eq:Eval}, \eqref{eq:LeftRightEval}.
In fact, for all 1-morphisms mentioned in the discussion of $\bi$, the choice of grading is unique. 

Also the 2-morphisms used in the discussion of $\bi$ are gradable, in the following sense. 
They are morphisms between gradable 1-morphisms with a unique choice of grading, i.\,e.\ they correspond to matrix factorisations over a bigraded polynomial ring $R$ of a homogeneous polynomial of bidegree $(2,0)$. 
It turns out that for all the matrix factorisations $(X,d_X)$ used in our discussion of $\bi$, the modules $X$ can be made into graded $R$-modules such that $d_X$ is homogeneous of degree $(1,0)$. 
By specifying the grading of $X$, one obtains  bigraded matrix factorisations, and hence 2-morphisms in $\bigra$. 
Indeed, for all 2-morphisms relevant for our discussion of $\bi$, the choice of grading of $X$ is unique but only up to grade shift.

Consider a matrix factorisation $[\pu,\qu]$ of Koszul type as defined in~\eqref{eq:Koszul1}--\eqref{eq:Koszul2} for which all the polynomials $p_i$ and $q_i$ are homogeneous with respect to the bigrading. 
This matrix factorisation can be graded by choosing any bigrading of the $\bigwedge^0$-part $M\subset K(\pu,\qu)$. The bigrading on the rest of $K(\pu,\qu)$ is then determined by the homogeneity of $d_{K(\pu,\qu)}$. Now 
$M$ is a rank-1 free module over the bigraded polynomial ring. Hence, it can be identified with the bigraded polynomial ring itself, which in turn has a natural bigrading as a module over itself. 
The choice of this bigrading then determines a bigrading on the entire module $K(\pu,\qu)$. We denote this naturally bigraded module by $K(\pu,\qu)$ as well. 
The only way to change the bigrading is to twist the rank-1 free graded module, $M\lmt M\{r,s\}$, where the degree $(m,n)$-part of $M\{r,s\}$ is given by the degree $(m+r,n+s)$-part of $M$: 
\begin{equation}
\big(M\{r,s\}\big)_{m,n}=M_{m+r,n+s}\,.
\end{equation}
The rest of $K(\pu,\qu)$ is then twisted accordingly, $K(\pu,\qu)\lmt K(\pu,\qu)\{r,s\}$. 
We denote the Koszul type matrix factorisation with the corresponding bigrading by
$[\pu,\qu]\{r,s\}$. In case of zero twist we write $[\pu,\qu]\{0,0\}=[\pu,\qu]$. 

Note that under the tensor product of Koszul type matrix factorisations, the twist behaves additively:
\begin{align}
\big[\pu,\qu\big]\{r,s\} \otimes \big[\pu^\prime,\qu^\prime\big]\{r^\prime,s^\prime\}
&=\left(\big[\pu,\qu\big]\otimes\big[\pu^\prime,\qu^\prime\big]\right)\{r+r^\prime,s+s^\prime\}\nonumber\\
&=\big[\pu,\qu\big]\{r+r^\prime,s+s^\prime\}\otimes\big[\pu^\prime,\qu^\prime\big]\nonumber\\
&=\big[\pu,\qu\big]\otimes\big[\pu^\prime,\qu^\prime\big]\{r+r^\prime,s+s^\prime\} \label{eq:twistsareadditive}
\,.
\end{align}

Next, we will spell out the gradings of the 2-morphisms appearing in the discussion of the full dualisability of $\bigra$. First of all, the matrix factorisations corresponding to the identity morphisms
\begin{align}
\tikzzbox{%
	\begin{tikzpicture}[thick,scale=1.0,color=black, baseline=0.9cm]
	\coordinate (p1) at (0,0);
	\coordinate (p2) at (0.5,0.2);
	\coordinate (p3) at (-0.5,0.4);
	\coordinate (p4) at (2.5,0);
	\coordinate (p5) at (1.5,0.2);
	\coordinate (p6) at (2.0,0.4);
	\coordinate (s) at ($(0,1.5)$);
	\coordinate (d) at ($(3,0)$);
	%
	\fill [orange!20!white, opacity=0.8] 
	($(p2)+(d)$) .. controls +(0,0.1) and +(0.5,0) ..  ($(p3)+(d)$)
	-- ($(p3)+(d)$) -- ($(p3)+(d)+(s)$) 
	-- ($(p3)+(d)+(s)$) .. controls +(0.5,0) and +(0,0.1) ..($(p2)+(d)+(s)$);
	%
	\draw[very thick, red!80!black] ($(p2)+(d)$) .. controls +(0,0.1) and +(0.5,0) ..  ($(p3)+(d)$); 
	\draw[very thick, red!80!black] ($(p2)+(d)+(s)$) .. controls +(0,0.1) and +(0.5,0) ..  ($(p3)+(d)+(s)$); 
	%
	\fill [orange!30!white, opacity=0.8] 
	($(p1)+(d)$) .. controls +(0.3,0) and +(0,-0.1) .. ($(p2)+(d)$)
	-- ($(p2)+(d)$) -- ($(p2)+(d)+(s)$)
	-- ($(p2)+(d)+(s)$) .. controls +(0,-0.1) and +(0.3,0) .. ($(p1)+(d)+(s)$); 
	%
	\draw[thin] ($(p1)+(d)$) -- ($(p1)+(d)+(s)$);
	\draw[thin] ($(p2)+(d)$) -- ($(p2)+(d)+(s)$);
	\draw[thin] ($(p3)+(d)$) -- ($(p3)+(d)+(s)$);
	%
	\draw[very thick, red!80!black] ($(p1)+(d)$) .. controls +(0.3,0) and +(0,-0.1) ..  ($(p2)+(d)$); 
	\draw[very thick, red!80!black] ($(p1)+(d)+(s)$) .. controls +(0.3,0) and +(0,-0.1) ..  ($(p2)+(d)+(s)$); 
	%
	\fill[red!80!black] ($(p4)+(s)$) circle (0pt) node[right] {{\small $\yu$}};
	\fill[red!80!black] ($(p6)+(s)$) circle (0pt) node[right] {{\small $\xu$}};
	\fill[red!80!black] (p4) circle (0pt) node[right] {{\small $\yu$}};
	\fill[red!80!black] (p6) circle (0pt) node[right] {{\small $\xu$}};
	\fill[red!80!black] ($(p2)+(d)$) circle (0pt) node[right] {{\small $\au$}};
	\fill[red!80!black] ($(p2)+(d)+(s)$) circle (0pt) node[right] {{\small $\bu$}};
	\end{tikzpicture}
}%
& = \id_{\coev_{\xu}}=\id_{{}^\dagger\tev_{\xu}}=\id_{\tev_{\xu}^\dagger}=\big[ \bu-\au, \, \yu-\xu \big] \\
\tikzzbox{%
	\begin{tikzpicture}[thick,scale=1.0,color=black, baseline=0.5cm, yscale=-1, xscale=-1]
	\coordinate (p1) at (0,0);
	\coordinate (p2) at (0.5,0.2);
	\coordinate (p3) at (-0.5,0.4);
	\coordinate (p4) at (2.5,0);
	\coordinate (p5) at (1.5,0.2);
	\coordinate (p6) at (2.0,0.4);
	\coordinate (s) at ($(0,-1.5)$);
	\coordinate (d) at ($(2.5,0)$);
	%
	\fill [orange!20!white, opacity=0.8] 
	($(p1)+(d)$) .. controls +(0.3,0) and +(0,-0.1) .. ($(p2)+(d)$)
	-- ($(p2)+(d)$) -- ($(p2)+(d)+(s)$)
	-- ($(p2)+(d)+(s)$) .. controls +(0,-0.1) and +(0.3,0) .. ($(p1)+(d)+(s)$); 
	\fill[red!80!black] (p4) circle (0pt) node[right] {{\small $\xu$}};
	%
	\draw[thin] ($(p1)+(d)$) -- ($(p1)+(d)+(s)$);
	\draw[thin] ($(p2)+(d)$) -- ($(p2)+(d)+(s)$);
	\draw[thin] ($(p3)+(d)$) -- ($(p3)+(d)+(s)$);
	%
	\draw[very thick, red!80!black] ($(p1)+(d)$) .. controls +(0.3,0) and +(0,-0.1) ..  ($(p2)+(d)$); 
	\draw[very thick, red!80!black] ($(p1)+(d)+(s)$) .. controls +(0.3,0) and +(0,-0.1) ..  ($(p2)+(d)+(s)$); 
	%
	\fill [orange!30!white, opacity=0.8] 
	($(p2)+(d)$) .. controls +(0,0.1) and +(0.5,0) ..  ($(p3)+(d)$)
	-- ($(p3)+(d)$) -- ($(p3)+(d)+(s)$) 
	-- ($(p3)+(d)+(s)$) .. controls +(0.5,0) and +(0,0.1) ..($(p2)+(d)+(s)$);
	%
	\draw[thin] ($(p2)+(d)$) -- ($(p2)+(d)+(s)$);
	\draw[thin] ($(p3)+(d)$) -- ($(p3)+(d)+(s)$);
	%
	\draw[very thick, red!80!black] ($(p2)+(d)$) .. controls +(0,0.1) and +(0.5,0) ..  ($(p3)+(d)$); 
	\draw[very thick, red!80!black] ($(p2)+(d)+(s)$) .. controls +(0,0.1) and +(0.5,0) ..  ($(p3)+(d)+(s)$); 
	%
	\fill[red!80!black] ($(p4)+(s)$) circle (0pt) node[right] {{\small $\xu$}};
	\fill[red!80!black] ($(p6)+(s)$) circle (0pt) node[right] {{\small $\yu$}};
	\fill[red!80!black] (p6) circle (0pt) node[right] {{\small $\yu$}};
	\fill[red!80!black] ($(p2)+(d)$) circle (0pt) node[left] {{\small $\au$}};
	\fill[red!80!black] ($(p2)+(d)+(s)$) circle (0pt) node[left] {{\small $\bu$}};
	\end{tikzpicture}
}%
&= \id_{\ev_{\xu}}=\id_{{}^\dagger\tcoev_{\xu}}=\id_{\tcoev_\xu^\dagger}=\big[ \bu-\au, \, \yu-\xu \big] \,
\end{align}
are not twisted. The matrix factorisations representing for instance the 2-morphisms between 
${}^\dagger \tev_\xu$ and $1_{\xu,\yu}\circ {}^\dagger \tev_\xu$ allow for a twist in terms of parameters $\rs,\qs\in\Q$: 
\begin{align}
\label{eq:sub4gr}
\tikzzbox{%
	\begin{tikzpicture}[thick,scale=1.0,color=black, baseline=3.3cm]
	\coordinate (p1) at (0,0);
	\coordinate (p2) at (0.5,0.2);
	\coordinate (p3) at (-0.5,0.4);
	\coordinate (p4) at (2.5,0);
	\coordinate (p5) at (1.5,0.2);
	\coordinate (p6) at (2.0,0.4);
	\coordinate (s) at ($(0,1.5)$);
	\coordinate (d) at ($(2,0)$);
	\coordinate (h) at ($(0,2.5)$);
	\coordinate (t) at ($(0.1,0)$);
	\coordinate (u) at ($(0,-1.1)$);
	%
	%
	\fill [orange!20!white, opacity=0.8] 
	($(p2)+(d)+(h)$) .. controls +(0,0.1) and +(0.5,0) ..  ($(p3)+(d)+(h)$) 
	-- ($(p3)+(h)$) -- ($(p3)+(s)+(h)$) -- ($(p3)+(d)+(s)+(h)$)
	-- ($(p3)+(d)+(s)+(h)$) .. controls +(0.5,0) and +(0,0.1) ..($(p2)+(d)+(s)+(h)$);
	%
	\fill[red!80!black] ($(p3)+(h)+(d)$) circle (2.5pt) node[below] {{\small $\xu'$}};
	\fill[red!80!black] ($(p3)+(s)+(0.8,1.2)$) circle (0pt) node {{\small $\bu$}};
	\fill[red!80!black] ($(p1)+(s)+(0.8,1.15)$) circle (0pt) node {{\small $\cu$}};
	%
	\draw[very thick, red!80!black] ($(p2)+(d)+(h)$) .. controls +(0,0.1) and +(0.5,0) ..  ($(p3)+(d)+(h)$) 
	-- ($(p3)+(h)$);
	\draw[very thick, red!80!black] ($(p2)+(d)+(h)+(s)$) .. controls +(0,0.1) and +(0.5,0) ..  ($(p3)+(d)+(h)+(s)$) 
	-- ($(p3)+(h)+(s)$);
	%
	\fill [orange!30!white, opacity=0.8] 
	($(p1)+(h)$) -- ($(p1)+(d)+(h)$) .. controls +(0.3,0) and +(0,-0.1) .. ($(p2)+(d)+(h)$)
	-- ($(p2)+(d)+(h)$) -- ($(p2)+(d)+(s)+(h)$)
	-- ($(p2)+(d)+(s)+(h)$) .. controls +(0,-0.1) and +(0.3,0) .. ($(p1)+(d)+(s)+(h)$)
	-- ($(p1)+(s)+(h)$); 
	%
	\draw[thin] ($(p1)+(h)$) -- ($(p1)+(h)+(s)$);
	\draw[thin] ($(p2)+(d)+(h)$) -- ($(p2)+(d)+(s)+(h)$);
	\draw[thin] ($(p3)+(h)$) -- ($(p3)+(h)+(s)$);
	%
	\draw[very thick, red!80!black] ($(p1)+(h)$) -- ($(p1)+(d)+(h)$) .. controls +(0.3,0) and +(0,-0.1) .. ($(p2)+(d)+(h)$); 
	\draw[very thick, red!80!black] ($(p1)+(h)+(s)$) -- ($(p1)+(d)+(h)+(s)$) .. controls +(0.3,0) and +(0,-0.1) .. ($(p2)+(d)+(h)+(s)$); 
	%
	\fill[red!80!black] ($(p1)+(h)+(t)$) circle (0pt) node[left] {{\small $\yu$}};
	\fill[red!80!black] ($(p1)+(h)+(s)+(t)$) circle (0pt) node[left] {{\small $\yu$}};
	\fill[red!80!black] ($(p3)+(h)+(t)$) circle (0pt) node[left] {{\small $\xu$}};
	\fill[red!80!black] ($(p3)+(h)+(s)+(t)$) circle (0pt) node[left] {{\small $\xu$}};
	\fill[red!80!black] ($(p2)+(d)+(h)-0.5*(t)$) circle (0pt) node[right] {{\small $\au$}};
	\fill[red!80!black] ($(p2)+(d)+(s)+(h)-0.5*(t)$) circle (0pt) node[right] {{\small $\du$}};
	\fill[red!80!black] ($(p1)+(h)+(d)$) circle (2.5pt) node[below] {{\small $\yu'$}};
	\end{tikzpicture}
}
&= \big[ \du-\cu, \, \yu'-\yu \big] \otimes \big[ \du-\au, \, \xu'-\yu' \big]  \\
&\qquad
\otimes \big[ \du-\bu, \, \xu-\xu' \big]\{\rs,\qs\} \, , \nonumber\\
\tikzzbox{%
	\begin{tikzpicture}[thick,scale=1.0,color=black, baseline=3.3cm]
	\coordinate (p1) at (0,0);
	\coordinate (p2) at (0.5,0.2);
	\coordinate (p3) at (-0.5,0.4);
	\coordinate (p4) at (2.5,0);
	\coordinate (p5) at (1.5,0.2);
	\coordinate (p6) at (2.0,0.4);
	\coordinate (s) at ($(0,1.5)$);
	\coordinate (d) at ($(2,0)$);
	\coordinate (h) at ($(0,2.5)$);
	\coordinate (t) at ($(0.1,0)$);
	\coordinate (u) at ($(0,-1.1)$);
	%
	%
	\fill [orange!20!white, opacity=0.8] 
	($(p2)+(d)+(h)$) .. controls +(0,0.1) and +(0.5,0) ..  ($(p3)+(d)+(h)$) 
	-- ($(p3)+(h)$) -- ($(p3)+(s)+(h)$) -- ($(p3)+(d)+(s)+(h)$)
	-- ($(p3)+(d)+(s)+(h)$) .. controls +(0.5,0) and +(0,0.1) ..($(p2)+(d)+(s)+(h)$);
	%
	\fill[red!80!black] ($(p3)+(h)+(d)+(s)$) circle (2.5pt) node[below] {{\small $\xu'$}};
	\fill[red!80!black] ($(p3)+(s)+(s)+(0.8,1.2)$) circle (0pt) node {{\small $\bu$}};
	\fill[red!80!black] ($(p1)+(s)+(s)+(0.8,1.15)$) circle (0pt) node {{\small $\cu$}};
	%
	\draw[very thick, red!80!black] ($(p2)+(d)+(h)$) .. controls +(0,0.1) and +(0.5,0) ..  ($(p3)+(d)+(h)$) 
	-- ($(p3)+(h)$);
	\draw[very thick, red!80!black] ($(p2)+(d)+(h)+(s)$) .. controls +(0,0.1) and +(0.5,0) ..  ($(p3)+(d)+(h)+(s)$) 
	-- ($(p3)+(h)+(s)$);
	%
	\fill [orange!30!white, opacity=0.8] 
	($(p1)+(h)$) -- ($(p1)+(d)+(h)$) .. controls +(0.3,0) and +(0,-0.1) .. ($(p2)+(d)+(h)$)
	-- ($(p2)+(d)+(h)$) -- ($(p2)+(d)+(s)+(h)$)
	-- ($(p2)+(d)+(s)+(h)$) .. controls +(0,-0.1) and +(0.3,0) .. ($(p1)+(d)+(s)+(h)$)
	-- ($(p1)+(s)+(h)$); 
	%
	\draw[thin] ($(p1)+(h)$) -- ($(p1)+(h)+(s)$);
	\draw[thin] ($(p2)+(d)+(h)$) -- ($(p2)+(d)+(s)+(h)$);
	\draw[thin] ($(p3)+(h)$) -- ($(p3)+(h)+(s)$);
	%
	\draw[very thick, red!80!black] ($(p1)+(h)$) -- ($(p1)+(d)+(h)$) .. controls +(0.3,0) and +(0,-0.1) .. ($(p2)+(d)+(h)$); 
	\draw[very thick, red!80!black] ($(p1)+(h)+(s)$) -- ($(p1)+(d)+(h)+(s)$) .. controls +(0.3,0) and +(0,-0.1) .. ($(p2)+(d)+(h)+(s)$); 
	%
	\fill[red!80!black] ($(p1)+(h)+(t)$) circle (0pt) node[left] {{\small $\yu$}};
	\fill[red!80!black] ($(p1)+(h)+(s)+(t)$) circle (0pt) node[left] {{\small $\yu$}};
	\fill[red!80!black] ($(p3)+(h)+(t)$) circle (0pt) node[left] {{\small $\xu$}};
	\fill[red!80!black] ($(p3)+(h)+(s)+(t)$) circle (0pt) node[left] {{\small $\xu$}};
	\fill[red!80!black] ($(p2)+(d)+(h)-0.5*(t)$) circle (0pt) node[right] {{\small $\du$}};
	\fill[red!80!black] ($(p2)+(d)+(s)+(h)-0.5*(t)$) circle (0pt) node[right] {{\small $\au$}};
	\fill[red!80!black] ($(p1)+(h)+(d)+(s)$) circle (2.5pt) node[below] {{\small $\yu'$}};
	\end{tikzpicture}
}
&= \big[ \cu-\du, \, \yu'-\yu \big] \otimes \big[ \au-\du, \, \xu'-\yu' \big]  \label{eq:sub4grp}\\
&\qquad 
\otimes \big[ \bu-\du, \, \xu-\xu' \big]\{-\rs,-\qs-2\} \, . 
\nonumber
\end{align}
Note that the relative twist is fixed by the fact that these 2-morphisms compose to $\id_{{}^\dagger\tev_{\xu}}$. Hence, we have only one twist parameter $(\rs,\qs)$ determining the bigrading of these 2-morphisms. 

The 2-morphisms corresponding to southern and northern hemispheres as well as saddles also allow for twists, in terms of parameters $\rsh,\rnh,\qsh,\qnh \in \Q$:
\begin{align}
\tikzzbox{%
	\begin{tikzpicture}[thick,scale=1.0,color=black, baseline=-0.9cm]
	\coordinate (p1) at (-2.75,0);
	\coordinate (p2) at (-1,0);
	\fill [orange!20!white, opacity=0.8] 
	(p1) .. controls +(0,-0.5) and +(0,-0.5) ..  (p2)
	-- (p2) .. controls +(0,0.5) and +(0,0.5) ..  (p1)
	;
	\fill [orange!30!white, opacity=0.8] 
	(p1) .. controls +(0,-0.5) and +(0,-0.5) ..  (p2)
	-- (p2) .. controls +(0,-2) and +(0,-2) ..  (p1)
	;
	\draw (p1) .. controls +(0,-2) and +(0,-2) ..  (p2); 
	\draw[very thick, red!80!black] (p1) .. controls +(0,0.5) and +(0,0.5) ..  (p2); 
	\draw[very thick, red!80!black] (p1) .. controls +(0,-0.5) and +(0,-0.5) ..  (p2); 
	%
	\fill[red!80!black] ($(p1)+(0.1,0)$) circle (0pt) node[left] {{\small $\au'$}};
	\fill[red!80!black] ($(p2)+(-0.05,0)$) circle (0pt) node[right] {{\small $\au\vphantom{\au'}$}};
	\fill[red!80!black] (-1.875,+0.38) circle (2.5pt) node[above] {{\small $\xu$}};
	\fill[red!80!black] (-1.875,-0.38) circle (2.5pt) node[below] {{\small $\yu$}};
	\end{tikzpicture}
}%
& 
= \coev_{\tev_\xu} 
= \tcoev_{\tcoev_\xu} 
= \big[ \au'-\au, \, \xu-\yu \big]\{\rsh,\qsh\}\, ,
\label{eq:DualityData2gr}
\\ 
\tikzzbox{%
	\begin{tikzpicture}[thick,scale=1.0,color=black, rotate=180, baseline=0.5cm]
	\coordinate (p1) at (-2.75,0);
	\coordinate (p2) at (-1,0);
	%
	\draw[very thick, red!80!black] (p1) .. controls +(0,-0.5) and +(0,-0.5) ..  (p2); 
	\fill[red!80!black] (-1.875,-0.38) circle (2.5pt) node[above] {{\small $\yu$}};
	\fill [orange!20!white, opacity=0.8] 
	(p1) .. controls +(0,-0.5) and +(0,-0.5) ..  (p2)
	-- (p2) .. controls +(0,0.5) and +(0,0.5) ..  (p1)
	;
	\fill [orange!30!white, opacity=0.8] 
	(p1) .. controls +(0,-0.5) and +(0,-0.5) ..  (p2)
	-- (p2) .. controls +(0,-2) and +(0,-2) ..  (p1)
	;
	\coordinate (q1) at (-1.875,0.38);
	\coordinate (q2) at (-2.67,-0.7);
	\draw (p1) .. controls +(0,-2) and +(0,-2) ..  (p2); 
	%
	\fill[red!80!black] ($(p1)+(0.1,0)$) circle (0pt) node[right] {{\small $\au'$}};
	\fill[red!80!black] ($(p2)+(-0.05,0)$) circle (0pt) node[left] {{\small $\au\vphantom{\au'}$}};
	\fill[red!80!black] (-1.875,+0.38) circle (2.5pt) node[below] {{\small $\xu$}};
	%
	\draw[very thick, red!80!black] (p1) .. controls +(0,0.5) and +(0,0.5) ..  (p2); 
	\end{tikzpicture}
}%
& 
= \tev_{\tev_\xu} 
= \ev_{\tcoev_\xu} 
= \big[ \yu-\xu, \, \au'-\au \big] \{\rnh,\qnh\}\, ,
\label{eq:DualityData3gr}
\\ 
\tikzzbox{%
	\begin{tikzpicture}[thick,scale=1.0,color=black, baseline=0.9cm]
	\coordinate (p1) at (0,0);
	\coordinate (p2) at (0.5,0.2);
	\coordinate (p3) at (-0.5,0.4);
	\coordinate (p4) at (2.5,0);
	\coordinate (p5) at (1.5,0.2);
	\coordinate (p6) at (2.0,0.4);
	\coordinate (s) at ($(0,1.5)$);
	\coordinate (d) at ($(3.8,0)$);
	\coordinate (h) at ($(0,2.5)$);
	\coordinate (t) at ($(0.1,0)$);
	\coordinate (u) at ($(0,-1.1)$);
	%
	\fill [orange!20!white, opacity=0.8] 
	(p3) .. controls +(0.5,0) and +(0,0.1) ..  (p2)
	-- (p2) .. controls +(0,0.9) and +(0,0.9) ..  (p5)
	-- (p5) .. controls +(0,0.1) and +(-0.3,0) .. (p6)
	-- (p6) -- ($(p6)+(s)$) -- ($(p3)+(s)$);
	%
	\fill[red!80!black] ($(p1)+(t)$) circle (0pt) node[left] {{\small $\yu'$}};
	\fill[red!80!black] ($(p2)-0.5*(t)$) circle (0pt) node[right] {{\small $\au'$}};
	\fill[red!80!black] ($(p3)+(t)$) circle (0pt) node[left] {{\small $\xu'$}};
	\fill[red!80!black] ($(p4)-0.5*(t)$) circle (0pt) node[right] {{\small $\yu$}};
	\fill[red!80!black] ($(p5)+0.5*(t)$) circle (0pt) node[left] {{\small $\au\vphantom{\au'}$}};
	\fill[red!80!black] ($(p6)-0.5*(t)$) circle (0pt) node[right] {{\small $\xu$}};
	\fill[red!80!black] ($(p4)+(s)-0.5*(t)$) circle (0pt) node[right] {{\small $\yu$}};
	\fill[red!80!black] ($(p3)+(s)+(1.25,0.2)$) circle (0pt) node {{\small $\bu$}};
	\fill[red!80!black] ($(p1)+(s)+(1.25,0.15)$) circle (0pt) node {{\small $\cu$}};
	\fill[red!80!black] ($(p1)+(s)+(t)$) circle (0pt) node[left] {{\small $\yu'$}};
	\fill[red!80!black] ($(p3)+(s)+(t)$) circle (0pt) node[left] {{\small $\xu'$}};
	\fill[red!80!black] ($(p6)+(s)-0.5*(t)$) circle (0pt) node[right] {{\small $\xu$}};
	%
	%
	\draw[very thick, red!80!black] (p2) .. controls +(0,0.1) and +(0.5,0) ..  (p3); 
	\draw[very thick, red!80!black] (p6) .. controls +(-0.3,0) and +(0,0.1) ..  (p5); 
	\draw[thin] (p6) -- ($(p6)+(s)$);
	%
	\fill [orange!30!white, opacity=0.8] 
	(p1) .. controls +(0.3,0) and +(0,-0.1) ..  (p2)
	-- (p2) .. controls +(0,0.9) and +(0,0.9) ..  (p5)
	-- (p5) .. controls +(0,-0.1) and +(-0.5,0) ..  (p4) 
	-- (p4) -- ($(p4)+(s)$) -- ($(p1)+(s)$);
	%
	\draw[very thick, red!80!black] (p1) .. controls +(0.3,0) and +(0,-0.1) ..  (p2); 
	\draw[very thick, red!80!black] (p5) .. controls +(0,-0.1) and +(-0.5,0) ..  (p4); 
	\draw[very thick, red!80!black] ($(p4)+(s)$) -- ($(p1)+(s)$);
	\draw[very thick, red!80!black] ($(p6)+(s)$) -- ($(p3)+(s)$);
	%
	\draw (p2) .. controls +(0,0.9) and +(0,0.9) ..  (p5);
	\draw[thin] (p1) -- ($(p1)+(s)$);
	\draw[thin] (p3) -- ($(p3)+(s)$);
	\draw[thin] (p4) -- ($(p4)+(s)$);
	\end{tikzpicture}
}%
& 
= 
\ev_{\tev_\xu} 
= \tev_{\tcoev_\xu} 
\nonumber \\ 
& 
= \big[ \cu-\au, \, \yu-\yu' \big] \otimes \big[ \bu-\au', \, \xu'-\xu \big] \label{eq:DualityData1gr}\\
&\qquad\otimes \big[ \au'-\au, \, \yu'-\xu \big]\{-\rs-\rsh,-\qs-\qsh-2\}\, ,
\nonumber
\\ 
\tikzzbox{%
	\begin{tikzpicture}[thick,scale=1.0,color=black, baseline=-0.9cm, yscale=-1]
	\coordinate (p1) at (0,0);
	\coordinate (p2) at (0.5,0.2);
	\coordinate (p3) at (-0.5,0.4);
	\coordinate (p4) at (2.5,0);
	\coordinate (p5) at (1.5,0.2);
	\coordinate (p6) at (2.0,0.4);
	\coordinate (s) at ($(0,1.5)$);
	\coordinate (d) at ($(3.8,0)$);
	\coordinate (h) at ($(0,2.5)$);
	\coordinate (t) at ($(0.1,0)$);
	\coordinate (u) at ($(0,-1.1)$);
	%
	\fill[red!80!black] ($(p1)+(t)$) circle (0pt) node[left] {{\small $\xu'$}};
	\fill[red!80!black] ($(p2)-0.5*(t)$) circle (0pt) node[right] {{\small $\au'$}};
	\fill[red!80!black] ($(p3)+(t)$) circle (0pt) node[left] {{\small $\yu'$}};
	\fill[red!80!black] ($(p4)-0.5*(t)$) circle (0pt) node[right] {{\small $\xu$}};
	\fill[red!80!black] ($(p5)+0.5*(t)$) circle (0pt) node[left] {{\small $\au\vphantom{\au'}$}};
	\fill[red!80!black] ($(p4)+(s)-0.5*(t)$) circle (0pt) node[right] {{\small $\xu$}};
	\fill[red!80!black] ($(p3)+(s)+(1.25,0.2)$) circle (0pt) node {{\small $\cu$}};
	\fill[red!80!black] ($(p1)+(s)+(1.25,0.15)$) circle (0pt) node {{\small $\bu$}};
	\fill[red!80!black] ($(p1)+(s)+(t)$) circle (0pt) node[left] {{\small $\xu'$}};
	\fill[red!80!black] ($(p3)+(s)+(t)$) circle (0pt) node[left] {{\small $\yu'$}};
	\fill[red!80!black] ($(p6)+(s)-0.5*(t)$) circle (0pt) node[right] {{\small $\yu$}};
	%
	\fill [orange!20!white, opacity=0.8] 
	(p1) .. controls +(0.3,0) and +(0,-0.1) ..  (p2)
	-- (p2) .. controls +(0,0.9) and +(0,0.9) ..  (p5)
	-- (p5) .. controls +(0,-0.1) and +(-0.5,0) ..  (p4) 
	-- (p4) -- ($(p4)+(s)$) -- ($(p1)+(s)$);
	%
	\draw[very thick, red!80!black] ($(p4)+(s)$) -- ($(p1)+(s)$);
	\draw[thin] (p1) -- ($(p1)+(s)$);
	%
	\fill [orange!30!white, opacity=0.8] 
	(p3) .. controls +(0.5,0) and +(0,0.1) ..  (p2)
	-- (p2) .. controls +(0,0.9) and +(0,0.9) ..  (p5)
	-- (p5) .. controls +(0,0.1) and +(-0.3,0) .. (p6)
	-- (p6) -- ($(p6)+(s)$) -- ($(p3)+(s)$);
	%
	\draw[very thick, red!80!black] (p1) .. controls +(0.3,0) and +(0,-0.1) ..  (p2); 
	\draw[very thick, red!80!black] (p2) .. controls +(0,0.1) and +(0.5,0) ..  (p3); 
	\draw[very thick, red!80!black] (p6) .. controls +(-0.3,0) and +(0,0.1) ..  (p5); 
	\draw[very thick, red!80!black] (p1) .. controls +(0.3,0) and +(0,-0.1) ..  (p2); 
	\draw[very thick, red!80!black] (p5) .. controls +(0,-0.1) and +(-0.5,0) ..  (p4); 
	\draw[very thick, red!80!black] ($(p6)+(s)$) -- ($(p3)+(s)$);
	%
	\draw[thin] (p6) -- ($(p6)+(s)$);
	\draw (p2) .. controls +(0,0.9) and +(0,0.9) ..  (p5);
	\draw[thin] (p3) -- ($(p3)+(s)$);
	\draw[thin] (p4) -- ($(p4)+(s)$);
	\fill[red!80!black] ($(p6)-0.5*(t)$) circle (0pt) node[right] {{\small $\yu$}};
	\end{tikzpicture}
}%
& 
= \tcoev_{\tev_\xu} 
= \coev_{\tcoev_\xu}
\nonumber \\ 
& = \big[ \cu-\au, \, \yu'-\yu \big] \otimes \big[ \bu-\au, \, \xu-\xu' \big] \nonumber \\
&\qquad\otimes \big[ \yu'-\xu' ,\, \au-\au' \big]\{\rs-\rnh,\qs-\qnh+2\} \, .\label{eq:DualityDatagr}
\end{align}
Here, the relative twists of  the southern hemisphere and the saddle are fixed by requiring the Zorro move~\eqref{eq:3dZorro} to hold in the graded setting. 
Similarly, the relative twists of the northern hemisphere and the upside-down saddle are fixed by the other Zorro move. 

Indeed, using the fact \eqref{eq:twistsareadditive} that the twist behaves additively under the tensor product, it is straightforward to verify that all relations required to show that $\bi$ is symmetric monoidal and that every object is fully dualisable also holds in $\bigra$. 
Moreover, also in the graded case there are precisely two trivialisations $S_\xu\cong 1_\xu$ of the Serre automorphism.

\section{Fully extended oriented TQFTs} 
\label{sec:FullyExtendedTQFT}

In this section we construct truncated affine Rozansky--Witten models as fully extended oriented TQFTs. 
We begin in Section~\ref{subsec:CobordismHypothesis} with a review of the latter, where we recall the classification in terms of fully dualisable objects and $\SO(2)$-homotopy fixed points, and how to explicitly compute the entire TQFT in terms of these elementary data. 
In Section~\ref{subsec:ExtendedAffineRW}, using our results from Section~\ref{sec:BicategoriesTruncatedRW}, we apply the general theory to construct ungraded TQFTs~$\zz_n$ with target category~$\bi$.
These are lifted in Section~\ref{subsec:ExtendedAffineRWWithCharges} to extended TQFTs~$\zzgra_n$ with target~$\bigra$, which incorporate flavour and R-charges.
In particular, we use the cobordism hypothesis to compute the graded vector spaces that~$\zzgra_n$ assigns to closed surfaces, cf.\ Corollary~\ref{cor:StateSpacesGradedCase} and~\eqref{eq:generatingfunction}.

\subsection{Cobordism hypothesis}
\label{subsec:CobordismHypothesis} 

The cobordism hypothesis \cite{BDpaper, l0905.0465, AyalaFrancis2017CH} identifies the fundamental building blocks of fully extended TQFTs and explains how to construct state spaces and partition functions from these data. 
Here we review what this means in the 2-dimensional oriented case, following \cite{spthesis, HSV, HV, Hesse}. 

\medskip 

As laid out in \cite[Sect.\,3.1--3.2]{spthesis}, there is a symmetric monoidal 2-category of oriented bordisms $\Bordor$. 
Its objects, 1- and 2-morphisms are 2-haloed compact 0-dimensional manifolds, 2-haloed compact 1-dimensional bordisms and diffeomorphism classes of compact 2-dimensional bordisms with corners, respectively, all with prescribed orientations. 
Horizontal and vertical composition is given by appropriate gluing of bordisms, and the monoidal structure is induced from disjoint union. 

We will have no need to deal with haloes explicitly. 
Hence we will treat objects of $\Bordor$ as finite disjoint unions of positively and negatively oriented points~$+$ and~$-$, respectively, while 1- and 2-morphisms are (represented by) oriented 1- and 2-dimensional bordisms. 
For example, the oriented interval $[0,1]$ can be lifted to the three distinct 1-morphisms 
\be 
\tikzzbox{%
	\begin{tikzpicture}[thick,scale=1.5, color=blue!50!black, baseline=-0.1cm]
	\coordinate (p1) at (0,0);
	\coordinate (p2) at (0.5,0);
	%
	\draw[very thick, red!80!black] (p1) -- (p2); 
	%
	\fill[red!80!black] ($(p2)+(0.15,0)$) circle (0pt) node {{\tiny $+$}};
	\fill[red!80!black] ($(p1)+(-0.15,0)$) circle (0pt) node {{\tiny $+$}};
	%
	%
	\end{tikzpicture}
} 
\colon +\lra +
\, , \quad 
\tikzzbox{%
	\begin{tikzpicture}[thick,scale=1.5,color=black, baseline=0.1cm]
	\coordinate (p1) at (0,0);
	\coordinate (p2) at (0,0.25);
	\draw[very thick, red!80!black] (p1) .. controls +(-0.3,0) and +(-0.3,0) ..  (p2); 
	\fill[red!80!black] (0.3,0) circle (0pt) node[left] {{\tiny$-$}};
	\fill[red!80!black] (0.3,0.25) circle (0pt) node[left] {{\tiny$+$}};
	\end{tikzpicture}
}
\colon +\sqcup - \lra \varnothing
\, , \quad 
\tikzzbox{%
	\begin{tikzpicture}[thick,scale=1.5,color=black, baseline=0.1cm]
	\coordinate (p1) at (0,0);
	\coordinate (p2) at (0,0.25);
	\draw[very thick, red!80!black] (p1) .. controls +(0.3,0) and +(0.3,0) ..  (p2); 
	\fill[red!80!black] (0.05,0) circle (0pt) node[left] {{\tiny$-$}};
	\fill[red!80!black] (0.05,0.25) circle (0pt) node[left] {{\tiny$+$}};
	\end{tikzpicture}
} 
\colon \varnothing \lra +\sqcup - \, , 
\ee 
where here and below we leave the orientation of bordisms implicit. 
Horizontally composing the latter two, we obtain the oriented circle~$S^1$, 
\be 
\tikzzbox{%
	\begin{tikzpicture}[thick,scale=1.5,color=black, baseline=0.1cm]
	\coordinate (p1) at (0,0);
	\coordinate (p2) at (0,0.25);
	\draw[very thick, red!80!black] (p1) .. controls +(-0.3,0) and +(-0.3,0) ..  (p2); 
	\fill[red!80!black] (0.3,0) circle (0pt) node[left] {{\tiny$-$}};
	\fill[red!80!black] (0.3,0.25) circle (0pt) node[left] {{\tiny$+$}};
	\end{tikzpicture}
} 
\circ 
\tikzzbox{%
	\begin{tikzpicture}[thick,scale=1.5,color=black, baseline=0.1cm]
	\coordinate (p1) at (0,0);
	\coordinate (p2) at (0,0.25);
	\draw[very thick, red!80!black] (p1) .. controls +(0.3,0) and +(0.3,0) ..  (p2); 
	\fill[red!80!black] (0.05,0) circle (0pt) node[left] {{\tiny$-$}};
	\fill[red!80!black] (0.05,0.25) circle (0pt) node[left] {{\tiny$+$}};
	\end{tikzpicture}
} 
= 
\tikzzbox{%
\begin{tikzpicture}[thick,scale=1.5,color=black, baseline=0.1cm]
\coordinate (p1) at (0,0);
\coordinate (p2) at (0,0.25);
\coordinate (g1) at (0,-0.05);
\coordinate (g2) at (0,0.3);		%
\draw[very thick, red!80!black] (p1) .. controls +(-0.2,0) and +(-0.2,0) ..  (p2); 
\draw[very thick, red!80!black] (p1) .. controls +(0.2,0) and +(0.2,0) ..  (p2); 
\end{tikzpicture}
}
\colon \varnothing \lra \varnothing \, . 
\ee 

Two examples of 2-morphisms in $\Bordor$ are
\be 
\label{eq:SaddleCap}
\tikzzbox{%
	\begin{tikzpicture}[thick,scale=1.0,color=black, baseline=-0.9cm, yscale=-1]
	\coordinate (p1) at (0,0);
	\coordinate (p2) at (0.5,0.2);
	\coordinate (p3) at (-0.5,0.4);
	\coordinate (p4) at (2.5,0);
	\coordinate (p5) at (1.5,0.2);
	\coordinate (p6) at (2.0,0.4);
	\coordinate (s) at ($(0,1.5)$);
	\coordinate (d) at ($(3.8,0)$);
	\coordinate (h) at ($(0,2.5)$);
	\coordinate (t) at ($(0.1,0)$);
	\coordinate (u) at ($(0,-1.1)$);
	%
	\fill[red!80!black] ($(p1)+(t)$) circle (0pt) node[left] {{\small $+$}};
	\fill[red!80!black] ($(p3)+(t)$) circle (0pt) node[left] {{\small $-$}};
	\fill[red!80!black] ($(p4)-0.5*(t)$) circle (0pt) node[right] {{\small $+$}};
	\fill[red!80!black] ($(p4)+(s)-0.5*(t)$) circle (0pt) node[right] {{\small $+$}};
	\fill[red!80!black] ($(p1)+(s)+(t)$) circle (0pt) node[left] {{\small $+$}};
	\fill[red!80!black] ($(p3)+(s)+(t)$) circle (0pt) node[left] {{\small $-$}};
	\fill[red!80!black] ($(p6)+(s)-0.5*(t)$) circle (0pt) node[right] {{\small $-$}};
	%
	\fill [orange!20!white, opacity=0.8] 
	(p1) .. controls +(0.3,0) and +(0,-0.1) ..  (p2)
	-- (p2) .. controls +(0,0.9) and +(0,0.9) ..  (p5)
	-- (p5) .. controls +(0,-0.1) and +(-0.5,0) ..  (p4) 
	-- (p4) -- ($(p4)+(s)$) -- ($(p1)+(s)$);
	%
	\draw[very thick, red!80!black] ($(p4)+(s)$) -- ($(p1)+(s)$);
	\draw[thin] (p1) -- ($(p1)+(s)$);
	%
	\fill [orange!30!white, opacity=0.8] 
	(p3) .. controls +(0.5,0) and +(0,0.1) ..  (p2)
	-- (p2) .. controls +(0,0.9) and +(0,0.9) ..  (p5)
	-- (p5) .. controls +(0,0.1) and +(-0.3,0) .. (p6)
	-- (p6) -- ($(p6)+(s)$) -- ($(p3)+(s)$);
	%
	\draw[very thick, red!80!black] (p1) .. controls +(0.3,0) and +(0,-0.1) ..  (p2); 
	\draw[very thick, red!80!black] (p2) .. controls +(0,0.1) and +(0.5,0) ..  (p3); 
	\draw[very thick, red!80!black] (p6) .. controls +(-0.3,0) and +(0,0.1) ..  (p5); 
	\draw[very thick, red!80!black] (p1) .. controls +(0.3,0) and +(0,-0.1) ..  (p2); 
	\draw[very thick, red!80!black] (p5) .. controls +(0,-0.1) and +(-0.5,0) ..  (p4); 
	\draw[very thick, red!80!black] ($(p6)+(s)$) -- ($(p3)+(s)$);
	%
	\draw[thin] (p6) -- ($(p6)+(s)$);
	\draw (p2) .. controls +(0,0.9) and +(0,0.9) ..  (p5);
	\draw[thin] (p3) -- ($(p3)+(s)$);
	\draw[thin] (p4) -- ($(p4)+(s)$);
	\fill[red!80!black] ($(p6)-0.5*(t)$) circle (0pt) node[right] {{\small $-$}};
	\end{tikzpicture}
}%
\colon 	
1_{+\sqcup -} \lra 
\tikzzbox{%
	\begin{tikzpicture}[thick,scale=1.5,color=black, baseline=0.1cm]
	\coordinate (p1) at (0,0);
	\coordinate (p2) at (0,0.25);
	\draw[very thick, red!80!black] (p1) .. controls +(0.3,0) and +(0.3,0) ..  (p2); 
	\fill[red!80!black] (0.05,0) circle (0pt) node[left] {{\tiny$-$}};
	\fill[red!80!black] (0.05,0.25) circle (0pt) node[left] {{\tiny$+$}};
	\end{tikzpicture}
} 
\circ
\tikzzbox{%
	\begin{tikzpicture}[thick,scale=1.5,color=black, baseline=0.1cm]
	\coordinate (p1) at (0,0);
	\coordinate (p2) at (0,0.25);
	\draw[very thick, red!80!black] (p1) .. controls +(-0.3,0) and +(-0.3,0) ..  (p2); 
	\fill[red!80!black] (0.3,0) circle (0pt) node[left] {{\tiny$-$}};
	\fill[red!80!black] (0.3,0.25) circle (0pt) node[left] {{\tiny$+$}};
	\end{tikzpicture}
}  
	\, , \quad 
	\tikzzbox{%
		\begin{tikzpicture}[thick,scale=1.0,color=black, rotate=180, baseline=0.5cm]
		\coordinate (p1) at (-2.75,0);
		\coordinate (p2) at (-1,0);
		%
		\draw[very thick, red!80!black] (p1) .. controls +(0,-0.5) and +(0,-0.5) ..  (p2); 
		%
		\fill [orange!20!white, opacity=0.8] 
		(p1) .. controls +(0,-0.5) and +(0,-0.5) ..  (p2)
		-- (p2) .. controls +(0,0.5) and +(0,0.5) ..  (p1)
		;
		\fill [orange!30!white, opacity=0.8] 
		(p1) .. controls +(0,-0.5) and +(0,-0.5) ..  (p2)
		-- (p2) .. controls +(0,-2) and +(0,-2) ..  (p1)
		;
		\coordinate (q1) at (-1.875,0.38);
		\coordinate (q2) at (-2.67,-0.7);
		\draw (p1) .. controls +(0,-2) and +(0,-2) ..  (p2); 
		%
		\draw[very thick, red!80!black] (p1) .. controls +(0,0.5) and +(0,0.5) ..  (p2); 
		\end{tikzpicture}
	}%
	\colon S^1 \lra \varnothing \, . 
\ee 
We observe that the graphical calculus in $\Bordor$ precisely captures the intuition of horizontally and vertically glueing surfaces with corners. 
Moreover, it follows that every object is fully dualisable, with $+^\dual = -$ and 
\begin{align}
\tikzzbox{%
	\begin{tikzpicture}[thick,scale=1.5,color=black, baseline=0.1cm]
	\coordinate (p1) at (0,0);
	\coordinate (p2) at (0,0.25);
	\draw[very thick, red!80!black] (p1) .. controls +(-0.3,0) and +(-0.3,0) ..  (p2); 
	\fill[red!80!black] (0.3,0) circle (0pt) node[left] {{\tiny$-$}};
	\fill[red!80!black] (0.3,0.25) circle (0pt) node[left] {{\tiny$+$}};
	\end{tikzpicture}
}
	& 
	  = \tev_+ 
	  = \ev_-
	  = {}^\dagger \!\coev_+ 
	  = \coev_+^\dagger
	  = {}^\dagger\! \tcoev_- 
	  = \tcoev_-^\dagger \, , 
	  \label{eq:StrictBord1} 
	  \\ 
\tikzzbox{%
	\begin{tikzpicture}[thick,scale=1.5,color=black, baseline=0.1cm]
	\coordinate (p1) at (0,0);
	\coordinate (p2) at (0,0.25);
	\draw[very thick, red!80!black] (p1) .. controls +(0.3,0) and +(0.3,0) ..  (p2); 
	\fill[red!80!black] (0.05,0) circle (0pt) node[left] {{\tiny$-$}};
	\fill[red!80!black] (0.05,0.25) circle (0pt) node[left] {{\tiny$+$}};
	\end{tikzpicture}
}  
	& 	
	  = \coev_+ 
	  = \tcoev_-
	  = {}^\dagger \!\tev_+ 
  	  = \tev_+^\dagger 
  	  = {}^\dagger\! \ev_-
	  = \ev_-^\dagger \, ,
	\label{eq:StrictBord2} 
\end{align}
while adjunction 2-morphisms are given by saddles and caps as in~\eqref{eq:SaddleCap} and their upside-down versions: 
\begin{align}
\tikzzbox{%
	\begin{tikzpicture}[thick,scale=1.0,color=black, baseline=0.9cm]
	\coordinate (p1) at (0,0);
	\coordinate (p2) at (0.5,0.2);
	\coordinate (p3) at (-0.5,0.4);
	\coordinate (p4) at (2.5,0);
	\coordinate (p5) at (1.5,0.2);
	\coordinate (p6) at (2.0,0.4);
	\coordinate (s) at ($(0,1.5)$);
	\coordinate (d) at ($(3.8,0)$);
	\coordinate (h) at ($(0,2.5)$);
	\coordinate (t) at ($(0.1,0)$);
	\coordinate (u) at ($(0,-1.1)$);
	%
	\fill [orange!20!white, opacity=0.8] 
	(p3) .. controls +(0.5,0) and +(0,0.1) ..  (p2)
	-- (p2) .. controls +(0,0.9) and +(0,0.9) ..  (p5)
	-- (p5) .. controls +(0,0.1) and +(-0.3,0) .. (p6)
	-- (p6) -- ($(p6)+(s)$) -- ($(p3)+(s)$);
	%
	\fill[red!80!black] ($(p1)+(t)$) circle (0pt) node[left] {{\small $-$}};
	\fill[red!80!black] ($(p3)+(t)$) circle (0pt) node[left] {{\small $+$}};
	\fill[red!80!black] ($(p4)-0.5*(t)$) circle (0pt) node[right] {{\small $-$}};
	\fill[red!80!black] ($(p6)-0.5*(t)$) circle (0pt) node[right] {{\small $+$}};
	\fill[red!80!black] ($(p4)+(s)-0.5*(t)$) circle (0pt) node[right] {{\small $-$}};
	\fill[red!80!black] ($(p1)+(s)+(t)$) circle (0pt) node[left] {{\small $-$}};
	\fill[red!80!black] ($(p3)+(s)+(t)$) circle (0pt) node[left] {{\small $+$}};
	\fill[red!80!black] ($(p6)+(s)-0.5*(t)$) circle (0pt) node[right] {{\small $+$}};
	%
	%
	\draw[very thick, red!80!black] (p2) .. controls +(0,0.1) and +(0.5,0) ..  (p3); 
	\draw[very thick, red!80!black] (p6) .. controls +(-0.3,0) and +(0,0.1) ..  (p5); 
	\draw[thin] (p6) -- ($(p6)+(s)$);
	%
	\fill [orange!30!white, opacity=0.8] 
	(p1) .. controls +(0.3,0) and +(0,-0.1) ..  (p2)
	-- (p2) .. controls +(0,0.9) and +(0,0.9) ..  (p5)
	-- (p5) .. controls +(0,-0.1) and +(-0.5,0) ..  (p4) 
	-- (p4) -- ($(p4)+(s)$) -- ($(p1)+(s)$);
	%
	\draw[very thick, red!80!black] (p1) .. controls +(0.3,0) and +(0,-0.1) ..  (p2); 
	\draw[very thick, red!80!black] (p5) .. controls +(0,-0.1) and +(-0.5,0) ..  (p4); 
	\draw[very thick, red!80!black] ($(p4)+(s)$) -- ($(p1)+(s)$);
	\draw[very thick, red!80!black] ($(p6)+(s)$) -- ($(p3)+(s)$);
	%
	\draw (p2) .. controls +(0,0.9) and +(0,0.9) ..  (p5);
	\draw[thin] (p1) -- ($(p1)+(s)$);
	\draw[thin] (p3) -- ($(p3)+(s)$);
	\draw[thin] (p4) -- ($(p4)+(s)$);
	\end{tikzpicture}
}%
	& 
	  = \ev_{\tev_+} 
	  = \tev_{\tcoev_+} \, , 
	  \label{eq:BordSaddleEv}
	  \\ 
\tikzzbox{%
	\begin{tikzpicture}[thick,scale=1.0,color=black, baseline=-0.9cm]
	\coordinate (p1) at (-2.75,0);
	\coordinate (p2) at (-1,0);
	\fill [orange!20!white, opacity=0.8] 
	(p1) .. controls +(0,-0.5) and +(0,-0.5) ..  (p2)
	-- (p2) .. controls +(0,0.5) and +(0,0.5) ..  (p1)
	;
	\fill [orange!30!white, opacity=0.8] 
	(p1) .. controls +(0,-0.5) and +(0,-0.5) ..  (p2)
	-- (p2) .. controls +(0,-2) and +(0,-2) ..  (p1)
	;
	\draw (p1) .. controls +(0,-2) and +(0,-2) ..  (p2); 
	\draw[very thick, red!80!black] (p1) .. controls +(0,0.5) and +(0,0.5) ..  (p2); 
	\draw[very thick, red!80!black] (p1) .. controls +(0,-0.5) and +(0,-0.5) ..  (p2); 
	\end{tikzpicture}
}%
	& 
	  = \coev_{\tev_+} 
	  = \tcoev_{\tcoev_+} \, , 
	\\
\tikzzbox{%
	\begin{tikzpicture}[thick,scale=1.0,color=black, rotate=180, baseline=0.5cm]
	\coordinate (p1) at (-2.75,0);
	\coordinate (p2) at (-1,0);
	%
	\draw[very thick, red!80!black] (p1) .. controls +(0,-0.5) and +(0,-0.5) ..  (p2); 
	%
	\fill [orange!20!white, opacity=0.8] 
	(p1) .. controls +(0,-0.5) and +(0,-0.5) ..  (p2)
	-- (p2) .. controls +(0,0.5) and +(0,0.5) ..  (p1)
	;
	\fill [orange!30!white, opacity=0.8] 
	(p1) .. controls +(0,-0.5) and +(0,-0.5) ..  (p2)
	-- (p2) .. controls +(0,-2) and +(0,-2) ..  (p1)
	;
	\coordinate (q1) at (-1.875,0.38);
	\coordinate (q2) at (-2.67,-0.7);
	\draw (p1) .. controls +(0,-2) and +(0,-2) ..  (p2); 
	%
	\draw[very thick, red!80!black] (p1) .. controls +(0,0.5) and +(0,0.5) ..  (p2); 
	\end{tikzpicture}
}%
	& 
	  = \tev_{\tev_+} 
	  = \ev_{\tcoev_+} \, , 
	  \\
\tikzzbox{%
	\begin{tikzpicture}[thick,scale=1.0,color=black, baseline=-0.9cm, yscale=-1]
	\coordinate (p1) at (0,0);
	\coordinate (p2) at (0.5,0.2);
	\coordinate (p3) at (-0.5,0.4);
	\coordinate (p4) at (2.5,0);
	\coordinate (p5) at (1.5,0.2);
	\coordinate (p6) at (2.0,0.4);
	\coordinate (s) at ($(0,1.5)$);
	\coordinate (d) at ($(3.8,0)$);
	\coordinate (h) at ($(0,2.5)$);
	\coordinate (t) at ($(0.1,0)$);
	\coordinate (u) at ($(0,-1.1)$);
	%
	\fill[red!80!black] ($(p1)+(t)$) circle (0pt) node[left] {{\small $+$}};
	\fill[red!80!black] ($(p3)+(t)$) circle (0pt) node[left] {{\small $-$}};
	\fill[red!80!black] ($(p4)-0.5*(t)$) circle (0pt) node[right] {{\small $+$}};
	\fill[red!80!black] ($(p4)+(s)-0.5*(t)$) circle (0pt) node[right] {{\small $+$}};
	\fill[red!80!black] ($(p1)+(s)+(t)$) circle (0pt) node[left] {{\small $+$}};
	\fill[red!80!black] ($(p3)+(s)+(t)$) circle (0pt) node[left] {{\small $-$}};
	\fill[red!80!black] ($(p6)+(s)-0.5*(t)$) circle (0pt) node[right] {{\small $-$}};
	%
	\fill [orange!20!white, opacity=0.8] 
	(p1) .. controls +(0.3,0) and +(0,-0.1) ..  (p2)
	-- (p2) .. controls +(0,0.9) and +(0,0.9) ..  (p5)
	-- (p5) .. controls +(0,-0.1) and +(-0.5,0) ..  (p4) 
	-- (p4) -- ($(p4)+(s)$) -- ($(p1)+(s)$);
	%
	\draw[very thick, red!80!black] ($(p4)+(s)$) -- ($(p1)+(s)$);
	\draw[thin] (p1) -- ($(p1)+(s)$);
	%
	\fill [orange!30!white, opacity=0.8] 
	(p3) .. controls +(0.5,0) and +(0,0.1) ..  (p2)
	-- (p2) .. controls +(0,0.9) and +(0,0.9) ..  (p5)
	-- (p5) .. controls +(0,0.1) and +(-0.3,0) .. (p6)
	-- (p6) -- ($(p6)+(s)$) -- ($(p3)+(s)$);
	%
	\draw[very thick, red!80!black] (p1) .. controls +(0.3,0) and +(0,-0.1) ..  (p2); 
	\draw[very thick, red!80!black] (p2) .. controls +(0,0.1) and +(0.5,0) ..  (p3); 
	\draw[very thick, red!80!black] (p6) .. controls +(-0.3,0) and +(0,0.1) ..  (p5); 
	\draw[very thick, red!80!black] (p1) .. controls +(0.3,0) and +(0,-0.1) ..  (p2); 
	\draw[very thick, red!80!black] (p5) .. controls +(0,-0.1) and +(-0.5,0) ..  (p4); 
	\draw[very thick, red!80!black] ($(p6)+(s)$) -- ($(p3)+(s)$);
	%
	\draw[thin] (p6) -- ($(p6)+(s)$);
	\draw (p2) .. controls +(0,0.9) and +(0,0.9) ..  (p5);
	\draw[thin] (p3) -- ($(p3)+(s)$);
	\draw[thin] (p4) -- ($(p4)+(s)$);
	\fill[red!80!black] ($(p6)-0.5*(t)$) circle (0pt) node[right] {{\small $-$}};
	\end{tikzpicture}
}%
	& 
	  = \tcoev_{\tev_+} 
	  = \coev_{\tev_+} \, .
	\label{eq:BordSaddleCoev}
\end{align}
Note that all but the first equalities in each line of \eqref{eq:StrictBord1}--\eqref{eq:BordSaddleCoev} are specific to the case of orientations. 
For example, these identities do not hold for the tangential structure of framings, i.\,e.\ in the associated 2-category $\Bordfr$ studied in detail in \cite{Pstragowski}; in this sense our oriented setting with $\Bordor$ is simpler. 
We also observe that every compact oriented surface can be (non-uniquely) decomposed into the pieces \eqref{eq:BordSaddleEv}--\eqref{eq:BordSaddleCoev}. 
This will be used repeatedly below, see e.\,g.\ \eqref{eq:TorusDecomposed} where a decomposition of the torus~$T^2$ is shown. 

\medskip 

Next we recall the type of TQFT that we will construct examples of. 
Let~$\B$ be a symmetric monoidal 2-category. 
A \textsl{2-dimensional (fully) extended oriented TQFT with values in~$\B$} is a symmetric monoidal 2-functor 
\be 
\zz \colon \Bordor \lra \B \, . 
\ee 
We denote the 2-category of such TQFTs as $\Fun(\Bordor,\B)$. 
The fundamental classification result for these TQFTs is in terms of $\SO(2)$-homotopy fixed points in the maximal sub-2-groupoid $(\Bfd)^\times$ of fully dualisable objects (cf.\ Section~\ref{subsubsec:SerreAutomorphisms} for the latter). 
These homotopy fixed points form the objects of a 2-groupoid $[(\Bfd)^\times]^{\SO(2)}$. 
Building on \cite{HSV, HV} the 2-dimensional oriented cobordism hypothesis was proven in \cite[Cor.\,5.9]{Hesse} as the equivalence of symmetric monoidal 2-categories 
\be 
\label{eq:OrCH1}
\Fun \big( \Bordor, \, \B \big) 
	\cong 
	\big[(\Bfd)^\times\big]^{\SO(2)} \, . 
\ee 
Instead of directly describing the right-hand side in terms of homotopy actions of $\SO(2)$ on~$\B$, i.\,e.\ monoidal 2-functors $\Pi_{\leqslant 2}(\SO(2)) \lra \Aut^{\textrm{sm}}(\B)$, we define an equivalent 2-groupoid $\Bhfp{\B}$, following \cite[Thm.\,4.3]{HV}.\footnote{The 2-groupoid $\Bhfp{\B}$ is identical to the one denoted $\textrm{2D}^1((\Bfd)^\times)$ in \cite[Sect.\,3.3.3]{RSpinLorantNils}.} 
This is also the 2-groupoid in terms of which we will study the cases when~$\B$ is one of the 2-categories of truncated affine Rozansky--Witten models introduced in Section~\ref{sec:BicategoriesTruncatedRW}: 
\begin{itemize}
	\item 
	Objects of $\Bhfp{\B}$ are pairs $(u,\lambda)$, where $u\in \Bfd$ and $\lambda \colon S_u \lra 1_u$ is a 2-isomorphism in $(\Bfd)^\times$. 
	\item 
	1-morphisms $(u,\lambda) \lra (u',\lambda')$ are 1-morphisms $X\colon u\lra u'$ in $(\Bfd)^\times$ such that the following diagram commutes, where $S_X$ is the 2-morphism component of the pseudonatural transformation~\eqref{eq:SerrePseudonaturalTransformationOnBfdCore} (see \cite[Eq.\,(3.27)]{RSpinLorantNils}): 
	\be 
	\label{eq:IsomorphismOfExtendedTQFTs}
	\begin{tikzpicture}[
	baseline=(current bounding box.base),
	descr/.style={fill=white,inner sep=3.5pt},
	normal line/.style={->}
	]
	\matrix (m) [matrix of math nodes, row sep=1em, column sep=5em, text height=1.5ex, text depth=0.1ex] {%
		X\otimes S_{u}&X\otimes 1_{u}&
		\\
		&&X
		\\
		S_{u'}\otimes X&1_{u'}\otimes X&
		\\
	};
	\path[font=\footnotesize] (m-1-1) edge[->] node[above,sloped] {$1_{X}\otimes \lambda$} (m-1-2);
	\path[font=\footnotesize] (m-3-1) edge[->] node[below,sloped] {$\lambda' \otimes 1_{X}$} (m-3-2);
	\path[font=\footnotesize] (m-1-1) edge[->] node[left] {$ S_{X} $} (m-3-1);
	\path[font=\footnotesize] (m-1-2) edge[->] node[above,sloped] {$ \cong $} (m-2-3);
	\path[font=\footnotesize] (m-3-2) edge[->] node[below,sloped] {$ \cong $} (m-2-3);
	\end{tikzpicture}
	\ee 
	\item 
	2-morphisms $X\lra Y$ in $\Bhfp{\B}$ are just 2-morphisms $X\lra Y$ in $(\Bfd)^\times$. 
	\item 
	Composition and units of $\Bhfp{\B}$ are induced from $(\Bfd)^\times$. 
\end{itemize}

Hence an object in $\Bhfp{\B}$ is a fully dualisable object $u\in\B$ together with a trivialisation~$\lambda$ of its Serre automorphism. 
Very roughly, the equivalence $[(\Bfd)^\times]^{\SO(2)} \cong \Bhfp{\B}$ comes about by combining the framed cobordism hypothesis (which states that fully extended framed TQFTs are classified by fully dualisable objects in the target category) with the fact that oriented bordisms can be obtained from framed bordisms through quotienting by the relation $S_+ \cong 1_+$ (in the 2-category $\Bordfr$, see \cite{Pstragowski}). 
Using this, the cobordism hypothesis~\eqref{eq:OrCH1} can be reformulated and rendered more explicitly to state that 2-dimensional extended oriented TQFTs with values in~$\B$ are classified by what they assign to the positive point~$+$, together with a trivialisation of the Serre automorphism: 

\begin{theorem}[\cite{HSV, HV, Hesse}]
	\label{thm:OrientedCobordismHypothesis}
	There is an equivalence of symmetric monoidal 2-categories 
	\begin{align}
	\Fun\big( \Bordor, \, \B \big) 
		& \stackrel{\cong}{\lra} \Bhfp{\B} 
		\nonumber
		\\ 
		\zz 
		& \lmt \Big( \zz(+), \, \lambda \colon  S_{\zz(+)}\stackrel{\cong}{\lra} 1_{\zz(+)} \Big) \, . 
	\end{align}
\end{theorem}

To understand that this is really a classification result, we should explain how to (re-)construct a TQFT merely from an object $(u,\lambda) \in \Bhfp{\B}$. 
To do so, we use the fact that, up to isomorphisms, symmetric monoidal 2-functors map adjunction data to adjunction data. 
Hence, up to isomorphisms which will not play a role in our applications in Sections~\ref{subsec:ExtendedAffineRW}--\ref{subsec:ExtendedAffineRWWithCharges}, given $u\in\Bfd$ and $\lambda\colon S_u\stackrel{{}_{\cong\,}}{\lra} 1_u$, we construct an extended TQFT $\zz\colon \Bordor \lra \B$ by setting  
\begin{align}
\zz( 
\tikzzbox{%
	\begin{tikzpicture}[thick,scale=1.5,color=black, baseline=0.06cm]
	\coordinate (p1) at (0,0);
	\coordinate (p2) at (0,0.25);
	\coordinate (g1) at (0,-0.05);
	\coordinate (g2) at (0,0.3);
	\draw[very thick, red!80!black] (p1) .. controls +(-0.3,0) and +(-0.3,0) ..  (p2); 
	\fill[red!80!black] (0.3,0) circle (0pt) node[left] {{\tiny$-$}};
	\fill[red!80!black] (0.3,0.25) circle (0pt) node[left] {{\tiny$+$}};
	\end{tikzpicture}
}
) 
= \zz(\tev_+) 
	& = \tev_u \, , 
	\\
\zz( 
\tikzzbox{%
	\begin{tikzpicture}[thick,scale=1.5,color=black, baseline=0.06cm]
	\coordinate (p1) at (0,0);
	\coordinate (p2) at (0,0.25);
	\coordinate (g1) at (0,-0.05);
	\coordinate (g2) at (0,0.3);
	\draw[very thick, red!80!black] (p1) .. controls +(0.3,0) and +(0.3,0) ..  (p2); 
	\fill[red!80!black] (00.07,0) circle (0pt) node[left] {{\tiny$-$}};
	\fill[red!80!black] (0.07,0.25) circle (0pt) node[left] {{\tiny$+$}};
	\end{tikzpicture}
}
)
= \zz(\coev_+) 
	& = \coev_u \, , 
	\label{eq:ZonCoev+}
	\\
\zz\Bigg( \,
\tikzzbox{%
	\begin{tikzpicture}[thick,scale=0.7,color=black, baseline=0.5cm]
	\coordinate (p1) at (0,0);
	\coordinate (p2) at (0.5,0.2);
	\coordinate (p3) at (-0.5,0.4);
	\coordinate (p4) at (2.5,0);
	\coordinate (p5) at (1.5,0.2);
	\coordinate (p6) at (2.0,0.4);
	\coordinate (s) at ($(0,1.5)$);
	\coordinate (d) at ($(3.8,0)$);
	\coordinate (h) at ($(0,2.5)$);
	\coordinate (t) at ($(0.1,0)$);
	\coordinate (u) at ($(0,-1.1)$);
	%
	\fill [orange!20!white, opacity=0.8] 
	(p3) .. controls +(0.5,0) and +(0,0.1) ..  (p2)
	-- (p2) .. controls +(0,0.9) and +(0,0.9) ..  (p5)
	-- (p5) .. controls +(0,0.1) and +(-0.3,0) .. (p6)
	-- (p6) -- ($(p6)+(s)$) -- ($(p3)+(s)$);
	%
	\draw[very thick, red!80!black] (p2) .. controls +(0,0.1) and +(0.5,0) ..  (p3); 
	\draw[very thick, red!80!black] (p6) .. controls +(-0.3,0) and +(0,0.1) ..  (p5); 
	\draw[thin] (p6) -- ($(p6)+(s)$);
	%
	\fill [orange!30!white, opacity=0.8] 
	(p1) .. controls +(0.3,0) and +(0,-0.1) ..  (p2)
	-- (p2) .. controls +(0,0.9) and +(0,0.9) ..  (p5)
	-- (p5) .. controls +(0,-0.1) and +(-0.5,0) ..  (p4) 
	-- (p4) -- ($(p4)+(s)$) -- ($(p1)+(s)$);
	%
	\draw[very thick, red!80!black] (p1) .. controls +(0.3,0) and +(0,-0.1) ..  (p2); 
	\draw[very thick, red!80!black] (p5) .. controls +(0,-0.1) and +(-0.5,0) ..  (p4); 
	\draw[very thick, red!80!black] ($(p4)+(s)$) -- ($(p1)+(s)$);
	\draw[very thick, red!80!black] ($(p6)+(s)$) -- ($(p3)+(s)$);
	%
	\draw (p2) .. controls +(0,0.9) and +(0,0.9) ..  (p5);
	\draw[thin] (p1) -- ($(p1)+(s)$);
	\draw[thin] (p3) -- ($(p3)+(s)$);
	\draw[thin] (p4) -- ($(p4)+(s)$);
	\end{tikzpicture}
}%
\, \Bigg)
= \zz(\ev_{\tev_+})
	& = \ev_{\tev_u} \, , 
	\\ 
\zz\Bigg( \,
\tikzzbox{%
	\begin{tikzpicture}[thick,scale=0.7,color=black, baseline=-0.5cm]
	\coordinate (p1) at (-2.75,0);
	\coordinate (p2) at (-1,0);
	\fill [orange!20!white, opacity=0.8] 
	(p1) .. controls +(0,-0.5) and +(0,-0.5) ..  (p2)
	-- (p2) .. controls +(0,0.5) and +(0,0.5) ..  (p1)
	;
	\fill [orange!30!white, opacity=0.8] 
	(p1) .. controls +(0,-0.5) and +(0,-0.5) ..  (p2)
	-- (p2) .. controls +(0,-2) and +(0,-2) ..  (p1)
	;
	\draw (p1) .. controls +(0,-2) and +(0,-2) ..  (p2); 
	\draw[very thick, red!80!black] (p1) .. controls +(0,0.5) and +(0,0.5) ..  (p2); 
	\draw[very thick, red!80!black] (p1) .. controls +(0,-0.5) and +(0,-0.5) ..  (p2); 
	\end{tikzpicture}
}%
\, \Bigg) 
= \zz ( \coev_{\tev_+} )
	& = \coev_{\tev_u}
		\\[-0.6cm]
	\zz\Bigg( \,
	\tikzzbox{%
		\begin{tikzpicture}[thick,scale=0.7,color=black, rotate=180, baseline=0.3cm]
		\coordinate (p1) at (-2.75,0);
		\coordinate (p2) at (-1,0);
		%
		\draw[very thick, red!80!black] (p1) .. controls +(0,-0.5) and +(0,-0.5) ..  (p2); 
		%
		\fill [orange!20!white, opacity=0.8] 
		(p1) .. controls +(0,-0.5) and +(0,-0.5) ..  (p2)
		-- (p2) .. controls +(0,0.5) and +(0,0.5) ..  (p1)
		;
		\fill [orange!30!white, opacity=0.8] 
		(p1) .. controls +(0,-0.5) and +(0,-0.5) ..  (p2)
		-- (p2) .. controls +(0,-2) and +(0,-2) ..  (p1)
		;
		\coordinate (q1) at (-1.875,0.38);
		\coordinate (q2) at (-2.67,-0.7);
		\draw (p1) .. controls +(0,-2) and +(0,-2) ..  (p2); 
		%
		\draw[very thick, red!80!black] (p1) .. controls +(0,0.5) and +(0,0.5) ..  (p2); 
		\end{tikzpicture}
	}%
	\, \Bigg) 
	= \zz ( \tev_{\tev_+} ) 
	& = \tev_{\tev_u} \, ,
		\\ 
	\zz\Bigg( \,
	\tikzzbox{%
		\begin{tikzpicture}[thick,scale=0.7,color=black, baseline=-0.7cm, yscale=-1]
		\coordinate (p1) at (0,0);
		\coordinate (p2) at (0.5,0.2);
		\coordinate (p3) at (-0.5,0.4);
		\coordinate (p4) at (2.5,0);
		\coordinate (p5) at (1.5,0.2);
		\coordinate (p6) at (2.0,0.4);
		\coordinate (s) at ($(0,1.5)$);
		\coordinate (d) at ($(3.8,0)$);
		\coordinate (h) at ($(0,2.5)$);
		\coordinate (t) at ($(0.1,0)$);
		\coordinate (u) at ($(0,-1.1)$);
		%
		\fill [orange!20!white, opacity=0.8] 
		(p1) .. controls +(0.3,0) and +(0,-0.1) ..  (p2)
		-- (p2) .. controls +(0,0.9) and +(0,0.9) ..  (p5)
		-- (p5) .. controls +(0,-0.1) and +(-0.5,0) ..  (p4) 
		-- (p4) -- ($(p4)+(s)$) -- ($(p1)+(s)$);
		%
		\draw[very thick, red!80!black] ($(p4)+(s)$) -- ($(p1)+(s)$);
		\draw[thin] (p1) -- ($(p1)+(s)$);
		%
		\fill [orange!30!white, opacity=0.8] 
		(p3) .. controls +(0.5,0) and +(0,0.1) ..  (p2)
		-- (p2) .. controls +(0,0.9) and +(0,0.9) ..  (p5)
		-- (p5) .. controls +(0,0.1) and +(-0.3,0) .. (p6)
		-- (p6) -- ($(p6)+(s)$) -- ($(p3)+(s)$);
		%
		\draw[very thick, red!80!black] (p1) .. controls +(0.3,0) and +(0,-0.1) ..  (p2); 
		\draw[very thick, red!80!black] (p2) .. controls +(0,0.1) and +(0.5,0) ..  (p3); 
		\draw[very thick, red!80!black] (p6) .. controls +(-0.3,0) and +(0,0.1) ..  (p5); 
		\draw[very thick, red!80!black] (p1) .. controls +(0.3,0) and +(0,-0.1) ..  (p2); 
		\draw[very thick, red!80!black] (p5) .. controls +(0,-0.1) and +(-0.5,0) ..  (p4); 
		\draw[very thick, red!80!black] ($(p6)+(s)$) -- ($(p3)+(s)$);
		%
		\draw[thin] (p6) -- ($(p6)+(s)$);
		\draw (p2) .. controls +(0,0.9) and +(0,0.9) ..  (p5);
		\draw[thin] (p3) -- ($(p3)+(s)$);
		\draw[thin] (p4) -- ($(p4)+(s)$);
		\end{tikzpicture}
	}%
	\, \Bigg) 
	= \zz ( \tcoev_{\tev_+} )
	& = \tcoev_{\tev_u} \, . 
\end{align}
As~$\zz$ sends units to units, this determines~$\zz$ on any compact oriented surface. 
For example, on a sphere we have 
\be 
\label{eq:ZonSphere}
\zz(S^2) 
	= \zz \Bigg(
	\tikzzbox{%
		\begin{tikzpicture}[thick,scale=0.5,color=black, rotate=180, baseline=-0.3cm]
		%
		%
		\coordinate (q1) at (-2.75,0.95);
		\coordinate (q2) at (-1,0.95);
		\fill [orange!20!white, opacity=0.8] 
		(q1) .. controls +(0,0.5) and +(0,0.5) ..  (q2)
		-- (q2) .. controls +(0,-0.5) and +(0,-0.5) ..  (q1)
		;
		\fill [orange!30!white, opacity=0.8] 
		(q1) .. controls +(0,0.5) and +(0,0.5) ..  (q2)
		-- (q2) .. controls +(0,1.25) and +(0,1.25) ..  (q1)
		;
		\draw (q1) .. controls +(0,1.25) and +(0,1.25) ..  (q2); 
		\draw[very thick, red!80!black] (q1) .. controls +(0,0.5) and +(0,0.5) ..  (q2); 
		\draw[very thick, red!80!black] (q1) .. controls +(0,-0.5) and +(0,-0.5) ..  (q2); 
		%
		%
		%
		\coordinate (p1) at (-2.75,0);
		\coordinate (p2) at (-1,0);
		%
		\draw[very thick, red!80!black] (p1) .. controls +(0,-0.5) and +(0,-0.5) ..  (p2); 
		%
		\fill [orange!20!white, opacity=0.8] 
		(p1) .. controls +(0,-0.5) and +(0,-0.5) ..  (p2)
		-- (p2) .. controls +(0,0.5) and +(0,0.5) ..  (p1)
		;
		\fill [orange!30!white, opacity=0.8] 
		(p1) .. controls +(0,-0.5) and +(0,-0.5) ..  (p2)
		-- (p2) .. controls +(0,-1.25) and +(0,-1.25) ..  (p1)
		;
		\coordinate (q1) at (-1.875,0.38);
		\coordinate (q2) at (-2.67,-0.7);
		\draw (p1) .. controls +(0,-1.25) and +(0,-1.25) ..  (p2); 
		%
		\draw[very thick, red!80!black] (p1) .. controls +(0,0.5) and +(0,0.5) ..  (p2); 
		\end{tikzpicture}
	}%
	\Bigg)
	= \zz( \tev_{\tev_+} \cdot \coev_{\tev_+} )
	= \tev_{\tev_u} \cdot \coev_{\tev_u} \, , 
\ee 
while on a torus we have
\begin{align}
\zz(T^2) 
& 
= \zz \left( 
	\tikzzbox{%
		\begin{tikzpicture}[thick,scale=1.0,color=black, baseline=-0.2cm]
		\coordinate (p1) at (0,0);
		\coordinate (p2) at (0.5,0.2);
		\coordinate (p3) at (-0.5,0.4);
		\coordinate (p4) at (2.5,0);
		\coordinate (p5) at (1.5,0.2);
		\coordinate (p6) at (2.0,0.4);
		\coordinate (s) at ($(0,1.5)$);
		\coordinate (s2) at ($(0,0.75)$);
		\coordinate (d) at ($(4,0)$);
		\coordinate (h) at ($(0,2.5)$);
		\coordinate (t) at ($(0.1,0)$);
		\coordinate (u) at ($(0,-1.1)$);
		%
		%
		%
		\fill [orange!20!white, opacity=0.8] 
		(p3) .. controls +(0.5,0) and +(0,0.1) ..  (p2)
		-- (p2) .. controls +(0,0.9) and +(0,0.9) ..  (p5)
		-- (p5) .. controls +(0,0.1) and +(-0.3,0) .. (p6)
		-- (p6) -- ($(p6)+(s)$) -- ($(p3)+(s)$);
		%
		\fill[red!80!black] ($(p1)+(t)$) circle (0pt) node[left] {{\small $-$}};
		\fill[red!80!black] ($(p3)+(t)$) circle (0pt) node[left] {{\small $+$}};
		\fill[red!80!black] ($(p4)-0.5*(t)$) circle (0pt) node[right] {{\small $-$}};
		\fill[red!80!black] ($(p6)-0.5*(t)$) circle (0pt) node[right] {{\small $+$}};
		\fill[red!80!black] ($(p4)+(s)-0.5*(t)$) circle (0pt) node[right] {{\small $-$}};
		\fill[red!80!black] ($(p1)+(s)+(t)$) circle (0pt) node[left] {{\small $-$}};
		\fill[red!80!black] ($(p3)+(s)+(t)$) circle (0pt) node[left] {{\small $+$}};
		\fill[red!80!black] ($(p6)+(s)-0.5*(t)$) circle (0pt) node[right] {{\small $+$}};
		%
		%
		\draw[very thick, red!80!black] (p2) .. controls +(0,0.1) and +(0.5,0) ..  (p3); 
		\draw[very thick, red!80!black] (p6) .. controls +(-0.3,0) and +(0,0.1) ..  (p5); 
		\draw[thin] (p6) -- ($(p6)+(s)$);
		%
		\fill [orange!30!white, opacity=0.8] 
		(p1) .. controls +(0.3,0) and +(0,-0.1) ..  (p2)
		-- (p2) .. controls +(0,0.9) and +(0,0.9) ..  (p5)
		-- (p5) .. controls +(0,-0.1) and +(-0.5,0) ..  (p4) 
		-- (p4) -- ($(p4)+(s)$) -- ($(p1)+(s)$);
		%
		\draw[very thick, red!80!black] (p1) .. controls +(0.3,0) and +(0,-0.1) ..  (p2); 
		\draw[very thick, red!80!black] (p5) .. controls +(0,-0.1) and +(-0.5,0) ..  (p4); 
		\draw[very thick, red!80!black] ($(p4)+(s)$) -- ($(p1)+(s)$);
		\draw[very thick, red!80!black] ($(p6)+(s)$) -- ($(p3)+(s)$);
		%
		\draw (p2) .. controls +(0,0.9) and +(0,0.9) ..  (p5);
		\draw[thin] (p1) -- ($(p1)+(s)$);
		\draw[thin] (p3) -- ($(p3)+(s)$);
		\draw[thin] (p4) -- ($(p4)+(s)$);
		%
		%
		%
		%
		\fill [orange!20!white, opacity=0.8] 
		($(p3)-(s2)$) .. controls +(0.5,0) and +(0,0.1) ..  ($(p2)-(s2)$)
		-- ($(p2)-(s2)$) .. controls +(0,-0.9) and +(0,-0.9) ..  ($(p5)-(s2)$)
		-- ($(p5)-(s2)$) .. controls +(0,0.1) and +(-0.3,0) .. ($(p6)-(s2)$)
		-- ($(p6)-(s2)$) -- ($(p6)-(s)-(s2)$) -- ($(p3)-(s)-(s2)$);
		%
		\fill[red!80!black] ($(p1)+(t)-(s2)$) circle (0pt) node[left] {{\small $-$}};
		\fill[red!80!black] ($(p3)+(t)-(s2)$) circle (0pt) node[left] {{\small $+$}};
		\fill[red!80!black] ($(p4)-0.5*(t)-(s2)$) circle (0pt) node[right] {{\small $-$}};
		\fill[red!80!black] ($(p6)-0.5*(t)-(s2)$) circle (0pt) node[right] {{\small $+$}};
		\fill[red!80!black] ($(p4)-(s2)-(s)-0.5*(t)$) circle (0pt) node[right] {{\small $-$}};
		\fill[red!80!black] ($(p1)-(s2)-(s)+(t)$) circle (0pt) node[left] {{\small $-$}};
		\fill[red!80!black] ($(p3)-(s2)-(s)+(t)$) circle (0pt) node[left] {{\small $+$}};
		\fill[red!80!black] ($(p6)-(s2)-(s)-0.5*(t)$) circle (0pt) node[right] {{\small $+$}};
		%
		%
		\draw[very thick, red!80!black] ($(p2)-(s2)$) .. controls +(0,0.1) and +(0.5,0) .. ($(p3)-(s2)$); 
		\draw[very thick, red!80!black] ($(p6)-(s2)$) .. controls +(-0.3,0) and +(0,0.1) .. ($(p5)-(s2)$); 
		\draw[very thick, red!80!black] ($(p6)-(s)-(s2)$) -- ($(p3)-(s)-(s2)$);	
		\draw[thin] ($(p6)-(s2)$) -- ($(p6)-(s)-(s2)$);
		%
		\fill [orange!30!white, opacity=0.8] 
		($(p1)-(s2)$) .. controls +(0.3,0) and +(0,-0.1) ..  ($(p2)-(s2)$)
		-- ($(p2)-(s2)$) .. controls +(0,-0.9) and +(0,-0.9) ..  ($(p5)-(s2)$)
		-- ($(p5)-(s2)$) .. controls +(0,-0.1) and +(-0.5,0) ..  ($(p4)-(s2)$) 
		-- ($(p4)-(s2)$) -- ($(p4)-(s)-(s2)$) -- ($(p1)-(s)-(s2)$);
		%
		\draw[very thick, red!80!black] ($(p1)-(s2)$) .. controls +(0.3,0) and +(0,-0.1) ..  ($(p2)-(s2)$); 
		\draw[very thick, red!80!black] ($(p5)-(s2)$) .. controls +(0,-0.1) and +(-0.5,0) ..  ($(p4)-(s2)$); 
		\draw[very thick, red!80!black] ($(p4)-(s)-(s2)$) -- ($(p1)-(s)-(s2)$); 
		%
		\draw ($(p2)-(s2)$) .. controls +(0,-0.9) and +(0,-0.9) ..  ($(p5)-(s2)$);
		\draw[thin] ($(p1)-(s2)$) -- ($(p1)-(s)-(s2)$);
		\draw[thin] ($(p3)-(s2)$) -- ($(p3)-(s)-(s2)$);
		\draw[thin] ($(p4)-(s2)$) -- ($(p4)-(s)-(s2)$);
		%
		%
		%
		%
		%
		\fill [orange!20!white, opacity=0.8] 
		($(p2)+(d)-(s)-(s2)$) .. controls +(0,0.1) and +(0.5,0) ..  ($(p3)+(d)-(s)-(s2)$)
		-- ($(p3)+(d)-(s)-(s2)$) -- ($(p3)+(d)+(s)$) 
		-- ($(p3)+(d)+(s)$) .. controls +(0.5,0) and +(0,0.1) ..($(p2)+(d)+(s)$); 
		%
		%
		\draw[very thick, red!80!black] ($(p2)+(d)-(s)-(s2)$) .. controls +(0,0.1) and +(0.5,0) ..  ($(p3)+(d)-(s)-(s2)$); 
		\draw[very thick, red!80!black] ($(p2)+(d)+(s)$) .. controls +(0,0.1) and +(0.5,0) ..  ($(p3)+(d)+(s)$); 
		%
		\fill [orange!30!white, opacity=0.8] 
		($(p1)+(d)-(s)-(s2)$) .. controls +(0.3,0) and +(0,-0.1) .. ($(p2)+(d)-(s)-(s2)$)
		-- ($(p2)+(d)-(s)-(s2)$) -- ($(p2)+(d)+(s)$)
		-- ($(p2)+(d)+(s)$) .. controls +(0,-0.1) and +(0.3,0) .. ($(p1)+(d)+(s)$);
		%
		\draw[very thick, red!80!black] ($(p1)+(d)-(s)-(s2)$) .. controls +(0.3,0) and +(0,-0.1) ..  ($(p2)+(d)-(s)-(s2)$); 
		\draw[very thick, red!80!black] ($(p1)+(d)+(s)$) .. controls +(0.3,0) and +(0,-0.1) ..  ($(p2)+(d)+(s)$); 
		%
		\draw[thin] ($(p1)+(d)-(s)-(s2)$) -- ($(p1)+(d)+(s)$);
		\draw[thin] ($(p3)+(d)-(s)-(s2)$) -- ($(p3)+(d)+(s)$);
		\draw[thin] ($(p2)+(d)-(s)-(s2)$) -- ($(p2)+(d)+(s)$);
		%
		\fill[red!80!black] ($(p3)+(s)+(d)+(t)$) circle (0pt) node[left] {{\small $+$}};
		\fill[red!80!black] ($(p1)+(s)+(d)+(t)$) circle (0pt) node[left] {{\small $-$}};
		\fill[red!80!black] ($(p3)-(s2)-(s)+(d)+(t)$) circle (0pt) node[left] {{\small $+$}};
		\fill[red!80!black] ($(p1)-(s2)-(s)+(d)+(t)$) circle (0pt) node[left] {{\small $-$}};
		%
		%
		%
		%
		\fill [orange!20!white, opacity=0.8] 
		($(p5)-(d)-(s)-(s2)$) .. controls +(0,0.1) and +(-0.3,0) .. ($(p6)-(d)-(s)-(s2)$)
		-- ($(p6)-(d)-(s)-(s2)$) -- ($(p6)-(d)+(s)$)
		-- ($(p6)-(d)+(s)$) .. controls +(-0.3,0) and +(0,0.1) .. ($(p5)-(d)+(s)$);
		%
		%
		\draw[thin] ($(p6)-(d)-(s)-(s2)$) -- ($(p6)-(d)+(s)$);
		%
		\draw[very thick, red!80!black] ($(p5)-(d)-(s)-(s2)$) .. controls +(0,0.1) and +(-0.3,0) ..  ($(p6)-(d)-(s)-(s2)$); 
		\draw[very thick, red!80!black] ($(p5)-(d)+(s)$) .. controls +(0,0.1) and +(-0.3,0) ..  ($(p6)-(d)+(s)$); 
		%
		\fill[red!80!black] ($(p4)+(s)-(d)-(t)$) circle (0pt) node[right] {{\small $-$}};
		\fill[red!80!black] ($(p6)+(s)-(d)-(t)$) circle (0pt) node[right] {{\small $+$}};
		\fill[red!80!black] ($(p4)-(s2)-(s)-(d)-(t)$) circle (0pt) node[right] {{\small $-$}};
		\fill[red!80!black] ($(p6)-(s2)-(s)-(d)-(t)$) circle (0pt) node[right] {{\small $+$}};
		colouring front: 
		\fill [orange!30!white, opacity=0.8] 
		($(p5)-(d)-(s)-(s2)$) .. controls +(0,-0.1) and +(-0.5,0) ..  ($(p4)-(d)-(s)-(s2)$)
		-- ($(p4)-(d)-(s)-(s2)$) -- ($(p4)-(d)+(s)$) 
		-- ($(p4)-(d)+(s)$) .. controls +(-0.5,0) and +(0,-0.1) ..($(p5)-(d)+(s)$); 
		%
		\draw[very thick, red!80!black] ($(p4)-(d)+(s)$) .. controls +(-0.5,0) and +(0,-0.1) ..($(p5)-(d)+(s)$);
		\draw[very thick, red!80!black] ($(p4)-(d)-(s)-(s2)$) .. controls +(-0.5,0) and +(0,-0.1) ..($(p5)-(d)-(s)-(s2)$);
		%
		\draw[thin] ($(p4)-(d)-(s)-(s2)$) -- ($(p4)-(d)+(s)$);
		\draw[thin] ($(p5)-(d)-(s)-(s2)$) -- ($(p5)-(d)+(s)$);
		%
		%
		%
		\coordinate (z) at (0,-1.0);
		%
		\fill [orange!20!white, opacity=0.8] 
		($(p5)-(d)-(s)-(s2)+(z)$) .. controls +(0,-1.5) and +(0,-1.5) .. ($(p2)+(d)-(s)-(s2)+(z)$) 
		-- ($(p2)+(d)-(s)-(s2)+(z)$)  .. controls +(0,0.4) and +(0,0.4) .. ($(p5)-(d)-(s)-(s2)+(z)$);
		%
		\fill [orange!30!white, opacity=0.8] 
		($(p5)-(d)-(s)-(s2)+(z)$) .. controls +(0,-1.5) and +(0,-1.5) .. ($(p2)+(d)-(s)-(s2)+(z)$) 
		-- ($(p2)+(d)-(s)-(s2)+(z)$)  .. controls +(0,-0.4) and +(0,-0.4) .. ($(p5)-(d)-(s)-(s2)+(z)$);
		%
		\draw[thin] ($(p5)-(d)-(s)-(s2)+(z)$) .. controls +(0,-1.5) and +(0,-1.5) .. ($(p2)+(d)-(s)-(s2)+(z)$) ;
		\draw[very thick, red!80!black] ($(p2)+(d)-(s)-(s2)+(z)$)  .. controls +(0,0.4) and +(0,0.4) .. ($(p5)-(d)-(s)-(s2)+(z)$);
		\draw[very thick, red!80!black] ($(p2)+(d)-(s)-(s2)+(z)$)  .. controls +(0,-0.4) and +(0,-0.4) .. ($(p5)-(d)-(s)-(s2)+(z)$);
		%
		%
		%
		\coordinate (z2) at (0,-0.25);
		%
		\fill [orange!20!white, opacity=0.8] 
		($(p5)-(d)+(s)+(s2)-(z2)$) .. controls +(0,1.5) and +(0,1.5) .. ($(p2)+(d)+(s)+(s2)-(z2)$) 
		-- ($(p2)+(d)+(s)+(s2)-(z2)$)  .. controls +(0,0.4) and +(0,0.4) .. ($(p5)-(d)+(s)+(s2)-(z2)$);
		%
		\draw[very thick, red!80!black] ($(p2)+(d)+(s)+(s2)-(z2)$)  .. controls +(0,0.4) and +(0,0.4) .. ($(p5)-(d)+(s)+(s2)-(z2)$);
		%
		\fill [orange!30!white, opacity=0.8] 
		($(p5)-(d)+(s)+(s2)-(z2)$) .. controls +(0,1.5) and +(0,1.5) .. ($(p2)+(d)+(s)+(s2)-(z2)$) 
		-- ($(p2)+(d)+(s)+(s2)-(z2)$)  .. controls +(0,-0.4) and +(0,-0.4) .. ($(p5)-(d)+(s)+(s2)-(z2)$);
		%
		\draw[thin] ($(p5)-(d)+(s)+(s2)-(z2)$) .. controls +(0,1.5) and +(0,1.5) .. ($(p2)+(d)+(s)+(s2)-(z2)$);
		\draw[very thick, red!80!black] ($(p2)+(d)+(s)+(s2)-(z2)$)  .. controls +(0,-0.4) and +(0,-0.4) .. ($(p5)-(d)+(s)+(s2)-(z2)$);
		\end{tikzpicture}
	}%
 \right)
\nonumber 
\\
& = \tev_{\tev_+} \cdot \Big[ 1_{\tev_u} \circ \big( \ev_{\tev_u} \cdot \: \tcoev_{\tev_u} \big) \circ 1_{\coev_u} \Big] \cdot \coev_{\tev_u} \, . 
\label{eq:TorusDecomposed}
\end{align}
Here we assumed that the adjunction data for the adjunction 1-morphisms $\tev_u, \tcoev_u, \ev_u, \coev_u$ are chosen in~$\B$ such that they satisfy the strict identities ${}^\dagger\! \tev_u = \coev_u = \tev_u^\dagger$ etc., analogous to the strict identities \eqref{eq:StrictBord1}--\eqref{eq:StrictBord2} in $\Bordor$. 
This is always possible and guarantees that e.\,g.\ the 2-morphism $\zz(\tev_{\tev_+}) = \tev_{\tev_u} \colon \tev_u \circ \tev_u^\dagger \lra 1_\one$ compiles, i.\,e.\ is compatible with~\eqref{eq:ZonCoev+}. 
If other adjunction data were chosen in~$\B$, the 2-isomorphisms $\tev_u^\dagger \cong \coev_u$ etc.\ (uniquely determined by the trivialisation~$\lambda$) have to be inserted by hand to make the formulas for~$\zz$ on surfaces compile. 
Note that in the cases $\B = \bi$ and $\B = \bigra$, our strict assumption on adjunction data is satisfied, see e.\,g.\ \eqref{eq:LeftRightEval}, \eqref{eq:LeftRightCoeval}, as every object~$\xu$ has an essentially unique trivialisation of its Serre automorphism (cf.\ Proposition~\ref{lem:SerreTrivial} and its bigraded version, as well as Theorem~\ref{cor:ExtendedTQFTZn} below), so in this sense we can treat the latter as the identity.

Along the same lines, and with the above assumption on adjunction data, we find that for any closed oriented surface~$\Sigma_g$ of genus $g\in\Z_{\geqslant 0}$: 
\be
\label{eq:PartitionFunctionFromExtendedTQFT}
\zz(\Sigma_g)
	= 
	\tev_{\tev_+} \cdot \Big[ 1_{\tev_u} \circ \big( \ev_{\tev_u} \cdot \: \tcoev_{\tev_u} \big)^g \circ 1_{\coev_u} \Big] \cdot \coev_{\tev_u} \, .  
\ee 

\begin{example}
	Consider the symmetric monoidal 2-category $\textrm{Alg}_\C$ of finite-dimensional $\C$-algebras, their finite-dimensional bimodules and bimodule maps. 
	The monoidal product is given by~$\otimes_\C$, horizontal composition is the relative tensor product over the intermediate algebra, and vertical composition is concatenation of $\C$-linear maps. 
	
	As explained in \cite[Sect.\,3.8]{spthesis}, an object $A\in \textrm{Alg}_\C$ is fully dualisable with trivialisable Serre automorphism iff~$A$ comes with the structure of a separable symmetric Frobenius algebra over~$\C$. 
	Its dual is the opposite algebra $A^\dual = A^{\textrm{op}}$, whose multiplication is that of~$A$ pre-composed with the swap map $a\otimes b \lmt b\otimes a$, with $\tev_A$ and $\tcoev_A$ given by~$A$ viewed as a $\C$-$(A\otimes_\C A^{\textrm{op}})$- and an $(A^{\textrm{op}}\otimes_\C A)$-$\C$-bimodule, respectively. 
	Since~$A$ is symmetric, we have $\tev_A^\dagger = {}_{\Ae}(A^*)_\C \cong {}_{\Ae} A_\C = \coev_A$, where $A^* := \Hom_\C(A,\C)$ and $\Ae := A\otimes_\C A^{\textrm{op}}$, hence we find that an extended TQFT~$\zz_A$ classified by~$A$ associates the 0-th Hochschild homology \textsl{and} cohomology to the oriented circle: 
	\begin{align}
	\textrm{HH}_0(A) & = A\otimes_{\Ae} A
						= \tev_A \circ \coev_A
						\cong \zz_A\big( \tev_+ \circ \coev_+ \big)
						= \zz_A(S^1)
						\nonumber
						\\
						& = \zz_A\big( \tev_+ \circ \tev_+^\dagger \big)
						\cong A \otimes_{\Ae} A^* 
						\cong \End_{\Ae}(A) 
						\nonumber
						\\
						& = \textrm{HH}^0(A) \, . 
	\end{align}
	Moreover, one finds that~$\zz_A$ sends the pair-of-pants to the induced (commutative) multiplication on $\textrm{HH}^0(A)$, while its values on the upside-down pair-of-pants, the cap and the cup are similarly induced from the Frobenius algebra structure of~$A$. 
	
	The above discussion for the 2-category $\textrm{Alg}_\C$ is naturally interpreted as that of oriented state sum models as fully extended oriented TQFTs. 
	Other standard examples of 2-dimensional TQFTs such as B-twisted sigma models and Landau--Ginzburg models also have well-established 2-categories associated to them which can be taken as targets for fully extended TQFTs, see e.\,g.\ \cite{cw1007.2679, BanksOnRozanskyWitten} and \cite{cm1208.1481, CMM}, respectively, for the symmetric monoidal and duality structures. 
	In the former case the Serre automorphism corresponds to tensoring with shifted canonical line bundles, and to a shift of matrix factorisations in the latter case. 
	In all cases the expression~\eqref{eq:PartitionFunctionFromExtendedTQFT} reproduces the partition functions on genus-$g$ surfaces obtained earlier by other methods. 
\end{example}

\begin{remark}
	\label{rem:ModuliSpaceOfFullyExtendedTQFTs}
	A natural interpretation of the cobordism hypothesis in Theorem~\ref{thm:OrientedCobordismHypothesis} is that the 2-groupoid~$\Bhfp{\B}$ contains information about the moduli space $\mathcal M_{\B_\infty}$ of extended oriented TQFTs with values in a 2-category~$\B$. 
	Indeed, we may think of~$\B$ as the truncation of an $(\infty,2)$-category~$\B_\infty$. 
	Then according to the $(\infty,2)$-version of the cobordism hypothesis \cite{l0905.0465, AyalaFrancis2017CH}, extended oriented TQFTs with values in~$\B_\infty$ are classified by an $\infty$-groupoid~$\Bhfp{\B}_\infty$ of $\SO(2)$-homotopy fixed points on fully dualisable objects in~$\B_\infty$, and our~$\Bhfp{\B}$ is its 2-categorical truncation. 
	But by the homotopy hypothesis an $\infty$-groupoid is to be identified with a topological space. 
	In particular, isomorphism classes of objects and isomorphism classes of 1-morphisms of~$\Bhfp{\B}$ correspond to connected components $\Pi_0(\mathcal M_{\B_\infty})$ and to morphisms in the fundamental groupoid $\Pi_1(\mathcal M_{\B_\infty})$, respectively, of the moduli space of extended oriented TQFTs with values in~$\B_\infty$. 
\end{remark}

\subsection{Truncated affine Rozansky--Witten models}
\label{subsec:ExtendedAffineRW}

In this section we apply the oriented cobordism hypothesis (Theorem~\ref{thm:OrientedCobordismHypothesis}) to the 2-category~$\bi$ of truncated affine Rozansky--Witten models, establishing that every $(x_1,\dots,x_n)\in\bi$ gives rise to a unique extended TQFT~$\zz_n$. 
We then go on, in Section~\ref{subsubsec:PartFuncStateSpaces}, to explicitly compute the $\Z_2$-graded vector spaces $\zz_n(\Sigma_g)$ associated to closed surfaces~$\Sigma_g$ of genus~$g$. 
From the perspective of 2-dimensional TQFT, the vector spaces $\zz_n(\Sigma_g)$ are ``partition functions'' for the ``spacetimes''~$\Sigma_g$, but since~$\zz_n$ is the truncation of a 3-dimensional field theory, they are naturally interpreted as ``state spaces'' of the latter. 
Finally in Section~\ref{subsubsec:ComFrobAlgGrothRing} we compute~$\zz_n$ on surfaces without corners like the pair-of-pants to identify the commutative Frobenius algebra classifying the closed TQFT obtained from~$\zz_n$ by restriction, and we discuss its relation to Grothendieck rings.

\subsubsection{Partition functions and state spaces}
\label{subsubsec:PartFuncStateSpaces}

According to Theorem~\ref{thm:FullyDualisable} and Proposition~\ref{lem:SerreTrivial}, every object in~$\bi$ is fully dualisable, and there are precisely two trivialisations of its Serre automorphism. 
One finds that both correspond to isomorphic TQFTs: 

\begin{theorem}
	\label{cor:ExtendedTQFTZn}
	For every $(x_1,\dots,x_n)\in\bi$ there is an extended oriented TQFT 
	\be 
	\zz_n \colon \Bordor \lra \bi
	\ee 
	with $\zz_n(+) = (x_1,\dots,x_n)$. 
	Such TQFTs are unique up to isomorphism. 
\end{theorem}
\begin{proof}
	Existence of~$\zz_n$ is immediate from the cobordism hypothesis. 
	To prove uniqueness we have to show that the two TQFTs corresponding (via Theorem~\ref{thm:OrientedCobordismHypothesis}) to $(\xu, [I_{1_\xu}])$ and $(\xu, [I_{1_\xu}[1]])$ in $\Bhfp{\bi}$ are isomorphic. 
	
	Note that in general the way a chosen trivialisation~$\lambda$ of a Serre automorphism~$S_u$ enters into the construction of the associated oriented TQFT is via identities like~\eqref{eq:RightAdjointFromSerre} and more generally \cite[Thm.\,3.9]{Pstragowski}, which expresses adjoints of evaluation 1-morphisms in terms of coevaluation 1-morphisms and a single factor of~$S_u$ (or~$S_u^{-1}$). 
	In particular, the left and right adjoints of adjunction 1-morphisms differ by~$S_u^{2}$ or~$S_u^{-2}$, and one finds that the associated adjunction 2-morphisms involve an even number of $\lambda$-factors (see e.\,g.\ \cite[(3.31)--(3.34)]{RSpinLorantNils} for more details). 
	In our case of~$u=\xu$ and $\lambda = [I_{1_\xu}]$ or $\lambda = [I_{1_\xu}[1]]$ these contributions are identities in both cases, because a double shift is trivial both under horizontal and vertical composition. 
\end{proof}

Since the 2-category~$\bi$ is under explicit control, it is straightforward to compute what the TQFT~$\zz_n$ does to bordisms, following the general discussion of Section~\ref{subsec:CobordismHypothesis}. 
Using the adjunction 1-morphisms~\eqref{eq:LeftRightEval}, \eqref{eq:LeftRightCoeval} we immediately see that to a circle, $\zz_n$ assigns 
\be 
\label{eq:ZnOnCircle}
\zz_n(S^1) = \Big( \au,\au',\xu,\xu' ; \, \big( \au-\au'\big) \cdot \big( \xu-\xu'\big) \Big) .
\ee 
Roughly, we may think of $\zz_n(S^1)$ as the homotopy category of matrix factorisations of $(\au-\au')\cdot(\xu-\xu')$ -- but as discussed in Section.~\ref{subsubsec:ComFrobAlgGrothRing} below, $\zz_n$ does not know about the entire structure of that category. 

To surfaces, the TQFT~$\zz_n$ associates isomorphism classes of matrix factorisations. 
For instance, $\zz_n(\Sigma_g)$ for a closed surface $\Sigma_g$ of genus $g$ is an isomorphism class of matrix factorisations of $0\in\C[\au,\au',\xu,\xu']$. 
Matrix factorisations of zero are isomorphic to the $\Z_2$-graded vector spaces given by their cohomology. 
Hence, $\zz_n(\Sigma_g)$ can be viewed as such a vector space, which should correspond to the state space of the underlying 3-dimensional theory on $\Sigma_g$. 

In the following Proposition~\ref{prop:StateSpaces}, we explicitly calculate $\zz_n(\Sigma_g)$, and indeed reproduce the state spaces of the affine Rozansky--Witten model with target manifold $T^*\C^n$. 
Since affine Rozansky--Witten models are free theories, their state spaces can be calculated in a straightforward manner, cf.\ \cite{RW1996}. 
(We will discuss the agreement of these state spaces with our results in slightly more detail in the graded setting in Section~\ref{subsec:ExtendedAffineRWWithCharges} below.)
\begin{proposition}
	\label{prop:StateSpaces}
	Let~$\Sigma_g$ be a closed surface of genus $g\in\Z_{\geqslant 0}$. 
	Then 
	\be 
	\label{eq:StateSpaceSigmag}
	\zz_n(\Sigma_g) \cong \big( \C \oplus \C[1] 
		 \big)^{\otimes 2ng}
	\otimes_\C \C[\au,\xu] 
	\ee 
	as $\Z_2$-graded vector spaces, where $\C[\au,\xu] = \C[a_1,\dots,a_n,x_1,\dots,x_n]$ has degree~0. 
\end{proposition}
\begin{proof}
	We first consider the case $g=0$, i.\,e.\ $\Sigma_0=S^2$. 
	By~\eqref{eq:DualityData2}, \eqref{eq:DualityData3} and~\eqref{eq:ZonSphere} we have $\zz_n(S^2) \cong [\yu-\xu,\au'-\au] \otimes [\au'-\au,\xu-\yu]$ as a matrix factorisation of zero over $\C[\au,\au',\xu,\yu]$. 
	Using Lemma~\ref{lem:elimination} this matrix factorisation can be simplified by eliminating the internal variable $\au'$. 
	This yields the matrix factorisation $[\yu-\xu,
		 \zeru 
	]$ over $\C[\au,\xu,\yu]$, whose cohomology is given by the $\Z_2$-graded vector space $\C[\au,\xu]$ in degree~$0$. 
	Hence $\zz_n(S^2) \cong \C[\au,\xu]$. 
	
	Next we consider $g\geqslant 1$. 
	Vertically composing $\tcoev_{\tev_\xu}$ in~\eqref{eq:DualityData} with $\ev_{\tev_\xu}$ in~\eqref{eq:DualityData1} (with $\bu \lmt \widetilde{\bu}$ and $\cu\lmt\widetilde{\cu}$) and then horizontally composing with 
	\be 
	\label{eq:HalfCylinders}
	\tikzzbox{%
		\begin{tikzpicture}[thick,scale=1.0,color=black, baseline=0.9cm]
		\coordinate (p1) at (0,0);
		\coordinate (p2) at (0.5,0.2);
		\coordinate (p3) at (-0.5,0.4);
		\coordinate (p4) at (2.5,0);
		\coordinate (p5) at (1.5,0.2);
		\coordinate (p6) at (2.0,0.4);
		\coordinate (s) at ($(0,1.5)$);
		\coordinate (d) at ($(3,0)$);
		%
		\fill [orange!20!white, opacity=0.8] 
		($(p2)+(d)$) .. controls +(0,0.1) and +(0.5,0) ..  ($(p3)+(d)$)
		-- ($(p3)+(d)$) -- ($(p3)+(d)+(s)$) 
		-- ($(p3)+(d)+(s)$) .. controls +(0.5,0) and +(0,0.1) ..($(p2)+(d)+(s)$);
		%
		\draw[very thick, red!80!black] ($(p2)+(d)$) .. controls +(0,0.1) and +(0.5,0) ..  ($(p3)+(d)$); 
		\draw[very thick, red!80!black] ($(p2)+(d)+(s)$) .. controls +(0,0.1) and +(0.5,0) ..  ($(p3)+(d)+(s)$); 
		%
		\fill [orange!30!white, opacity=0.8] 
		($(p1)+(d)$) .. controls +(0.3,0) and +(0,-0.1) .. ($(p2)+(d)$)
		-- ($(p2)+(d)$) -- ($(p2)+(d)+(s)$)
		-- ($(p2)+(d)+(s)$) .. controls +(0,-0.1) and +(0.3,0) .. ($(p1)+(d)+(s)$); 
		%
		\draw[thin] ($(p1)+(d)$) -- ($(p1)+(d)+(s)$);
		\draw[thin] ($(p2)+(d)$) -- ($(p2)+(d)+(s)$);
		\draw[thin] ($(p3)+(d)$) -- ($(p3)+(d)+(s)$);
		%
		\draw[very thick, red!80!black] ($(p1)+(d)$) .. controls +(0.3,0) and +(0,-0.1) ..  ($(p2)+(d)$); 
		\draw[very thick, red!80!black] ($(p1)+(d)+(s)$) .. controls +(0.3,0) and +(0,-0.1) ..  ($(p2)+(d)+(s)$); 
		%
		\fill[red!80!black] ($(p4)+(s)$) circle (0pt) node[right] {{\small $\yu$}};
		\fill[red!80!black] ($(p6)+(s)$) circle (0pt) node[right] {{\small $\xu$}};
		\fill[red!80!black] (p4) circle (0pt) node[right] {{\small $\yu$}};
		\fill[red!80!black] (p6) circle (0pt) node[right] {{\small $\xu$}};
		\fill[red!80!black] ($(p2)+(d)$) circle (0pt) node[right] {{\small $\du$}};
		\fill[red!80!black] ($(p2)+(d)+(s)$) circle (0pt) node[right] {{\small $\eu$}};
		\end{tikzpicture}
	}%
	= \big[ \eu-\du, \, \yu-\xu \big] 
	\, , \quad 
	\tikzzbox{%
		\begin{tikzpicture}[thick,scale=1.0,color=black, baseline=0.5cm, yscale=-1, xscale=-1]
		\coordinate (p1) at (0,0);
		\coordinate (p2) at (0.5,0.2);
		\coordinate (p3) at (-0.5,0.4);
		\coordinate (p4) at (2.5,0);
		\coordinate (p5) at (1.5,0.2);
		\coordinate (p6) at (2.0,0.4);
		\coordinate (s) at ($(0,-1.5)$);
		\coordinate (d) at ($(2.5,0)$);
		%
		\fill [orange!20!white, opacity=0.8] 
		($(p1)+(d)$) .. controls +(0.3,0) and +(0,-0.1) .. ($(p2)+(d)$)
		-- ($(p2)+(d)$) -- ($(p2)+(d)+(s)$)
		-- ($(p2)+(d)+(s)$) .. controls +(0,-0.1) and +(0.3,0) .. ($(p1)+(d)+(s)$); 
		\fill[red!80!black] (p4) circle (0pt) node[right] {{\small $\xu'$}};
		%
		\draw[thin] ($(p1)+(d)$) -- ($(p1)+(d)+(s)$);
		\draw[thin] ($(p2)+(d)$) -- ($(p2)+(d)+(s)$);
		\draw[thin] ($(p3)+(d)$) -- ($(p3)+(d)+(s)$);
		%
		\draw[very thick, red!80!black] ($(p1)+(d)$) .. controls +(0.3,0) and +(0,-0.1) ..  ($(p2)+(d)$); 
		\draw[very thick, red!80!black] ($(p1)+(d)+(s)$) .. controls +(0.3,0) and +(0,-0.1) ..  ($(p2)+(d)+(s)$); 
		%
		\fill [orange!30!white, opacity=0.8] 
		($(p2)+(d)$) .. controls +(0,0.1) and +(0.5,0) ..  ($(p3)+(d)$)
		-- ($(p3)+(d)$) -- ($(p3)+(d)+(s)$) 
		-- ($(p3)+(d)+(s)$) .. controls +(0.5,0) and +(0,0.1) ..($(p2)+(d)+(s)$);
		%
		\draw[thin] ($(p2)+(d)$) -- ($(p2)+(d)+(s)$);
		\draw[thin] ($(p3)+(d)$) -- ($(p3)+(d)+(s)$);
		%
		\draw[very thick, red!80!black] ($(p2)+(d)$) .. controls +(0,0.1) and +(0.5,0) ..  ($(p3)+(d)$); 
		\draw[very thick, red!80!black] ($(p2)+(d)+(s)$) .. controls +(0,0.1) and +(0.5,0) ..  ($(p3)+(d)+(s)$); 
		%
		\fill[red!80!black] ($(p4)+(s)$) circle (0pt) node[right] {{\small $\xu'$}};
		\fill[red!80!black] ($(p6)+(s)$) circle (0pt) node[right] {{\small $\yu'$}};
		\fill[red!80!black] (p6) circle (0pt) node[right] {{\small $\yu'$}};
		\fill[red!80!black] ($(p2)+(d)$) circle (0pt) node[left] {{\small $\du'$}};
		\fill[red!80!black] ($(p2)+(d)+(s)$) circle (0pt) node[left] {{\small $\eu'$}};
		\end{tikzpicture}
	}%
	= \big[ \eu'-\du', \, \yu'-\xu' \big] 
	\ee 
	we find 
	\begin{align}
	& \zz_n \left( 
	\tikzzbox{%
		\begin{tikzpicture}[thick,scale=1.0,color=black, baseline=-0.2cm]
		\coordinate (p1) at (0,0);
		\coordinate (p2) at (0.5,0.2);
		\coordinate (p3) at (-0.5,0.4);
		\coordinate (p4) at (2.5,0);
		\coordinate (p5) at (1.5,0.2);
		\coordinate (p6) at (2.0,0.4);
		\coordinate (s) at ($(0,1.5)$);
		\coordinate (s2) at ($(0,0.75)$);
		\coordinate (d) at ($(4,0)$);
		\coordinate (h) at ($(0,2.5)$);
		\coordinate (t) at ($(0.1,0)$);
		\coordinate (u) at ($(0,-1.1)$);
		%
		%
		%
		\fill [orange!20!white, opacity=0.8] 
		(p3) .. controls +(0.5,0) and +(0,0.1) ..  (p2)
		-- (p2) .. controls +(0,0.9) and +(0,0.9) ..  (p5)
		-- (p5) .. controls +(0,0.1) and +(-0.3,0) .. (p6)
		-- (p6) -- ($(p6)+(s)$) -- ($(p3)+(s)$);
		%
		\fill[red!80!black] ($(p1)+(t)$) circle (0pt) node[left] {{\small $-$}};
		\fill[red!80!black] ($(p3)+(t)$) circle (0pt) node[left] {{\small $+$}};
		\fill[red!80!black] ($(p4)-0.5*(t)$) circle (0pt) node[right] {{\small $-$}};
		\fill[red!80!black] ($(p6)-0.5*(t)$) circle (0pt) node[right] {{\small $+$}};
		\fill[red!80!black] ($(p4)+(s)-0.5*(t)$) circle (0pt) node[right] {{\small $-$}};
		\fill[red!80!black] ($(p1)+(s)+(t)$) circle (0pt) node[left] {{\small $-$}};
		\fill[red!80!black] ($(p3)+(s)+(t)$) circle (0pt) node[left] {{\small $+$}};
		\fill[red!80!black] ($(p6)+(s)-0.5*(t)$) circle (0pt) node[right] {{\small $+$}};
		%
		%
		\draw[very thick, red!80!black] (p2) .. controls +(0,0.1) and +(0.5,0) ..  (p3); 
		\draw[very thick, red!80!black] (p6) .. controls +(-0.3,0) and +(0,0.1) ..  (p5); 
		\draw[thin] (p6) -- ($(p6)+(s)$);
		%
		\fill [orange!30!white, opacity=0.8] 
		(p1) .. controls +(0.3,0) and +(0,-0.1) ..  (p2)
		-- (p2) .. controls +(0,0.9) and +(0,0.9) ..  (p5)
		-- (p5) .. controls +(0,-0.1) and +(-0.5,0) ..  (p4) 
		-- (p4) -- ($(p4)+(s)$) -- ($(p1)+(s)$);
		%
		\draw[very thick, red!80!black] (p1) .. controls +(0.3,0) and +(0,-0.1) ..  (p2); 
		\draw[very thick, red!80!black] (p5) .. controls +(0,-0.1) and +(-0.5,0) ..  (p4); 
		\draw[very thick, red!80!black] ($(p4)+(s)$) -- ($(p1)+(s)$);
		\draw[very thick, red!80!black] ($(p6)+(s)$) -- ($(p3)+(s)$);
		%
		\draw (p2) .. controls +(0,0.9) and +(0,0.9) ..  (p5);
		\draw[thin] (p1) -- ($(p1)+(s)$);
		\draw[thin] (p3) -- ($(p3)+(s)$);
		\draw[thin] (p4) -- ($(p4)+(s)$);
		%
		%
		%
		%
		\fill [orange!20!white, opacity=0.8] 
		($(p3)-(s2)$) .. controls +(0.5,0) and +(0,0.1) ..  ($(p2)-(s2)$)
		-- ($(p2)-(s2)$) .. controls +(0,-0.9) and +(0,-0.9) ..  ($(p5)-(s2)$)
		-- ($(p5)-(s2)$) .. controls +(0,0.1) and +(-0.3,0) .. ($(p6)-(s2)$)
		-- ($(p6)-(s2)$) -- ($(p6)-(s)-(s2)$) -- ($(p3)-(s)-(s2)$);
		%
		\fill[red!80!black] ($(p1)+(t)-(s2)$) circle (0pt) node[left] {{\small $-$}};
		\fill[red!80!black] ($(p3)+(t)-(s2)$) circle (0pt) node[left] {{\small $+$}};
		\fill[red!80!black] ($(p4)-0.5*(t)-(s2)$) circle (0pt) node[right] {{\small $-$}};
		\fill[red!80!black] ($(p6)-0.5*(t)-(s2)$) circle (0pt) node[right] {{\small $+$}};
		\fill[red!80!black] ($(p4)-(s2)-(s)-0.5*(t)$) circle (0pt) node[right] {{\small $-$}};
		\fill[red!80!black] ($(p1)-(s2)-(s)+(t)$) circle (0pt) node[left] {{\small $-$}};
		\fill[red!80!black] ($(p3)-(s2)-(s)+(t)$) circle (0pt) node[left] {{\small $+$}};
		\fill[red!80!black] ($(p6)-(s2)-(s)-0.5*(t)$) circle (0pt) node[right] {{\small $+$}};
		%
		%
		\draw[very thick, red!80!black] ($(p2)-(s2)$) .. controls +(0,0.1) and +(0.5,0) .. ($(p3)-(s2)$); 
		\draw[very thick, red!80!black] ($(p6)-(s2)$) .. controls +(-0.3,0) and +(0,0.1) .. ($(p5)-(s2)$); 
		\draw[very thick, red!80!black] ($(p6)-(s)-(s2)$) -- ($(p3)-(s)-(s2)$);	
		\draw[thin] ($(p6)-(s2)$) -- ($(p6)-(s)-(s2)$);
		%
		\fill [orange!30!white, opacity=0.8] 
		($(p1)-(s2)$) .. controls +(0.3,0) and +(0,-0.1) ..  ($(p2)-(s2)$)
		-- ($(p2)-(s2)$) .. controls +(0,-0.9) and +(0,-0.9) ..  ($(p5)-(s2)$)
		-- ($(p5)-(s2)$) .. controls +(0,-0.1) and +(-0.5,0) ..  ($(p4)-(s2)$) 
		-- ($(p4)-(s2)$) -- ($(p4)-(s)-(s2)$) -- ($(p1)-(s)-(s2)$);
		%
		\draw[very thick, red!80!black] ($(p1)-(s2)$) .. controls +(0.3,0) and +(0,-0.1) ..  ($(p2)-(s2)$); 
		\draw[very thick, red!80!black] ($(p5)-(s2)$) .. controls +(0,-0.1) and +(-0.5,0) ..  ($(p4)-(s2)$); 
		\draw[very thick, red!80!black] ($(p4)-(s)-(s2)$) -- ($(p1)-(s)-(s2)$); 
		%
		\draw ($(p2)-(s2)$) .. controls +(0,-0.9) and +(0,-0.9) ..  ($(p5)-(s2)$);
		\draw[thin] ($(p1)-(s2)$) -- ($(p1)-(s)-(s2)$);
		\draw[thin] ($(p3)-(s2)$) -- ($(p3)-(s)-(s2)$);
		\draw[thin] ($(p4)-(s2)$) -- ($(p4)-(s)-(s2)$);
		%
		%
		%
		%
		%
		\fill [orange!20!white, opacity=0.8] 
		($(p2)+(d)-(s)-(s2)$) .. controls +(0,0.1) and +(0.5,0) ..  ($(p3)+(d)-(s)-(s2)$)
		-- ($(p3)+(d)-(s)-(s2)$) -- ($(p3)+(d)+(s)$) 
		-- ($(p3)+(d)+(s)$) .. controls +(0.5,0) and +(0,0.1) ..($(p2)+(d)+(s)$); 
		%
		%
		\draw[very thick, red!80!black] ($(p2)+(d)-(s)-(s2)$) .. controls +(0,0.1) and +(0.5,0) ..  ($(p3)+(d)-(s)-(s2)$); 
		\draw[very thick, red!80!black] ($(p2)+(d)+(s)$) .. controls +(0,0.1) and +(0.5,0) ..  ($(p3)+(d)+(s)$); 
		%
		\fill [orange!30!white, opacity=0.8] 
		($(p1)+(d)-(s)-(s2)$) .. controls +(0.3,0) and +(0,-0.1) .. ($(p2)+(d)-(s)-(s2)$)
		-- ($(p2)+(d)-(s)-(s2)$) -- ($(p2)+(d)+(s)$)
		-- ($(p2)+(d)+(s)$) .. controls +(0,-0.1) and +(0.3,0) .. ($(p1)+(d)+(s)$);
		%
		\draw[very thick, red!80!black] ($(p1)+(d)-(s)-(s2)$) .. controls +(0.3,0) and +(0,-0.1) ..  ($(p2)+(d)-(s)-(s2)$); 
		\draw[very thick, red!80!black] ($(p1)+(d)+(s)$) .. controls +(0.3,0) and +(0,-0.1) ..  ($(p2)+(d)+(s)$); 
		%
		\draw[thin] ($(p1)+(d)-(s)-(s2)$) -- ($(p1)+(d)+(s)$);
		\draw[thin] ($(p3)+(d)-(s)-(s2)$) -- ($(p3)+(d)+(s)$);
		\draw[thin] ($(p2)+(d)-(s)-(s2)$) -- ($(p2)+(d)+(s)$);
		%
		\fill[red!80!black] ($(p3)+(s)+(d)+(t)$) circle (0pt) node[left] {{\small $+$}};
		\fill[red!80!black] ($(p1)+(s)+(d)+(t)$) circle (0pt) node[left] {{\small $-$}};
		\fill[red!80!black] ($(p3)-(s2)-(s)+(d)+(t)$) circle (0pt) node[left] {{\small $+$}};
		\fill[red!80!black] ($(p1)-(s2)-(s)+(d)+(t)$) circle (0pt) node[left] {{\small $-$}};
		%
		%
		%
		%
		\fill [orange!20!white, opacity=0.8] 
		($(p5)-(d)-(s)-(s2)$) .. controls +(0,0.1) and +(-0.3,0) .. ($(p6)-(d)-(s)-(s2)$)
		-- ($(p6)-(d)-(s)-(s2)$) -- ($(p6)-(d)+(s)$)
		-- ($(p6)-(d)+(s)$) .. controls +(-0.3,0) and +(0,0.1) .. ($(p5)-(d)+(s)$);
		%
		%
		\draw[thin] ($(p6)-(d)-(s)-(s2)$) -- ($(p6)-(d)+(s)$);
		%
		\draw[very thick, red!80!black] ($(p5)-(d)-(s)-(s2)$) .. controls +(0,0.1) and +(-0.3,0) ..  ($(p6)-(d)-(s)-(s2)$); 
		\draw[very thick, red!80!black] ($(p5)-(d)+(s)$) .. controls +(0,0.1) and +(-0.3,0) ..  ($(p6)-(d)+(s)$); 
		%
		\fill[red!80!black] ($(p4)+(s)-(d)-(t)$) circle (0pt) node[right] {{\small $-$}};
		\fill[red!80!black] ($(p6)+(s)-(d)-(t)$) circle (0pt) node[right] {{\small $+$}};
		\fill[red!80!black] ($(p4)-(s2)-(s)-(d)-(t)$) circle (0pt) node[right] {{\small $-$}};
		\fill[red!80!black] ($(p6)-(s2)-(s)-(d)-(t)$) circle (0pt) node[right] {{\small $+$}};
		colouring front: 
		\fill [orange!30!white, opacity=0.8] 
		($(p5)-(d)-(s)-(s2)$) .. controls +(0,-0.1) and +(-0.5,0) ..  ($(p4)-(d)-(s)-(s2)$)
		-- ($(p4)-(d)-(s)-(s2)$) -- ($(p4)-(d)+(s)$) 
		-- ($(p4)-(d)+(s)$) .. controls +(-0.5,0) and +(0,-0.1) ..($(p5)-(d)+(s)$); 
		%
		\draw[very thick, red!80!black] ($(p4)-(d)+(s)$) .. controls +(-0.5,0) and +(0,-0.1) ..($(p5)-(d)+(s)$);
		\draw[very thick, red!80!black] ($(p4)-(d)-(s)-(s2)$) .. controls +(-0.5,0) and +(0,-0.1) ..($(p5)-(d)-(s)-(s2)$);
		%
		\draw[thin] ($(p4)-(d)-(s)-(s2)$) -- ($(p4)-(d)+(s)$);
		\draw[thin] ($(p5)-(d)-(s)-(s2)$) -- ($(p5)-(d)+(s)$);
		\end{tikzpicture}
	}%
	\right) 
	=
	\tikzzbox{%
		\begin{tikzpicture}[thick,scale=1.0,color=black, baseline=3.9cm]
		\coordinate (p1) at (0,0);
		\coordinate (p2) at (0.5,0.2);
		\coordinate (p3) at (-0.5,0.4);
		\coordinate (q2) at (-1.0,0.2);
		\coordinate (p4) at (2.5,0);
		\coordinate (p5) at (1.5,0.2);
		\coordinate (p6) at (2.0,0.4);
		\coordinate (s) at ($(0,2.5)$);
		\coordinate (d) at ($(2.5,0)$);
		\coordinate (h) at ($(0,2.5)$);
		\coordinate (t) at ($(0.1,0)$);
		\coordinate (u) at ($(0,-1.1)$);
		%
		\fill [orange!20!white, opacity=0.8] 
		($(q2)+(h)+(s)$) .. controls +(0,0.1) and +(-0.3,0) ..  ($(p3)+(h)+(s)$)
		-- ($(p3)+(s)$) .. controls +(-0.3,0) and +(0,0.1) .. ($(q2)+(s)$);
		\fill [orange!20!white, opacity=0.8] 
		($(p2)+(d)+(h)$) .. controls +(0,0.1) and +(0.5,0) ..  ($(p3)+(d)+(h)$) 
		-- ($(p3)+(h)$) -- ($(p3)+(s)+(h)$) -- ($(p3)+(d)+(s)+(h)$)
		-- ($(p3)+(d)+(s)+(h)$) .. controls +(0.5,0) and +(0,0.1) .. ($(p2)+(d)+(s)+(h)$);
		%
		\draw[very thick, red!80!black] ($(p2)+(d)+(h)$) .. controls +(0,0.1) and +(0.5,0) .. ($(p3)+(d)+(h)$) 
		-- ($(p3)+(h)$) .. controls +(-0.3,0) and +(0,0.1) .. ($(q2)+(h)$);
		\draw[very thick, red!80!black] ($(p2)+(d)+(h)+(s)$) .. controls +(0,0.1) and +(0.5,0) .. ($(p3)+(d)+(h)+(s)$) 
		-- ($(p3)+(h)+(s)$) .. controls +(-0.3,0) and +(0,0.1) .. ($(q2)+(h)+(s)$);
		%
		%
		\fill[red!80!black] ($(p3)+(d)+(s)$) circle (2.5pt) node[above] {{\small $\xu$}};
		\fill[red!80!black] ($(p3)+(s)$) circle (2.5pt) node[below] {};
		\fill[red!80!black] ($(p3)+(s)+(0.05,0)$) circle (0pt) node[above] {{\small $\xu'$}};
		\fill[red!80!black] ($(p3)+(h)+(s)+(0.05,0)$) circle (0pt) node[above] {{\small $\xu'$}};
		\fill[red!80!black] ($(p3)+(s)+(1.25,2.8)$) circle (0pt) node {{\small $\widetilde\bu$}};
		\fill[red!80!black] ($(p3)+(1.25,2.74)$) circle (0pt) node {{\small $\bu$}};
		\fill[red!80!black] ($(p1)+(1.25,2.3)$) circle (0pt) node {{\small $\cu$}};
		%
		%
		\fill [orange!30!white, opacity=0.8] 
		($(q2)+(h)+(s)$) .. controls +(0,-0.1) and +(-0.5,0) ..  ($(p1)+(h)+(s)$)
		-- ($(p1)+(s)$) .. controls +(-0.5,0) and +(0,-0.1) .. ($(q2)+(s)$);
		\fill [orange!30!white, opacity=0.8] 
		($(p1)+(h)$) -- ($(p1)+(d)+(h)$) .. controls +(0.3,0) and +(0,-0.1) .. ($(p2)+(d)+(h)$)
		-- ($(p2)+(d)+(h)$) -- ($(p2)+(d)+(s)+(h)$)
		-- ($(p2)+(d)+(s)+(h)$) .. controls +(0,-0.1) and +(0.3,0) .. ($(p1)+(d)+(s)+(h)$)
		-- ($(p1)+(s)+(h)$); 
		%
		\draw[thin] ($(p2)+(d)+(h)$) -- ($(p2)+(d)+(h)+(s)$);
		\draw[thin] ($(q2)+(h)$) -- ($(q2)+(h)+(s)$);
		\draw[very thick, red!80!black] 
		($(q2)+(h)$) .. controls +(0,-0.1) and +(-0.5,0) .. ($(p1)+(h)$) -- ($(p1)+(d)+(h)$) .. controls +(0.3,0) and +(0,-0.1) .. ($(p2)+(d)+(h)$); 
		\draw[very thick, red!80!black] 
		($(q2)+(h)+(s)$) .. controls +(0,-0.1) and +(-0.5,0) .. ($(p1)+(h)+(s)$) -- ($(p1)+(d)+(h)+(s)$) .. controls +(0.3,0) and +(0,-0.1) .. ($(p2)+(d)+(h)+(s)$); 
		%
		%
		\fill[color=white] ($1/2*(d)+1/2*(s)+(h)$) circle (0.5);
		\draw[thin] ($1/2*(d)+1/2*(s)+(h)$) circle (0.5);
		%
		\fill[red!80!black] ($(p2)+(d)+(h)-0.5*(t)$) circle (0pt) node[right] {{\small $\du$}};
		\fill[red!80!black] ($(p2)+(d)+(s)+(h)-0.5*(t)$) circle (0pt) node[right] {{\small $\eu$}};
		\fill[red!80!black] ($(q2)+(h)+0.75*(t)$) circle (0pt) node[left] {{\small $\du'$}};
		\fill[red!80!black] ($(q2)+(s)+(h)+0.75*(t)$) circle (0pt) node[left] {{\small $\eu'$}};
		\fill[red!80!black] ($(p3)+(h)+(d)+(s)$) circle (2.5pt) node[above] {{\small $\xu$}};
		\fill[red!80!black] ($(p1)+(h)+(d)+(s)$) circle (2.5pt) node[below] {{\small $\yu$}};
		\fill[red!80!black] ($(p1)+(d)+(s)$) circle (2.5pt) node[below] {{\small $\yu$}};
		\fill[red!80!black] ($(p3)+(h)+(s)$) circle (2.5pt) node[below] {};
		\fill[red!80!black] ($(p1)+(h)+(s)$) circle (2.5pt) node[below] {};
		\fill[red!80!black] ($(p1)+(h)+(s)+(0.05,0.1)$) circle (0pt) node[below] {{\small $\yu'$}};
		\fill[red!80!black] ($(p1)+(s)$) circle (2.5pt) node[below] {};
		\fill[red!80!black] ($(p1)+(s)+(0.05,0.1)$) circle (0pt) node[below] {{\small $\yu'$}};
		\fill[red!80!black] ($(p1)+(s)+(1.25,2.25)$) circle (0pt) node {{\small $\widetilde\cu$}};
		\end{tikzpicture} 
	}    
	\nonumber
	\\ 
	& \quad = \big[\widetilde{\cu}-\au,\,\yu-\yu'\big]\otimes\big[\widetilde{\bu}-\au',\,\xu'-\xu\big]\otimes\big[\au'-\au,\,\yu'-\xu\big]\nonumber\\
	&\qquad\qquad\otimes\big[\cu-\au,\,\yu'-\yu\big]\otimes\big[\bu-\au,\,\xu-\xu'\big]\otimes\big[\yu'-\xu',\,\au-\au'\big]\nonumber\\
	&\qquad\qquad\otimes\big[\eu-\du,\,\xu-\yu\big]\otimes\big[\eu'-\du',\,\yu'-\xu'\big] \, . \label{eq:g>1step1}
\end{align}
Here we used that all compositions are given by the tensor product of matrix factorisations. The first line on the right-hand side of \eqref{eq:g>1step1} corresponds to the saddle, the second line to the upside-down saddle, and the last line to the two half-cylinders in the decomposition. The variables $\au$ and $\au'$ in \eqref{eq:g>1step1} are internal.
Eliminating them using 
Lemma~\ref{lem:elimination}, we obtain that~\eqref{eq:g>1step1} is isomorphic to
\begin{align}\label{eq:g>1step2}
&
\big[\widetilde{\cu}-\cu,\,\yu-\yu'\big]\otimes\big[\widetilde{\bu}-\cu,\,\yu'-\xu\big]\otimes\big[\bu-\cu,\,\xu-\xu'\big]\otimes
\big[\yu'-\xu',\,\cu-\widetilde{\bu}\big]\nonumber\\
&\quad\otimes\big[\eu-\du,\,\xu-\yu\big]\otimes\big[\eu'-\du',\,\yu'-\xu'\big]\, . 
\end{align}
Next, we repeatedly apply the fact (proved as Lemma~\ref{lem:addition} in the appendix) that 
\begin{equation}\nonumber
\big[\pu,\,\qu\big]\otimes\big[\pu',\,\qu'\big]\,\cong\,
\big[\pu\pm\pu',\,\qu\big]\otimes\big[\pu',\,\qu'\mp\qu\big]\,,
\end{equation}
whenever $\pu,\qu,\pu',\qu'$ are lists of variables of the same length. 
Furthermore, since in general $(X,d_X) \cong (X,-d_X)$, and hence in particular $[\au,\xu]\cong[-\au,-\xu]$, we find with~\eqref{eq:tpshift} that \eqref{eq:g>1step2} is isomorphic to
\begin{align}
	& \big[ \widetilde{\cu}-\cu, \, \yu-\yu' \big] 
			\otimes \big[ \widetilde{\bu}-\bu, \, \xu'-\xu \big] 
			\otimes \big[ \eu - \du, \, \xu-\yu \big] \otimes \big[ \eu' - \du', \, \yu'-\xu' \big]
			\nonumber 
	\\
	& \quad\qquad 
			\otimes \big[ 
				 \zeru 
			, \, \yu' -\xu \big] 
			\otimes \big[ 
				 \zeru 
			, \, \widetilde{\bu} - \widetilde{\cu} \big]
			 \, .
			\label{eq:Middle}
	\end{align}
	Next, we pre-compose this with the 2-morphism
	\begin{align}\label{eq:s24p}
	&\tikzzbox{%
		\begin{tikzpicture}[thick,scale=1.0,color=black, baseline=-1.5cm]
		\coordinate (p1) at (0,0);
		\coordinate (p2) at (0.5,0.2);
		\coordinate (p3) at (-0.5,0.4);
		\coordinate (p4) at (2.5,0);
		\coordinate (p5) at (1.5,0.2);
		\coordinate (p6) at (2.0,0.4);
		\coordinate (s) at ($(0,1.5)$);
		\coordinate (d) at ($(3.8,0)$);
		\coordinate (h) at ($(0,2.5)$);
		\coordinate (t) at ($(0.1,0)$);
		\coordinate (u) at ($(0,-1.1)$);
		%
		\fill [orange!20!white, opacity=0.8] 
		($(p5)+(u)$) .. controls +(0,-2) and +(0,-2) .. ($(p2)+(d)+(u)$) 
		-- ($(p2)+(d)+(u)$) .. controls +(0,0.4) and +(0,0.4) .. ($(p5)+(u)$);
		%
		\fill [orange!30!white, opacity=0.8] 
		($(p5)+(u)$) .. controls +(0,-2) and +(0,-2) .. ($(p2)+(d)+(u)$) 
		-- ($(p2)+(d)+(u)$) .. controls +(0,-0.4) and +(0,-0.4) .. ($(p5)+(u)$);
		%
		\draw ($(p5)+(u)$) .. controls +(0,-2) and +(0,-2) ..  ($(p2)+(d)+(u)$);
		\draw[very thick, red!80!black] ($(p5)+(u)$) .. controls +(0,0.4) and +(0,0.4) ..  ($(p2)+(d)+(u)$);
		\draw[very thick, red!80!black] ($(p5)+(u)$) .. controls +(0,-0.4) and +(0,-0.4) ..  ($(p2)+(d)+(u)$);
		\fill[red!80!black] ($(p5)+(u)+(2.2,0.25)$) circle (2.5pt) node[above] {{\small $\xu$}};
		\fill[red!80!black] ($(p5)+(u)+(2.2,-0.255)$) circle (2.5pt) node[below] {{\small $\yu$}};
		\fill[red!80!black] ($(p5)+(u)+(0.6,0.25)$) circle (2.5pt) node[above] {{\small $\xu'$}};
		\fill[red!80!black] ($(p5)+(u)+(0.6,-0.255)$) circle (2.5pt) node[below] {{\small $\yu'$}};
		%
		\fill[red!80!black] ($(p5)+(u)+0.5*(t)$) circle (0pt) node[left] {{\small $\du'$}};
		\fill[red!80!black] ($(p2)+(d)+(u)-0.5*(t)$) circle (0pt) node[right] {{\small $\du\vphantom{\du'}$}};
		\fill[red!80!black] ($(p5)+(u)+(1.4,0.6)$) circle (0pt) node {{\small ${\bu}$}};
		\fill[red!80!black] ($(p5)+(u)+(1.4,-0.55)$) circle (0pt) node {{\small ${\cu}$}};
		\end{tikzpicture}
	}%
	\cong\big[\bu-\du,\,\xu'-\xu\big]\otimes\big[\cu-\du,\,\yu-\yu'\big]\otimes\big[\du'-\du,\,\xu'-\yu'\big]
	\end{align}
	obtained by composing 
	\begin{align}
	& 
	\tikzzbox{%
		\begin{tikzpicture}[thick,scale=1.0,color=black, baseline=1.0cm, yscale=-1, xscale=-1]
		\coordinate (p1) at (0,0);
		\coordinate (p2) at (0.5,0.2);
		\coordinate (p3) at (-0.5,0.4);
		\coordinate (p4) at (2.5,0);
		\coordinate (p5) at (1.5,0.2);
		\coordinate (p6) at (2.0,0.4);
		\coordinate (s) at ($(0,-2.5)$);
		\coordinate (d) at ($(2.5,0)$);
		%
		\fill [orange!20!white, opacity=0.8] 
		($(p1)+(d)$) .. controls +(0.3,0) and +(0,-0.1) .. ($(p2)+(d)$)
		-- ($(p2)+(d)$) -- ($(p2)+(d)+(s)$)
		-- ($(p2)+(d)+(s)$) .. controls +(0,-0.1) and +(0.3,0) .. ($(p1)+(d)+(s)$); 
		\fill[red!80!black] (p4) circle (0pt) node[right] {{\small $\xu'$}};
		%
		\draw[thin] ($(p1)+(d)$) -- ($(p1)+(d)+(s)$);
		%
		\draw[very thick, red!80!black] ($(p1)+(d)$) .. controls +(0.3,0) and +(0,-0.1) ..  ($(p2)+(d)$); 
		\draw[very thick, red!80!black] ($(p1)+(d)+(s)$) .. controls +(0.3,0) and +(0,-0.1) ..  ($(p2)+(d)+(s)$); 
		%
		\fill [orange!30!white, opacity=0.8] 
		($(p2)+(d)$) .. controls +(0,0.1) and +(0.5,0) ..  ($(p3)+(d)$)
		-- ($(p3)+(d)$) -- ($(p3)+(d)+(s)$) 
		-- ($(p3)+(d)+(s)$) .. controls +(0.5,0) and +(0,0.1) ..($(p2)+(d)+(s)$);
		%
		\draw[thin] ($(p2)+(d)$) -- ($(p2)+(d)+(s)$);
		\draw[thin] ($(p3)+(d)$) -- ($(p3)+(d)+(s)$);
		%
		\draw[very thick, red!80!black] ($(p2)+(d)$) .. controls +(0,0.1) and +(0.5,0) ..  ($(p3)+(d)$); 
		\draw[very thick, red!80!black] ($(p2)+(d)+(s)$) .. controls +(0,0.1) and +(0.5,0) ..  ($(p3)+(d)+(s)$); 
		%
		\fill[red!80!black] ($(p4)+(s)$) circle (0pt) node[right] {{\small $\xu'$}};
		\fill[red!80!black] ($(p6)+(s)$) circle (0pt) node[right] {{\small $\yu'$}};
		\fill[red!80!black] (p6) circle (0pt) node[right] {{\small $\yu'$}};
		\fill[red!80!black] ($(p2)+(d)$) circle (0pt) node[left] {{\small $\au'$}};
		\fill[red!80!black] ($(p2)+(d)+(s)$) circle (0pt) node[left] {{\small $\du'$}};
		\end{tikzpicture}
	}%
	\tikzzbox{%
		\begin{tikzpicture}[thick,scale=1.0,color=black, baseline=3.9cm]
		\coordinate (p1) at (0,0);
		\coordinate (p2) at (0.5,0.2);
		\coordinate (p3) at (-0.5,0.4);
		\coordinate (p4) at (2.5,0);
		\coordinate (p5) at (1.5,0.2);
		\coordinate (p6) at (2.0,0.4);
		\coordinate (s) at ($(0,2.5)$);
		\coordinate (d) at ($(2,0)$);
		\coordinate (h) at ($(0,2.5)$);
		\coordinate (t) at ($(0.1,0)$);
		\coordinate (u) at ($(0,-1.1)$);
		%
		%
		\fill [orange!20!white, opacity=0.8] 
		($(p2)+(d)+(h)$) .. controls +(0,0.1) and +(0.5,0) ..  ($(p3)+(d)+(h)$) 
		-- ($(p3)+(h)$) -- ($(p3)+(s)+(h)$) -- ($(p3)+(d)+(s)+(h)$)
		-- ($(p3)+(d)+(s)+(h)$) .. controls +(0.5,0) and +(0,0.1) ..($(p2)+(d)+(s)+(h)$);
		%
		\draw[very thick, red!80!black] ($(p2)+(d)+(h)$) .. controls +(0,0.1) and +(0.5,0) ..  ($(p3)+(d)+(h)$) 
		-- ($(p3)+(h)$);
		\draw[very thick, red!80!black] ($(p2)+(d)+(h)+(s)$) .. controls +(0,0.1) and +(0.5,0) ..  ($(p3)+(d)+(h)+(s)$) 
		-- ($(p3)+(h)+(s)$);
		%
		\fill [orange!30!white, opacity=0.8] 
		($(p1)+(h)$) -- ($(p1)+(d)+(h)$) .. controls +(0.3,0) and +(0,-0.1) .. ($(p2)+(d)+(h)$)
		-- ($(p2)+(d)+(h)$) -- ($(p2)+(d)+(s)+(h)$)
		-- ($(p2)+(d)+(s)+(h)$) .. controls +(0,-0.1) and +(0.3,0) .. ($(p1)+(d)+(s)+(h)$)
		-- ($(p1)+(s)+(h)$); 
		%
		\draw[thin] ($(p1)+(h)$) -- ($(p1)+(h)+(s)$);
		\draw[thin] ($(p2)+(d)+(h)$) -- ($(p2)+(d)+(s)+(h)$);
		\draw[thin] ($(p3)+(h)$) -- ($(p3)+(h)+(s)$);
		%
		\draw[very thick, red!80!black] ($(p1)+(h)$) -- ($(p1)+(d)+(h)$) .. controls +(0.3,0) and +(0,-0.1) .. ($(p2)+(d)+(h)$); 
		\draw[very thick, red!80!black] ($(p1)+(h)+(s)$) -- ($(p1)+(d)+(h)+(s)$) .. controls +(0.3,0) and +(0,-0.1) .. ($(p2)+(d)+(h)+(s)$); 
		%
		\fill[red!80!black] ($(p1)+(h)+(t)$) circle (0pt) node[left] {{\small $\yu'$}};
		\fill[red!80!black] ($(p1)+(h)+(s)+(t)$) circle (0pt) node[left] {{\small $\yu'$}};
		\fill[red!80!black] ($(p3)+(h)+(t)$) circle (0pt) node[left] {{\small $\xu'$}};
		\fill[red!80!black] ($(p3)+(h)+(s)+(t)$) circle (0pt) node[left] {{\small $\xu'$}};
		\fill[red!80!black] ($(p2)+(d)+(h)-0.5*(t)$) circle (0pt) node[right] {{\small $\au$}};
		\fill[red!80!black] ($(p2)+(d)+(s)+(h)-0.5*(t)$) circle (0pt) node[right] {{\small $\du$}};
		\fill[red!80!black] ($(p3)+(h)+(d)+(s)$) circle (2.5pt) node[below] {{\small $\xu$}};
		\fill[red!80!black] ($(p3)+(s)+(0.8,2.3)$) circle (0pt) node {{\small $\bu$}};
		\fill[red!80!black] ($(p1)+(s)+(0.8,2.25)$) circle (0pt) node {{\small $\cu$}};
		\fill[red!80!black] ($(p1)+(h)+(d)+(s)$) circle (2.5pt) node[below] {{\small $\yu$}};
		\end{tikzpicture}
	}  
	\label{eq:CylinderWithExtraDots}
	\\ 
	& \quad 
	= \big[ \du'-\au', \, \yu' - \xu' \big] 
			\otimes \big[ \bu-\au, \, \xu'-\xu\big] 
			\otimes \big[ \du - \au, \, \xu-\yu \big] 
			\otimes \big[ \cu - \au, \, \yu-\yu' \big] \, 
		\nonumber	
	\end{align}
	(the right diagram corresponds to the inverse of~\eqref{eq:sub4}) with
	\be
	\label{eq:ZnOnCup}
	\tikzzbox{%
		\begin{tikzpicture}[thick,scale=1.0,color=black, baseline=-0.7cm]
		\coordinate (p1) at (-2.75,0);
		\coordinate (p2) at (-1,0);
		\fill [orange!20!white, opacity=0.8] 
		(p1) .. controls +(0,-0.5) and +(0,-0.5) ..  (p2)
		-- (p2) .. controls +(0,0.5) and +(0,0.5) ..  (p1)
		;
		\fill [orange!30!white, opacity=0.8] 
		(p1) .. controls +(0,-0.5) and +(0,-0.5) ..  (p2)
		-- (p2) .. controls +(0,-2) and +(0,-2) ..  (p1)
		;
		\draw (p1) .. controls +(0,-2) and +(0,-2) ..  (p2); 
		\draw[very thick, red!80!black] (p1) .. controls +(0,0.5) and +(0,0.5) ..  (p2); 
		\draw[very thick, red!80!black] (p1) .. controls +(0,-0.5) and +(0,-0.5) ..  (p2); 
		%
		\fill[red!80!black] ($(p1)+(0.1,0)$) circle (0pt) node[left] {{\small $\au'$}};
		\fill[red!80!black] ($(p2)+(-0.05,0)$) circle (0pt) node[right] {{\small $\au\vphantom{\au'}$}};
		\fill[red!80!black] (-1.875,+0.38) circle (2.5pt) node[above] {{\small $\xu'$}};
		\fill[red!80!black] (-1.875,-0.38) circle (2.5pt) node[below] {{\small $\yu'$}};
		\end{tikzpicture}
	}%
	= \big[ \au'-\au, \, \xu'-\yu'\big] \, .
	\ee 
	We obtain that~$\zz_n$ sends the decomposition 
	\be  
	\tikzzbox{%
		\begin{tikzpicture}[thick,scale=1.0,color=black, baseline=-0.2cm]
		\coordinate (p1) at (0,0);
		\coordinate (p2) at (0.5,0.2);
		\coordinate (p3) at (-0.5,0.4);
		\coordinate (p4) at (2.5,0);
		\coordinate (p5) at (1.5,0.2);
		\coordinate (p6) at (2.0,0.4);
		\coordinate (s) at ($(0,1.5)$);
		\coordinate (s2) at ($(0,0.75)$);
		\coordinate (d) at ($(4,0)$);
		\coordinate (h) at ($(0,2.5)$);
		\coordinate (t) at ($(0.1,0)$);
		\coordinate (u) at ($(0,-1.1)$);
		%
		%
		%
		\fill [orange!20!white, opacity=0.8] 
		(p3) .. controls +(0.5,0) and +(0,0.1) ..  (p2)
		-- (p2) .. controls +(0,0.9) and +(0,0.9) ..  (p5)
		-- (p5) .. controls +(0,0.1) and +(-0.3,0) .. (p6)
		-- (p6) -- ($(p6)+(s)$) -- ($(p3)+(s)$);
		%
		\fill[red!80!black] ($(p1)+(t)$) circle (0pt) node[left] {{\small $-$}};
		\fill[red!80!black] ($(p3)+(t)$) circle (0pt) node[left] {{\small $+$}};
		\fill[red!80!black] ($(p4)-0.5*(t)$) circle (0pt) node[right] {{\small $-$}};
		\fill[red!80!black] ($(p6)-0.5*(t)$) circle (0pt) node[right] {{\small $+$}};
		\fill[red!80!black] ($(p4)+(s)-0.5*(t)$) circle (0pt) node[right] {{\small $-$}};
		\fill[red!80!black] ($(p1)+(s)+(t)$) circle (0pt) node[left] {{\small $-$}};
		\fill[red!80!black] ($(p3)+(s)+(t)$) circle (0pt) node[left] {{\small $+$}};
		\fill[red!80!black] ($(p6)+(s)-0.5*(t)$) circle (0pt) node[right] {{\small $+$}};
		%
		%
		\draw[very thick, red!80!black] (p2) .. controls +(0,0.1) and +(0.5,0) ..  (p3); 
		\draw[very thick, red!80!black] (p6) .. controls +(-0.3,0) and +(0,0.1) ..  (p5); 
		\draw[thin] (p6) -- ($(p6)+(s)$);
		%
		\fill [orange!30!white, opacity=0.8] 
		(p1) .. controls +(0.3,0) and +(0,-0.1) ..  (p2)
		-- (p2) .. controls +(0,0.9) and +(0,0.9) ..  (p5)
		-- (p5) .. controls +(0,-0.1) and +(-0.5,0) ..  (p4) 
		-- (p4) -- ($(p4)+(s)$) -- ($(p1)+(s)$);
		%
		\draw[very thick, red!80!black] (p1) .. controls +(0.3,0) and +(0,-0.1) ..  (p2); 
		\draw[very thick, red!80!black] (p5) .. controls +(0,-0.1) and +(-0.5,0) ..  (p4); 
		\draw[very thick, red!80!black] ($(p4)+(s)$) -- ($(p1)+(s)$);
		\draw[very thick, red!80!black] ($(p6)+(s)$) -- ($(p3)+(s)$);
		%
		\draw (p2) .. controls +(0,0.9) and +(0,0.9) ..  (p5);
		\draw[thin] (p1) -- ($(p1)+(s)$);
		\draw[thin] (p3) -- ($(p3)+(s)$);
		\draw[thin] (p4) -- ($(p4)+(s)$);
		%
		%
		%
		%
		\fill [orange!20!white, opacity=0.8] 
		($(p3)-(s2)$) .. controls +(0.5,0) and +(0,0.1) ..  ($(p2)-(s2)$)
		-- ($(p2)-(s2)$) .. controls +(0,-0.9) and +(0,-0.9) ..  ($(p5)-(s2)$)
		-- ($(p5)-(s2)$) .. controls +(0,0.1) and +(-0.3,0) .. ($(p6)-(s2)$)
		-- ($(p6)-(s2)$) -- ($(p6)-(s)-(s2)$) -- ($(p3)-(s)-(s2)$);
		%
		\fill[red!80!black] ($(p1)+(t)-(s2)$) circle (0pt) node[left] {{\small $-$}};
		\fill[red!80!black] ($(p3)+(t)-(s2)$) circle (0pt) node[left] {{\small $+$}};
		\fill[red!80!black] ($(p4)-0.5*(t)-(s2)$) circle (0pt) node[right] {{\small $-$}};
		\fill[red!80!black] ($(p6)-0.5*(t)-(s2)$) circle (0pt) node[right] {{\small $+$}};
		\fill[red!80!black] ($(p4)-(s2)-(s)-0.5*(t)$) circle (0pt) node[right] {{\small $-$}};
		\fill[red!80!black] ($(p1)-(s2)-(s)+(t)$) circle (0pt) node[left] {{\small $-$}};
		\fill[red!80!black] ($(p3)-(s2)-(s)+(t)$) circle (0pt) node[left] {{\small $+$}};
		\fill[red!80!black] ($(p6)-(s2)-(s)-0.5*(t)$) circle (0pt) node[right] {{\small $+$}};
		%
		%
		\draw[very thick, red!80!black] ($(p2)-(s2)$) .. controls +(0,0.1) and +(0.5,0) .. ($(p3)-(s2)$); 
		\draw[very thick, red!80!black] ($(p6)-(s2)$) .. controls +(-0.3,0) and +(0,0.1) .. ($(p5)-(s2)$); 
		\draw[very thick, red!80!black] ($(p6)-(s)-(s2)$) -- ($(p3)-(s)-(s2)$);	
		\draw[thin] ($(p6)-(s2)$) -- ($(p6)-(s)-(s2)$);
		%
		\fill [orange!30!white, opacity=0.8] 
		($(p1)-(s2)$) .. controls +(0.3,0) and +(0,-0.1) ..  ($(p2)-(s2)$)
		-- ($(p2)-(s2)$) .. controls +(0,-0.9) and +(0,-0.9) ..  ($(p5)-(s2)$)
		-- ($(p5)-(s2)$) .. controls +(0,-0.1) and +(-0.5,0) ..  ($(p4)-(s2)$) 
		-- ($(p4)-(s2)$) -- ($(p4)-(s)-(s2)$) -- ($(p1)-(s)-(s2)$);
		%
		\draw[very thick, red!80!black] ($(p1)-(s2)$) .. controls +(0.3,0) and +(0,-0.1) ..  ($(p2)-(s2)$); 
		\draw[very thick, red!80!black] ($(p5)-(s2)$) .. controls +(0,-0.1) and +(-0.5,0) ..  ($(p4)-(s2)$); 
		\draw[very thick, red!80!black] ($(p4)-(s)-(s2)$) -- ($(p1)-(s)-(s2)$); 
		%
		\draw ($(p2)-(s2)$) .. controls +(0,-0.9) and +(0,-0.9) ..  ($(p5)-(s2)$);
		\draw[thin] ($(p1)-(s2)$) -- ($(p1)-(s)-(s2)$);
		\draw[thin] ($(p3)-(s2)$) -- ($(p3)-(s)-(s2)$);
		\draw[thin] ($(p4)-(s2)$) -- ($(p4)-(s)-(s2)$);
		%
		%
		%
		%
		%
		\fill [orange!20!white, opacity=0.8] 
		($(p2)+(d)-(s)-(s2)$) .. controls +(0,0.1) and +(0.5,0) ..  ($(p3)+(d)-(s)-(s2)$)
		-- ($(p3)+(d)-(s)-(s2)$) -- ($(p3)+(d)+(s)$) 
		-- ($(p3)+(d)+(s)$) .. controls +(0.5,0) and +(0,0.1) ..($(p2)+(d)+(s)$); 
		%
		%
		\draw[very thick, red!80!black] ($(p2)+(d)-(s)-(s2)$) .. controls +(0,0.1) and +(0.5,0) ..  ($(p3)+(d)-(s)-(s2)$); 
		\draw[very thick, red!80!black] ($(p2)+(d)+(s)$) .. controls +(0,0.1) and +(0.5,0) ..  ($(p3)+(d)+(s)$); 
		%
		\fill [orange!30!white, opacity=0.8] 
		($(p1)+(d)-(s)-(s2)$) .. controls +(0.3,0) and +(0,-0.1) .. ($(p2)+(d)-(s)-(s2)$)
		-- ($(p2)+(d)-(s)-(s2)$) -- ($(p2)+(d)+(s)$)
		-- ($(p2)+(d)+(s)$) .. controls +(0,-0.1) and +(0.3,0) .. ($(p1)+(d)+(s)$);
		%
		\draw[very thick, red!80!black] ($(p1)+(d)-(s)-(s2)$) .. controls +(0.3,0) and +(0,-0.1) ..  ($(p2)+(d)-(s)-(s2)$); 
		\draw[very thick, red!80!black] ($(p1)+(d)+(s)$) .. controls +(0.3,0) and +(0,-0.1) ..  ($(p2)+(d)+(s)$); 
		%
		\draw[thin] ($(p1)+(d)-(s)-(s2)$) -- ($(p1)+(d)+(s)$);
		\draw[thin] ($(p3)+(d)-(s)-(s2)$) -- ($(p3)+(d)+(s)$);
		\draw[thin] ($(p2)+(d)-(s)-(s2)$) -- ($(p2)+(d)+(s)$);
		%
		\fill[red!80!black] ($(p3)+(s)+(d)+(t)$) circle (0pt) node[left] {{\small $+$}};
		\fill[red!80!black] ($(p1)+(s)+(d)+(t)$) circle (0pt) node[left] {{\small $-$}};
		\fill[red!80!black] ($(p3)-(s2)-(s)+(d)+(t)$) circle (0pt) node[left] {{\small $+$}};
		\fill[red!80!black] ($(p1)-(s2)-(s)+(d)+(t)$) circle (0pt) node[left] {{\small $-$}};
		%
		%
		%
		%
		\fill [orange!20!white, opacity=0.8] 
		($(p5)-(d)-(s)-(s2)$) .. controls +(0,0.1) and +(-0.3,0) .. ($(p6)-(d)-(s)-(s2)$)
		-- ($(p6)-(d)-(s)-(s2)$) -- ($(p6)-(d)+(s)$)
		-- ($(p6)-(d)+(s)$) .. controls +(-0.3,0) and +(0,0.1) .. ($(p5)-(d)+(s)$);
		%
		%
		\draw[thin] ($(p6)-(d)-(s)-(s2)$) -- ($(p6)-(d)+(s)$);
		%
		\draw[very thick, red!80!black] ($(p5)-(d)-(s)-(s2)$) .. controls +(0,0.1) and +(-0.3,0) ..  ($(p6)-(d)-(s)-(s2)$); 
		\draw[very thick, red!80!black] ($(p5)-(d)+(s)$) .. controls +(0,0.1) and +(-0.3,0) ..  ($(p6)-(d)+(s)$); 
		%
		\fill[red!80!black] ($(p4)+(s)-(d)-(t)$) circle (0pt) node[right] {{\small $-$}};
		\fill[red!80!black] ($(p6)+(s)-(d)-(t)$) circle (0pt) node[right] {{\small $+$}};
		\fill[red!80!black] ($(p4)-(s2)-(s)-(d)-(t)$) circle (0pt) node[right] {{\small $-$}};
		\fill[red!80!black] ($(p6)-(s2)-(s)-(d)-(t)$) circle (0pt) node[right] {{\small $+$}};
		colouring front: 
		\fill [orange!30!white, opacity=0.8] 
		($(p5)-(d)-(s)-(s2)$) .. controls +(0,-0.1) and +(-0.5,0) ..  ($(p4)-(d)-(s)-(s2)$)
		-- ($(p4)-(d)-(s)-(s2)$) -- ($(p4)-(d)+(s)$) 
		-- ($(p4)-(d)+(s)$) .. controls +(-0.5,0) and +(0,-0.1) ..($(p5)-(d)+(s)$); 
		%
		\draw[very thick, red!80!black] ($(p4)-(d)+(s)$) .. controls +(-0.5,0) and +(0,-0.1) ..($(p5)-(d)+(s)$);
		\draw[very thick, red!80!black] ($(p4)-(d)-(s)-(s2)$) .. controls +(-0.5,0) and +(0,-0.1) ..($(p5)-(d)-(s)-(s2)$);
		%
		\draw[thin] ($(p4)-(d)-(s)-(s2)$) -- ($(p4)-(d)+(s)$);
		\draw[thin] ($(p5)-(d)-(s)-(s2)$) -- ($(p5)-(d)+(s)$);
		%
		%
		%
		\coordinate (z) at (0,-1.0);
		%
		\fill [orange!20!white, opacity=0.8] 
		($(p5)-(d)-(s)-(s2)+(z)$) .. controls +(0,-1.5) and +(0,-1.5) .. ($(p2)+(d)-(s)-(s2)+(z)$) 
		-- ($(p2)+(d)-(s)-(s2)+(z)$)  .. controls +(0,0.4) and +(0,0.4) .. ($(p5)-(d)-(s)-(s2)+(z)$);
		%
		\fill [orange!30!white, opacity=0.8] 
		($(p5)-(d)-(s)-(s2)+(z)$) .. controls +(0,-1.5) and +(0,-1.5) .. ($(p2)+(d)-(s)-(s2)+(z)$) 
		-- ($(p2)+(d)-(s)-(s2)+(z)$)  .. controls +(0,-0.4) and +(0,-0.4) .. ($(p5)-(d)-(s)-(s2)+(z)$);
		%
		\draw[thin] ($(p5)-(d)-(s)-(s2)+(z)$) .. controls +(0,-1.5) and +(0,-1.5) .. ($(p2)+(d)-(s)-(s2)+(z)$) ;
		\draw[very thick, red!80!black] ($(p2)+(d)-(s)-(s2)+(z)$)  .. controls +(0,0.4) and +(0,0.4) .. ($(p5)-(d)-(s)-(s2)+(z)$);
		\draw[very thick, red!80!black] ($(p2)+(d)-(s)-(s2)+(z)$)  .. controls +(0,-0.4) and +(0,-0.4) .. ($(p5)-(d)-(s)-(s2)+(z)$);
		\end{tikzpicture}
	}%
	\ee 
	of the torus with one disc cut out to the isomorphism class of the tensor product of the matrix factorisations 
	\eqref{eq:Middle}, \eqref{eq:s24p}:
	\begin{align}
	\, & \big[ \widetilde{\cu}-\cu, \, \yu-\yu' \big] 
	\otimes \big[ \widetilde{\bu}-\bu, \, \xu'-\xu \big] 
	\otimes \big[ \eu - \du, \, \xu-\yu \big] 
	\otimes \big[ \eu' - \du', \, \yu'-\xu' \big]
	\nonumber 
	\\
	& \quad\qquad 
	\otimes \big[ 
		 \zeru 
	, \, \yu' -\xu \big] 
	\otimes \big[ 
		 \zeru 
	, \, \widetilde{\bu} - \widetilde{\cu} \big]
	\nonumber
	\\ 
	\, & \quad\qquad 
	\otimes \big[\bu-\du,\,\xu'-\xu\big]\otimes\big[\cu-\du,\,\yu-\yu'\big]\otimes\big[\du'-\du,\,\xu'-\yu'\big]\,.
	\label{eq:g>1step3}
\end{align}
The first four tensor factors in \eqref{eq:g>1step3} can now be used to eliminate the internal variables 
by means of Lemma~\ref{lem:elimination}. One is left with the tensor factors in the last two lines of \eqref{eq:g>1step3}, with the replacements $\bu\lmt\widetilde{\bu}$, $\cu\lmt\widetilde{\cu}$, 
$\du\lmt{\eu}$ and $\du'\lmt{\eu}'$. 
Note that the third line of \eqref{eq:g>1step3} comes from the 2-morphism \eqref{eq:s24p} associated to the southern hemisphere with four marked points on the equator. Since it is only changed by the replacement of variables above, it reproduces the same 2-morphism. The matrix factorisations in the second line of \eqref{eq:g>1step3} can both be brought into the form 
	 $[\zeru,\zeru] \cong (\C\oplus\C[1])^{\otimes n}$
by applying Lemma~\ref{lem:addition}.
Thus, \eqref{eq:g>1step3} is isomorphic to 
\begin{align}\label{eq:g>1step4}
	  & 
	  	 \big( \C \oplus \C[1] \big)^{\otimes 2n}
	   \otimes_\C 
	\tikzzbox{%
		\begin{tikzpicture}[thick,scale=1.0,color=black, baseline=-1.5cm]
		\coordinate (p1) at (0,0);
		\coordinate (p2) at (0.5,0.2);
		\coordinate (p3) at (-0.5,0.4);
		\coordinate (p4) at (2.5,0);
		\coordinate (p5) at (1.5,0.2);
		\coordinate (p6) at (2.0,0.4);
		\coordinate (s) at ($(0,1.5)$);
		\coordinate (d) at ($(3.8,0)$);
		\coordinate (h) at ($(0,2.5)$);
		\coordinate (t) at ($(0.1,0)$);
		\coordinate (u) at ($(0,-1.1)$);
		%
		\fill [orange!20!white, opacity=0.8] 
		($(p5)+(u)$) .. controls +(0,-2) and +(0,-2) .. ($(p2)+(d)+(u)$) 
		-- ($(p2)+(d)+(u)$) .. controls +(0,0.4) and +(0,0.4) .. ($(p5)+(u)$);
		%
		\fill [orange!30!white, opacity=0.8] 
		($(p5)+(u)$) .. controls +(0,-2) and +(0,-2) .. ($(p2)+(d)+(u)$) 
		-- ($(p2)+(d)+(u)$) .. controls +(0,-0.4) and +(0,-0.4) .. ($(p5)+(u)$);
		%
		\draw ($(p5)+(u)$) .. controls +(0,-2) and +(0,-2) ..  ($(p2)+(d)+(u)$);
		\draw[very thick, red!80!black] ($(p5)+(u)$) .. controls +(0,0.4) and +(0,0.4) ..  ($(p2)+(d)+(u)$);
		\draw[very thick, red!80!black] ($(p5)+(u)$) .. controls +(0,-0.4) and +(0,-0.4) ..  ($(p2)+(d)+(u)$);
		\fill[red!80!black] ($(p5)+(u)+(2.2,0.25)$) circle (2.5pt) node[above] {{\small $\xu$}};
		\fill[red!80!black] ($(p5)+(u)+(2.2,-0.255)$) circle (2.5pt) node[below] {{\small $\yu$}};
		\fill[red!80!black] ($(p5)+(u)+(0.6,0.25)$) circle (2.5pt) node[above] {{\small $\xu'$}};
		\fill[red!80!black] ($(p5)+(u)+(0.6,-0.255)$) circle (2.5pt) node[below] {{\small $\yu'$}};
		%
		\fill[red!80!black] ($(p5)+(u)+0.5*(t)$) circle (0pt) node[left] {{\small $\eu'$}};
		\fill[red!80!black] ($(p2)+(d)+(u)-0.5*(t)$) circle (0pt) node[right] {{\small $\eu\vphantom{\eu'}$}};
		\fill[red!80!black] ($(p5)+(u)+(1.4,0.6)$) circle (0pt) node {{\small $\widetilde{\bu}$}};
		\fill[red!80!black] ($(p5)+(u)+(1.4,-0.55)$) circle (0pt) node {{\small $\widetilde{\cu}$}};
		\end{tikzpicture}
	}%
	\, . 
	\end{align} 
	Post-composing this with 
	\be 
	\tikzzbox{%
		\begin{tikzpicture}[thick,scale=1.0,color=black, baseline=1.2cm, yscale=-1]
		\coordinate (p1) at (0,0);
		\coordinate (p2) at (0.5,0.2);
		\coordinate (p3) at (-0.5,0.4);
		\coordinate (p4) at (2.5,0);
		\coordinate (p5) at (1.5,0.2);
		\coordinate (p6) at (2.0,0.4);
		\coordinate (s) at ($(0,1.5)$);
		\coordinate (d) at ($(3.8,0)$);
		\coordinate (h) at ($(0,2.5)$);
		\coordinate (t) at ($(0.1,0)$);
		\coordinate (u) at ($(0,-1.1)$);
		%
		\fill [orange!30!white, opacity=0.8] 
		($(p5)+(u)$) .. controls +(0,-2) and +(0,-2) .. ($(p2)+(d)+(u)$) 
		-- ($(p2)+(d)+(u)$) .. controls +(0,-0.4) and +(0,-0.4) .. ($(p5)+(u)$);
		%
		\draw[very thick, red!80!black] ($(p5)+(u)$) .. controls +(0,-0.4) and +(0,-0.4) ..  ($(p2)+(d)+(u)$);
		\fill[red!80!black] ($(p5)+(u)+(1.4,-0.6)$) circle (0pt) node {{\small $\widetilde{\bu}$}};
		\fill[red!80!black] ($(p5)+(u)+(2.2,-0.255)$) circle (2.5pt) node[above] {{\small $\xu$}};
		\fill[red!80!black] ($(p5)+(u)+(0.6,-0.255)$) circle (2.5pt) node[above] {{\small $\xu'$}};
		%
		%
		\fill [orange!30!white, opacity=0.8] 
		($(p5)+(u)$) .. controls +(0,-2) and +(0,-2) .. ($(p2)+(d)+(u)$) 
		-- ($(p2)+(d)+(u)$) .. controls +(0,0.4) and +(0,0.4) .. ($(p5)+(u)$);
		%
		\draw ($(p5)+(u)$) .. controls +(0,-2) and +(0,-2) ..  ($(p2)+(d)+(u)$);
		\draw[very thick, red!80!black] ($(p5)+(u)$) .. controls +(0,0.4) and +(0,0.4) ..  ($(p2)+(d)+(u)$);
		\fill[red!80!black] ($(p5)+(u)+(2.2,0.25)$) circle (2.5pt) node[below] {{\small $\yu$}};
		\fill[red!80!black] ($(p5)+(u)+(0.6,0.25)$) circle (2.5pt) node[below] {{\small $\yu'$}};
		%
		\fill[red!80!black] ($(p5)+(u)+0.5*(t)$) circle (0pt) node[left] {{\small $\eu'$}};
		\fill[red!80!black] ($(p2)+(d)+(u)-0.5*(t)$) circle (0pt) node[right] {{\small $\eu\vphantom{\eu'}$}};
		\fill[red!80!black] ($(p5)+(u)+(1.4,0.55)$) circle (0pt) node {{\small $\widetilde{\cu}$}};
		\end{tikzpicture}
	}%
	\cong \big[ \widetilde{\cu} - \eu', \, \yu'-\yu \big] 
			\otimes \big[ \eu-\eu', \, \yu-\yu \big] 
			\otimes \big[ \widetilde{\bu} -\eu', \, \xu-\xu' \big] \, , 
	\ee 
	we obtain $\zz_n(\Sigma_1) = \zz_n(T^2) \cong 
		 (\C\oplus\C[1])^{\otimes 2n} \otimes_\C \C[\au,\xu]$ 
	by the type of reasoning we used to establish $\zz_n(S^2) \cong \C[\au,\xu]$. 

	Finally, we can obtain the 2-morphism corresponding to the genus-$g$ surface $\Sigma_g$ by composing the $g$-th power of~\eqref{eq:g>1step1} with the 2-morphisms corresponding to the hemispheres. Making repeated use of the fact that the 2-morphism~\eqref{eq:g>1step1} composes with the one associated to the southern hemisphere to~\eqref{eq:g>1step3}, we produce the formula for $\zz_n(\Sigma_g)$ in~\eqref{eq:StateSpaceSigmag}. 
\end{proof}

\begin{remark}
	\label{rem:ModuliSpaceOfTruncatedAffineRWmodels}
	We may apply the cobordism hypothesis to obtain some information on the moduli space~$\mathcal M_\bi$ of truncated affine Rozansky--Witten models. 
	Indeed, according to Remark~\ref{rem:ModuliSpaceOfFullyExtendedTQFTs} and because we have that $\bi^\times \cong \Bhfp{\bi}$ (as implied by Theorem~\ref{cor:ExtendedTQFTZn}), the connected components and the fundamental groupoid of~$\mathcal M_\bi$ are encoded in $\Pi_0(\bi^\times)$ and $\Pi_1(\bi^\times)$. 
	
	To determine $\Pi_0(\bi^\times)$, we note that if for positive integers~$m$ and~$n$ we have $(x_1,\dots,x_m) \cong (x_1,\dots,x_n)$ in~$\bi$, then $\zz_m \cong \zz_n$ as TQFTs. 
	Hence because $\zz_m(\Sigma_g) \ncong \zz_n(\Sigma_g)$ for $m\neq n$ by Proposition~\ref{prop:StateSpaces}, we have $\zz_m \ncong \zz_n$ for $m\neq n$, which in turn implies $(x_1,\dots,x_m) \ncong (x_1,\dots,x_n)$. 
	Thus 
	\be 
	\label{eq:Pi0ModuliSpace}
	\Pi_0(\mathcal M_\bi) 
		\cong \Pi_0(\bi^\times)
		\cong \N \, . 
	\ee 
	
	To determine $\Pi_1(\bi^\times)$, we must consider all invertible 1-morphisms in~$\bi$ and ask when two parallel such 1-morphisms $(\au;W), (\bu;V) \colon \xu \lra \yu$ are isomorphic. 
	Because of~\eqref{eq:Pi0ModuliSpace} invertibility implies $\xu\cong\yu$, while an isomorphism $(\au;W) \cong (\bu;V)$ in~$\bi$ by definition gives $\pi_0(\hmf(\C[\au,\xu,\yu], W)^\omega) \cong \pi_0(\hmf(\C[\bu,\xu,\yu], V)^\omega)$. 
	If this were to lift to an equivalence $\hmf(\C[\au,\xu,\yu], W)^\omega \cong \hmf(\C[\bu,\xu,\yu], V)^\omega$, then \cite[Thm.\,1.4]{Kalck2021} would imply that~$W$ and~$V$ differ by an even number of squares, and by Knörrer periodicity we would then find that the fundamental groups of~$\mathcal M_\bi$ are all trivial. 
\end{remark}

\subsubsection{Commutative Frobenius algebras and Grothendieck rings}
\label{subsubsec:ComFrobAlgGrothRing}

From every fully extended 2-dimensional TQFT~$\zz$ one obtains a closed TQFT by restricting~$\zz$ to circles and bordisms between them. 
In particular, from the extended TQFT~$\zz_n$ of Theorem~\ref{cor:ExtendedTQFTZn} we obtain a closed oriented TQFT 
\be 
\zz_n^{\textrm{cl}} := \zz_n \Big|_{\Bordor(\varnothing,\varnothing)} \colon \Bord_{2,1}^{\textrm{or}} \lra \bi(\xu,\xu) \, . 
\ee 

Recall that 2-dimensional oriented closed TQFTs with values in a given symmetric monoidal category $(\mathcal V,\otimes,\one)$ are equivalent to commutative Frobenius algebras $(A,\mu,\eta,\beta)$ in~$\mathcal V$, see e.\,g.\ \cite{Kockbook}. 
If $\mathcal V = \Vect_{\C}$ with the standard tensor product, those algebras are precisely ordinary commutative Frobenius algebras over~$\C$. 
In general, $A$ is an object in~$\mathcal V$ together with a commutative associative multiplication $\mu\colon A\otimes A\lra A$, a unit $\eta\colon \one \lra A$, and a non-degenerate pairing $\beta \colon A \otimes A \lra \one$ that is compatible with~$\mu$. 
Applying this classification result to the closed TQFT~$\zz_n^{\textrm{cl}}$, we can equivalently describe it as follows: 

\begin{proposition}
	\label{prop:CommutativeFrobeniusAlgebraFromZn}
	The closed TQFT~$\zz_n^{\textrm{cl}}$ is classified by the commutative Frobenius algebra $(A,\mu,\eta,\beta)$ in $\bi(\xu,\xu)$ with 
	\begin{align}
	A = \zz_n\big(
	\tikzzbox{
		\begin{tikzpicture}[thick,scale=1.5,color=black, baseline=0.065cm]
		\coordinate (p1) at (0,0);
		\coordinate (p2) at (0,0.25);
		\coordinate (g1) at (0,-0.05);
		\coordinate (g2) at (0,0.3);		
		\draw[very thick, red!80!black] (p1) .. controls +(-0.2,0) and +(-0.2,0) ..  (p2); 
		\draw[very thick, red!80!black] (p1) .. controls +(0.2,0) and +(0.2,0) ..  (p2); 
		\end{tikzpicture}
	}
	\big) 
	& \cong \Big( \au,\du,\xu,\yu ; \, \big( \au-\du \big) \cdot \big( \xu-\yu\big) \Big) , 
	\\
	\mu = \zz_n\Big(
	\tikzzbox{
		\begin{tikzpicture}[thick,scale=0.8,color=black, baseline=0.2cm]
		\coordinate (p1) at (-0.55,0);
		\coordinate (p2) at (-0.2,0);
		\coordinate (p3) at (0.2,0);
		\coordinate (p4) at (0.55,0);
		\coordinate (p5) at (0.175,0.8);
		\coordinate (p6) at (-0.175,0.8);
		\draw[very thick, red!80!black] (p1) .. controls +(0,0.1) and +(0,0.1) ..  (p2); 
		\draw[very thick, red!80!black] (p3) .. controls +(0,0.1) and +(0,0.1) ..  (p4); 
		%
		\fill [orange!35!white, opacity=0.8] 
		(p1) .. controls +(0,0.1) and +(0,0.1) ..  (p2)
		-- (p2) .. controls +(0,0.35) and +(0,0.35) ..  (p3)
		-- (p3) .. controls +(0,0.1) and +(0,0.1) ..  (p4)
		-- (p4) .. controls +(0,0.5) and +(0,-0.5) ..  (p5)
		-- (p5) .. controls +(0,-0.1) and +(0,-0.1) ..  (p6)
		-- (p6) .. controls +(0,-0.5) and +(0,0.5) ..  (p1)
		;
		\fill [orange!20!white, opacity=0.8] 
		(p5) .. controls +(0,-0.1) and +(0,-0.1) .. (p6)
		-- (p6) .. controls +(0,0.1) and +(0,0.1) .. (p5);
		\fill [orange!30!white, opacity=0.8] 
		(p1) .. controls +(0,-0.1) and +(0,-0.1) .. (p2)
		-- (p2) .. controls +(0,0.1) and +(0,0.1) .. (p1);
		\fill [orange!30!white, opacity=0.8] 
		(p3) .. controls +(0,-0.1) and +(0,-0.1) .. (p4)
		-- (p4) .. controls +(0,0.1) and +(0,0.1) .. (p3);
		\draw (p2) .. controls +(0,0.35) and +(0,0.35) ..  (p3); 
		\draw (p4) .. controls +(0,0.5) and +(0,-0.5) ..  (p5); 
		\draw (p6) .. controls +(0,-0.5) and +(0,0.5) ..  (p1); 
		\draw[very thick, red!80!black] (p1) .. controls +(0,-0.1) and +(0,-0.1) ..  (p2); 
		\draw[very thick, red!80!black] (p3) .. controls +(0,-0.1) and +(0,-0.1) ..  (p4); 
		\draw[very thick, red!80!black] (p5) .. controls +(0,0.1) and +(0,0.1) ..  (p6); 
		\draw[very thick, red!80!black] (p5) .. controls +(0,-0.1) and +(0,-0.1) ..  (p6); 
		\end{tikzpicture}
	}
	\Big)
	& \cong \big[ \au''-\au, \, \yu-\yu' \big]  \otimes \big[ \au''-\au' ,\, \xu'-\xu \big] \otimes \big[ \au'-\au, \, \yu'-\xu \big] 
	\nonumber
	\\ & \qquad \otimes \big[ \du''-\du',\, \yu'-\xu' \big]  \otimes \big[ \au''- \du, \, \xu-\yu \big] 
	\, , \label{eq:ZnPoP}
	\\
	\eta = \zz_n
	\big(
	\tikzzbox{%
		\begin{tikzpicture}[thick,scale=0.25,color=black, rotate=180, baseline=-0.4cm]
		%
		%
		\coordinate (q1) at (-2.75,0.95);
		\coordinate (q2) at (-1,0.95);
		\fill [orange!20!white, opacity=0.8] 
		(q1) .. controls +(0,0.5) and +(0,0.5) ..  (q2)
		-- (q2) .. controls +(0,-0.5) and +(0,-0.5) ..  (q1)
		;
		\fill [orange!30!white, opacity=0.8] 
		(q1) .. controls +(0,0.5) and +(0,0.5) ..  (q2)
		-- (q2) .. controls +(0,1.25) and +(0,1.25) ..  (q1)
		;
		\draw (q1) .. controls +(0,1.25) and +(0,1.25) ..  (q2); 
		\draw[very thick, red!80!black] (q1) .. controls +(0,0.5) and +(0,0.5) ..  (q2); 
		\draw[very thick, red!80!black] (q1) .. controls +(0,-0.5) and +(0,-0.5) ..  (q2); 
		\end{tikzpicture}
	}%
	\big) 
	& \cong \big[ \au-\du,\, \xu-\yu \big] 
	\, , 
	\\
	\beta = \zz_n\Big(
	\tikzzbox{%
		\begin{tikzpicture}[very thick,scale=0.8,color=blue!50!black, baseline=0.1cm]
		\coordinate (p1) at (-0.55,0);
		\coordinate (p2) at (-0.2,0);
		\coordinate (p3) at (0.2,0);
		\coordinate (p4) at (0.55,0);
		\draw[very thick, red!80!black] (p1) .. controls +(0,0.15) and +(0,0.15) ..  (p2); 
		\draw[very thick, red!80!black] (p3) .. controls +(0,0.15) and +(0,0.15) ..  (p4); 
		%
		\fill [orange!35!white, opacity=0.8] 
		(p1) .. controls +(0,-0.15) and +(0,-0.15) .. (p2)
		-- (p2) .. controls +(0,0.35) and +(0,0.35) ..  (p3)
		-- (p3) .. controls +(0,-0.15) and +(0,-0.15) .. (p4)
		-- (p4) .. controls +(0,0.9) and +(0,0.9) ..  (p1)
		;
		\draw[thick, black] (p2) .. controls +(0,0.35) and +(0,0.35) ..  (p3); 
		\draw[thick, black] (p4) .. controls +(0,0.9) and +(0,0.9) ..  (p1);
		\draw[very thick, red!80!black] (p1) .. controls +(0,-0.15) and +(0,-0.15) ..  (p2); 
		\draw[very thick, red!80!black] (p3) .. controls +(0,-0.15) and +(0,-0.15) ..  (p4); 
		\end{tikzpicture}
	}
	\Big)
	& \cong \big[ \du-\au, \, \yu-\yu' \big]  \otimes \big[ \du-\au' ,\, \xu'-\xu \big] \otimes \big[ \au'-\au, \, \yu'-\xu \big] 
	\nonumber
	\\ & \qquad \otimes \big[ \yu'-\xu',\,\du-\du' \big]
	\, . \label{eq:ZnPairing}
	\end{align}
\end{proposition}

\begin{proof}
	We already noticed the expressions for 
	$
	\zz_n(\!
	\tikzzbox{
		\begin{tikzpicture}[thick,scale=1.5,color=black, baseline=0.065cm]
		\coordinate (p1) at (0,0);
		\coordinate (p2) at (0,0.25);
		\coordinate (g1) at (0,-0.05);
		\coordinate (g2) at (0,0.3);		%
		\draw[very thick, red!80!black] (p1) .. controls +(-0.2,0) and +(-0.2,0) ..  (p2); 
		%
		\draw[very thick, red!80!black] (p1) .. controls +(0.2,0) and +(0.2,0) ..  (p2); 
		\end{tikzpicture}
	}
	\!)
	$
	and
	$
	\zz_n(
	\tikzzbox{%
		\begin{tikzpicture}[thick,scale=0.25,color=black, rotate=180, baseline=-0.4cm]
		%
		%
		\coordinate (q1) at (-2.75,0.95);
		\coordinate (q2) at (-1,0.95);
		\fill [orange!20!white, opacity=0.8] 
		(q1) .. controls +(0,0.5) and +(0,0.5) ..  (q2)
		-- (q2) .. controls +(0,-0.5) and +(0,-0.5) ..  (q1)
		;
		\fill [orange!30!white, opacity=0.8] 
		(q1) .. controls +(0,0.5) and +(0,0.5) ..  (q2)
		-- (q2) .. controls +(0,1.25) and +(0,1.25) ..  (q1)
		;
		\draw (q1) .. controls +(0,1.25) and +(0,1.25) ..  (q2); 
		\draw[very thick, red!80!black] (q1) .. controls +(0,0.5) and +(0,0.5) ..  (q2); 
		\draw[very thick, red!80!black] (q1) .. controls +(0,-0.5) and +(0,-0.5) ..  (q2); 
		\end{tikzpicture}
	}%
	) 
	$ 
	in~\eqref{eq:ZnOnCircle} and~\eqref{eq:ZnOnCup}, respectively; here we simply re-named some of the variables. 
	To compute the action of~$\zz_n$ on the pair-of-pants, we decompose it as 
	\be  
	\tikzzbox{
		\begin{tikzpicture}[thick,scale=2,color=black, baseline=0.75cm]
		\coordinate (p1) at (-0.55,0);
		\coordinate (p2) at (-0.2,0);
		\coordinate (p3) at (0.2,0);
		\coordinate (p4) at (0.55,0);
		\coordinate (p5) at (0.175,0.8);
		\coordinate (p6) at (-0.175,0.8);
		\draw[very thick, red!80!black] (p1) .. controls +(0,0.1) and +(0,0.1) ..  (p2); 
		\draw[very thick, red!80!black] (p3) .. controls +(0,0.1) and +(0,0.1) ..  (p4); 
		%
		\fill [orange!35!white, opacity=0.8] 
		(p1) .. controls +(0,0.1) and +(0,0.1) ..  (p2)
		-- (p2) .. controls +(0,0.35) and +(0,0.35) ..  (p3)
		-- (p3) .. controls +(0,0.1) and +(0,0.1) ..  (p4)
		-- (p4) .. controls +(0,0.5) and +(0,-0.5) ..  (p5)
		-- (p5) .. controls +(0,-0.1) and +(0,-0.1) ..  (p6)
		-- (p6) .. controls +(0,-0.5) and +(0,0.5) ..  (p1)
		;
		\fill [orange!20!white, opacity=0.8] 
		(p5) .. controls +(0,-0.1) and +(0,-0.1) .. (p6)
		-- (p6) .. controls +(0,0.1) and +(0,0.1) .. (p5);
		\fill [orange!30!white, opacity=0.8] 
		(p1) .. controls +(0,-0.1) and +(0,-0.1) .. (p2)
		-- (p2) .. controls +(0,0.1) and +(0,0.1) .. (p1);
		\fill [orange!30!white, opacity=0.8] 
		(p3) .. controls +(0,-0.1) and +(0,-0.1) .. (p4)
		-- (p4) .. controls +(0,0.1) and +(0,0.1) .. (p3);
		\draw (p2) .. controls +(0,0.35) and +(0,0.35) ..  (p3); 
		\draw (p4) .. controls +(0,0.5) and +(0,-0.5) ..  (p5); 
		\draw (p6) .. controls +(0,-0.5) and +(0,0.5) ..  (p1); 
		\draw[very thick, red!80!black] (p1) .. controls +(0,-0.1) and +(0,-0.1) ..  (p2); 
		\draw[very thick, red!80!black] (p3) .. controls +(0,-0.1) and +(0,-0.1) ..  (p4); 
		\draw[very thick, red!80!black] (p5) .. controls +(0,0.1) and +(0,0.1) ..  (p6); 
		\draw[very thick, red!80!black] (p5) .. controls +(0,-0.1) and +(0,-0.1) ..  (p6); 
		\end{tikzpicture}
	}
	\;
	=
	\;\;
	\tikzzbox{%
		\begin{tikzpicture}[thick,scale=1.0,color=black, baseline=0.9cm]
		\coordinate (p1) at (0,0);
		\coordinate (p2) at (0.5,0.2);
		\coordinate (p3) at (-0.5,0.4);
		\coordinate (p4) at (2.5,0);
		\coordinate (p5) at (1.5,0.2);
		\coordinate (p6) at (2.0,0.4);
		\coordinate (s) at ($(0,1.5)$);
		\coordinate (d) at ($(3.25,0)$);
		\coordinate (d2) at ($(-3.25,0)$);
		\coordinate (h) at ($(0,2.5)$);
		\coordinate (t) at ($(0.1,0)$);
		\coordinate (u) at ($(0,-1.1)$);
		%
		\fill [orange!20!white, opacity=0.8] 
		(p3) .. controls +(0.5,0) and +(0,0.1) ..  (p2)
		-- (p2) .. controls +(0,0.9) and +(0,0.9) ..  (p5)
		-- (p5) .. controls +(0,0.1) and +(-0.3,0) .. (p6)
		-- (p6) -- ($(p6)+(s)$) -- ($(p3)+(s)$);
		%
		\draw[very thick, red!80!black] (p2) .. controls +(0,0.1) and +(0.5,0) ..  (p3); 
		\draw[very thick, red!80!black] (p6) .. controls +(-0.3,0) and +(0,0.1) ..  (p5); 
		\draw[thin] (p6) -- ($(p6)+(s)$);
		%
		\fill [orange!30!white, opacity=0.8] 
		(p1) .. controls +(0.3,0) and +(0,-0.1) ..  (p2)
		-- (p2) .. controls +(0,0.9) and +(0,0.9) ..  (p5)
		-- (p5) .. controls +(0,-0.1) and +(-0.5,0) ..  (p4) 
		-- (p4) -- ($(p4)+(s)$) -- ($(p1)+(s)$);
		%
		\draw[very thick, red!80!black] (p1) .. controls +(0.3,0) and +(0,-0.1) ..  (p2); 
		\draw[very thick, red!80!black] (p5) .. controls +(0,-0.1) and +(-0.5,0) ..  (p4); 
		\draw[very thick, red!80!black] ($(p4)+(s)$) -- ($(p1)+(s)$);
		\draw[very thick, red!80!black] ($(p6)+(s)$) -- ($(p3)+(s)$);
		%
		\draw (p2) .. controls +(0,0.9) and +(0,0.9) ..  (p5);
		\draw[thin] (p1) -- ($(p1)+(s)$);
		\draw[thin] (p3) -- ($(p3)+(s)$);
		\draw[thin] (p4) -- ($(p4)+(s)$);
		%
		%
		\fill [orange!20!white, opacity=0.8] 
		($(p2)+(d)$) .. controls +(0,0.1) and +(0.5,0) ..  ($(p3)+(d)$)
		-- ($(p3)+(d)$) -- ($(p3)+(d)+(s)$) 
		-- ($(p3)+(d)+(s)$) .. controls +(0.5,0) and +(0,0.1) ..($(p2)+(d)+(s)$);
		%
		\draw[very thick, red!80!black] ($(p2)+(d)$) .. controls +(0,0.1) and +(0.5,0) ..  ($(p3)+(d)$); 
		\draw[very thick, red!80!black] ($(p2)+(d)+(s)$) .. controls +(0,0.1) and +(0.5,0) ..  ($(p3)+(d)+(s)$); 
		%
		\fill [orange!30!white, opacity=0.8] 
		($(p1)+(d)$) .. controls +(0.3,0) and +(0,-0.1) .. ($(p2)+(d)$)
		-- ($(p2)+(d)$) -- ($(p2)+(d)+(s)$)
		-- ($(p2)+(d)+(s)$) .. controls +(0,-0.1) and +(0.3,0) .. ($(p1)+(d)+(s)$); 
		%
		\draw[thin] ($(p1)+(d)$) -- ($(p1)+(d)+(s)$);
		\draw[thin] ($(p2)+(d)$) -- ($(p2)+(d)+(s)$);
		\draw[thin] ($(p3)+(d)$) -- ($(p3)+(d)+(s)$);
		%
		\draw[very thick, red!80!black] ($(p1)+(d)$) .. controls +(0.3,0) and +(0,-0.1) ..  ($(p2)+(d)$); 
		\draw[very thick, red!80!black] ($(p1)+(d)+(s)$) .. controls +(0.3,0) and +(0,-0.1) ..  ($(p2)+(d)+(s)$); 
		%
		
		%
		\fill [orange!20!white, opacity=0.8] 
		($(p5)+(d2)$) .. controls +(0,0.1) and +(-0.3,0) ..  ($(p6)+(d2)$)
		-- ($(p6)+(d2)$) -- ($(p6)+(d2)+(s)$) 
		-- ($(p6)+(d2)+(s)$) .. controls +(-0.3,0) and +(0,0.1) ..($(p5)+(d2)+(s)$);
		%
		\draw[thin] ($(p6)+(d2)$) -- ($(p6)+(d2)+(s)$);
		\draw[very thick, red!80!black] ($(p5)+(d2)$) .. controls +(0,0.1) and +(-0.3,0) ..  ($(p6)+(d2)$); 
		\draw[very thick, red!80!black] ($(p5)+(d2)+(s)$) .. controls +(0,0.1) and +(-0.3,0) ..  ($(p6)+(d2)+(s)$); 
		%
		\fill [orange!30!white, opacity=0.8] 
		($(p4)+(d2)$) .. controls +(-0.5,0) and +(0,-0.1) .. ($(p5)+(d2)$)
		-- ($(p5)+(d2)$) -- ($(p5)+(d2)+(s)$)
		-- ($(p5)+(d2)+(s)$) .. controls +(0,-0.1) and +(-0.5,0) .. ($(p4)+(d2)+(s)$); 
		%
		\draw[thin] ($(p4)+(d2)$) -- ($(p4)+(d2)+(s)$);
		\draw[thin] ($(p5)+(d2)$) -- ($(p5)+(d2)+(s)$);
		%
		\draw[very thick, red!80!black] ($(p4)+(d2)$) .. controls +(-0.5,0) and +(0,-0.1) ..  ($(p5)+(d2)$); 
		\draw[very thick, red!80!black] ($(p4)+(d2)+(s)$) .. controls +(-0.5,0) and +(0,-0.1) ..  ($(p5)+(d2)+(s)$); 
		\end{tikzpicture}
	}%
	\, . 
	\ee 
	Hence 
	$
	\zz_n(
	\tikzzbox{
		\begin{tikzpicture}[thick,scale=0.5,color=black, baseline=0.05cm]
		\coordinate (p1) at (-0.55,0);
		\coordinate (p2) at (-0.2,0);
		\coordinate (p3) at (0.2,0);
		\coordinate (p4) at (0.55,0);
		\coordinate (p5) at (0.175,0.8);
		\coordinate (p6) at (-0.175,0.8);
		\draw[very thick, red!80!black] (p1) .. controls +(0,0.1) and +(0,0.1) ..  (p2); 
		\draw[very thick, red!80!black] (p3) .. controls +(0,0.1) and +(0,0.1) ..  (p4); 
		%
		\fill [orange!35!white, opacity=0.8] 
		(p1) .. controls +(0,0.1) and +(0,0.1) ..  (p2)
		-- (p2) .. controls +(0,0.35) and +(0,0.35) ..  (p3)
		-- (p3) .. controls +(0,0.1) and +(0,0.1) ..  (p4)
		-- (p4) .. controls +(0,0.5) and +(0,-0.5) ..  (p5)
		-- (p5) .. controls +(0,-0.1) and +(0,-0.1) ..  (p6)
		-- (p6) .. controls +(0,-0.5) and +(0,0.5) ..  (p1)
		;
		\fill [orange!20!white, opacity=0.8] 
		(p5) .. controls +(0,-0.1) and +(0,-0.1) .. (p6)
		-- (p6) .. controls +(0,0.1) and +(0,0.1) .. (p5);
		\fill [orange!30!white, opacity=0.8] 
		(p1) .. controls +(0,-0.1) and +(0,-0.1) .. (p2)
		-- (p2) .. controls +(0,0.1) and +(0,0.1) .. (p1);
		\fill [orange!30!white, opacity=0.8] 
		(p3) .. controls +(0,-0.1) and +(0,-0.1) .. (p4)
		-- (p4) .. controls +(0,0.1) and +(0,0.1) .. (p3);
		\draw (p2) .. controls +(0,0.35) and +(0,0.35) ..  (p3); 
		\draw (p4) .. controls +(0,0.5) and +(0,-0.5) ..  (p5); 
		\draw (p6) .. controls +(0,-0.5) and +(0,0.5) ..  (p1); 
		\draw[very thick, red!80!black] (p1) .. controls +(0,-0.1) and +(0,-0.1) ..  (p2); 
		\draw[very thick, red!80!black] (p3) .. controls +(0,-0.1) and +(0,-0.1) ..  (p4); 
		\draw[very thick, red!80!black] (p5) .. controls +(0,0.1) and +(0,0.1) ..  (p6); 
		\draw[very thick, red!80!black] (p5) .. controls +(0,-0.1) and +(0,-0.1) ..  (p6); 
		\end{tikzpicture}
	}
	)
	$ 
	is given by first horizontally composing~\eqref{eq:DualityData1} and~\eqref{eq:HalfCylinders}, and then vertically post-composing with 
	\be 
	\label{eq:AuxIso}
	\big[ \du''-\eu',\, \yu'-\xu' \big] \otimes \big[ \au''-\cu,\, \yu-\yu' \big] \otimes \big[ \au''-\eu,\, \xu-\yu \big] \otimes \big[ \au''-\bu,\, \xu'-\xu \big] \, , 
	\ee 
	which (up to re-naming of variables) is the inverse of~\eqref{eq:CylinderWithExtraDots}. 
	A straightforward computation then shows that this gives the expression~\eqref{eq:ZnPoP}, viewed as a 2-morphism 
	\be 
	\big( \au'-\du' \big) \cdot \big( \xu'-\yu' \big) 
		+ \big( \du-\au \big) \cdot \big( \xu-\yu \big)
		\lra 
		\big( \du''-\au'' \big) \cdot \big( \xu''-\yu'' \big) 
	\ee 
	in~$\bi$ (after re-naming $\xu\lmt \xu''$ and $\yu\lmt\yu''$). 
	The expression for 
	$
	\zz_n(
	\tikzzbox{%
		\begin{tikzpicture}[very thick,scale=0.5,color=blue!50!black, baseline=0cm]
		\coordinate (p1) at (-0.55,0);
		\coordinate (p2) at (-0.2,0);
		\coordinate (p3) at (0.2,0);
		\coordinate (p4) at (0.55,0);
		\draw[very thick, red!80!black] (p1) .. controls +(0,0.15) and +(0,0.15) ..  (p2); 
		\draw[very thick, red!80!black] (p3) .. controls +(0,0.15) and +(0,0.15) ..  (p4); 
		%
		\fill [orange!35!white, opacity=0.8] 
		(p1) .. controls +(0,-0.15) and +(0,-0.15) .. (p2)
		-- (p2) .. controls +(0,0.35) and +(0,0.35) ..  (p3)
		-- (p3) .. controls +(0,-0.15) and +(0,-0.15) .. (p4)
		-- (p4) .. controls +(0,0.9) and +(0,0.9) ..  (p1)
		;
		\draw[thick, black] (p2) .. controls +(0,0.35) and +(0,0.35) ..  (p3); 
		\draw[thick, black] (p4) .. controls +(0,0.9) and +(0,0.9) ..  (p1);
		\draw[very thick, red!80!black] (p1) .. controls +(0,-0.15) and +(0,-0.15) ..  (p2); 
		\draw[very thick, red!80!black] (p3) .. controls +(0,-0.15) and +(0,-0.15) ..  (p4); 
		\end{tikzpicture}
	}
	) 
	$ 
	in~\eqref{eq:ZnPairing} is obtained similarly, by composing with~\eqref{eq:DualityData3} with suitably re-named variables. 
\end{proof}

The commutative Frobenius algebras of Proposition~\ref{prop:CommutativeFrobeniusAlgebraFromZn} can be partially described by ordinary algebras over~$\C$, as we shall discuss next. 
First, as already noted around~\eqref{eq:ZnOnCircle}, we may think of $\zz_n(S^1) = (\au,\du,\xu,\yu ; ( \au-\du\big) \cdot \big( \xu-\yu))$ as the associated (non-semisimple) homotopy category of matrix factorisations
\be 
\A_n 
	:= 
	\textrm{hmf}\Big( \C[\au,\du,\xu,\yu] ; \, \big( \au-\du\big) \cdot \big( \xu-\yu\big) \Big)^\omega 
	\cong 
	\textrm{mod}^{\Z_2}\big( \C[\au,\xu] \big)
	\, .
\ee 
(For the equivalence to $\Z_2$-graded $\C[\au,\xu]$-modules we recall the argument in the proof of Proposition~\ref{lem:SerreTrivial}.)

The algebra structure in Proposition~\ref{prop:CommutativeFrobeniusAlgebraFromZn} is compatible with the standard monoidal structure on~$\A_n$. 
To see this, note that 
$
\zz_n(
\tikzzbox{%
	\begin{tikzpicture}[thick,scale=0.25,color=black, rotate=180, baseline=-0.4cm]
	%
	%
	\coordinate (q1) at (-2.75,0.95);
	\coordinate (q2) at (-1,0.95);
	\fill [orange!20!white, opacity=0.8] 
	(q1) .. controls +(0,0.5) and +(0,0.5) ..  (q2)
	-- (q2) .. controls +(0,-0.5) and +(0,-0.5) ..  (q1)
	;
	\fill [orange!30!white, opacity=0.8] 
	(q1) .. controls +(0,0.5) and +(0,0.5) ..  (q2)
	-- (q2) .. controls +(0,1.25) and +(0,1.25) ..  (q1)
	;
	\draw (q1) .. controls +(0,1.25) and +(0,1.25) ..  (q2); 
	\draw[very thick, red!80!black] (q1) .. controls +(0,0.5) and +(0,0.5) ..  (q2); 
	\draw[very thick, red!80!black] (q1) .. controls +(0,-0.5) and +(0,-0.5) ..  (q2); 
	\end{tikzpicture}
}%
) 
\in \A_n
$ 
is represented by the unit object, and a straightforward computation shows that (where~$\A_n'$ is a copy of~$\A_n$ with all variables primed, and similarly for~$\A_n''$)
\begin{align}
\zz_n(
\tikzzbox{
	\begin{tikzpicture}[thick,scale=0.5,color=black, baseline=0.05cm]
	\coordinate (p1) at (-0.55,0);
	\coordinate (p2) at (-0.2,0);
	\coordinate (p3) at (0.2,0);
	\coordinate (p4) at (0.55,0);
	\coordinate (p5) at (0.175,0.8);
	\coordinate (p6) at (-0.175,0.8);
	\draw[very thick, red!80!black] (p1) .. controls +(0,0.1) and +(0,0.1) ..  (p2); 
	\draw[very thick, red!80!black] (p3) .. controls +(0,0.1) and +(0,0.1) ..  (p4); 
	%
	\fill [orange!35!white, opacity=0.8] 
	(p1) .. controls +(0,0.1) and +(0,0.1) ..  (p2)
	-- (p2) .. controls +(0,0.35) and +(0,0.35) ..  (p3)
	-- (p3) .. controls +(0,0.1) and +(0,0.1) ..  (p4)
	-- (p4) .. controls +(0,0.5) and +(0,-0.5) ..  (p5)
	-- (p5) .. controls +(0,-0.1) and +(0,-0.1) ..  (p6)
	-- (p6) .. controls +(0,-0.5) and +(0,0.5) ..  (p1)
	;
	\fill [orange!20!white, opacity=0.8] 
	(p5) .. controls +(0,-0.1) and +(0,-0.1) .. (p6)
	-- (p6) .. controls +(0,0.1) and +(0,0.1) .. (p5);
	\fill [orange!30!white, opacity=0.8] 
	(p1) .. controls +(0,-0.1) and +(0,-0.1) .. (p2)
	-- (p2) .. controls +(0,0.1) and +(0,0.1) .. (p1);
	\fill [orange!30!white, opacity=0.8] 
	(p3) .. controls +(0,-0.1) and +(0,-0.1) .. (p4)
	-- (p4) .. controls +(0,0.1) and +(0,0.1) .. (p3);
	\draw (p2) .. controls +(0,0.35) and +(0,0.35) ..  (p3); 
	\draw (p4) .. controls +(0,0.5) and +(0,-0.5) ..  (p5); 
	\draw (p6) .. controls +(0,-0.5) and +(0,0.5) ..  (p1); 
	\draw[very thick, red!80!black] (p1) .. controls +(0,-0.1) and +(0,-0.1) ..  (p2); 
	\draw[very thick, red!80!black] (p3) .. controls +(0,-0.1) and +(0,-0.1) ..  (p4); 
	\draw[very thick, red!80!black] (p5) .. controls +(0,0.1) and +(0,0.1) ..  (p6); 
	\draw[very thick, red!80!black] (p5) .. controls +(0,-0.1) and +(0,-0.1) ..  (p6); 
	\end{tikzpicture}
}
) \colon \A_n' \times \A_n & \lra \A_n'' \nonumber
	\\
	\big( P(\au',\du',\xu',\yu'), \, Q(\au,\du,\xu,\yu) \big) & \lmt P(\du'',\du,\xu'',\yu'') \otimes_{\C[\du]} Q(\du,\au'',\xu'',\yu'') \, , 
\end{align}
which is indeed the relative tensor product of matrix factorisations. 

Since~$\bi$ (and hence~$\zz_n$) sees only isomorphism classes of matrix factorisations, 
$
\zz_n(
\tikzzbox{
	\begin{tikzpicture}[thick,scale=0.5,color=black, baseline=0.05cm]
	\coordinate (p1) at (-0.55,0);
	\coordinate (p2) at (-0.2,0);
	\coordinate (p3) at (0.2,0);
	\coordinate (p4) at (0.55,0);
	\coordinate (p5) at (0.175,0.8);
	\coordinate (p6) at (-0.175,0.8);
	\draw[very thick, red!80!black] (p1) .. controls +(0,0.1) and +(0,0.1) ..  (p2); 
	\draw[very thick, red!80!black] (p3) .. controls +(0,0.1) and +(0,0.1) ..  (p4); 
	%
	\fill [orange!35!white, opacity=0.8] 
	(p1) .. controls +(0,0.1) and +(0,0.1) ..  (p2)
	-- (p2) .. controls +(0,0.35) and +(0,0.35) ..  (p3)
	-- (p3) .. controls +(0,0.1) and +(0,0.1) ..  (p4)
	-- (p4) .. controls +(0,0.5) and +(0,-0.5) ..  (p5)
	-- (p5) .. controls +(0,-0.1) and +(0,-0.1) ..  (p6)
	-- (p6) .. controls +(0,-0.5) and +(0,0.5) ..  (p1)
	;
	\fill [orange!20!white, opacity=0.8] 
	(p5) .. controls +(0,-0.1) and +(0,-0.1) .. (p6)
	-- (p6) .. controls +(0,0.1) and +(0,0.1) .. (p5);
	\fill [orange!30!white, opacity=0.8] 
	(p1) .. controls +(0,-0.1) and +(0,-0.1) .. (p2)
	-- (p2) .. controls +(0,0.1) and +(0,0.1) .. (p1);
	\fill [orange!30!white, opacity=0.8] 
	(p3) .. controls +(0,-0.1) and +(0,-0.1) .. (p4)
	-- (p4) .. controls +(0,0.1) and +(0,0.1) .. (p3);
	\draw (p2) .. controls +(0,0.35) and +(0,0.35) ..  (p3); 
	\draw (p4) .. controls +(0,0.5) and +(0,-0.5) ..  (p5); 
	\draw (p6) .. controls +(0,-0.5) and +(0,0.5) ..  (p1); 
	\draw[very thick, red!80!black] (p1) .. controls +(0,-0.1) and +(0,-0.1) ..  (p2); 
	\draw[very thick, red!80!black] (p3) .. controls +(0,-0.1) and +(0,-0.1) ..  (p4); 
	\draw[very thick, red!80!black] (p5) .. controls +(0,0.1) and +(0,0.1) ..  (p6); 
	\draw[very thick, red!80!black] (p5) .. controls +(0,-0.1) and +(0,-0.1) ..  (p6); 
	\end{tikzpicture}
}
)
$ 
induces the monoidal product for~$\A_n$ only on isomorphism classes of objects. 
If~$\bi$ were the truncation of a 3-category associated to a 3-dimensional theory~$\mathcal T$, the 3-morphisms of the latter would in particular include the morphisms of~$\A_n$ and complete its monoidal structure. 
This monoidal structure would then be interpreted as the fusion of topological line defects (on the identity surface defect) in~$\mathcal T$. 
Moreover, the braiding of line defects would give rise to a braided monoidal structure on~$\A_n$. 

From the perspective of our truncated 2-category~$\bi$ and the TQFT~$\zz_n$ taking values in it, what is visible of the monoidal category~$\A_n$ is its Grothendieck ring $K_0(\A_n)$. 
By definition, this is the abelian group generated by isomorphism classes of objects in~$\A_n$ subject to relations coming from short exact sequences in~$\A_n$, together with multiplication 
\be 
\label{eq:ProductOnGroth}
\big( [P], \, [Q] \big) 
	\lmt \big[ \zz_n(
	\tikzzbox{
		\begin{tikzpicture}[thick,scale=0.5,color=black, baseline=0.05cm]
		\coordinate (p1) at (-0.55,0);
		\coordinate (p2) at (-0.2,0);
		\coordinate (p3) at (0.2,0);
		\coordinate (p4) at (0.55,0);
		\coordinate (p5) at (0.175,0.8);
		\coordinate (p6) at (-0.175,0.8);
		\draw[very thick, red!80!black] (p1) .. controls +(0,0.1) and +(0,0.1) ..  (p2); 
		\draw[very thick, red!80!black] (p3) .. controls +(0,0.1) and +(0,0.1) ..  (p4); 
		%
		\fill [orange!35!white, opacity=0.8] 
		(p1) .. controls +(0,0.1) and +(0,0.1) ..  (p2)
		-- (p2) .. controls +(0,0.35) and +(0,0.35) ..  (p3)
		-- (p3) .. controls +(0,0.1) and +(0,0.1) ..  (p4)
		-- (p4) .. controls +(0,0.5) and +(0,-0.5) ..  (p5)
		-- (p5) .. controls +(0,-0.1) and +(0,-0.1) ..  (p6)
		-- (p6) .. controls +(0,-0.5) and +(0,0.5) ..  (p1)
		;
		\fill [orange!20!white, opacity=0.8] 
		(p5) .. controls +(0,-0.1) and +(0,-0.1) .. (p6)
		-- (p6) .. controls +(0,0.1) and +(0,0.1) .. (p5);
		\fill [orange!30!white, opacity=0.8] 
		(p1) .. controls +(0,-0.1) and +(0,-0.1) .. (p2)
		-- (p2) .. controls +(0,0.1) and +(0,0.1) .. (p1);
		\fill [orange!30!white, opacity=0.8] 
		(p3) .. controls +(0,-0.1) and +(0,-0.1) .. (p4)
		-- (p4) .. controls +(0,0.1) and +(0,0.1) .. (p3);
		\draw (p2) .. controls +(0,0.35) and +(0,0.35) ..  (p3); 
		\draw (p4) .. controls +(0,0.5) and +(0,-0.5) ..  (p5); 
		\draw (p6) .. controls +(0,-0.5) and +(0,0.5) ..  (p1); 
		\draw[very thick, red!80!black] (p1) .. controls +(0,-0.1) and +(0,-0.1) ..  (p2); 
		\draw[very thick, red!80!black] (p3) .. controls +(0,-0.1) and +(0,-0.1) ..  (p4); 
		\draw[very thick, red!80!black] (p5) .. controls +(0,0.1) and +(0,0.1) ..  (p6); 
		\draw[very thick, red!80!black] (p5) .. controls +(0,-0.1) and +(0,-0.1) ..  (p6); 
		\end{tikzpicture}
	}
	) (P,Q) 
	\big] \, . 
\ee 
It follows from Proposition~\ref{prop:CommutativeFrobeniusAlgebraFromZn} that this product is associative and unital with respect to the unit 
$
[\zz_n(
\tikzzbox{%
	\begin{tikzpicture}[thick,scale=0.25,color=black, rotate=180, baseline=-0.4cm]
	%
	%
	\coordinate (q1) at (-2.75,0.95);
	\coordinate (q2) at (-1,0.95);
	\fill [orange!20!white, opacity=0.8] 
	(q1) .. controls +(0,0.5) and +(0,0.5) ..  (q2)
	-- (q2) .. controls +(0,-0.5) and +(0,-0.5) ..  (q1)
	;
	\fill [orange!30!white, opacity=0.8] 
	(q1) .. controls +(0,0.5) and +(0,0.5) ..  (q2)
	-- (q2) .. controls +(0,1.25) and +(0,1.25) ..  (q1)
	;
	\draw (q1) .. controls +(0,1.25) and +(0,1.25) ..  (q2); 
	\draw[very thick, red!80!black] (q1) .. controls +(0,0.5) and +(0,0.5) ..  (q2); 
	\draw[very thick, red!80!black] (q1) .. controls +(0,-0.5) and +(0,-0.5) ..  (q2); 
	\end{tikzpicture}
}%
)]
$. 

In addition to the unital associative structure, $K_0(\A_n)$ has a coassociative comultiplication 
\begin{align} 
\big[ P \equiv P(\du,\au,\xu,\yu)  \big]
& \lmt \big[ \zz_n(
\tikzzbox{
	\begin{tikzpicture}[thick,scale=0.5,color=black, baseline=-0.3cm, yscale=-1]
	\coordinate (p1) at (-0.55,0);
	\coordinate (p2) at (-0.2,0);
	\coordinate (p3) at (0.2,0);
	\coordinate (p4) at (0.55,0);
	\coordinate (p5) at (0.175,0.8);
	\coordinate (p6) at (-0.175,0.8);
	\draw[very thick, red!80!black] (p1) .. controls +(0,0.1) and +(0,0.1) ..  (p2); 
	\draw[very thick, red!80!black] (p3) .. controls +(0,0.1) and +(0,0.1) ..  (p4); 
	%
	\fill [orange!35!white, opacity=0.8] 
	(p1) .. controls +(0,0.1) and +(0,0.1) ..  (p2)
	-- (p2) .. controls +(0,0.35) and +(0,0.35) ..  (p3)
	-- (p3) .. controls +(0,0.1) and +(0,0.1) ..  (p4)
	-- (p4) .. controls +(0,0.5) and +(0,-0.5) ..  (p5)
	-- (p5) .. controls +(0,-0.1) and +(0,-0.1) ..  (p6)
	-- (p6) .. controls +(0,-0.5) and +(0,0.5) ..  (p1)
	;
	\fill [orange!20!white, opacity=0.8] 
	(p5) .. controls +(0,-0.1) and +(0,-0.1) .. (p6)
	-- (p6) .. controls +(0,0.1) and +(0,0.1) .. (p5);
	\fill [orange!30!white, opacity=0.8] 
	(p1) .. controls +(0,-0.1) and +(0,-0.1) .. (p2)
	-- (p2) .. controls +(0,0.1) and +(0,0.1) .. (p1);
	\fill [orange!30!white, opacity=0.8] 
	(p3) .. controls +(0,-0.1) and +(0,-0.1) .. (p4)
	-- (p4) .. controls +(0,0.1) and +(0,0.1) .. (p3);
	\draw (p2) .. controls +(0,0.35) and +(0,0.35) ..  (p3); 
	\draw (p4) .. controls +(0,0.5) and +(0,-0.5) ..  (p5); 
	\draw (p6) .. controls +(0,-0.5) and +(0,0.5) ..  (p1); 
	\draw[very thick, red!80!black] (p1) .. controls +(0,-0.1) and +(0,-0.1) ..  (p2); 
	\draw[very thick, red!80!black] (p3) .. controls +(0,-0.1) and +(0,-0.1) ..  (p4); 
	\draw[very thick, red!80!black] (p5) .. controls +(0,0.1) and +(0,0.1) ..  (p6); 
	\draw[very thick, red!80!black] (p5) .. controls +(0,-0.1) and +(0,-0.1) ..  (p6); 
	\end{tikzpicture}
}
) (P) 
\big] 
\nonumber 
\\ & \qquad\qquad 
= 
\Big[ 
	 P(\du',\au',\xu',\yu') \otimes \Big( \big[ \au''-\au', \, \zeru \big] \otimes \big[ \xu'-\xu, \, \zeru \big] \Big) 
\nonumber 
\\ 
& 
\qquad\qquad\qquad
\otimes \big[ \du''-\au'', \, \xu''-\yu'' \big] 
\Big] 
\end{align}
where the last expression is a consequence of a straightforward computation along the lines of the proof of Proposition~\ref{prop:CommutativeFrobeniusAlgebraFromZn}, and by the same argument this comultiplication is compatible with the multiplication in~\eqref{eq:ProductOnGroth} in the sense that 
\be 
\Big[\zz_n
\Big(
\tikzzbox
{
	\begin{tikzpicture}[thick,scale=0.45, color=black, yscale=-1, baseline=-0.5cm]
	\coordinate (p1) at (-0.55,0);
	\coordinate (p2) at (-0.2,0);
	\coordinate (p3) at (0.45,0);
	\coordinate (p4) at (0.8,0);
	\coordinate (q1) at (0.2,1.6);
	\coordinate (q2) at (0.55,1.6);
	\coordinate (q3) at (1.2,1.6);
	\coordinate (q4) at (1.55,1.6);
	%
	\fill [orange!35!white, opacity=0.8] 
	(p1) -- (p2)
	-- (p2) .. controls +(0.7,1.55) and +(-0.2,0.85) ..  (p3)
	-- (p3) -- (p4)
	-- (p4) .. controls +(0,0.5) and +(0,-0.5) .. (q4)
	-- (q4) .. controls +(0,0.1) and +(0,0.1) .. (q3)
	-- (q3) .. controls +(-0.7,-1.55) and +(0.2,-0.85) .. (q2)
	-- (q2) .. controls +(0,0.1) and +(0,0.1) .. (q1)
	-- (q1) .. controls +(0,-0.5) and +(0,0.5) ..  (p1)
	;
	%
	\fill [orange!38] 
	(p1) .. controls +(0,0.1) and +(0,0.1) ..  (p2) -- 
	(p2) .. controls +(0,-0.1) and +(0,-0.1) ..  (p1)
	;
	%
	\fill [orange!38] 
	(p3) .. controls +(0,0.1) and +(0,0.1) ..  (p4) -- 
	(p4) .. controls +(0,-0.1) and +(0,-0.1) ..  (p3)
	;
	\draw (p2) .. controls +(0.7,1.55) and +(-0.2,0.85) ..  (p3); 
	\draw (p4) .. controls +(0,0.5) and +(0,-0.5) .. (q4); 
	\draw (q3) .. controls +(-0.7,-1.55) and +(0.2,-0.85) .. (q2); 
	\draw (q1) .. controls +(0,-0.5) and +(0,0.5) ..  (p1); 
	\draw[very thick, color=red!80!black] (p1) .. controls +(0,0.1) and +(0,0.1) ..  (p2); 
	\draw[very thick, color=red!80!black] (p2) .. controls +(0,-0.1) and +(0,-0.1) ..  (p1); 
	\draw[very thick, color=red!80!black] (p3) .. controls +(0,0.1) and +(0,0.1) .. (p4); 
	\draw[very thick, color=red!80!black] (p4) .. controls +(0,-0.1) and +(0,-0.1) ..  (p3); 
	\draw[very thick, color=red!80!black] (q1) .. controls +(0,0.1) and +(0,0.1) ..  (q2); 
	\draw[very thick, color=red!80!black, opacity=0.2] (q2) .. controls +(0,-0.1) and +(0,-0.1) ..  (q1); 
	\draw[very thick, color=red!80!black] (q3) .. controls +(0,0.1) and +(0,0.1) .. (q4); 
	\draw[very thick, color=red!80!black, opacity=0.2] (q4) .. controls +(0,-0.1) and +(0,-0.1) ..  (q3); 
	\end{tikzpicture}
} 
\Big)(P,Q) \Big]
=
\Big[\zz_n
\Big(
\tikzzbox
{
	\begin{tikzpicture}[thick,scale=0.45, color=black, yscale=-1, baseline=-0.5cm]
	\coordinate (p1) at (-0.55,0);
	\coordinate (p2) at (-0.2,0);
	\coordinate (p3) at (0.2,0);
	\coordinate (p4) at (0.55,0);
	\coordinate (p5) at (0.175,0.8);
	\coordinate (p6) at (-0.175,0.8);
	\coordinate (q1) at (-0.55,1.6);
	\coordinate (q2) at (-0.2,1.6);
	\coordinate (q3) at (0.2,1.6);
	\coordinate (q4) at (0.55,1.6);
	%
	\fill [orange!35!white, opacity=0.8] 
	(p1) -- (p2)
	-- (p2) .. controls +(0,0.35) and +(0,0.35) ..  (p3)
	-- (p3) -- (p4)
	-- (p4) .. controls +(0,0.5) and +(0,-0.5) ..  (p5)
	-- (p5) .. controls +(0,0.5) and +(0,-0.5) .. (q4)
	-- (q4) .. controls +(0,0.1) and +(0,0.1) ..  (q3)
	-- (q3) .. controls +(0,-0.35) and +(0,-0.35) ..  (q2)
	-- (q2) .. controls +(0,0.1) and +(0,0.1) ..  (q1)
	-- (q1) .. controls +(0,-0.5) and +(0,0.5) .. (p6)
	-- (p6) .. controls +(0,-0.5) and +(0,0.5) ..  (p1)
	;
	%
	\fill [orange!38] 
	(p1) .. controls +(0,0.1) and +(0,0.1) ..  (p2) -- 
	(p2) .. controls +(0,-0.1) and +(0,-0.1) ..  (p1)
	;
	%
	\fill [orange!38] 
	(p3) .. controls +(0,0.1) and +(0,0.1) ..  (p4) -- 
	(p4) .. controls +(0,-0.1) and +(0,-0.1) ..  (p3)
	;
	\draw (p2) .. controls +(0,0.35) and +(0,0.35) ..  (p3); 
	\draw (p4) .. controls +(0,0.5) and +(0,-0.5) .. (p5); 
	\draw (p5) .. controls +(0,0.5) and +(0,-0.5) .. (q4); 
	\draw (q3) .. controls +(0,-0.35) and +(0,-0.35) ..  (q2); 
	\draw (q1) .. controls +(0,-0.5) and +(0,0.5) .. (p6); 
	\draw (p6) .. controls +(0,-0.5) and +(0,0.5) ..  (p1); 
	\draw[very thick, color=red!80!black] (p1) .. controls +(0,0.1) and +(0,0.1) ..  (p2); 
	\draw[very thick, color=red!80!black] (p2) .. controls +(0,-0.1) and +(0,-0.1) ..  (p1); 
	\draw[very thick, color=red!80!black] (p3) .. controls +(0,0.1) and +(0,0.1) .. (p4); 
	\draw[very thick, color=red!80!black] (p4) .. controls +(0,-0.1) and +(0,-0.1) ..  (p3); 
	\draw[very thick, color=red!80!black] (q1) .. controls +(0,0.1) and +(0,0.1) ..  (q2); 
	\draw[very thick, color=red!80!black, opacity=0.2] (q2) .. controls +(0,-0.1) and +(0,-0.1) ..  (q1); 
	\draw[very thick, color=red!80!black] (q3) .. controls +(0,0.1) and +(0,0.1) .. (q4); 
	\draw[very thick, color=red!80!black, opacity=0.2] (q4) .. controls +(0,-0.1) and +(0,-0.1) ..  (q3); 
	\end{tikzpicture}
} 
\Big)(P,Q) \Big]
=
\Big[\zz_n\Big(
\tikzzbox
{
	\begin{tikzpicture}[thick,scale=0.45, color=black, yscale=-1, baseline=-0.5cm]
	\coordinate (p1) at (0.55,0);
	\coordinate (p2) at (0.2,0);
	\coordinate (p3) at (-0.45,0);
	\coordinate (p4) at (-0.8,0);
	\coordinate (q1) at (-0.2,1.6);
	\coordinate (q2) at (-0.55,1.6);
	\coordinate (q3) at (-1.2,1.6);
	\coordinate (q4) at (-1.55,1.6);
	%
	\fill [orange!35!white, opacity=0.8] 
	(p1) -- (p2)
	-- (p2) .. controls +(-0.7,1.55) and +(0.2,0.85) ..  (p3)
	-- (p3) -- (p4)
	-- (p4) .. controls +(0,0.5) and +(0,-0.5) .. (q4)
	-- (q4) .. controls +(0,0.1) and +(0,0.1) ..  (q3)
	-- (q3) .. controls +(0.7,-1.55) and +(-0.2,-0.85) .. (q2)
	-- (q2) .. controls +(0,0.1) and +(0,0.1) ..  (q1)
	-- (q1) .. controls +(0,-0.5) and +(0,0.5) ..  (p1)
	;
	%
	\fill [orange!38] 
	(p1) .. controls +(0,0.1) and +(0,0.1) ..  (p2) -- 
	(p2) .. controls +(0,-0.1) and +(0,-0.1) ..  (p1)
	;
	%
	\fill [orange!38] 
	(p3) .. controls +(0,0.1) and +(0,0.1) ..  (p4) -- 
	(p4) .. controls +(0,-0.1) and +(0,-0.1) ..  (p3)
	;
	\draw (p2) .. controls +(-0.7,1.55) and +(0.2,0.85) ..  (p3); 
	\draw (p4) .. controls +(0,0.5) and +(0,-0.5) .. (q4); 
	\draw (q3) .. controls +(0.7,-1.55) and +(-0.2,-0.85) .. (q2); 
	\draw (q1) .. controls +(0,-0.5) and +(0,0.5) ..  (p1); 
	\draw[very thick, color=red!80!black] (p1) .. controls +(0,0.1) and +(0,0.1) ..  (p2); 
	\draw[very thick, color=red!80!black] (p2) .. controls +(0,-0.1) and +(0,-0.1) ..  (p1); 
	\draw[very thick, color=red!80!black] (p3) .. controls +(0,0.1) and +(0,0.1) .. (p4); 
	\draw[very thick, color=red!80!black] (p4) .. controls +(0,-0.1) and +(0,-0.1) ..  (p3); 
	\draw[very thick, color=red!80!black] (q1) .. controls +(0,0.1) and +(0,0.1) ..  (q2); 
	\draw[very thick, color=red!80!black, opacity=0.2] (q2) .. controls +(0,-0.1) and +(0,-0.1) ..  (q1); 
	\draw[very thick, color=red!80!black] (q3) .. controls +(0,0.1) and +(0,0.1) .. (q4); 
	\draw[very thick, color=red!80!black, opacity=0.2] (q4) .. controls +(0,-0.1) and +(0,-0.1) ..  (q3); 
	\end{tikzpicture}
} 
\Big)(P,Q) \Big]
\, . 
\ee 
However, the counit $\beta \cdot (\eta \circ 1_A)$ of the Frobenius algebra classifying~$\zz_n^{\textrm{cl}}$ does not induce a counit for the complexification of $K_0(\A_n)$ (viewed as an algebra either over~$\C$ or over $\C[\xu,\yu]$). 
The reason is that $\beta \cdot (\eta \circ 1_A)$ applied to an object in~$\A_n$ produces a matrix factorisation of zero, and not an element in~$\C$ or $\C[\xu,\yu]$. 
This is consistent with the fact that the complexification of $K_0(\A_n)$ is infinite-dimensional while Frobenius algebras in $\Vect_\C$ are necessarily finite-dimensional. 

In summary, the extended TQFT~$\zz_n$ induces an associative unital coassociative non-counital algebra structure on the Grothendieck ring of~$\A_n$, where multiplication and comultiplication are compatible as in the case of Frobenius algebras. 
The lack of a counit, or equivalently the lack of a non-degenerate pairing, is expected for a truncation of a 3-dimensional twisted sigma model with non-compact target $T^*\C^n$. 

Finally, we mention that while~$\zz_n$ does not know about the full structure of~$\A_n$, it does see more than its Grothendieck ring. 
For example, by combining Proposition~\ref{prop:StateSpaces} with the discussion in \cite[Sect.\,2.6.2]{CDGG} we explicitly verify that the state space assigned to the torus is the \textsl{full} Hochschild homology of the non-semisimple category~$\A_n$: 
\be 
\zz_n(T^2) \cong \textrm{HH}_\bullet(\A_n) \, . 
\ee

\subsection{Including flavour and R-charges}
\label{subsec:ExtendedAffineRWWithCharges}

The discussion of Section~\ref{subsec:ExtendedAffineRW} carries over immediately to the graded setting. 
As discussed in Section~\ref{subsec:GradedCase}, 
also every object in $\bigra$ is fully dualisable, and there are precisely two trivialisations of its Serre automorphism, both of which lead to isomorphic TQFTs. 
Thus, for every object $\xu\in\bigra$, there is an (up to isomorphism) unique extended oriented TQFT
\be 
	\zzgra_n \colon \Bordor \lra \bigra
\ee 
with $\zzgra_n(+) = (x_1,\dots,x_n)$. Since all the structure morphisms of $\bigra$ are lifts of those in $\bi$, the gradings immediately push through all the calculations of Section~\ref{subsec:ExtendedAffineRW}. 
In particular,
\be 
\zzgra_n(S^1) 
	= 
	\Big( \au,\au',\xu,\xu' ; \, \big( \au-\au'\big) \cdot \big( \xu-\xu'\big) \Big) ,
\ee 
where now $\xu,\xu'$ carry bidegree $(1,-1)$, and $\au,\au'$ have bidegree $(1,1)$.
Modulo a caveat analogous to the ungraded case, we may think of $\zzgra_n(S^1)$ as the homotopy category of bigraded matrix factorisations of $(\au-\au')\cdot(\xu-\xu')$. 

To a closed surface~$\Sigma_g$ of genus~$g$, the TQFT~$\zzgra_n$ associates an isomorphism class of bigraded matrix factorisation of zero, or equivalently, the ($\Z_2\times\Q\times\Q$)-graded vector space given by its cohomology. 
This vector space corresponds to the state space of the underlying 3-dimensional Rozansky--Witten theory. 

Making use of the fact that the gradings behave additively under the tensor product, cf.~\eqref{eq:twistsareadditive}, it is straightforward to incorporate the bigrading into the calculations in the proof of Proposition~\ref{prop:StateSpaces}, leading to: 

\begin{corollary}
	\label{cor:StateSpacesGradedCase}
	Let $\Sigma_g$ be a closed oriented surface of genus $g$, then 
	\begin{align}
	\zzgra_n(\Sigma_g) 
		\cong & \Big( 
			 \big(\C \oplus \C[1]\{0,1\}\big)^{\otimes n}
		\otimes_\C 
			 \big(\C \oplus \C[1]\{0,-1\}\big)^{\otimes n}
		\{-\rnh-\rsh,-\qnh-\qsh\} \Big)^{\otimes g}\nonumber\\
		& \qquad\qquad\qquad\qquad\qquad\qquad \otimes_\C \C[\au,\xu] \{\rnh+\rsh,\qnh+\qsh\}\,.\label{eq:StateSpacegrSigmag}
	\end{align}
\end{corollary}
This space indeed agrees with the state space of the Rozansky--Witten theory with target manifold $T^*\C^n$ from \cite{RW1996}, cf.\ Appendix~\ref{sec:appalgebras} and \cite[Sect.\,2]{CDGG}.
For fixed degrees, the corresponding subspace of the state space is finite-dimensional. 
We furthermore note that for the case $g=1$ the choices for the parameters $\rnh, \rsh, \qnh, \qsh$ drop out,\footnote{A more general discussion of state spaces as well as indices can be found in \cite{CDGG}, where also flat background connections combining twisted mass parameters with the flavour symmetry are considered. 
Here, the torus is again special in the sense that for generic background connections the state space becomes trivial.}
and that the state space associated to the torus 
\be 
\zzgra_n(T^2) \cong \textrm{HH}_\bullet(\A^{\textrm{gr}}_n ) 
\ee 
is the full Hochschild homology of the $(\Z_2\times\Q\times\Q)$-graded $\C[\au,\xu]$-modules
\be 
\A^{\textrm{gr}}_n 
	:= 
	\textrm{hmf}^{\textrm{gr}}\Big( \C[\au,\du,\xu,\yu] ; \, \big( \au-\du\big) \cdot \big( \xu-\yu\big) \Big)^\omega 
	\cong 
	\textrm{mod}^{\textrm{gr}}\big( \C[\au,\xu] \big)
\, , 
\ee 
where the $\au$- and $\xu$-variables have bidegrees $(1,1)$ and $(1,-1)$, respectively, as discussed in Section~\ref{subsec:GradedCase}. 

In the physics literature, these spaces are often described in terms of their generating functions
\be
Z_{\Sigma_g}(s,t,u):=\textrm{tr}_{\zzgra_n(\Sigma_g)}\Big(s^F\,t^R\,u^Q\Big)\,,
\ee
where $F$ is the fermion number operator counting the homological $\Z_2$-degree, while~$R$ and~$Q$ are the flavour and R-charge operators, respectively. 
Such indices can often be evaluated by path integral computations in rather general situations, going beyond the free field case. 
From~\eqref{eq:StateSpacegrSigmag} one reads off the generating function
\be\label{eq:generatingfunction}
Z_{\Sigma_g}(s,t,u)
=
\Big(
	 \big(1+su\big)^n\,\big(1+\tfrac{s}{u}\big)^n
\, t^{-\rnh-\rsh}\,u^{-\qnh-\qsh}
\Big)^g\,\frac{t^{\rnh+\rsh}\,u^{\qnh+\qsh}}
	 {\big(1-tu\big)^n\,\big(1-\frac{t}{u}\big)^n}\,.
\ee
Setting $s=-t$ (i.\,e.\ counting the odd homological degree with a minus sign in the generating function and adding the homological degree to the R-charge)
we obtain the more familiar form
\be\label{eq:generatingfunction2}
Z_{\Sigma_g}(t,u)
:=Z_{\Sigma_g}(-t,t,u)
	= \Big(
		 \big(1-tu\big)^n\,\big(1-\tfrac{t}{u}\big)^n
	\, t^{-\rnh-\rsh}\,u^{-\qnh-\qsh}\Big)^{g-1} \, ,
\ee
which after setting $\rnh+\rsh=-1$ and $\qnh+\qsh=0$ agrees with the formulas of e.\,g.\ \cite[Sect.\,2]{GHNPPS}.

\medskip 

The discussion of the commutative Frobenius structure in Section~\ref{subsubsec:ComFrobAlgGrothRing} for the ungraded case carries over in a straightforward manner to the graded case. 
Inclusion of the gradings yields the commutative Frobenius algebra $(A^\textrm{gr},\mu^\textrm{gr},\eta^\textrm{gr},\beta^\textrm{gr})$ in $\bigra(\xu,\xu)$, where 
	\begin{align}
	A^\textrm{gr} = & \zzgra_n\big(
	\tikzzbox{
		\begin{tikzpicture}[thick,scale=1.5,color=black, baseline=0.065cm]
		\coordinate (p1) at (0,0);
		\coordinate (p2) at (0,0.25);
		\coordinate (g1) at (0,-0.05);
		\coordinate (g2) at (0,0.3);		%
		\draw[very thick, red!80!black] (p1) .. controls +(-0.2,0) and +(-0.2,0) ..  (p2); 
		%
		\draw[very thick, red!80!black] (p1) .. controls +(0.2,0) and +(0.2,0) ..  (p2); 
		\end{tikzpicture}
	}
	\big) 
	 \cong \Big( \au,\du,\xu,\yu ; \, \big( \au-\du \big) \cdot \big( \xu-\yu\big) \Big) , 
	\\
	\mu^\textrm{gr} = &\zzgra_n\Big(
	\tikzzbox{
		\begin{tikzpicture}[thick,scale=0.8,color=black, baseline=0.2cm]
		\coordinate (p1) at (-0.55,0);
		\coordinate (p2) at (-0.2,0);
		\coordinate (p3) at (0.2,0);
		\coordinate (p4) at (0.55,0);
		\coordinate (p5) at (0.175,0.8);
		\coordinate (p6) at (-0.175,0.8);
		\draw[very thick, red!80!black] (p1) .. controls +(0,0.1) and +(0,0.1) ..  (p2); 
		\draw[very thick, red!80!black] (p3) .. controls +(0,0.1) and +(0,0.1) ..  (p4); 
		%
		\fill [orange!35!white, opacity=0.8] 
		(p1) .. controls +(0,0.1) and +(0,0.1) ..  (p2)
		-- (p2) .. controls +(0,0.35) and +(0,0.35) ..  (p3)
		-- (p3) .. controls +(0,0.1) and +(0,0.1) ..  (p4)
		-- (p4) .. controls +(0,0.5) and +(0,-0.5) ..  (p5)
		-- (p5) .. controls +(0,-0.1) and +(0,-0.1) ..  (p6)
		-- (p6) .. controls +(0,-0.5) and +(0,0.5) ..  (p1)
		;
		\fill [orange!20!white, opacity=0.8] 
		(p5) .. controls +(0,-0.1) and +(0,-0.1) .. (p6)
		-- (p6) .. controls +(0,0.1) and +(0,0.1) .. (p5);
		\fill [orange!30!white, opacity=0.8] 
		(p1) .. controls +(0,-0.1) and +(0,-0.1) .. (p2)
		-- (p2) .. controls +(0,0.1) and +(0,0.1) .. (p1);
		\fill [orange!30!white, opacity=0.8] 
		(p3) .. controls +(0,-0.1) and +(0,-0.1) .. (p4)
		-- (p4) .. controls +(0,0.1) and +(0,0.1) .. (p3);
		\draw (p2) .. controls +(0,0.35) and +(0,0.35) ..  (p3); 
		\draw (p4) .. controls +(0,0.5) and +(0,-0.5) ..  (p5); 
		\draw (p6) .. controls +(0,-0.5) and +(0,0.5) ..  (p1); 
		\draw[very thick, red!80!black] (p1) .. controls +(0,-0.1) and +(0,-0.1) ..  (p2); 
		\draw[very thick, red!80!black] (p3) .. controls +(0,-0.1) and +(0,-0.1) ..  (p4); 
		\draw[very thick, red!80!black] (p5) .. controls +(0,0.1) and +(0,0.1) ..  (p6); 
		\draw[very thick, red!80!black] (p5) .. controls +(0,-0.1) and +(0,-0.1) ..  (p6); 
		\end{tikzpicture}
	}
	\Big)
	 \cong \big[ \au''-\au, \, \yu-\yu' \big]  \otimes \big[ \au''-\au' ,\, \xu'-\xu \big] \otimes \big[ \au'-\au, \, \yu'-\xu \big] 
	\nonumber
	\\ & \qquad\qquad\qquad\quad \otimes \big[ \du''-\du',\, \yu'-\xu' \big]  \otimes \big[ \au''- \du, \, \xu-\yu \big] \{-\rsh,-\qsh-2\}
	\, ,
	\\
	\eta^\textrm{gr} = &\zzgra_n
	\big(
	\tikzzbox{%
		\begin{tikzpicture}[thick,scale=0.25,color=black, rotate=180, baseline=-0.4cm]
		%
		%
		\coordinate (q1) at (-2.75,0.95);
		\coordinate (q2) at (-1,0.95);
		\fill [orange!20!white, opacity=0.8] 
		(q1) .. controls +(0,0.5) and +(0,0.5) ..  (q2)
		-- (q2) .. controls +(0,-0.5) and +(0,-0.5) ..  (q1)
		;
		\fill [orange!30!white, opacity=0.8] 
		(q1) .. controls +(0,0.5) and +(0,0.5) ..  (q2)
		-- (q2) .. controls +(0,1.25) and +(0,1.25) ..  (q1)
		;
		\draw (q1) .. controls +(0,1.25) and +(0,1.25) ..  (q2); 
		\draw[very thick, red!80!black] (q1) .. controls +(0,0.5) and +(0,0.5) ..  (q2); 
		\draw[very thick, red!80!black] (q1) .. controls +(0,-0.5) and +(0,-0.5) ..  (q2); 
		\end{tikzpicture}
	}%
	\big) 
	 \cong \big[ \au-\du,\, \xu-\yu \big] \{\rsh,\qsh\}
	\, , 
	\\
	\beta^\textrm{gr} = &\zzgra_n\Big(
	\tikzzbox{%
		\begin{tikzpicture}[very thick,scale=0.8,color=blue!50!black, baseline=0.1cm]
		\coordinate (p1) at (-0.55,0);
		\coordinate (p2) at (-0.2,0);
		\coordinate (p3) at (0.2,0);
		\coordinate (p4) at (0.55,0);
		\draw[very thick, red!80!black] (p1) .. controls +(0,0.15) and +(0,0.15) ..  (p2); 
		\draw[very thick, red!80!black] (p3) .. controls +(0,0.15) and +(0,0.15) ..  (p4); 
		%
		\fill [orange!35!white, opacity=0.8] 
		(p1) .. controls +(0,-0.15) and +(0,-0.15) .. (p2)
		-- (p2) .. controls +(0,0.35) and +(0,0.35) ..  (p3)
		-- (p3) .. controls +(0,-0.15) and +(0,-0.15) .. (p4)
		-- (p4) .. controls +(0,0.9) and +(0,0.9) ..  (p1)
		;
		\draw[thick, black] (p2) .. controls +(0,0.35) and +(0,0.35) ..  (p3); 
		\draw[thick, black] (p4) .. controls +(0,0.9) and +(0,0.9) ..  (p1);
		\draw[very thick, red!80!black] (p1) .. controls +(0,-0.15) and +(0,-0.15) ..  (p2); 
		\draw[very thick, red!80!black] (p3) .. controls +(0,-0.15) and +(0,-0.15) ..  (p4); 
		\end{tikzpicture}
	}
	\Big)
	 \cong \big[ \du-\au, \, \yu-\yu' \big]  \otimes \big[ \du-\au' ,\, \xu'-\xu \big] \otimes \big[ \au'-\au, \, \yu'-\xu \big] 
	\nonumber
	\\ & \qquad\qquad\qquad\quad \otimes \big[ \yu'-\xu',\,\du-\du' \big]\{\rnh-\rsh,\qnh-\qsh-2\}
	\, .
	\end{align}
This structure can also be expressed in terms of Grothendieck rings in complete analogy to the case of Section~\ref{subsubsec:ComFrobAlgGrothRing}.

\appendix

\section{Two technical lemmas}\label{sec:applemmas}

In dealing with matrix factorisations we repeatedly apply the following two lemmas. 
The first one deals with the elimination of internal variables:

\begin{lemma}\label{lem:elimination}
	Fix $\xu=(x_1,\ldots,x_n)$, $\au=(a_1,\ldots,a_k)$, $\bu=(b_1,\ldots,b_k)$, $W\in \C[\xu,\au,\bu]$ and $\pu=(p_1,\ldots,p_k) \in \C[\xu,\au,\bu]^{\times k}$. 
	Let $(X,d_X)$ be a matrix factorisation of~$W$ such that $\partial_{b_i}(W+\sum_i(b_i-a_i)p_i)=0$ for all $1\leq i\leq k$.
	Then the infinite-rank matrix factorisation $[\bu-\au,\pu]\otimes (X,d_X)$ over $\C[\xu,\au]$, 
	in which~$\bu$ appears as internal variables, is isomorphic to  
	\begin{equation}
	\big[\bu-\au,\,\pu\big]\otimes (X,d_X) \cong (X,d_X)(\xu,\au,\au)\,,
	\end{equation}
	where $(X,d_X)(\xu,\au,\au)$ is obtained by setting $\bu=\au$ in $(X,d_X)$. 
	In particular, it is a matrix factorisation of $W(\xu,\au,\au)$ over $\C[\xu,\au]$. 
\end{lemma}

\begin{proof}
	We set $(Y,d_Y) := [\bu-\au,\pu]\otimes (X,d_X)$, and we denote the components of $d_X$ and $d_Y$ by $d_X^i\colon X^i\lra X^{i+1}$ and $d_Y^i\colon Y^i\lra Y^{i+1}$. 
	Moreover, we consider the ring $R:=\C[\xu,\au]/(W(\xu,\au,\au))$, and the modules $S_{X}^i := X(\xu,\au,\au)^i\otimes_{\C[\xu,\au]}R$, $S^i_Y := Y^i\otimes_{\C[\xu,\au,\bu]}R[\bu]$. 
	Then the $R$-module 
	\begin{equation}
	M:=S_Y^0\Big/\!\big(d_Y^1(S_Y^1),b_1-a_1,\ldots,b_k-a_k\big)
	\end{equation}
	has an $R$-free Koszul type resolution
	\begin{equation}
	\cdots\longrightarrow F_2\stackrel{f_2}{\longrightarrow} F_1\stackrel{f_1}{\longrightarrow} F_0\longrightarrow M\,,
	\end{equation}
	which after finitely many steps turns into the two-periodic exact sequence
	\begin{equation}
	\cdots\longrightarrow S_Y^1\stackrel{d_Y^1}{\longrightarrow} S_Y^0\stackrel{d_Y^0}{\longrightarrow} S_Y^1
	\stackrel{d_Y^1}{\longrightarrow} S_Y^0\longrightarrow \cdots
	\end{equation}
	On the other hand, $M$ is isomorphic to the $R$-module
	\begin{equation}
	N=S^X_0\Big/\!\big(d_X^1(\xu,\au,\au)(S_X^1)\big)\,,
	\end{equation}
	obtained by setting $\bu=\au$ in $M$. 
	This module has an $R$-free resolution 
	\begin{equation}
	\cdots\longrightarrow E_2\stackrel{e_2}{\longrightarrow} E_1\stackrel{e_1}{\longrightarrow} E_0\longrightarrow N\,,
	\end{equation}
	which is two-periodic from the start and defined by $(X,d_X)(\xu,\au,\au)$:
	\begin{equation}
	E_i:=S_X^i\,,\quad
	e_i:=d^i_X(\xu,\au,\au)\,.
	\end{equation}
	The isomorphism between $M$ and $N$ now lifts to the resolutions $F_\bullet$ and $E_\bullet$ and defines an isomorphism between the matrix factorisations $(Y,d_Y)=[\bu-\au,\pu]\otimes (X,d_X)$ and $(X,d_X)(\xu,\au,\au)$. 
	(For the general relation between modules over hypersurface rings and matrix factorisations, see \cite{eisenbud}. 
	This type of argument has been used to determine fusion of matrix factorisations in \cite{Brunner:2007qu}, see also \cite{Brunner:2021cga}.)
\end{proof}

The second lemma can be used to add and subtract terms in Koszul factors of a tensor product of Koszul matrix factorisations:

\begin{lemma}\label{lem:addition}
Let $[\pu,\qu]$ and $[\pu',\qu']$ be two Koszul matrix factorisations of the same size, i.\,e.\ the lists of variables $\pu,\qu,\pu',\qu'$ all have the same length, say~$k$. 
Then
\begin{equation}\nonumber
\big[\pu,\,\qu\big]\otimes\big[\pu',\,\qu'\big] 
	\cong 
	\big[\pu\pm\pu',\,\qu\big]\otimes\big[\pu',\,\qu'\mp\qu\big]\,.
\end{equation}
\end{lemma}
\begin{proof}
Let us first prove the statement for rank-1 Koszul matrix factorisations, i.\,e.~for $k=1$. In this case, $[\pu,\qu]\otimes[\pu',\qu']\cong(X,d_X)$ with $X=\C[\pu,\qu,\pu',\qu']^{\oplus 4}$ and 
\be\label{eq:dxtp}
d_X=\left(
\begin{array}{cccc}
&&q_1&q_1'\\
&&-p_1'&p_1\\
p_1&-q_1'&&\\
p_1'&q_1&&
\end{array}
\right).
\ee
Conjugating $d_X$ with the matrix
\be\label{eq:dxT}
T=
\left(
\begin{array}{cccc}
1&0&&\\
0&1&&\\
&&1&\pm 1\\
&&0&1
\end{array}
\right)
\ee
yields
\begin{equation}\label{eq:dxconj}
T\,d_X\,T^{-1}=\left(
\begin{array}{cccc}
&&q_1&q_1'\mp q_1\\
&&-p_1'&p_1\pm p_1'\\
p_1\pm p_1'&-(q_1'\mp q_1)&&\\
p_1'&q_1&&
\end{array}
\right) .
\end{equation}
This has the same form as $d_X$ in \eqref{eq:dxtp} where $p_1$ is replaced by $p_1\pm p_1'$ and $q_1'$ by $q_1'\mp q_1$. Hence,
\be\label{eq:dxcong}
\big[\pu,\,\qu\big]\otimes\big[\pu',\,\qu'\big] 
	\cong 
	(X,d_X) 
	\cong 
	\big[\pu\pm\pu',\,\qu\big]\otimes\big[\pu',\,\qu'\mp\qu\big]\,.
\ee
This also implies the statement for higher rank. For this we note that 
 $[\pu,\qu]=\bigotimes_{i=1}^k[p_i,q_i]$, and use the fact that up to isomorphism the tensor product of matrix factorisations is commutative and associative. 
\end{proof}

\section{Super algebras, twists and symmetries}
\label{sec:appalgebras}

In this appendix we collect a few formulas that show how the topological theory considered in the main part of the paper arises as a topological subsector of a supersymmetric 3-dimensional $\mathcal N=4$ theory of free hypermultiplets. 
We use standard physics notation and follow the original work \cite{RW1996} for our arguments.

\subsection{Algebras}

The 3-dimensional $\mathcal N=4$ super Poincar\'e algebra is
\begin{equation}
\big\{ Q_\alpha^{A \dot{A}} , Q_\beta^{B\dot{B}} \big\}
	=
	\epsilon^{AB}\epsilon^{\da \db} P_{\alpha \beta} \, . 
\end{equation}
The Lorentz symmetry  is $\SU(2)_E$, and the indices $\alpha, \beta$ refer to it. 
The R-symmetry is a product of two copies of $\SU(2)$ denoted as $\SU(2)_H \times \SU(2)_C$. 
The upper undotted index $A$ refers to $\SU(2)_H$, while $\da$ refers to $\SU(2)_C$. 
The supercharges transform as (${\textbf 2}, {\textbf 2}, {\textbf 2})$ under $\SU(2)_E\times \SU(2)_H \times \SU(2)_C$. 

We consider a theory of hypermultiplets. 
A hypermultiplet contains four scalars in the representation $({\textbf 1},   {\textbf 2}, {\textbf 1})$  of $\SU(2)_E\times \SU(2)_H \times \SU(2)_C$ as well as fermions in $({\textbf 2},   {\textbf 1}, {\textbf 2})$.
In addition, there is an $\textrm{Sp}(n)$ flavour symmetry for $n$ hypers, under which scalars and fermions in the same hypermultiplet transform in the same way. 
Geometrically, the hypers take value in a hyper K\"ahler manifold, which in the case of the current paper is $T^*\C^n$, where the K\"ahler forms transform in the adjoint of $\SU(2)_H$ and are invariant under $\SU(2)_C$.

There are two different topological twists, using either of the two $\SU(2)$ R-symmetries. 
We pick the $\SU(2)_C$ (Rozansky--Witten) twist and  modify the Lorentz symmetry to become the diagonal
\begin{equation}
\SU(2)_{E}' \subset \SU(2)_E \times \SU(2)_C \, .
\end{equation}
As a consequence, we obtain  a pair of scalar supercharges $Q^A$, $A\in\{1,2\}$, and a pair of vector supercharges $Q_\mu^A$, where $\mu \in\{0,1,2\}$ is the 3-dimensional vector index. 
The scalars in the hypermultiplet remain scalars with respect to $\SU(2)_E'$, and the fermions become scalars and vectors. 

Following Rozansky and Witten \cite{RW1996}, we pick a complex structure, thereby breaking $\SU(2)_H$ to $\textrm{U}(1)$, and denote the scalars in it by $\phi^I $ and $\bar\phi^{\bar{I}}$. 
Here, $I$ takes values $1,\dots,2n$ for a target manifold of real dimension $4n$. 
Furthermore, the scalar originating from the fermions is denoted $\eta^I$, the vector $\chi_\mu^I$. 
Combining also the two scalar supercharges into a holomorphic and an anti-holomorphic one, one obtains for the $\bar{Q}$-variation of the fields a flat space version of \cite[Eq.\,(2.22)]{RW1996}:
\begin{equation}
\delta \phi^I=0 \, , \quad 
	\delta \eta^I=0 \, ,  \quad 
	\delta \chi_\mu^I \sim \partial_\mu \phi^I \, , \quad 
	\delta \bar\phi^{\bar{I}}\sim \eta^I \, .
\end{equation}

\subsection{State spaces}
\label{subsec:StateSpecesAppendix}

To obtain the state space ${\mathcal H}_\Sigma$ associated to a surface $\Sigma$, we quantise on $\Sigma \times \R$ where $\R$ is the time direction, and we follow the procedure of \cite[Sect.\,5]{RW1996}. 
Such state spaces have recently  been computed for  more general theories (and both twists) by considering the supersymmetric quantum mechanics along the lines of \cite{BFK}.
The state space is a Fock space, carrying a representation of the algebra given by quantising the Poisson brackets given in \cite{RW1996}. 
Since one restricts to a topological subsector, one only needs to take care of the zero modes of all fields. 
If $\Sigma$ is $S^2$, the only contribution to the cohomology comes from the bosons $\phi^I$. For a single hypermultiplet, $I$ takes two values $1,2$. 
In the main text $x$ corresponds to the zero mode of $\phi^2$, and $a$ to the zero mode of $\phi^1$. 
For $n$ hypermultiplets, the state space $\mathcal H_{S^2}$ is simply a polynomial ring in $2n$ variables, which we denote $\C[\xu,\au]$. 

For surfaces $\Sigma_g$ of genus $g\geqslant 1$, we pick a basis of harmonic 1-forms $\omega^{(\alpha)}$ on $\Sigma_g$ and expand the vectors $\chi_\mu^I$ along $\Sigma_g$ as $\chi_\mu^I (x) = \chi_\alpha^I \omega^{(\alpha)}_\mu$ 
where $\mu\in\{1,2\}$ and $\alpha \in \{ 1, \dots, 2g \}$. 
The $\chi_\alpha^I$ are constant fermionic coefficients. 
To quantise, it is furthermore necessary to pick a polarisation.
We take the canonical commutation relations from \cite{RW1996}, 
\be
\big\{ \chi_\alpha^I, \chi_\beta^J \big\} 
	= 
	\epsilon^{IJ} (L^{-1})_{\alpha \beta} \, ,
\ee
where $L$ is the intersection form on $\Sigma_g$. 
The fermionic part of the state space can thus be regarded as being generated by $2g$ fermions $\chi_\alpha^I$, where $\alpha$ takes $g$ values. 
	 The state space (for a single hypermultiplet) associated to $\Sigma_g$ thus is ${\cal H}_{\Sigma_g} = (\C \oplus \C[1])^{\otimes 2g} \otimes_\C \C[\au,\xu]$. 
This space is naturally $\Z_2$-graded, splitting into bosonic and fermionic subspaces.

\subsection{Gradings}

There are two global symmetries  that act on the fields of the twisted theory. The first one originates from the $\SU(2)_H$ R-symmetry of the untwisted theory. 
Since we have chosen a complex structure, this R-symmetry is broken, leaving a $\textrm{U}(1)$ R-symmetry, whose associated charges we denote~$q_R$.  
Under it, the scalars $\phi^I$ have eigenvalue $+1$, whereas the conjugate scalars have eigenvalue $-1$. The fields $\eta^I$ and $\chi_\mu^I$ originating from the fermions have charge $0$, since the fermions of the original theory transformed trivially under $\SU(2)_H$. 
Furthermore, there is a flavour symmetry with a $\textrm{U}(1)$ subgroup whose charges we denote~$q_F$.  
Decomposing the initial $\mathcal N=4$ algebra with respect to a $\mathcal N=(2,2)$ subalgebra, the two resulting chiral multiplets transform with opposite flavour charge. 
Hence $\phi^1$ and $\phi^2$ have opposite charge. 
We summarise these charges as follows:
\begin{center}
	\setlength{\tabcolsep}{9pt}
	\begin{tabular}{lrrrrrrrr}
		\toprule[1.5pt]
		&$\phi^1$& $\bar\phi^1$& $\phi^2$& $\bar\phi^2$ &  $\chi_\mu^1$ & $\chi_\mu^2$ & $\eta^1$ & $\eta^2$
		\\
		\midrule[1.pt]
		&  &&&&&&&
		\\[-13pt]
		$q_R$ & 1 & $-1$ & 1& $-1$&0 &0 &0& 0
		\\
		$q_F$ & 1 & 1 & $-1$ & $-1$ & 1 & $-1$ & 1 & $-1$
		\\[1pt]
		\bottomrule[1.5pt]
	\end{tabular}
\end{center}

Note that the opposite flavour charges and equal R-charges assigned to $\phi^1$ and $\phi^2$ directly match with the opposite flavour charges of variables $x$ and $a$ in the main text. 
Furthermore, the gradings in  \eqref{eq:StateSpacegrSigmag} are  directly compatible with the ones originating from the assignment of R- and flavour charges, if we identify the two factors corresponding to the fermions in \eqref{eq:StateSpacegrSigmag} with contributions from $\chi^1_\alpha$ and $\chi^2_\alpha$.

\end{document}